\documentclass[iop,revtex4]{emulateapj} 
\usepackage{lscape}
\usepackage{epsfig,natbib} 
\usepackage{graphicx}
\usepackage{verbatim} 
\usepackage{float}  
\usepackage{natbib} 
\newcommand{\ek}{\emph{Kepler\ }}
\newcommand{\eke}{\emph{Kepler}}

\newcommand{\ms}{\mbox{m\,s$^{-1}~$}}
\newcommand{\kms}{\mbox{km\,s$^{-1}~$}}
\newcommand{\kmse}{\mbox{km\,s$^{-1}$}}

\newcommand{\kse}{\mbox{km\,s$^{-1}$}}
\newcommand{\mse}{\mbox{m\,s$^{-1}$}}

\newcommand{\msun}{M$_{\odot}~$}

\newcommand{\rsun}{R$_{\odot}~$}
\newcommand{\rsune}{R$_{\odot}$}
\newcommand{\mjup}{M$_{\rm JUP}~$}
\newcommand{\mjupe}{M$_{\rm JUP}$}

\newcommand{\mearth}{$M_\earth$~}
\newcommand{\mearthe}{$M_\earth$}
\newcommand{\rearth}{$R_\earth$~}
\newcommand{\rearthe}{$R_\earth$}

\newcommand{\msinie}{$M \sin i$}
\newcommand{\vsinie}{$V \sin i$}

\newcommand{\feh}{\ensuremath{[\mbox{Fe}/\mbox{H}]}}
\newcommand{\fehe}{\ensuremath{[\mbox{Fe}/\mbox{H}]~}}
\newcommand{\rphk}{\ensuremath{log~R'_{\mbox{\scriptsize HK}}}}

\newcommand{\msini}{\ensuremath{M \sin i}}

\newcommand{\teff}{\ensuremath{T_{\rm eff}}}

\newcommand{\logg}{\ensuremath{\log{g}}}
\newcommand{\vsini}{\ensuremath{v \sin{i}}}
\newcommand{\gcc}{\mbox{g\,cm$^{-3}$}} 

\shortauthors{Marcy {et~al.}}
\shorttitle{Kepler Planet Masses}
\submitted{Accepted by ApJS} 
\begin{document}
\pagenumbering{arabic}

\title{Masses, Radii, and Orbits of Small Kepler Planets: \\  The
  Transition from Gaseous to Rocky Planets   \altaffilmark{$\dagger$}}

\author{
Geoffrey~W.~Marcy\altaffilmark{1}  ,  
Howard~Isaacson\altaffilmark{1},  
Andrew~W.~Howard\altaffilmark{29},  
Jason~F.~Rowe\altaffilmark{2}, 
Jon~M.~Jenkins\altaffilmark{3}, 
Stephen~T.~Bryson\altaffilmark{2}, 
David~W.~Latham\altaffilmark{8}, 
Steve~B.~Howell\altaffilmark{2}, 
Thomas~N.~Gautier III\altaffilmark{6}, 
Natalie~M.~Batalha\altaffilmark{2}
% ...
}   
\altaffiltext{1}{University of California, Berkeley, CA 94720}  
\altaffiltext{2}{NASA Ames Research Center, Moffett Field, CA 94035}
\altaffiltext{3}{SETI Institute/NASA Ames Research Center, Moffett Field, CA 94035}
% ...
\altaffiltext{6}{Jet Propulsion Laboratory/Caltech, Pasadena, CA 91109}
% ...
\altaffiltext{8}{Harvard Smithsonian Center for Astrophysics, 60 Garden Street, Cambridge, MA 02138}
% ...
\altaffiltext{29}{University of  Hawaii, Honolulu, HI}
% ...

\begin{abstract}

We report on the masses, sizes, and orbits of the planets orbiting 22
\ek stars.  There are 49 planet candidates around these stars,
including 42 detected through transits and 7 revealed by precise
Doppler measurements of the host stars.  Based on an analysis of the
\ek brightness measurements, along with high-resolution imaging and
spectroscopy, Doppler spectroscopy, and (for 11 stars)
asteroseismology, we establish low false-positive probabilities for
all of the transiting planets (41 of 42 have a false-positive
probability under 1\%),  and we constrain their sizes and masses.
Most of the transiting planets are smaller than 3$\times$ the size of Earth.  For 16
planets, the Doppler signal was securely detected, providing a direct
measurement of the planet's mass.  For the other 26 planets we provide
either marginal mass measurements or upper limits to their masses and
densities; in many cases we can rule out a rocky composition.  We
identify 6 planets with densities above 5 \gcc, suggesting a mostly
rocky interior for them.   Indeed, the only planets that are compatible
with a purely rocky composition are smaller than $\sim$2 \rearthe.
Larger planets evidently contain a larger fraction of low-density
material (H, He, and H$_2$O).
\end{abstract}

\clearpage

\section{Introduction}
\label{sec:intro}
Our Solar System contains no planets with radii between Earth and
Neptune (3.9 \rearthe), a size gap that differs from the apparent
distribution of small planets in the Milky Way and requires
adjustments to the core-accretion model to explain.  For example,
Uranus and Neptune, with equatorial radii of 4.01 and 3.88 \rearthe, respectively,
and their massive rocky cores would presumably have grown to Saturn or
even Jupiter size through runaway accretion had the protoplanetary
disk not disappeared when it did \citep{Pollack1996, Goldreich2004,
  Rogers_Seager2010b, Morbidelli2013}.  Still, based on formation models of our own Solar
System, the size domain of 1--4 \rearth was expected to be nearly
deserted \citep{Ida_Lin2010, Mordasini2012a}.  It is not.  Instead,
 most of the observed planets around other stars have radii in the range
of 1--4 \rearth \citep{Borucki2011, Batalha2013}.

This great population of sub-Neptune-size exoplanets had first been
revealed by precise Doppler surveys of solar-mass stars within 50 pc.
Such surveys find planet counts increase toward smaller masses, at
least within the range of 1000 \mearth down to $\sim$5 \mearth
\citep{Howard2010b, Mayor2011}.  Independently, the NASA \ek telescope
finds that 85\% of its transiting planet ``candidates'' have radii
less than 4 \rearth \citep{Batalha2013}.  Since more than 80\% of
these small planet candidates are actually planets
\citep{Morton_Johnson2011, Fressin2013}, the population of
sub-4-\rearth planets is assuredly large (but see
\citealt{Santerne2012} for the confirmation rate of Jupiter-size
planets).  No detection bias would favor the discovery of small
planets over the large ones (for a given orbital period), and indeed
the small planets enjoy a smaller rate of false-positive scenarios.  Thus, in both the solar vicinity probed
by Doppler surveys and in the \ek field of view (slightly above the
plane of the Milky Way), an {\em overwhelming majority} of planets
orbiting within 1 au of solar-type stars are smaller than Uranus and
Neptune (i.e., $<\,\sim$4 \rearthe).

For planets orbiting close to their host star, the great occurrence of
small planets is particularly well determined.  Within 0.25 au of
solar-type stars, the number of planets rises rapidly moving from 15
\rearth to 2 \rearthe, based on analyses of \ek data that correct for
detection biases due to photometric noise, orbital inclination, and
the completeness of the \ek planet-search detection pipeline
\citep{Howard2012, Fressin2013, Petigura2013}.  Further corrections
for photometric SNR and detection completeness show that the
occurrence of planets remains at a (high) constant level for sizes
from 2 to 1 \rearthe, with $\sim$15\% of FGK stars having a planet of
1--3 \rearth within 0.25 au \citep{Fressin2013, Petigura2013}.

With no Solar System analogs, the chemical compositions, interior
structures, and formation processes for 1--4 \rearth planets, including
their gravitational interactions with other planets, present profound
questions \citep{Seager2007, Fortney07, Zeng_Seager08, Rogers2011,
  Zeng_Sasselov2013, Lissauer2011, Lissauer2012, Fabrycky2012}.  Determining chemical
composition is one step toward a deeper understanding, but at this planet-size scale,
the relative amounts of rock, water, and H and He gas remain poorly
known.  Most likely, the admixture of those three ingredients changes
as a function of planet mass, but differs among planets at a given
mass, as well.

Beyond the question of their characteristics, these 1--4 \rearth
planets pose a great challenge for the theory of planet
formation:~like Venus, Earth, Uranus, and Neptune, they likely contain
a ratio of rock to light material that is much greater than cosmic
abundances, and therefore their formation must have required some
complex processing in the protoplanetary disk.  However, new ideas are
emerging about the formation of such Neptune-mass-and-smaller planets,
most of which are variations on the theme of core-accretion theory
\citep{Chiang_Laughlin2013, Mordasini2012a, Hansen_Murray2013}.  Particularly intriguing
is the notion that taken as an ensemble, the hundreds of \ek exoplanet
candidates reflect the mass densities of protoplanetary disks during
the period of planet formation, leading to a theory that within 0.5 au
of their host stars, sub-Neptunes formed in situ, i.e., without
migration \citep{Chiang_Laughlin2013, Hansen_Murray2013}.  The predicted
relations between mass, radius, and incident flux agree with those
observed \citep{Lopez2012, Lopez2013, Weiss2013}.  These models and their
associated predictions of in situ mini-Neptune and super-Earth
formation can be further tested with accurate measurements of planet
masses and radii.

Measuring masses for transiting planets that already have measured
radii can constrain the mean molecular weight, internal chemical
composition, and hence formation mechanisms for 1--4 \rearth planets
\citep{Seager2007, Zeng_Seager08, Zeng_Sasselov2013, Rogers2011,
  Chiang_Laughlin2013}.  Only a handful of small planets have mass
measurements, and those with radii above 2 \rearth often have low
densities inconsistent with pure rocky composition.  Two
well-studied examples are GJ 436 b and GJ 1214 b \citep{Maness07,
  Gillon2007, Torres2008, Charbonneau2009} with radii of 4.21 and 2.68
\rearthe, masses of 23.2 and 6.55 \mearthe, and resulting bulk densities of 1.69 and 1.87 \gcc,
respectively.  Their densities are slightly higher than those of
Uranus and Neptune (1.27 and 1.63 \gcc), but still well below
Earth's (5.5 \gcc). The sub-Earth bulk densities indicate that the two
exoplanets contain significant amounts of low-density material (by volume), presumably H, He, and
water \citep{Figueira2009, Rogers_Seager2010a, Batygin2013}.  At the larger end of
the small-planet spectrum is Kepler-18 c, with a radius of 5.5 \rearth
but a mass of only 17.3 \mearthe, implying a low density of 0.59
$\pm$ 0.07 \gcc \citep{Cochran2011}. Similarly, HAT-P-26, with 6.3
\rearthe, has low density of 0.40 $\pm$ 0.1 \gcc \citep{Hartman2011}.

The several other $\sim$2--4 \rearth exoplanets with secure masses and
radii support this trend, including the five inner planets around
Kepler-11, GJ 3470 b, 55 Cnc e, and Kepler-68 b \citep{Lissauer2013,
  Bonfils2012, Endl2012, Demory2013, Demory2011, Gilliland2013}.  All
of these planets have densities less than 5 \gcc and some under 1
\gcc, indicating a significant amount of light material by volume (H, He, water)
mixed with some rock and Fe.  (The uncertainties for 55 Cnc e admit
the possibility this 2.1 \rearth planet could be pure rock.)  Perhaps
these securely measured lower-than-rock densities are representative
of planets of size 2.0---4.5 \rearth in general, and hence
representative of the chemical composition of such planets.

Most tellingly, the five planets with radii less 2 \rearthe, namely
CoRoT 7b, Kepler-10b, Kepler-36b, KOI-1843.03, and Kepler-78b all have
measured densities of 6--10 \gcc \citep{Queloz2009, Batalha2011,
  Carter2012, Rappaport2013,Sanchis-Ojeda2013, Pepe2013,Howard2013}.  
 Thus, below 2 \rearth some planets have
densities consistent with pure solid rock and iron-nickel.  The
dichotomy of planet densities has been considered theoretically as due
to accumulation and photo-evaporation of volatiles
\citep{Chiang_Laughlin2013, Hansen_Murray2013, Lopez2012, Lopez2013}.

To quantify this transition to rocky planets, one may use the extant
empirical relation between density and planet mass that has been
discovered for the planets smaller than 5 \rearthe:~$\rho = 1.3
M_p^{-0.60} F^{-0.09}$, where $\rho$ is in \gcc, $M_p$ is in \mearthe,
and $F$ is the incident stellar flux on the planet in erg s$^{-1}$
cm$^{-2}$ \citep{Weiss2013}. The Weiss et al.~relation shows that
planets with masses over $\sim$2 \mearth (equivalently, with radii
over 1.5 \rearthe) have typical densities less than 5.5 \gcc and hence
typically contain significant amounts of light material (H, He, and
water).  {\it Thus, the transition from planets containing significant
  light material to those that are rocky occurs at planet radii near
  1.0--2.5 \rearthe, i.e., masses near 1--3 \mearth}, based
tentatively on the handful of planets in that size domain located
within 0.2 au.  This suggestion of a transition to rocky planets
below masses of 3 \mearth is a major result from current \ek exoplanet
observations.

However, the Weiss et al.~relation, and the predicted transition to
rocky planets below 2 \rearthe, is based on the measured masses and
radii of only a handful of planets.  It surely requires both
confirmation and quantification, by measuring the masses and radii of
more small exoplanets.  Those additional small exoplanets would also
greatly inform models of planet formation, based on correlations
between the volatile or rocky nature of the planets and the
metallicities of their host stars \citep{Buchhave2012, Latham2012,
  Johnson2007}.

Here we report measured masses, radii, and densities (or upper limits
on those values) for 42 transiting planet candidates contained within
22 bright \ek Objects of Interest (KOIs) from \cite{Batalha2013}.  We
carried out multiple Doppler-shift measurements of the host
stars using the Keck 1 telescope. From the spectroscopy and Doppler
measurements, we compute self-consistent measurements of stellar and
planet radii, employing either stellar structure models or
asteroseismology measurements from the \ek photometry. We also search
for (and report) 7 additional non-transiting planets revealed by the
precise radial velocities, for a total of 49 planets.

\section{Vetting and Selection of 22 Target KOIs}
\label{sec:vetting}

This paper contains the results of extensive precise-RV measurements
of KOIs, made by the \ek team.  The intense RV follow-up observations
described here were carried out on 22 KOIs chosen through a careful
vetting process.  The initial identification of the KOIs from the
photometry was an extensive, iterative program carried out by the \ek
team during the nominal NASA mission from launch 2009 March to 2012
November.  The identification process has been described elsewhere,
notably by \cite{Caldwell2010, Jenkins2010a, Jenkins2010b,
  VanCleve2009}, and \cite{Argabright2008}, with an overview in
\cite{Borucki2010}.  The $\sim$2300 KOIs identified in these searches are
listed in \cite{Batalha2013}.

\subsection{Data Validation: TCERT}
\label{sec:dv}

The selection of the 22 KOIs for this study involved several major
stages of pruning of the candidates, starting with the ``threshold
crossing events'' (TCEs) that are the series of repeated dimmings
found for a particular star by the \ek ``Transit Planet Search'' (TPS)
pipeline.    Working within the \ek TCE review team
(``TCERT''), we vetted the TCEs to distinguish planet candidates from
false positives and to measure more accurately the properties of the
planets and their host stars.  Detailed descriptions of the components
of this TCE vetting can be found in \cite{Gautier2010, Borucki2011,
  Batalha2013}.

A TCE was elevated to KOI status (planet candidate) based on simple
(often eye-ball) criteria involving the inspection of each \ek light
curve, using long cadence photometry, overplotted on a model of a
transiting planet, and noting a lack of eclipsing 
binary signatures such as secondary eclipses and ``odd-even'' alternate 
variability of successive transit depths.  This TCERT-based identification 
of the KOIs involved only \ek data, not outside observations.

Actual transiting planets should exhibit photometry that is well fit,
within errors, by a transiting planet model.  They should also show
an astrometric displacement (if any) during transit that is consistent with the
hypothesis that the intended target star is the source of the
photometric variations during transit.  Such ``Data
Validation'' (DV) techniques are described in \cite{Batalha_fp,
  Batalha2011, Bryson2013}.  These DV tests have undergone
improvements and automation during the past three years \citep{dv,
  Bryson2013}. All 22 KOIs in this work passed their DV tests,
conferring KOI status on them as continued planet candidates.  Details
on the nature of DV criteria for each KOI are given below in
Section~\ref{sec:obs}.

The TCERT identification of KOIs as ``planet candidates'' made them
worthy of follow-up observations with other telescopes, designed both
to weed out false positives and to better measure the planet
properties through superior knowledge of host star properties, notably
radii.  Various types of follow-up observations of some, but not all,
of the $\sim$2300 KOIs had been carried out by the time of our
selection process of the 22 KOIs studied here.  Publication of those
KOIs are in \cite{Borucki2011, Borucki2012, Batalha2013}.

\subsection{Follow-Up Observation Program: KFOP}
\label{sec:followup}
We activated in May 2009 the \ek ``Follow-up Observation Program''
(KFOP) with the goals of vetting the KOIs for false positives and
improving the measurement of the planet radii. The goals were to
characterize all of the KOIs, as resources permitted, using a variety
of ground-based telescopes.  Each of the KOIs first had their \ek
light curves and astrometric integrity scrutinized again, polishing
the TCERT vetting.  Here we summarize the key KFOP observational efforts
that were carried out on $\sim$1000 KOIs from which the 22 KOIs
presented here were selected.

In brief, each of the $\sim$1000 KOIs had its light curve further scrutinized
and its position further measured (Section ~\ref{sec:dv}) to alert us to angularly
nearby stars (within 2$''$) in the photometric aperture. As
described below, we carried
out adaptive optics (AO) imaging and speckle interferometry
(Section~\ref{sec:ao} and Section~\ref{sec:speckle}) to hunt for
neighboring stars.  KOIs having a neighboring star within 2$''$ and
brighter than 1\% of the primary star are not amenable for follow-up
spectroscopy due to the light from both stars entering the slit.
Roughly 20\% of the KOIs were deemed not suitable for spectroscopy due
a close stellar neighbor.

KOIs meeting those criteria were observed with high-resolution,
low-SNR echelle spectroscopy to measure atmospheric stellar
parameters, magnetic activity, and rotational Doppler broadening,
designed to detect binaries and, importantly, to assess suitability
for precise RV measurements.  Only single, FGKM-type stars with narrow
lines (\vsini $<$ 10 \kms) are suitable for the highest precision RV
measurements.  Below is a summary of the nature of these KFOP vetting
actions on over 1000 KOIs, prioritized by brightness, leading to the selection of the 22 KOIs in this study.
KOIs brighter than Kp$\sim$14.5 mag received most of the KFOP
observational resources, while those fainter than Kp=15 mag were only rarely
observed.

\subsubsection{Follow-Up Reconnaissance Spectroscopy}
\label{sec:recon}

We carried out ``reconnaissance'' high-resolution spectroscopy on
$\sim$1000 KOIs with spectral resolution, R$\sim$50,000, and $SNR =
20-100$ per pixel.  The dual goals were searching for false positives
and refining the stellar parameters. We obtained one or two such
reconnaissance spectra using one of four facilities: the McDonald
Observatory 2.7 m, the Tillinghast 1.5 m on Mt. Hopkins, the Lick
Observatory 3 m, and the 2.6-m Nordic Optical Telescope.
 
Of greatest importance was to detect angularly nearby stars that,
themselves, might be eclipsed or transited by a companion star or
planet, the light from which would be diluted by the primary star
mimicking a transiting planet around it.  With a typical spectrometer
slit width of 1$''$, stellar companions within 0{\farcs}5 would
send light into the slit, permitting their detection if bright enough (see
below).  A cross correlation of each spectrum was performed, usually
with a best-matched synthetic template, to detect stellar companions separated by more
than $\sim$10 \kms in radial velocity and brighter than $\sim$5\% of
the primary in optical flux.

Also, a second reconnaissance spectrum was obtained to detect radial
velocity (``RV'') variation above a threshold of $\sim$0.5 \kms,
indicating the presence of a binary.  We selected the 22
KOIs in this paper by rejecting all KOIs that showed such RV variation
from binary motion.  The fraction of KOIs rejected by reconnaissance
spectroscopy was roughly 5\%, leaving 95\% as surviving planet
candidates.   The absence of a secondary spectrum and RV
variations (confirmed later by the precise RVs with 2 \ms precision)
for all 22 KOIs rules out a large portion of parameter space for
possible false positives in the form of an angularly nearby star that
may be the source of the periodic dimming.  As described in Section ~
\ref{sec:fpp1}, a further analysis of the Keck-HIRES spectra taken later
with {\it high SNR} further ruled out stellar companions within 0{\farcs}5
down to optical flux levels of 1\% that of the primary star.
  
The reconnaissance spectra were also analyzed to measure the
properties of the host star more precisely than was available in the
\ek Input Catalog (KIC). The spectra were analyzed by comparing each
one to a library of theoretical stellar spectra, e.g.,
\cite{Buchhave2012}.  This ``recon'' analysis (later refined by
Buchhave et al. as ``SPC'' analysis) was done with grid step sizes
between individual library spectra of 250 K for \teff, 0.5 dex for
\logg, 1 \kms for \vsini, and 0.5 dex in metallicity ([m/H]).  This
``recon'' spectroscopy analysis yielded approximate values of \teff
~(within 200 K), \logg ~ (within 0.10 dex), and \vsini ~(within 2 \kms) for
the primary star of the KOI, valuable for deciding whether the KOI was
suitable for follow-up precise RV observations. Only stars cooler than
6100 K on the main sequence (\logg $>$ 4.0) with \vsini $<$ 5 \kms were deemed
suitable for the RV measurements of highest precision near $\sim$2
\ms.  All relevant details about the reconnaissance spectroscopy for
each KOI are given in Section~\ref{sec:obs}.

\subsubsection{Speckle Imaging}
\label{sec:speckle}

Speckle imaging of each of the 22 KOIs was obtained using the
two-color DSSI speckle camera at the WIYN 3.5 m telescope on Kitt
Peak, with technical details given in \citet{Howell2011,
  Horch2009}. The speckle camera simultaneously obtained 3,000 images
of 40 msec duration in two filters: $V$ (5620/400\AA) and $R$
(6920/400\AA). These data yielded a final speckle image for each
filter. Section~\ref{sec:obs} describes the results of the speckle
observation for each KOI noting if any other sources appeared.

The speckle data for each star allowed detection of a companion star
within the $2{\farcs}76 \times 2{\farcs}76$  field of view centered on the
target.  The speckle observations could detect, or rule out,
companions between 0{\farcs}05 and 1{\farcs}5 from each KOI. The
speckle images were all obtained with the WIYN telescope during seeing
of 0{\farcs}6--1{\farcs}0 .  The threshold for detection of companion stars
was a delta magnitude of 3.8 mag in the R band and 4.1 mag in V band
(within the sensitivity annulus from 0{\farcs}05--1{\farcs}5),
relative to the brightness of the KOI target star. For Kepler-97 the
detection threshold was compromised by a stellar companion
0{\farcs}36 away from the primary and 2.7 magnitudes fainter at optical
wavelengths (3.2 mag fainter in K band). This companion is farther
than the maximum possible separation of a false positive star, based
on centroid astrometry in and out of transit.

These speckle observations were used to select the 22 KOIs studied
here, by rejecting all KOIs (roughly 5\%) that showed such a stellar
companion.  Thus, these initial speckle observations showed that none
of the 22 KOIs (except Kepler-97) in this work had a detected companion
by speckle.  {\em This selection process using speckle imaging by
  which the 22 KOIs were chosen surely favors single stars rather than
  binaries or multiples.  Hence it eliminates, a priori, a major
  domain of false positives (i.e. neighboring stars with transiting
  companions)}.

\subsubsection{AO Imaging}
\label{sec:ao}

Near-infrared adaptive optics (AO) imaging was obtained for $\sim$300 KOIs
of the 1000 KOIs brighter than Kp=14 mag.  The goal, as with the
speckle observations described above, was to detect stellar companions
that might be the source of the periodic dimming (a false positive).
Seeing-limited imaging, obtained with various telescopes at both
optical and IR wavelengths, revealed companions located more
than 2$''$ from the primary KOI star.  Seeing-limited J-band Images
from UKIRT were particularly useful \citep{Lawrence2007}.  (After
selection of the 22 KOIs here, we also examined optical seeing-limited
images from the Keck-HIRES guide camera, and we provide those images
here.)  Any seeing-limited images showing a companion stars within 2$''$
 was rejected as a useful candidate for high precision RV work,
i.e. rejected for inclusion in this study.  Thus all 22 KOIs in this
paper were selected with a prior AO image, as well as the speckle
imaging described above.

The strength of AO imaging is the ability to detect companions located
between 0{\farcs}05--2{\farcs}0 of the KOI primary star with detection limits
6 - 8 magnitudes fainter than the primary (depending on the telescope
and AO camera).  The goal was to detect angularly nearby stars, either
background or gravitationally bound, that might potentially have an
eclipsing companion or a transiting planet that might mimic a
transiting planet around the primary star, i.e., a false positive.

Four different AO instruments were used in the near IR on four
different telescopes, namely the Keck 2 telescope on Mauna Kea
(NIRC2-AO), the MMT telescope on Mt. Hopkins (ARIES), the 5m telescope
on Mt. Palomar (PHARO), and the 3m telescope at Lick Observatory
(IRCAL), each described briefly below \citep{Hayward2001, Troy2003,
  Adams2012}.  After the 22 KOIs were selected based on an absence of
stellar companions found with the reconnaissance AO observations with
those four AO instruments, we carried out subsequent AO
imaging with the Keck 2 telescope and NIRC2 camera, generally superior
to the other imaging.

All 22 KOIs were observed with the Keck NIRC2-AO system
\citep{Wizinowich2004, Johansson2008}.  We employed a natural guide
star rather than the laser guide star as the Galactic field is rich
with useful 13th mag guide stars in the \ek field.  We obtained all
images on two nights, 2013 June 13/14 and 14/15.  We used the $K'$
filter (wavelength coverage 1.9--2.3 $\mu$), except for the brightest
five KOIs for which we employed a narrow-band Bracket-gamma filter to
avoid saturation and to achieve flatter wavefronts. On both nights the
natural seeing (before AO) was $\sim$0{\farcs}2 (FWHM) in $K'$-band.

For each KOI, we obtained 15 images with NIRC2-AO, employing a pattern of
 three dither positions (using the three best quadrants on the
 detector) and 5 exposures at each position. The images were
 sky-subtracted, flat-fielded, and co-added to yield a final AO image.
  All final Keck AO images have a PSF with a FWHM of 0{\farcs}05 $\pm$ 0{\farcs}01, 
  with a field of view of 2".  

The detection thresholds from these Keck AO images for each KOI are
shown in Figures \ref{fig:koi41_fig1} - \ref{fig:koi1925_fig2}.  The
detection thresholds uniformly yielded 5-sigma detection of
delta-$K'$-magnitude = 6 mag as close as 0{\farcs}2 from the KOI.  At
0{\farcs}4, the detection threshold is delta-$K'$-magnitude = 8 mag.
Excellent spatial resolution and sensitivity permit the detection of a
large fraction of the background stars that could mimic a transiting
planet.  This spatial resolution also permits detection of a
significant fraction of the widely separated {\em bound stellar
  companions}, given the typical distances ($\sim$200 pc) to these
magnitude 10--13 solar-type stars, as described in
Section~\ref{sec:fpp1}.  None of the 22 KOIs showed a stellar
companion within 1", except for Kepler-97, described below.

The AO-vetting of the $\sim$2000 KOIs was done with four
smaller telescopes.  The MMT ARIES camera achieves near
diffraction-limited imaging, with typical PSF FWHM of 0{\farcs}25 in
the J-band and 0{\farcs}14 in the Ks band, yielding Strehl ratios of
0.3 in Ks and 0.05 in J band\citep{Adams2012}.  While guiding on the
primary star, a set of 16 images, on a four-point dither pattern was
acquired for each KOI. Full details and a description of calibration
and reduction of the images is described by \cite{Adams2012}
  
Some KOIs were vetted with the Palomar 5m ``PHARO'' adaptive
optics camera, observed in both the Ks and J infrared bands using a
5-point dither pattern with integration times between 1.4 and 70
seconds, depending on the target brightness. The AO system used the
primary star itself, not a laser, to guide and correct the images,
achieving a best resolution of 0{\farcs}05 at J and 0{\farcs}09 in the
Ks band, with Strehl ratios of 0.10--0.15 in J and 0.35--0.5 in Ks.
Typical detection thresholds were 7 mag at a separation of 0{\farcs}5 
and 9.3 mag at 1{\farcs}0.
  
The remaining KOIs were AO-vetted with the Lick Observatory 3-m
telescope and high-resolution camera, $''$IRCAL$''$.  Observations
were made in Natural Guide Star(NGS) mode, allowing the AO system to
guide on the target star.  The Lick IRCAL AO system, built by Claire
Max and James R. Graham, is described in detail at
\begin{verbatim}astro.berkeley.edu/~jrg/ircal/spie/ircal.html \end{verbatim}

This mode of observing allows stars as faint as $K p$ = 13.5 to be
 observed.  Detection limits down to a delta-J-magnitude of 6 mag, as
 close as 0{\farcs}1 are typically achieved.  Background sky emission
 in the J-band is typically 16.0 magnitudes per square arcsecond. The
 K-band background sky emission at Lick Observatory is 10.3 magnitudes
 per square arcsecond, making K-band observing difficult. Typically
 only J-band images were taken at Lick Observatory.  For details about
 the IRCAL AO system, see
\begin{verbatim}mtham.ucolick.org/techdocs/instruments/\end{verbatim} 
\cite{Adams2012} provides an excellent description of AO imaging
of KOIs.

The AO imaging described above with five telescopes revealed a
companion star within 6$''$ of nine targets: Kepler-103, 95, 109, 48,113, 96, 131, 97, and 407.
The following KOIs have no detected stellar
companion within 6$''$: Kepler-100, 93, 102, 94, 106, 25, 37, 68, 98, 99, 406, 408, and 409. 
Of the nine KOIs with a detected companion,
only one, Kepler-97, has a companion within 1$''$. See Table \ref{tab:fpp_tbl} for details. 

 None of the nine stars having stellar companions
reside angularly within the maximum exclusion radius found from
astrometry in and out of transit (see Section~\ref{sec:fpp2} and Table \ref{tab:fpp_tbl}).
 {\em Thus we find that none of
the neighboring stars can arguably be causing the dimming as a false
positive.}   For Kepler-97, the observed neighboring star resides 0{\farcs}38 away
while the exclusion radius is 0{\farcs}20, suggesting that the
companion does not cause the apparent transit in the photometry.
  
Toward selection of the 22 KOIs in this study, any KOI with a
neighboring star located within 2$''$ that had more than 1\% the
flux of the primary star at optical wavelengths was rejected as a
suitable candidate for precise RV measurements due to the
contamination of light from that nearby star and due to the possible
false positive.  Of the 22 KOIs in this paper, only Kepler-97 has a
companion within 2$''$, and it is less than 1\% of the brightness
of the primary star at optical wavelengths, as described in Section~
\ref{sec:obs}.  The collection of imaging data and the associated
search for companions, as well as the photometry and RV
measurements, are found in Figures \ref{fig:koi41_fig1} -
\ref{fig:koi1925_fig2}.

\subsection{Selecting the 22 KOIs}
\label{sec:choose22}

In the last three years of the four-year \ek mission, the TCERT
committee systematically shifted its prioritization to select
smaller-radii planets for precise-RV follow-up observations.
Initially, the criteria had emphasized verifying the planet nature of
KOIs. The effort had favored large planets with sizes above 4 \rearth
and short-period orbits that might yield a detectable RV variation in
the host star, to check the existence of \ek transiting planets.
Detections of the RV signatures of the large planets around Kepler 4,
5, 6, 7, and 8 followed from this conservative prioritization. After
the successes of the first six months of the \ek mission, the criteria
shifted toward verifying and measuring the masses of the smaller
planets, 2--5 \rearthe, and of planets in multi-planet systems, if
they were likely to be detected with RVs.  Resulting RV detections
included Kepler 10, 18, 20, 22, 25, and 68 yielding constraints on the
masses of the planets.

During the second and third years of the 4-year \ek mission,
i.e., 2010 and 2011, the TCERT prioritization shifted toward
planets having smaller radii, below
3 \rearth and down to 1.0 \rearthe. Obviously such small planets are
expected to have low masses, inducing small RV amplitudes in their
host star. We carried out careful, optimized selection of
suitable KOIs for RV work.

One selection criterion was a brightness limit, Kp $<$ 13.5 mag,
 to permit Poisson-limited signal-to-noise ratios near 100 per pixel
 within a 45 minute exposure with the Keck-HIRES spectrometer.  Such
 exposures yield a photon-limited Doppler precision of 1.5 \ms.  Another
 selection criteria was \teff~$<$ 6100 K (based on reconnaissance
 spectra) to promote numerous, narrow spectral lines that
 contribute Doppler information. 
Another criterion was small
 rotational Doppler broadening of the spectral lines,
 \vsini~$<$ 5 \kmse, based on reconnaissance spectra,
 to limit broadening of the lines that degrades Doppler precision. 

In the face of pervasive astrophysical ``jitter'' of $\sim$1 \ms for G and 
K dwarfs \citep{Isaacson2010}, the \ek TCERT committee selected
KOIs for which an estimated planet mass might be sufficient to induce
an RV amplitude greater than 1 \ms.  To anticipate the RV amplitude, we used a nominal mass
based on the planet radius taken from the KIC, coupled with a rough
 estimate of planet density for that radius.
The density assumptions were simplistically
 based on the planets in our solar system along with the few known
 small exoplanets, notably GJ436b, GJ1214b, and Kepler-10b. We simply
 assumed a rocky constitution and density of $\sim$5.5 \gcc for
 planets smaller than 2 \rearthe. We assumed densities of 2 \gcc for
 planets of 2--5 \rearthe, and we assumed densities of 1 \gcc for
 planets larger than 5 \rearthe. These densities allowed the
 TCERT to choose planets that might meet the
 criteria above, including a prospective RV amplitude above 1
 \ms. The selection process was imperfect and biased as the assumed
 stellar parameters and planet densities were only approximately known
 and the target KOIs were selected based on RV detectability.

Here we report on the 22 KOIs selected by the process
 described above, with a preference for small planets suitable for
 detection and mass determination by precise RV measurements.  All 22 KOIs are
 identified in \cite{Batalha2013}, but the follow-up observations,
 their analysis, asteroseismology, and the RVs have not
 been published to date, except for Kepler-68 for which we
 provide an update to its long-period, non-transiting
 planet. This sample of 22 KOIs contains neither a random selection
 of KOIs nor a defined distribution of any parameters. They were
 selected during the first three years of ever-evolving criteria, as described
 above. 

Importantly, the planet masses were unknown at the time of target
selection, except for estimates based on measured planet radii and
guesses of density. {\it Thus, for each of the selected planet
  candidates, the subsequently measured planet mass provides an
  unbiased sampling of planet masses for its particular planet
  radius.}  We could not have selected planet candidates biased toward
high or low planet masses for a given planet radius, as we had no such
mass indicator. 

\section{Stellar Characterization}
\label{sec:starchar}
For each of the 22 KOIs, we obtained an optical ``template'' spectrum
using the Keck telescope and HIRES echelle spectrometer
\citep{Vogt1994} with no iodine gas in the light path.  Each spectrum
spanned wavelengths from 3600--8000 \AA, with a spectral resolution of
$R$=60,000 and typical SNR per pixel of 100--200.  These template
spectra were analyzed with the standard LTE spectrum synthesis code,
SME \citep{Valenti96, Valenti2005, Fischer2005} to yield values of
\teff, \logg, and \feh~ yielding formal uncertainties (of roughly 50
K, 0.1 dex, and 0.05 dex, respectively, with slight differences in
precision due to SNR and spectral type).  We augment the formal
uncertainties to account for addition contributions to errors seen in
56 transiting planet hosts for which constraints on stellar properties
stem from analysis of the light curves \citep{Torres2012}.  We added
dispersions in quadrature of $\sigma_{T_{\rm eff}}$ = 59K,
$\sigma_{[Fe/H]}$ = 0.062 dex.  Values of \logg~ are somewhat more
  uncertain and may be systematically in error for \teff~ $>$ 6100 K,
  due to poor sensitivity of the magnesium b triplet lines to surface
  gravity \citep{Torres2012}.

 For 11 of the 22 KOIs an asteroseismic signal was detected in the \ek
 photometry, namely for Kepler-100, 93, 103, 95, 109, 25, 37, 68, 406, 408 and 409.
 For those 11 KOIs the output stellar parameters
 from the SME analysis \teff, \logg, and \feh, were fed into the
 asteroseismology analysis as priors. The asteroseismology analysis
 yielded a more precise measure of stellar radius and mass, and hence
 of surface gravity.  This surface gravity was fed back, frozen, in
 the SME analysis of the spectrum, allowing a redetermination of
 \teff~ and \fehe without the usual covariances with \logg.  The
 resulting values of \teff~ and \fehe were then fed back to an
 asteroseismology analysis as before, achieving an iterative
 convergence quickly \citep{Huber2013, Gilliland2013}. The resulting
 uncertainties in stellar radius are between 2 and 4\%
 \citep{Huber2013}. Stellar parameters for these 11 KOIs with
 asteroseismology are reported in Table~\ref{tab:stellar_pars_tbl}.

For the remaining 11 KOIs that offered no asteroseismology signal, we
determined the stellar mass and radius from the SME spectrum analysis
combined with the Yonsei-Yale stellar structure models \citep{Yi2001,
  Demarque04}. The SME output values of \teff, \logg, and \feh ~ map
to a stellar mass and radius.  For the mild subgiants, the output SME
stellar parameters may correspond to regions of the HR diagram where
some convergence of the evolutionary tracks occurs, leaving greater
uncertainties in the resulting stellar mass and radius, e.g.,
\cite{Batalha2011}.  Any such uncertainties are duly noted and
included in the subsequent analysis of the properties of the planets.

The determinations of stellar masses and radii for the 22
KOIs (with or without asteroseismology) are employed as priors in a
self-consistent Markov-Chain Monte Carlo (MCMC) analysis of the \ek
transit light curves and Keck RVs.  Final stellar parameters are
determined by self-consistent fits of the \ek light curve and RVs to a
model of a planet transiting its host star (see below). 
The output stellar masses and radii differ from input values by
typically less than 10\%. The \ek transit light curve shape and
orbital period (notably transit duration) implicitly further constrain
the stellar density and hence further constrain stellar radius and
mass.  By solving for all stellar (and planet) parameters
simultaneously, and by constraining the fit with priors on \teff,
\logg, and metallicity, along with Yonsei-Yale stellar isochrones, we
obtain final values of stellar radius and mass, along with planet
parameters. Excellent discussions of the iterative convergence of
spectroscopic and asteroseismology results, along with self-consistent
light curve analysis, are provided by \cite{Torres2012, Gilliland2013,
  Borucki2013}.  The final values of all stellar parameters are listed
in Table~\ref{tab:stellar_pars_tbl}.  In the following sections, these
stellar parameters are used, along with the \ek photometry, RVs, and
stellar structure models, to derive the properties of the 42 planet
candidates, listed in Table~\ref{tab:orbital_pars_tbl}, and the false
positive probabilities (FPP) listed in Table~\ref{tab:fpp_tbl} and
discussed in Section~\ref{sec:fpp1}.

\section{Keck-HIRES Precise Velocity Measurements}
\label{sec:preciseRVs}
 We observed the 22 KOIs with the HIRES spectrometer at the Keck
 Observatory from 2009 July to 2013 August, obtaining 20--50 RV
 measurements for each star. The setup used for the RV observations
 was the same as used by the California Planet Search (CPS), including
 a slit width of 0{\farcs}87, yielding a resolving power of $R \approx
 60,000$ between wavelengths 3600 and 8000 \AA \citep{Marcy_Butler92,
   Marcy08}.  For those bright ($V<10$) FGKM stars in the CPS, the
 photon-limited RV precision of $\sim$1.5 \ms matched the typical RV
 fluctuations (jitter) from complex gas flows in the photosphere, also
 $\sim$1.5 \ms, on time scales from minutes to years
 \citep{Howard2010b}.

For the KOIs observed here, typical errors were slightly higher.  The
typical exposure times were 20 to 45 minutes (for Kp = 10--13 mag),
resulting in a signal to noise (SNR) ratio between 70 and 200 per
pixel, depending on the brightness of the target. As a benchmark, at
Kp = 13.0 mag, the typical exposure was 45 minutes, giving SNR=75 per
pixel, and each pixel spanned $\sim$1.3 \kmse. With such exposures,
photon statistics of the observed spectrum, along with the comparable
SNR of the comparison template spectrum, limited the RV precision to
$\sim$2 \mse, slightly greater than typical jitter of $\sim$1 \ms and
systematic errors, also of $\sim$1 \ms.  Indeed, KOIs yielding
non-detections typically have an RMS of the RVs of $\sim$ 3 \ms, as
shown in Tables~\ref{tab:rvs_k00041}--\ref{tab:rvs_k01925}.  We note
that at SNR=70, uncertainties in wavelength scale are estimated to be
less than 0.5 \ms due to the wavelength information contained in
thousands of iodine lines, making wavelength errors a minor source of
error compared to the astrophysical jitter of 1.5 \mse.

The raw reduction of the CCD images followed the standard pipeline of
 the CPS group, but with the addition of sky subtraction, made 
necessary by the faint stars and longer exposure times.
The spectra were obtained with
 the iodine absorption cell in front of the entrance slit of the
 spectrometer, superimposing iodine lines directly on the stellar
 absorption line spectrum, providing both the observatory-frame
 wavelength scale and the instrumental profile of the HIRES
 spectrometer at each wavelength \citep{Marcy92}.  

The Doppler analysis is the same as that used by the CPS group
\citep{Johnson2010}. ``Template'' spectra obtained without iodine gas
in the beam are used in the forward modeling of spectra taken through
iodine to solve simultaneously for the wavelength scale, the
instrumental profile, and the RV in each of 718 segments of length 80
pixels corresponding to $\sim$2.0 \AA, depending on position along
each spectral order. The internal uncertainty in the final RV
measurement for each exposure is the weighted uncertainty in the mean
RV of those 718 segments, the weights of which are determined
dynamically by the RV scatter of each segment relative to the mean RV
of the other segments. The resulting weights reflect the actual RV
performance quality of each spectrum segment. The template spectra are
also used in spectroscopic analysis to determine stellar parameters,
as described in Section~\ref{sec:starchar}.

 The typical long exposures of 10--45 minutes and modest SNR of the
 stellar spectra imply that night sky emission lines and scattered
 moonlight may significantly contaminate the spectra.  To measure and
 remove the contaminating light we use the C2 decker on HIRES which
 projects to 0{\farcs}87$\times$14{\farcs}0 on the sky. 
The C2 decker collects
 both the stellar light and night-sky light simultaneously. The star
 is guided at the center of the slit while the sky light passes
 through the entire 14$''$ length of the slit.  The sky
 contamination is thus simultaneously recorded with the stellar
 spectrum at each wavelength in the regions above and below each
 spectral order, beyond the wings of the PSF of the star image
 projected onto the CCD detector.  The ``sky pixels'' located above and
 below each spectral order provide a direct measure of the spectrum of
 the sky and we subtract that sky light on a column by column basis
 (wavelength by wavelength). When the seeing is greater than 1{\farcs}5
 (which occurs less than 10\% of the time at Mauna Kea), we do not use
 the C2 decker but instead use a smaller 
slit of dimensions 0{\farcs}87 x 3{\farcs}5 (B5
 decker) and we observe only bright stars, Kp $<$ 11 mag, with
 exposure times of $\sim$10 min to avoid sky contamination. 

Observations of KOIs acquired in 2009 did not employ the C2 decker.
With no ability to perform sky subtraction, those observations have
additional RV errors from scattered moonlight.  We quantified these RV
errors by studying the contamination seen in long-slit spectra and by
comparing the scatter in the RVs during 2009 (no sky subtraction) to
the RVs obtained in later years (with sky subtraction), permitting us
to compute the additional RV uncertainties incurred in 2009.  In typical
gibbous moon conditions with light clouds, the moonlight contributed
1--2\% of the light of a Kp = 13 mag star (Rayleigh scattering causing a
wavelength dependence) within a projected $\sim$3{\farcs}5 extraction
width of each spectral order.  Under such gibbous conditions, the
moonlit sky at Mauna Kea is apparently 19th mag per square arcsec in
$V$ band.  Increasing amounts of cirrus clouds will scatter more
moonlight into the slit but will transmit less star light, thereby
increasing the relative amount of contamination of the stellar
spectrum.

We find that RV errors of up to 10 \ms occurred during 2009,
depending on the amount of contamination and the relative radial
velocity of the stellar spectrum and the scattered solar spectrum from
the moon.  Employing sky subtraction with the C2 decker yield RV
precision as if no sky contamination occurred; the observed RV
scatter does not depend on the phase or presence of the moon.  For
stars brighter than Kp = 11 mag the sky subtraction made no difference
in RV precision as moon light was apparently negligible.

Plots of the RVs for each of the 42 transiting planet candidates,
phased to the final orbit (see Section~\ref{sec:planetchar}), are
shown in Figures \ref{fig:koi41_fig2}--\ref{fig:koi1925_fig2}.  The
measured RVs for each of the 22 KOIs are listed in Tables
\ref{tab:rvs_k00041}--\ref{tab:rvs_k01925}.  In those tables, the
first column contains the barycentric Julian date when the star light
arrived at the solar system barycenter (BJD) based on the measured
photon-weighted mid-time of the exposure.  The second column contains
the relative RV (with no defined RV zero point) in the frame of the
barycenter of the solar system.  Only the changes with time in the RVs
are physically meaningful for a given star, not the individual RV
values.  The absolute radial velocities can be determined
relative to the solar system barycenter, but only with an accuracy of
$\sim$50 \ms \citep{Chubak2012}. The third column contains the 
time-series RV
uncertainty, which includes both the internal uncertainty (from the
uncertainty in the mean Doppler shift of 718 spectral segments) and an
approximate jitter of 2 \ms (from photospheric and instrumental
sources) based on hundreds of stars of similar FGK spectral type
\citep{Isaacson2010}.  

The actual RV jitter has values between 1--3 \ms for individual
stars, but the actual photospheric fluid flows for any particular star
and the detailed systematic RV errors are both difficult to estimate
with any accuracy better than 1 \mse. The jitter is added in quadrature
to the internal uncertainty for each RV measurement, to yield a final
RV uncertainty. The actual uncertainties are surely non-Gaussian from
both the photospheric hydrodynamics and from systematic errors in the
Doppler analysis, and they are likely to be temporally coherent with
separate power spectra.  Such error distributions are difficult to
characterize precisely.  Still, it is marvelous that the Doppler-shift
errors for 13th magnitude stars located hundreds of light years away
are less than human jogging speed.

\section{Planet Characterization}
\label{sec:planetchar}

We determine the physical and orbital properties of the 42 transiting
planet candidates around the 22 KOIs by simultaneously fitting
\ek\ photometry and Keck RVs with an analytical model of a transiting
planet \citep{Mandel2002}.  To build these models, we started with an
adopted stellar density as determined by either
the SME analysis of the high-resolution Keck spectrum of the star or
the accompanying asteroseismology analysis (both described in
Section~\ref{sec:starchar}).  The models assume Keplerian orbits with no
gravitational interactions between the planets of the multiple-planet
systems.  This non-interaction assumption is adequate to yield
parameters as accurate as the limited time series permits, as any
precession or secular resonances will create detectable effects (by
RVs) only after a decade, even for periods as short as weeks.  The
parameters in the model include the stellar density (initially from
the SME or asteroseismology analysis), the RV gamma (center of mass
velocity), a mean photometric flux, an RV zero-point, the time of one
transit ($T0$), orbital period ($P$), impact parameter ($b$), the
scaled planet radius ($R_{\rm PL}/R_*$), and the RV amplitude ($K$).

We use the parameterization of limb-darkening \citep{Mandel2002} with
coefficients calculated by \cite{limbdarkening} for the \ek\ bandpass.
We simultaneously fit all measurements with a model using a
Markov-Chain-Monte-Carlo (MCMC) routine.  To determine planet mass and
radius the Markov-Chains from the stellar modeling are combined with
the Markov-chains from the transit model.  For each Markov chain in
the transit model, we pick a stellar model from the stellar evolution
model Markov Chain and calculate planet radius and mass.  This
produces a posterior distribution for radius and mass from which we
measure the median and uncertainties.

The final values of the
planet parameters in Table~\ref{tab:orbital_pars_tbl} are the values
at which the posterior distribution is a maximum, often termed the
``mode'' of the distribution.

We considered both eccentric and circular orbits for the models of all
transiting planets.  A comparison of the chi-square statistic from the
best-fitting models for circular and eccentric orbits showed that in
no cases was a non-zero eccentricity demanded, or even compelling.
The best-fitting RV semi-amplitude, $K$, for all transiting planets in this study is
 less than 6.1 \ms (for Kepler-94b), only a factor of two or three larger than
 the RV errors.  This modest signal-to-noise ratio for the RVs limits their
 capability to detect eccentricities securely.  We found that
 models with non-zero eccentricities open the door for peculiar and undefended
 Keplerian orbits that predict high acceleration during periastron passages where no RVs were
 obtained.  These models predict wild, brief departures (during periastron)
 of the RVs from the measured standard deviation and thus violate Occam's
 Razor that favors the simplest possible model that satisfies the RV
 data.  {\em Therefore, all models of the transiting planets were computed
 with a circular orbit.}  Only the RVs for Kepler-94b exhibit some evidence of an
 eccentricity near $e=$0.2, but the non-zero eccentricity is not
 compelling (see Section~\ref{subsec:koi104summary}).

For the 7 non-transiting planets, non-zero eccentricities are commonly
demanded by the non-sinusoidal and large (many sigma) RV variations.
The RVs for three non-transiting planets revealed evidence of non-zero
eccentricities, namely for KOIs Kepler-94c, Kepler-25d, Kepler-68d.
Table~\ref{tab:orbital_pars_tbl} lists the best-fitting orbital parameters for those four
planets. The derived eccentricities of 0.38 $\pm$ 0.05, 0.18 $\pm$ 0.10,
and 0.10 $\pm$ 0.04  respectively.  The best-fit values
of $\omega$ are 157 $\pm$ 6 deg, 51 $\pm$ 70 deg, 347 $\pm$ 100 deg, respectively.

In all models, we allowed the value of the RV amplitude to be negative
as well as positive, corresponding to both negative and positive
values of planet mass.  Obviously negative mass is not physically
allowed.  But fluctuations in the RV measurements due to errors may
result in RVs that are anti-correlated with the ephemeris of the
planet as dictated by the photometric light curve.  Fluctuations can
spuriously cause the RVs to be slightly positive when orbital phase
dictates they should be negative, and vice versa.  In such cases, the
derived negative mass, and the posterior distribution of masses, is a
statistically important measure of the possible masses of the planet,
especially useful when included with the ensemble of masses of other
planets and their posterior mass distributions.  By allowing planet
masses to float negative, we account for the natural fluctuations in
planet mass from RV errors.  

For all planet candidates, especially those that yielded less than
2-sigma detections of the RV signal ($K$ less than 2-sigma from zero),
we also compute the 95th percentile upper limit to the planet mass.
To compute this for each planet, we integrated the posterior mass
distribution from the MCMC analysis to determine the mass at the 95th
percentile.  This 95th percentile serves as a useful metric of an
``upper limit'' to the planet mass, and there remains a 5\%
probability that the actual planet mass is higher.  In many cases, the
posterior mass distribution formally permits the planet's mass to be
zero, or even negative.  In such cases, the physically acceptable
upper limit, such as that computed from the 95th percentile, offers a
useful upper bound on the actual mass of the planet.  For such
planets, we determine both metrics of planet mass for these
non-detections.  The planet mass at the peak of the posterior
distribution can be positive or negative, which is useful for
statistical treatment of the planets as an ensemble. The 95th
percentile upper limit is positive, useful for constraining planet
mass, density, and chemical composition.  In
Table~\ref{tab:orbital_pars_tbl}, column 4 gives the planet mass at
the peak of the posterior distribution and column 5 give the 95th
percentile upper limit.

For each KOI, we plot the RVs as a function of time, the phase folded 
RVs for each transiting and non-transiting planet, and the phase 
folded \ek photometry. (Figures \ref{fig:koi41_fig1} -\ref{fig:koi1925_fig1}). 
The errors for each RV measurement include the internal 
error and 2.0 \ms of jitter, which is added in quadrature to obtain 
the final error. In each phase folded RV plot, the best fit RV curve  
is over plotted on top of the RVs. Each blue point is the average of 
the RVs that fall within one of the two quadrature ranges, 
$0.25 \pm 0.125$ and 0.75 $\pm$ 0.125,
 of phase set from the transits for which RV excursions
 are expected to be maximum. The value of the binned point consists of the 
weighted average of the RVs within the bin. The times of 
observation are also weighted, causing the blue RV point to be 
slightly offset from 0.25 or 0.75, based on the average phase of 
the RVs in each bin. The error of the binned RV is the standard 
deviation of the RVs within the bin divided by the square root of 
the number of binned RVs.

In summary, we fit the photometry and RVs with a \citealt{Mandel2002}
model by adopting the star's properties based on spectroscopy
(SME) and on asteroseismology, if available. Model parameters are
determined by the chi-squared statistic, and we compute posterior
distributions for the properties of the planet and the star using
MCMC.  We derive planet radius, mass, orbital period, ephemeris, and
stellar parameters, including the mean stellar density, in the final
solution. The final stellar parameters for each star are in
Table~\ref{tab:stellar_pars_tbl}.  The final planetary parameters are
listed in Table~\ref{tab:orbital_pars_tbl}, including stellar density
from the model and unbiased planet masses and densities that can be
negative.  The associated 1-sigma uncertainty for each parameter is
computed by integrating the posterior distribution of a parameter to
34\% of its area on either side of the peak, with values listed in
Table~\ref{tab:orbital_pars_tbl}.

\section{False Positive Assessments}
\label{sec:fpp1}

As has been well documented \citep{Torres2011}, a series of periodic
photometric dimmings consistent with a transiting planet may actually
be the result of various astrophysical phenomena that involve no
planet at all.  Such ``false positive'' scenarios involve the light
from some angularly nearby star located within the ($\sim$15$''$
diameter) \ek software aperture that dims with a duration and
periodicity consistent with an orbiting object passing in front of the
target star.  The light from that nearby star may be located within
the software aperture of the target star or located just outside that
aperture so that the wings of its PSF encroach into the aperture,
polluting the brightness measurements.  The amount of pollution may
vary with the quarterly roll of the spacecraft, as each star
experiences small changes in both the relative position of its aperture
and in its differential aberration from the changing velocity vector
of the spacecraft.  The polluting nearby star may be physically
unrelated to the target star (in the background or foreground) or it
may be gravitationally bound, and the cause of its dimming could be a
transiting planet, brown dwarf, star, cloud, or other construct.

By considering all astrophysical false-positive scenarios in the
direction of the \ek field of view, and in the absence of follow-up
measurements, the false positive probability (FPP) for \ek planet
candidates smaller than Jupiter is $\sim$10\%
\citep{Morton_Johnson2011, Morton2012, Fressin2013}.  However, the 22
KOIs here were selected after follow-up observations had already been
done, notably spectroscopy, high-resolution imaging, and careful
astrometry, removing many of the apparent false positives, as
described in Section~\ref{sec:choose22}.  We thus expect a false
positive rate for the 42 planet candidates studied here to be well
below 10\%.

  For Jupiter-size planet candidates the false positive rate is
  higher, near $\sim$35\% \citep{Santerne2012, Fressin2013} because
  both brown dwarfs and M dwarfs are roughly the size of Jupiter,
  allowing them to masquerade as giant planets. Also, gas giant
  planets are geometrically more likely to transit with only some
  fraction of the planet's apparent disk covering the star's disk.
  Such ``grazing incidence transits'' with impact parameter, $b>$0.9,
  cause ``V-shaped'' light curves that resemble those caused by
  eclipsing binaries (for the same reason).  Thus, the V-shaped light
  curves from gas giants forces the \ek TCERT planet validation effort
  to retain both the true planets and the background eclipsing
  binaries, thereby increasing the occurrence of false-positives.
  However, none of the transiting planets in this work are nearly as
  large as Jupiter.

The detailed assessment of the FPP for any individual planet candidate
requires careful analysis.  This ``planet-validation'' process can be
aided by the corroborating detection of the planet with some other
technique such as with RVs or transit-timing measurements. Validation
may also be accomplished by estimating the probability that the planet
is real (from measured occurrence rates) and comparing it to the sum
of the probabilities of all false-positive scenarios that are
consistent with the observations.

 \subsection{Follow-up Observations Constrain False Positives}
 \label{sec:fpp_fop}
 
 To tighten the estimates of the false-positive probabilities for the
 42 transiting planet candidates in this paper, we performed a wide
 variety of follow-up observations, described in
 Section~\ref{sec:vetting} and its subsections.  The follow-up
 observations include AO imaging, speckle
 interferometry, and high-resolution spectroscopy, all capable of
 detecting angularly nearby stars that might be the source of the
 dimming that mimics a transiting planet around the target star.  The
 AO and speckle techniques detect companions beyond a few tenths of an
 arcsec (detailed below) while the spectroscopy detects nearby stars
 (from secondary absorption lines or asymmetries in spectral lines) located within a few
 tenths of an arcsec for relative RVs  $>$10 \kms.  Thus these 
  techniques are useful to detect stellar companions located 
 within a few arcsec of the target star.   The non-detections were 
 taken into account, along with the exclusion radius, in the 
 calculation of the FPP using the method of \cite{Morton2012}.
 
 For all 22 KOIs in this paper, we have obtained AO
 imaging and speckle interferometry.  Figures \ref{fig:koi41_fig1} --
 \ref{fig:koi1925_fig1} (middle panel) show the detectability
 thresholds for companion stars to all 22 KOIs from these two
 techniques.  The AO and speckle techniques typically rule out stellar
 companions as close as $\sim$0{\farcs}1 of the target star
 (especially for Keck AO), depending on wavelength and technique (see
 Figures \ref{fig:koi41_fig2} -- \ref{fig:koi1925_fig2}, bottom
 panel).  The spectroscopic technique (see below) becomes effective
 for companions located closer than $\sim$0{\farcs}4 (half of the slit
 width), complementing AO and speckle.  Thus this suite of techniques
 offers good coverage of companion stars located at a wide range of
 orbital separations, except for 5--20 au (RV offset too small to
 support spectral separation, angular offset too small for AO or
 Speckle to resolve) within which the techniques are not robust
 at the typical distances of these targets of 100-200 pc.  All of the
 non-detections of stellar companions contributed to the FPP values
 listed in Table~\ref{tab:fpp_tbl}.

False positives can be caused by a background eclipsing binary or by
star spots \citep{Buchhave2011, Queloz2001}.  Such effects cause the
profiles of the absorption lines from the observed composite spectrum
to vary in shape as a function of ``orbital'' phase.  We searched for
changes in the shapes of line profiles by computing the usual ``line
bisector'', i.e., the relative Doppler shift of the profile near the
line core  to that in the wings.  For all 22 KOIs, the line bisectors
varied by no more than the noise ($\sim$30 \ms) and were not
correlated in time with the observed RV variations.   Thus, we rule
out all eclipsing binaries and star spots that would have caused such
bisector changes.

To further aid in constraining potential stellar false positive
scenarios, we also evaluate whether there is a linear trend present in
the RV data.  A hierarchical triple system, for example, would likely
cause the primary to display a long-term acceleration, so the absence
of a trend would help rule out these scenarios.  We find that all the
RV time series have linear trends less than 5m s$^{-1}$ amplitude over
the course of observation for all KOIs except Kepler-93, Kepler-97, and Kepler-407
which have trends of 39, 11, and 300m s$^{-1}$ amplitude. The slight
curvature in the RVs for Kepler-407 allow us to place limits on the mass 
and period of the companion.

The FPP calculations included the detectability of physically close-in
($<$5au) companion stars to the target star.  We analyzed the
high-resolution (R=60,000), high signal-to-noise (SNR$\approx$150)
optical spectra of all 22 KOIs for the presence of absorption lines
from any second star besides the identified \ek target star, as
described and tested in detail in Kolbl(2014, in prep).  In brief, the 
entrance slit of the Keck-HIRES spectrometer had a width of 0{\farcs}87, 
allowing the light from any neighboring stars located within
0{\farcs}4 to enter the slit.  This offers detectability of companion
stars complementary to that of AO and speckle
interferometry.  The algorithm fits the observed spectrum with the
closest-matching member (in a chi-square sense) of our library of 640
AFGKM-type spectra stored on disk, spanning a wide range of \teff,
\logg, and metallicities.  After proper Doppler shifting, artificial
rotational broadening, continuum normalization, and also flux
dilution (due to a possible secondary star), that best-fitting primary
star spectrum is subtracted from the observed spectrum.

The code then takes the residuals to that spectral fit and performs
the same chi-squared search for a ``second'' spectrum that best fits
those residuals.  This approach stems from an Occam's razor
perspective, rather than immediately doing a self-consistent
two-spectrum fit.  If one spectrum adequately fits
the spectrum, without ``need'' to invoke a second spectrum, then the
spectrum can only be deemed single.  A low value of chi-squared for the
fit of any library spectrum (indeed a subset of them)
to the residuals serves to indicate the presence of a second spectrum.
We establish a detection threshold by injecting fake spectra into the
observed spectrum and executing the algorithm above to determine the
value of chi-square for any relative Doppler shift, $\Delta RV$
between the companion star and the primary star that causes a 3-sigma
detection of the secondary star.  Figures \ref{fig:koi41_fig1}-
\ref{fig:koi1925_fig1} (bottom panels) show the resulting plot of
chi-squared vs $\Delta RV$ in search of a clear minimum that would
signify the presence of a second spectrum.  The blue and red lines
show how low chi-square would be for a companion having 0.3\% and
1.0\%, respectively, of the optical flux of the primary, based on the
injection of fake secondary spectra.  None of the 22 KOIs shows
evidence of a second star within 0{\farcs}4, at flux thresholds of
$\sim$0.3\% of the flux of the primary star.  There is a blind spot
for $\Delta RV < 10 \kms$ for which the absorption lines overlap,
preventing effective detection of any companion stars.  Thus,
companion stars orbiting inward of $\sim$5 au are detectable by this
technique, but companions farther out will have too small a $\Delta
RV$ to be seen.  Even for companions orbiting within $\sim$5 au, 
the orbital phase might result in a radial velocity less than 10 \kms 
relative to the primary star.
Secondary stars with rotational \vsini~ $>$ 10 \kms suffer from 
degraded detectability due to enhanced line broadening. 
Normal FGKM secondaries only rarely have high \vsini~except 
tidally locked close binaries. Similarly, secondary stars with spectral 
types earlier than F5 or white dwarfs have few spectral lines and 
suffer from poor detectability. We note that short period stellar 
binaries are unlikely given the clean transit signature and the 
dynamical instability of the planet that would result.  

Four KOIs were also observed with the lucky imaging technique, 
namely Kepler-100, Kepler-102, Kepler-37, and Kepler-409. We used the 
lucky imaging camera AstraLux \cite{Hormuth07} mounted at the 
2.2m telescope at Calar Alto Observatory (Almer\'ia, Spain). 
Good quality on-site seeing of 0{\farcs}7-0{\farcs}9 during the 
observations combined with short exposure times lead to 
diffraction limited images of the four targets in a 
$24\times 24$ arcsec field of view. A total of 30,000 frames 
of 0.030s each were acquired for Kepler-37 and Kepler-409, 
and 40000 frames were acquired for Kepler-100 (with 0.083 s 
of exposure time) and Kepler-102 ($t_{\rm exp}=0.068$ s).
 The basic reduction, frame selection and image combination 
 were carried out with the AstraLux 
 pipeline\footnote{www.mpia.mpg.de/ASTRALUX} \citep{Hormuth07}. 
 During the reduction process, the images are resampled 
 from the original pixel scale of 47.18 mas pixel$^{-1}$ to 23.59 
 mas pixel$^{-1}$. The plate scale was measured with the 
 \textit{ccmap} package of IRAF by matching the XY positions 
 of 66 stars identified in an AstraLux image with their counterparts 
 in the \cite{yanny94} catalog of the Hubble Space Telescope.
 The results from Lucky imaging were not used in the False
 Alarm Probability calculations.

We computed the sensitivity curves of our reduced images with 
a 10\% of selection rate (which optimizes the instrument/telescope 
configuration), and obtained the limiting magnitudes at different 
angular separations \citep[see additional details in][]{lillo-box12}. 
The specific limits for each observed KOI are shown in Figures \ref{fig:koi41_fig1}-
\ref{fig:koi1925_fig1} (bottom panels). 
Thus, we can assure that no objects are found within such sensitivity limits.

Light curves in the infrared may also be used to inform false positive
probabilities \citep{Cochran2011, Ballard2013} (D\'esert 2013,
submitted), but we did
not use them here, deferring such analysis for later papers
(i.e. Ballard et al. 2013 in preparation and Sarah
Ballard, personal communication). These additional constraints that
often rule out some false-positive scenarios serve to reduce the false
positive probability below that reported formally here.

\subsection{Computing Formal False Positive Probabilities}
\label{sec:fpp2}

 For each planet candidate presented in this paper, we incorporate all
 these follow-up constraints---imaging, spectroscopic analysis, and RV
 trend analysis---into the FPP-calculating procedure described in
 detail in Morton (2012).  This procedure combines information about
 predicted occurrence rates and distributions of false positives with
 the observed shape of the light curve and follow-up observations in
 order to calculate the relative probabilities of an observed signal
 being by a caused by true planet or any of a number of false positive
 scenarios.  The analysis parametrizes each phase-folded dimming
 profile with three parameters, its duration, depth, and the ratio of
 the total duration to the duration of ``ingress'' and ``egress'',
 using the geometrical approximation of a trapezoid.  The distribution
 of properties of the stars and their companions (stellar or
 planetary) toward the target star within the Milky Way Galaxy inform
 the probabilities of all the false positive scenarios by using their
 corresponding light curves, also approximated as trapezoids.

In addition to the follow-up observations and analysis discussed
above, these FPP calculations also take into account detailed
measurements of an angular ``exclusion radius''.  The maximum possible
angular distance from the target star that a false-positive-causing
dimming star could be while remaining consistent with the lack of
astrometric displacement detected between the times in and out of
``transit''. Any neighboring stars within the photometric aperture
that both dim adequately to cause the observed overall dimming and are
located farther than this exclusion radius from the \ek target star
would cause an astrometric shift in photometric difference image
centroid position between times ``in-transit'' and ``out-of-transit''.
The \ek Project data validation (DV) process routinely checks for such
displacements, ruling them false positives.

We have carefully measured this maximum exclusion radius for all stars
using the method described in detail by \cite{Bryson2013}. The
exclusion radii are listed in Table~\ref{tab:fpp_tbl} in column 2, in
arcsec from the target star. The exclusion radii range from 0{\farcs}01 - 4$''$ , 
with a median value of 0{\farcs}30 .  For 10 bright target
stars, namely Kepler-100, 93, 102, 25, 37, 68, 96, 131, 408, and 409
no formal exclusion radius could be computed due to saturation of the
\ek CCD detector.  For them, a reasonable exclusion radius of 4$''$
was adopted corresponding to a position displacement of a full \ek
pixel that can be excluded despite the saturation.

The exclusion radius for each of the 22 KOIs establishes a circular
area on the sky (i.e., a solid angle) centered on the target star
within which there could be a background star that either is an
eclipsing binary (BGEB) or has a transiting planet (BGPL), mimicking a
transiting planet around the target star.  That circular area could
also contain an eclipsing binary star gravitationally bound to the
target star, constituting a hierarchical triple system (HEB).  We also
consider the possibility that the target star itself is an eclipsing
binary; however, this scenario is completely ruled out for every KOI
by the observed lack of large RV variations.  Note that for each
scenario including some sort of eclipsing binary, \textit{we also
  include eclipsing binaries in elliptical orbits that show only a
  secondary eclipse.}

We list the results of these calculations in Table~\ref{tab:fpp_tbl}.  For each KOI,
we list the probability that it is caused by each false positive
scenario, with their sum being the total FPP.  Also included in Table~\ref{tab:fpp_tbl}
 is the assumed planet occurrence rate for each KOI (between 2/3
$R_p$ and 4/3 $R_p$) that is a factor in the probability calculations.
For probabilities smaller than 0.0001, we simply state in the table as
``$<$1e-4,ÕÕ as quoting exact probabilities smaller than this seems
unrealistic to us given unavoidable uncertainties in the input
assumptions.  Table~\ref{tab:fpp_tbl} also lists the exclusion radii that are used and
whether there was any detected companion, and its separation(s) in the
event of detections.

The exclusion radius and false-positive scenarios are further informed
by AO imaging and speckle interferometry for all 22
KOIs.  The detection thresholds for companions from those two types of
observations are shown in Figures \ref{fig:koi41_fig1}-
\ref{fig:koi1925_fig2}.  All companion stars detected are listed in
Table~\ref{tab:fpp_tbl} in column 4 and described for individual
targets in Section~\ref{sec:obs}.  The majority of the target stars
have no detected neighboring star in any of our high-resolution
images, including Keck AO images with resolution of 0{\farcs}05
 (see Table~\ref{tab:fpp_tbl}).  Such ``single'' stars with no
detected stellar neighbor enjoy a severely limited set of plausible
false positive scenarios.  For those target stars that have a
neighboring companion, listed in Table~\ref{tab:fpp_tbl} column 4, the
companions all reside angularly outside the maximum exclusion radius
derived from centroid astrometry described above.  Thus, those
detected companions are not able to explain the transit-like dimming
of the target star, supporting the planet hypothesis.  We also use the
non-detections of companion stars in a spectroscopic search for
secondary lines (see the next subsection).

Table~\ref{tab:fpp_tbl} presents the results of our formal
false-positive calculations.  The first two columns gives the name of the
planets.  The third column gives the angular exclusion
radius described above.  The fourth column gives the angular separation
of any neighboring stars detected by AO, speckle imaging, or seeing
limited imaging.  When no such neighbor was found, column 4 contains
``single''.  The fifth column gives the probability that the target
might itself be an eclipsing binary.  All of these probabilities are
less than 0.0001 because it is geometrically difficult for a companion
star to eclipse the primary star yielding the same short transit
duration and small transit depth accomplished by a planet.  Columns 6,
7, and 8 give the probabilities that the observed light curve is
produced by a gravitationally bound (Hierarchical) eclipsing binary
(HEB), a background eclipsing binary (BGEB), and a background star
orbited by a planet (BGPL), respectively.  Column 9 gives the
estimated prior probability for the candidate planet within a 30\%
range in period and size.  This 30\% is arbitrary and not particularly
conservative.  But the resulting blend probabilities are so small for
the planets here that increasing the radius range would have little effect
on the conclusions about the validation of the planets, except for the
two cases with FPP above 2\%, which would incur an increased FPP.

Column 10 gives the sum of the
false-positive probabilities (in columns 5--8), constituting the final
false-positive probability (FPP).  More precisely, the FPP is the sum
of probabilities in columns 5--8 divided by the sum of those columns
plus the probability that it's a planet, which is 1-FPP$\approx$1.

The false-positive probabilities (FPPs) are less than 1\% for
 all transiting planet candidates here, except for KOI-1612.01, which
 has a FPP of 0.021, respectively,  as shown in column 9 of
 Table~\ref{tab:fpp_tbl}.  {\it We thus find
 41 transiting planets to be formally ``validated'' as highly likely
to be real, with a false-positive probability less than 1\%.}
The remaining planet candidate, KOI-1612.01, has an
FPP of 2.1\%, making that planet's existence highly likely also, 
but not satisfying the strict 1\% FPP criterion.

We note here that Kepler-93, Kepler-97, and Kepler-407 have detected linear trends.
The FPP calculations whose results are presented in the table are
based on the assumption that there is no detected linear trend up to
the limit provided ($<$5 \ms over the relevant time baselines for all
except these three).  The fact that these three systems have detected
trends means there is a companion there, which changes the
probabilities involved in the calculation in a way that has not been
quantified.  However, the FP scenario that would get a boost from this
is the HEB scenario, whose likelihood is quite small already for all
three of these systems.  Thus even a large prior boost for the HEB
scenario would be unlikely to qualitatively change the results.

In addition, Kepler-97 has a close companion detected at just outside
the exclusion radius provided by the centroid analysis.  This changes
the false positive calculation in a similar manner, meaning the priors
for the various false positive scenarios will be higher, as there is a
known companion present.  However, given that there is also a detected
RV trend in Kepler-97, this close companion is likely to be physically
associated, meaning the HEB scenario should receive the prior boost
rather than the BGEB or BGPL.  And as mentioned above, the HEB
scenario has a low likelihood for this KOI (i.e., the signal shape is
not compatible with the vast majority of simulated HEBs).  As a
result, we still validate all three of these planets, though with the
caveat that the true FPPs of these systems have not been as
confidently quantified as those for which neither a trend nor a
companion was detected.

\subsection{Gravitationally Bound Stars with Transiting Planets}
\label{sec:BoundPL}

In our calculation of the FPP we do not deem as false positives those
scenarios that involve a gravitationally bound companion star
transited by a planet.  In such systems a real planet does exist,
albeit orbiting an unresolved bound companion star.  We have
considered carefully whether to deem such planets as ``real'' or
``false positives.''  We find no easy answer.  One useful thought
experiment involves a bound companion star that is the nearly the same
brightness as the primary star.  It makes little sense to deem a
planet a ``false positive'' simply because it orbits a slightly
fainter companion star.  Indeed, the true planet-host star may be the
``secondary star'' only in some wavelength bandpasses, thus rendering
the planet ``real'' depending on which bandpass one considers, which
is clearly absurd.

Continuing the thought experiment to companion stars that are
progressively fainter than the primary star leads to no ``break
point'' at which the planet around that companion should be suddenly
deemed a ``false positive.''  {\em Therefore, all transiting planets
  orbiting any (perhaps unknown) bound star in the stellar system is
  considered a real ``planet.''}  But if a planet does orbit a cooler,
smaller secondary star the planet is likely to be larger than was
inferred from a model of the planet orbiting the hotter, larger
primary star, to yield a predicted transit depth as deep as that
observed.  Planet occurrence decreases toward larger sizes
\citep{Howard2012, Fressin2013, Petigura2013}, and the contribution of
light from any secondary star to the photometric aperture is much
smaller than that by the primary star.  Therefore, the probability
that the transit light curve is caused by a larger planet transiting a
fainter secondary star in a binary system is less than 50\% and can be
estimated, as follows.

Roughly 50\% of FGK stars have a companion star and among those
binaries, there is a roughly 50\% probability that the transiting
planet orbits the secondary star.  Thus the probability that a KOI
consists of a transiting planet orbiting a secondary star is
0.5$\times$0.5 = 0.25.  This reasoning ignores the planet occurrence 
as a function of host star mass, which could be higher or lower than 
that of the primary. As mentioned, the required larger planet
(with lower occurrence) and the dilution from the primary star (less
detectable) make it less likely that the observed dimming is actually
from the secondary star.  {\it Thus the probability that a given
  planet candidate is actually orbiting a secondary star is probably under
  25\%.}  Thus, less than 25\% of the planets reported here
actually orbit a bound companion star.  In such cases the planet is
likely to be larger than given here in Table~\ref{tab:orbital_pars_tbl}.

\subsection{False Positive Probabilities Above 1\%}
\label{sec:fpp_high}

The one KOI with a formal FPP above 1\%, KOI-1612.01,
merits more detailed attention.  

KOI-1612.01 exhibits a V-shaped transit light curve, yielding a
nominal FPP of 2.1\% (\ref{tab:fpp_tbl}) stemming primarily from a
background eclipsing binary scenario.  With its exclusion radius of
2{\farcs}1, this planet candidate must remain a candidate, as the
background eclipsing binary scenario remains viable with a probability
of 2\%.   The FPP above 1\% excludes KOI-1612.01 from receiving a Kepler number

The high FPP for KOI-116.01 stems from its transit duration of only
1.4 hours, even though the orbital period is 2.4 d orbit around
a star of radius 1.23 \rsune.  This short observed transit duration is
shorter than expected for such a transit and star, unless the impact
parameter is near unity, which is unlikely.  The star field is indeed
crowded, and the astrometric centroid diagnostics are understandably
ambiguous.  This light curve of such short duration could instead be
caused by a eclipsing binary around a smaller star in the
background. Moreover, the transit light curve for KOI-1612.01 is shallow,
with a fractional depth of only 0.00003 (see Figure
\ref{fig:koi1612_fig2}, bottom), leaving fractional noise high
enough to allow a wide variety of false-positive scenarios.  

We note that many systems, such as 108.01 and 108.02, should have FPP
reduced by factors of over 5 due to their occurrence in a system of
multiple planets.  Such systems apparently have their orbital planes 
oriented nearly edge-on to our line of sight.  Given one transiting planet, such
alignment increases the probability of a transit by any other planets in the
system \citep{Lissauer2012, Lissauer2013}.
We do not apply this FPP benefit for multi-transiting-planets here.
For a complete assessment of the statistical boost to enable
validation of multi-planet transiting systems, with hundreds of
planets validated, see Lissauer et al. (2014, submitted), and Rowe et
  al. (2014, submitted).  

\section{Individual KOIs: Observations, Analysis, and Planet Properties}
\label{sec:obs}

We provide here detailed descriptions of the observations made for
each KOI individually.  We include descriptions of the sequence of
follow-up observations organized in chronological order to highlight
the reasoning during the reconnaissance work that led to the final
measurements with precise RVs and AO made at the Keck 1
and Keck 2 telescopes, respectively.  The goal is to highlight the
evolution of our understanding of each planet, from preliminary
estimates of its radius and orbital period, to the final posterior
distributions of the masses, radii, densities, and orbital parameters.

The ground-based follow-up observations were made by the \ek FOP team
from 2009--2012, \citep{Gautier2010, Batalha2013}, as described in
Section~ \ref{sec:vetting} with a few additional observations made
during the era after the end of the nominal \ek mission deemed the
``Community Follow-up Observing Program'' (CFOP). Detailed
descriptions of the origin of stellar parameters and planet parameters
were previously given in Sections~\ref{sec:starchar} and
\ref{sec:planetchar}, respectively. Nearly all of the observations
described here are publicly available on the CFOP website:

\url{cfop.ipac.caltech.edu/home/login.php}

\subsection{KOI-41, Kepler-100} 
\label{sec:k41}

Three transiting planet candidates were detected in \ek photometry 
around Kepler-100 and reported in \cite{Borucki2011}
and \cite{Batalha2013}.
These are now understood to have orbital periods of 6.9, 12.8, and
35.3 d and planet radii of 1.3, 2.2, and 1.6 \rearthe, respectively.
Due to their order of identification, the KOI numbers are not in
ascending order of orbital period.  A ``recon'' spectrum (see
\ref{sec:recon}) was taken at the McDonald 2.7m telescope in 2009
June, followed by another at the Tillinghast 1.5m telescope that same
month, and both were analyzed with an early version of the
spectroscopic analysis package, ``SPC'' described by
\cite{Buchhave2012} that had \teff~discrimination only accurate to
250 K and \logg \ to 0.2 dex. The resulting values of \teff~and \logg~were 
found to be consistent with those in the \ek Input Catalog (KIC).
No large RV variations (over 1 \kse) were found, and the rotational
Doppler broadening, \vsini, was found to be less than 4 \kms.  These
stellar parameters based on ``recon'' spectroscopy, along with the
relative brightness of Kepler-100, made it a good candidate for high
signal-to-noise, high-resolution spectroscopy to better constrain
stellar parameters and to carry out precise Doppler measurements.  

A ``template'' spectrum was obtained with Keck-HIRES (no iodine in the
beam).  Its analysis with SME gave values of \teff~ and \feh, used to
constrain the asteroseismology analysis, coupled with stellar
interiors modeling, to yield stellar mass and radius (described in
detail in Section~\ref{sec:starchar}).  A subsequent iteration between
SME and asteroseismology brought a quick convergence of stellar mass
and radius. The final stellar parameters are \teff~= 5825 $\pm$ 75 K,
\logg~= 4.13 $\pm$ 0.03, \vsini~= 3.7 $\pm$ 1.0 \kms and \feh~= 0.02
$\pm$ 0.10, (see Table~\ref{tab:stellar_pars_tbl}).

High-resolution imaging was acquired in 2010 May with the ARIES AO
system on the MMT, under seeing of 0{\farcs}1 in the Ks band and 0{\farcs}2
in the J-band.  Speckle imaging was taken at the WIYN telescope in
June 2010.  (citations for all instruments are given at the beginning
of Section~\ref{sec:followup}.)  Neither imaging technique found any
nearby companion stars within 6$''$ of the primary.

Keck AO imaging was performed on 2013 June 13/14 and 14/15, using a
Bracket-gamma filter to avoid saturation, as we had to use for all
stars brighter than Kp = 11 mag. The images of Kepler-100 have a PSF
with a FWHM of 0{\farcs}05 (described in detail in
Section~\ref{sec:ao}).  A full detectability
curve at $K'$ band (2.2 $\mu$), giving the detection threshold of any
neighboring stars as a function of angular separation, is shown in
Figure~\ref{fig:koi41_fig1}.  No companion was seen.

Figure~\ref{fig:koi41_fig1} shows a seeing limited image with the
spectrometer slit causing the vertical line (upper
left panel), an AO image(upper right), the detection threshold
in delta magnitudes achieved with each imaging method (middle panel),
and the limits on companions as determined by searching this star's
HIRES spectrum for secondary lines (lower panel).

 The low rotational line broadening (\vsini) and the lack of nearby
 stellar companions made Kepler-100 a high quality target for precise
 radial velocity (RV) measurements with Keck-HIRES.  They began 2009
 July 29 and span 1221 days (Figure~\ref{fig:koi41_fig2}, top panel),
 including 44 precise RVs. The RVs show a weak correlation with the
 ephemeris of the 6.9 d transiting planet identified by \eke, limiting
 the planet mass to be less than 9.6 \mearth at the one sigma
 level. This upper limit to the planet mass corresponds to a bulk
 density of the planet of 20.5 \gcc. Since this density is greater than
 any plausible solid material, taking into account gravitational compression,
 the limit is not physically meaningful. Additional RV observations
 would likely push this limiting mass lower.

The RVs also do not correlate with the ephemeris of the 12.8 d
transiting planet from \ek photometry, but the upper limits to the
planet mass are physically meaningful. The one sigma upper limit to
the mass is 4.8 \mearthe, corresponding to a bulk density of 2.0
\gcc. If this planet had consisted of mostly dense material, such as
iron with an expected density near 10 \gcc, the RV amplitude would
have been 5x larger, making such an RV signal easily detectable. But
such was not the case.  Instead, this planet must consist of a
significant admixture of lighter materials, consistent with the trend
described in Section~\ref{sec:discussion} in which
planets with radii greater than 2.0 \rearth typically have densities
less than 3.0 \gcc.  The RVs also do not correlate with the 35 day
planet ephemeris. The peak value in the posterior distribution of the
planet mass is below zero, meaning that the measured mass is
consistent with zero. The one sigma upper limit for the planet mass is
1.7 \mearth and 1.9 \gcc.  Figure~\ref{fig:koi41_fig2} (bottom right
panel) shows the phased RVs for each planet.

 The full set of RVs are plotted in Figure~\ref{fig:koi41_fig2} (top) and
 listed in Table~\ref{tab:rvs_k00041}, along with chromospheric 
 \rphk~values measured from the same spectra as the RVs, making them
 simultaneously obtained.

\subsection{KOI-69, Kepler-93}  

\ek photometry identified Kepler-93 as having one transiting
planet candidate with a period of 4.7 d and a radius of 1.5
\rearthe. A second orbiting companion is indicated (as described below)
by the precise RVs that exhibit a linear trend of +11 \ms $yr^{-1}$. 
During the past 4 years, the RVs show no departure from a straight
line at the 1 \ms level.

Follow up recon spectroscopy commenced in 2009 August at McDonald 2.7m
and the 2.6m Nordic Optical telescope.  Moderate signal-to-noise
spectra were acquired, yielding spectral parameters found with SPC
analysis that were consistent with the values in the KIC. The \vsini~
was measured to be $<$ 2 \kms, consistent with a slowly rotating main
sequence star G5V, and no RV variation was seen between the two
spectra at the $~$500\ms level (confirmed by later precise RVs).  SME
analysis on the Keck-HIRES template spectrum showed the star to be
slightly cooler than the KIC and recon-determined parameters.  Final
stellar parameters including stellar mass and radius were calculated
using asteroseismology with SME values used as input parameters, along
with the usual iteration between the two. The final stellar parameters
are \teff~= 5669 $\pm$ 75 K, \logg~= 4.47$\pm$ 0.03 and \feh~= 0.02 
$\pm$ 0.10 (see Section~\ref{sec:starchar}).  
Table ~\ref{tab:stellar_pars_tbl} lists
the final stellar parameters including stellar mass and radius.
 
 Speckle imaging was first obtained in 2009 October and a binary
 companion was suggested 0{\farcs}05 away with a delta magnitude of 1.4
 mag. Two additional speckle images were taken and the secondary star
 was not detected.  A warning about the possible existence of this
 companion appears in Table 3 of \cite{Borucki2011} but it should be
 disregarded.  Adaptive optics imaging in 2010 May at MMT-ARIES
 confirmed no secondary stars were present from 1$''$--6$''$ of the
 primary, but the putative 0{\farcs}05 companion would have fallen within
 the inner working angle of ARIES.  The field of view near this KOI is
 shown in an image from the Keck-HIRES guide camera, which is a seeing
 limited and shown in Figure ~\ref{fig:koi69_fig1} (top left). Keck AO
 imaging on 2013 June 13/14 and 14/15 yielded a FWHM of 0{\farcs}05 
 Figure ~\ref{fig:koi69_fig1} (top right),
 with a full detectability curve given in Figure~\ref{fig:koi69_fig1} (middle).  
 No companion was seen.  Limits on companions
 from the high-resolution imaging are displayed in Figure~\ref{fig:koi69_fig1}
 (bottom).

Keck-HIRES precise RVs span 1132 days, from 2009 July to 2012
September (Figure~\ref{fig:koi69_fig2}, top). The most prominent
feature in the RVs is the linear trend of 11.2 $\pm$ 1.5  \ms $yr^{-1}$.
Since we see no curvature, the orbital period of the companion causing
the linear trend is much longer than the time baseline of nearly four
years. Given the line-of-sight component of acceleration during 4
years, we can place lower limits on the mass and period ($M > $3 \mjup
and $P > $5 yr), based on the duration and magnitude of the RV trend.
The linear RV trend suggests the period is much longer than 10 years,
indicating that the orbiting companion has a mass at least tens of
Jupiter masses.  It could be an M-dwarf or brown dwarf orbiting beyond
5 au, or perhaps a compact stellar object, all being too dark to be
revealed in the Keck-AO images with resolution of 0{\farcs}05.  This
non-transiting companion has a designation of Kepler-93c and these values
are listed in Table ~\ref{tab:orbital_pars_tbl}.  It remains possible
that the transiting planet orbits the unresolved massive companion
rather than the primary star.  If so, it would have to be Jupiter-size
to yield the observed transit depth, with a period of 4.7 d, which
would be a rarity.  A more detailed analysis of Kepler-93 is carried
out by Ballard et al., 2013, submitted.

We place upper limits on the mass of the transiting planet at 4.4
\mearthe, which corresponds to a bulk density of 7.2 \gcc. This is
only an upper limit because the median value of the posterior
distribution of the MCMC analysis of the planet mass is only slightly
above zero, and the value is consistent with zero at the two sigma
level.  The phase folded RV curve shows the K-amplitude of
1.05 $\pm$ 0.8\ms (Figure \ref{fig:koi69_fig2}, bottom right), not
significant at better than 1-sigma.  The full set of RVs is plotted
versus time in Figure \ref{fig:koi69_fig2} (top) and listed along with
\rphk~activity values in Table ~\ref{tab:rvs_k00069}.

\subsection{KOI-82, Kepler-102}  

The five planets in this system have periods of 5.28, 7.1, 10.3, 16.1,
and 27.5 d and have corresponding radii of 0.5, 0.6, 1.0, 2.2 and 0.9
\rearthe. Discovered according to \ek photometric detectability, their
KOI numbers are not in order of increasing orbital period.  Due to
their very small transit depths of only 41 and 65 ppm, 82.04 and 82.05
were detected last, but have the shortest periods. The Kepler-102 letters
are assigned to each planet by increasing orbital period.

 Recon spectroscopy was first acquired with the Lick 3m telescope in
 2009 August.  Two spectra taken four days apart gave stellar values
 in agreement with the KIC value for \teff, but the value of \logg~=
 5.0 conflicted with the KIC \logg~value of 4.0. The final stellar
 parameters determined with SME analysis combined with stellar models
 and light-curve fits are \teff~= 4903 $\pm$ 74 K, \logg~= 4.61
 $\pm$ 0.03 and \feh~= 0.08 $\pm$ 0.07.  The final \logg~value and
 stellar mass changed the KIC value of stellar radius from 1.8 \rsun
 to 0.74 \rsune.  The initial values of planet radii were likewise
 reduced by a factor of $\sim$2.5.

Due to the intense interest in the system of five small planets, and the recon-derived
stellar parameters that differed greatly from the KIC values, high-resolution 
spectra were taken at Keck/HIRES and FIES to confirm the stellar properties. 
Several trials of SME analysis and SPC analysis were made to robustly confirm 
the \teff, \logg, and \feh. In the end all analyses agreed within one
sigma errors, given above.

 Standard follow-up with speckle imaging in 2010 June and ARIES AO observations in 
  2010 September found no stellar companions within their detection
 domains, supporting the notion that the transits occur on the primary
 star or an unresolved bound companion star. Figure \ref{fig:koi82_fig1}, top left,
 shows a seeing limited image of the field of view from the Keck-HIRES 
 guide camera.   Figure  \ref{fig:koi82_fig1} (middle panel) shows the 
 limiting magnitudes achieved with each imaging method.  
Keck AO imaging on 2013 June 13/14 and 14/15 yielded a PSF with FWHM of 0{\farcs}05', 
and detectability 7 magnitudes fainter than the primary star at $K'$,
 for all separations more than 0{\farcs}2.  The full detectability
 curve is given in Figure \ref{fig:koi82_fig1}.  The absence of any
 stellar companion in the Keck AO images greatly reduces the probability of any stellar
 neighbor that might pose an alternative to the transiting planet interpretation.
This sense is confirmed with the detailed false positive analysis (See
Section~\ref{sec:fpp1}). Confirmation of Kepler-102e was performed independently 
by \cite{Wang2013}  using pixel centroid offsets, transit depth
measurements and a UKIRT constrast curve.

With five transiting planets having orbital periods less than 25 d and
planet radii less than 2.5 \rearthe, Kepler-102 appears to be a
densely packed, rich system of small planets.  The RV time baseline of
nearly 900 days shows only a 4.3 \ms scatter.  We fit the RVs to the
orbits specified by the transit ephemerides, using circular models (as
with all transiting planets in this paper), see (Figure
\ref{fig:koi82_fig2}, top).  We fit all five planets simultaneously
with circular orbits. The 16-day planet (Kepler-102e) shows clear
coherence with the RVs.  We measure the planet mass at 8.9 $\pm$ 2.0
\mearth (radius 2.22 \rearthe), and a density of 4.68 $\pm$ 1.1
\gcc. The RV semi-amplitude, $K$, is 2.77 $\pm$ 0.6 \ms, the largest
for any planet in the system.  The four remaining planets in the
system do not appear in the RVs convincingly.  Their induced RV
variations are apparently at or below their respective RV detection
thresholds, and we report two sigma upper limits to their masses in
Table ~\ref{tab:orbital_pars_tbl}.  With the three year time baseline,
we constrain the presence of non-transiting planets out to 5 au with
masses down to $\sim$1 \mjupe. No detectable trends or periodic
signals are seen in the RV residuals to the five planet fit.  The full
set of RVs are listed, along with their \rphk~activity values, in
Table~\ref{tab:rvs_k00082}.

\subsection{KOI-104, Kepler-94}  
\label{subsec:koi104summary}

The \ek photometry revealed a single transiting planet candidate
orbiting Kepler-94 with a period of 2.50 days and radius
(later found to be) 3.51 \rearthe.  It resides in a multiple system,
as precise RVs reveal the presence of a non-transiting orbiting
companion with a period of 820 $\pm$ 5 days and
\msini~ of 9.8 $\pm$ 0.6 \mjupe. Before the non-transiting planet was
detected, we acquired the usual suite of follow up observations. Recon
spectroscopy was first done with the 2.6m Nordic Optical Telescope in
2009 August.  The three spectra that were acquired showed no radial
velocity variation above the errors of ~200 \ms, indicating the host
star to be a slowly rotating cool dwarf.

Initial photometric analysis based on KIC stellar parameters yielded a
star and planet radius smaller by 20\% than the final planet
radius. This slowly rotating K-type host star was recognized as ideal
for precise RV spectroscopy.  SME analysis of the HIRES template
spectrum and comparison to Yonsei-Yale stellar models refined the
stellar properties to be \teff~= 4781 $\pm$ 98 K, \logg~= 4.59 $\pm$ 0.04
and \feh~= +0.34 $\pm$ 0.07, in agreement with the stellar parameters determined
from the recon analysis. The final planet radius is 3.51 $\pm$ 0.15 \rearthe,
stemming from our standard MCMC modeling of light-curve and RVs and after
adopting the final stellar radius.  Kepler-94 was analyzed by
\cite{Muirhead2012b}, but the resulting parameters were found to be 
uncertain with their method, which is best suited for mid- to late M dwarfs
having \teff~$<$ 4000 K.  The final stellar parameters can be found in
Table ~\ref{tab:stellar_pars_tbl}.

The ``V-shaped light curve'' led the KFOP to adopt
caution regarding the planet interpretation. The \ek photometric
diagnostics suggested a chance that this was a blend of two or more
stars. Speckle imaging revealed no companions within its
limits. Subsequent AO observations taken at Palomar further limited
the presence of stellar companions to within 1{\farcs}0 away from the
primary.  The concerns regarding the V-shaped transit and possible
blend of other stars initially yielded a false positive probability
of ~10\%, prior to the acquisition of the Keck AO images.
The Keck AO images of Kepler-94 greatly reduced the probability
of background and bound stellar companions, as described in detail in
Section~\ref{sec:fpp_fop}, yielding a final false positive probability of
$<0.0001$, validating the planet.

Moreover, the RV periodicity is in phase with the \ek ephemeris of this 2.51 d
planet, adding further support to the reality of the
transiting planet, and limiting the plausible false-positive
scenarios that would insidiously mimic both the light curve and RVs of
a transiting planet.  For a detailed description of the FPP
calculation, see Section~\ref{sec:fpp2} and \ref{sec:BoundPL}.

A seeing limited image of the field of view from the Keck-HIRES guider
is found in Figure~\ref{fig:koi104_fig1}(top left) while 
Figure~\ref{fig:koi104_fig1}(middle) shows the limiting magnitudes 
achieved with each high-resolution imaging method.  No stellar companions 
were found. Keck AO imaging on 2013 June 13/14 and 14/15 yielded a 
FWHM of 0{\farcs}05. No companion was seen.

The first Keck-HIRES RV was acquired in 2010 June. The long baseline
of the RVs, spanning 800 days (Figure \ref{fig:koi104_fig2}, top),
was vital for mapping out the orbit of the non-transiting planet.  The
precise RVs are in phase with the transiting planet in a circular
orbit, and we measure the minimum planet mass (\msinie) to be 10.8
$\pm$ 1.4 \mearthe.  The transiting planet mass, combined with the
planet radius measurement from \ek yields a bulk density of 1.45 $\pm$
0.26\gcc, which is consistent with theoretical expectations by
\cite{Lopez2012}.  This is a low density planet composed of a large
fraction of non-rocky material, perhaps H and He. While the planet
radius and mass quoted above stems from a circular orbit model, the
transiting planet appears to be in a non-circular orbit, despite its
short period. It is one of the few cases in which a non-zero
eccentricity is called for, allowing a better fit to the RVs than the
circular orbit. The higher RV variation is due to the non-transiting
planet with its period of 820 d, and we measure \msini~of that planet to be 9.8 $\pm$ 0.6
\mjupe. This non-transiting planet has a non-circular orbit, and we measure the
eccentricity to be $e$=0.38 $\pm$ 0.05. The phase folded RV curves are
shown in Figure~\ref{fig:koi104_fig2} (bottom right).

The eccentricity of the short period planet is unusual. Perhaps the
non-transiting planet is pumping the eccentricity of the transiting
planet. It is worth noting that there is only one transiting planet,
perhaps because the non-transiting planet has scattered other planets
out of the original protoplanetary plane.  The most likely true mass
of the non-transiting planet 10 - 20 \mjupe, accounting for likely
orbital inclinations, and the actual planet mass could be higher.  The
full set of RVs are listed, along with \rphk~values, in Table
~\ref{tab:rvs_k00104}.  It could be fruitful to search carefully for
TTVs in the transit times of the inner planet.

\subsection{KOI-108, Kepler-103} 

The two planets identified by \ek photometry around Kepler-103 
have periods of 15.97 and 179.6 d, with planet radii of 3.37 and 5.14 \rearthe,
respectively. Follow-up spectroscopy at the McDonald 2.7m in 2009 August
were found to be in marginal agreement with the KIC estimates of
\teff~ and \logg, and rotational \vsini $<$ 4 \kmse.  These stellar
properties, and its brightness at $K p$ = 12.3, warranted high
resolution spectroscopy at the Keck Observatory.  SME analysis 
was conducted on a Keck-HIRES template
spectrum. The results were used as initial guesses in the
asteroseismology analysis, which determined the final stellar
parameters, notably increasing the stellar radius by 25\% from
from the KIC radius to 1.43 \rsune.  The derived stellar parameters
are \teff~= 5845 $\pm$ 88 K, \logg~= 4.16 $\pm$ 0.04, and \feh~= 0.07 $\pm$ 0.1.
The lack of RV measurements in 2012 is due to the refined
stellar parameters, which increased the planet radii to $>$3.0
\rearthe, diminishing its priority.

Keck AO imaging on 2013 June 13/14 and 14/15 yielded images with PSFs
having a FWHM of 0{\farcs}05.  The detectability curve for neighboring
stars is given in Figure \ref{fig:koi108_fig1}.  No companion was
seen.  Adaptive optics imaging was also acquired in the J-band at
Palomar Observatory, and two nearby stars were found separated by
2{\farcs}44 and 4{\farcs}87 from the primary, both being 7.2
magnitudes fainter than the primary. No companion stars were found
within the limits of speckle imaging, taken in 2010 June, which probes
from 0{\farcs}05 to 1{\farcs}4 from the primary.  During Keck-HIRES
observations, the 4{\farcs}87 stellar companion was purposely kept out
of the HIRES slit, however the 2{\farcs}76 companion was not visible
on the HIRES guider and some of its light may have inadvertently gone
into the slit of the spectrometer.  The HIRES slit is only 0{\farcs}87
wide, so for most orientations of the rotating HIRES field of view
(due to use of the image rotator) that nearby star would not fall in
the slit. If it did, then the flux would be $<$1\% the brightness of
the primary at optical wavelengths. For all of these reasons, we
suspect any contamination from the 2{\farcs}76 companion is negligible
in the spectroscopy of the primary star, given the SNR of
$\sim$100. Figure~\ref{fig:koi108_fig1} (top left) shows a seeing
limited image of the field of view of the HIRES guider camera.  Figure
\ref{fig:koi108_fig1} (middle) shows the limiting magnitudes achieved
with each high-resolution imaging method.

The second transit of the 180d planet was found only after a full year
of \ek data was analyzed. Upon discovery of the first transit event of
the 180d planet, centroid analysis was conducted.  Initially centroid
motion seemed apparent, indicating a source location for the transit
located 0{\farcs}7 from the primary star, albeit at the two sigma
level. After more \ek photometry was processed, the astrometric 
displacement was not confirmed, and the centroid
analysis was found to be consistent with a transit on the target star.

The time baseline for the RVs spans 735 days (Figure
~\ref{fig:koi108_fig2}, top). The RV signal does not strongly correlate with the
orbital period of 16 days for the inner planet. The nominal mass of
the 15 d planet is measured to be 9.7 $\pm$ 8.6 \mearthe,
corresponding to a bulk density of 1.38 $\pm$ 1.4 \gcc~when combined
with the planet radius of 3.37 \rearthe. The RV signal has a
semi-amplitude of 2.32 $\pm$ 2.1 \mse, providing only an upper limit
to the mass. The 95th percentile of the posterior mass distribution gives
the upper limit to the planet mass at 30 \mearthe. Since RV
measurements are planned to occur at the predicted times of
quadrature, the phase coverage of the RVs is poor and the eccentricity
is not well constrained, leading us to conduct these fits using
circular orbits.

For the 180 d planet, orbital analysis of the RVs yields a peak of the
posterior distribution of masses to be at 3.85 \mse, but with a large uncertainty
of 2.7 \mse.  Poor phase coverage of the RVs for this long period planet make
upper mass limit only marginally useful. When the RV is measured at both
quadratures, both high and low, the mass limits will be more robust.  
The RVs are listed, along with their \rphk~activity values, in 
Table~\ref{tab:rvs_k00108}. This system of two planets should be 
examined carefully for TTV signals.

\subsection{KOI-116, Kepler-106}  
 
\ek identified a quartet of candidate transiting planets around
    Kepler-106, with orbital periods of 6.2, 13.6, 24.0, and 43.8
    d and planet radii of 0.8, 2.5, 0.95 and 2.6 \rearthe,
    respectively. The two largest planets were identified within the
    first three months of \ek data.  Kepler-106b required quarters 1--6
    of \ek data, and the Kepler-106d required quarters 1--8. The longer
    time baselines of the \ek photometry were required to find such
    small signals.

Ground based follow-up observing started with acquisition of recon
spectra using the 2.6m Nordic Optical Telescope in 2009 August and the
McDonald 2.7m in 2009 September. Analysis of these spectra gave \teff~
in agreement with the KIC, but \logg~was found to be nearer 4.5 (than
4.0 from the KIC), placing this star on the main sequence, not slightly
evolved. The main effect of this change in gravity is the decrease of
the stellar radius by 50\% to 1.04 \rsun. The planet radii were
likewise decreased from $~5$ \rearth to their final values stated
above and in Table ~\ref{tab:orbital_pars_tbl}.  SME analysis of a
HIRES template spectrum was used in combination with Yonsei-Yale
stellar isochrones to determine the final stellar parameters of 
\teff~= 5858 $\pm$ 114 K, \logg~= 4.407 $\pm$ 0.14, and 
\feh~= -0.12 $\pm$ 0.1. (Table ~\ref{tab:stellar_pars_tbl}).

Keck AO imaging on 2013 June 13/14 and 14/15 yielded a FWHM of 0{\farcs}05 , 
with a full detectability curve given in Figure
\ref{fig:koi116_fig1}.  No companion was seen, ruling out companions
8 mag fainter in $K'$ band (or brighter) located beyond 0{\farcs}4 
from the primary star.  Speckle imaging at WIYN taken in 2010 June and
AO observation taken at ARIES in Ks band in late 2010, also showed no
companion stars within their detection limits. Figure
\ref{fig:koi116_fig1} (top left) shows a seeing limited image of the
field of view of the HIRES guide camera, and Figure
\ref{fig:koi116_fig1} (top right) shows the Keck AO image.  Figure
\ref{fig:koi116_fig1} (middle) shows the limiting magnitudes achieved
with each high-resolution imaging method.

While precise RV monitoring started in 2010, the discovery of the
fourth planet, and the adjustment of all four planet radii to values
below 2.5 \rearthe, motivated an increased RV cadence in the 2012 observing
season.  The RVs obtained during 1073 d  (Figure \ref{fig:koi116_fig2}, top) exhibit
no evidence of non-transiting planets nor monotonic trends.   We computed a
self-consistent four-planet model fit to the RV, assuming circular
orbits (as with all transiting planets in this paper) and the \ek
orbital ephemeris.  Those fits show no evidence of the 6 d and
24 d planets in phased plots of the RVs (each plot having the
remaining three planets removed).

The phased RVs for the planet (Kepler-106c) with a period of 13.57 d
and radius 2.5 \rearth do correlate with the RVs predicted from the
ephemeris, as shown in Figure \ref{fig:koi116_fig2}, lower right.  That RV
signal yields a planet mass of 10.4 $\pm$ 3 \mearthe, a 3-sigma
detection of the planet.  This planet mass and radius corresponds to
a planet density of 3.3 $\pm$ 1.6 \gcc.

Finally, Kepler-106e with $P$=43 d and $R_{p}$=2.6 \rearthe, shows only a
weak correlation with the RVs, enabling a constraint on the planet
mass of 11.2 $\pm$ 6 \mearthe, below a 2-sigma detection, and density
= 3.1 $\pm$ 2.1 \gcc.  While we cannot place useful upper limits on
the masses or densities of the two sub-Earth radii planets, the RVs
are consistent with the four planet system. The phase-folded RV curves
for each planet are shown in Figure~\ref{fig:koi116_fig2} (bottom
right).  The full set of RVs are listed, along with their \rphk~activity 
values in Table~\ref{tab:rvs_k00116}.

\subsection{KOI-122, Kepler-95} 
Kepler-95 was identified by \ek to have a single transiting
planet with orbital period 11.5 d and radius 3.4 \rearthe. The first
follow-up observations were two recon spectra taken in 2009 August at
the McDonald 2.7m.  The resulting measurements of \teff~agree with the
KIC value, but the \logg~value differs significantly, yielding an
implied stellar radius of 1.41 \rsune, 75\% larger than the KIC value.
We obtained a Keck-HIRES spectrum, and the SME analysis combined with
asteroseismology \citep{Huber2013} provides \logg~= 4.17 $\pm$ 0.04,
\teff~= 5699 $\pm$74 K and \feh~= +0.30 $\pm$ 0.1 (see Table
~\ref{tab:stellar_pars_tbl}).   The corresponding
adjustment to the planet radius (originally 1.9 \rearthe), moved this
KOI out of the \ek TCERT prioritized range of planet radii, i.e. above
3 \rearthe.  This planet probably would not have been included in this
survey, had we known the radii precisely to begin with.

Follow-up speckle imaging with the WIYN telescope in 2009 August found
no stellar companions, but AO imaging at Palomar in 2010 June revealed
a single companion star 4{\farcs}1 from the primary that is 6.5
magnitudes fainter in the J-band. Centroid analysis of this target
rules out the possibility of the transit occurring on that neighboring
star. Figure~\ref{fig:koi122_fig1} (top left) shows a seeing limited
image of the field of view of the HIRES guide camera, in which the
4{\farcs}1 neighboring stars does not appear, presumably because it is so
faint in the optical.  Figure~\ref{fig:koi122_fig1} (middle) shows the
limiting magnitudes achieved with each high-resolution imaging method.
Keck AO imaging on 2013 June 13/14 and 14/15 yielded a FWHM of 0{\farcs}05, 
with a full detectability curve given in Figure
\ref{fig:koi122_fig1}.  No additional companions were found beside
that 4{\farcs}1 neighbor.

The precise RV time baseline from Keck-HIRES spans 1078 days and
exhibits an RMS of 5.1 \ms (see Figure~\ref{fig:koi122_fig2},
top). After fitting the ephemeris of the single planet, of radius =
3.4 \rearthe, to the RVs we detect the planet mass to be 13.0 $\pm$ 2.9
\mearth with a corresponding density of 1.7 $\pm$ 0.4 \gcc .  The RV
detection is statistically significant, and the low density of 1.7
\gcc~requires the planet to consist of a large fraction of volatiles
by volume.  This planet density is consistent with theoretical
expectations for a planet of its size, 3-4 \rearthe, as noted in in Table
4 of \cite{Lopez2012}.  The full set of RVs are listed, along with
their \rphk~activity values in Table~\ref{tab:rvs_k00122}.

\subsection{KOI-123, Kepler-109} 

The Kepler-109 planetary system consists of two transiting
planets with orbital periods of 6.5 d (Kepler-109b) and 21 d
(Kepler-109c), and radii of 2.4 and 2.5 \rearthe, respectively. Two
recon-level spectra were acquired at the McDonald 2.7m in 2009 August,
separated by 18 days, and they showed no RV variation of more than
500 \mse.  The measured \teff~and \logg~values agreed, within errors,
with the KIC values. The low projected rotational velocity of \vsini~=
4 \kms made this a good target for follow-up with precise RVs. The
template spectrum from Keck-HIRES was used to determine the stellar
parameters with SME, which was used as inputs to the asteroseismology
analysis. The final stellar parameters are \teff~= 5952 $\pm$ 75 K,
\logg~= 4.21 $\pm$ 0.04, and \feh~= -0.08 $\pm$ 0.1 (see Table
~\ref{tab:stellar_pars_tbl}).  The stellar parameters from the KIC,
recon spectra, and SME/asteroseismology are all in agreement, and no
large modifications to the stellar radius nor to the planet radii were
needed.

In 2009 September AO imaging at Mt. Palomar revealed two
stellar companions in the J-band, located 2{\farcs}03 and 5{\farcs}27 away
from the primary target star, with delta magnitudes of 7.4 and 8.1,
respectively. Speckle imaging was acquired in 2010 June, and no
additional companions were found. The two companions found with AO are
beyond the detection limits of speckle. Centroid analysis of the \ek
photometry excludes the possibility that the transits fall on either
of the known companion stars. Figure~\ref{fig:koi123_fig1} (top left)
shows a seeing limited image of the field of view of the HIRES guide
camera that reaches to Vmag = 21.  It shows four neighboring stars,
the two brightest being those seen in the Palomar AO imaging. More stringent limits on
companions were placed with Keck AO images we obtained on 2013 June 13/14 and
14/15 yielding a FWHM of 0{\farcs}05 and a detection threshold of 8
magnitudes at $K'$ beyond 0{\farcs}4 from the primary.  The absence of
neighboring stars between 0{\farcs}1 and 2{\farcs}0  reduces the probability
of any background and gravitationally bound stars that might
masquerade as the two transiting planets.  Figure
\ref{fig:koi123_fig1} (middle) shows the threshold magnitudes achieved
with each high-resolution imaging method.    

The sparsely populated $\sim$1100 d time baseline of RV 
measurements has an RMS of 7.1 \ms
(see Figure~\ref{fig:koi123_fig2}, top), and the RVs do not correlate
in phase with either transiting planet (see Figure~\ref{fig:koi123_fig2}, bottom right).  
We believe that the 7.1 \ms scatter
may be due to a combination of the noise, stellar jitter, instrumental
errors, and perhaps non-transiting planets not identified with our
limited number of RVs.

We provide mass limits from our usual MCMC analysis for each
planet. The nominal planet masses are 1.3 $\pm$ 5.4 \mearth and 2.2
$\pm$ 7.8 \mearth for the inner and outer planets respectively. The
corresponding density upper limits of 0.3 $\pm$ 2 \gcc and 0.6 $\pm$
2.3 \gcc suggest that compositions of pure rocky material are very
unlikely in both cases, as the densities are inconsistent with rocky
interiors.  If they were mostly rocky, both planets would have masses
easily detectable by the existing RVs.  The full set of RVs are
listed, along with their \rphk~ activity values in Table
~\ref{tab:rvs_k00123}.

\subsection{KOI-148, Kepler-48} 

This four planet system, has three transiting planets with orbital
periods of 4.8, 9.7, and 43 d with radii of 1.9, 2.7 and 2.0 \rearthe,
respectively.  The RVs reveal a non-transiting planet with an orbital
period of 982 $\pm$ 10 d which we name Kepler-48e. It has a minimum mass
of \msini = 2.1 $\pm$ 0.08 \mjupe. Follow-up observations of this
system began with the acquisition of three recon spectra from the Lick
3m and McDonald 2.7m. The spectral analysis of these recon spectra
conducted with SPC found this star to be a slowly rotating main
sequence K0 star, confirming the KIC parameters and adding a
measurement of \vsini~$<$ 2 \kmse. When we started collecting precise
RVs with Keck-HIRES, SME analysis was conducted on the template
spectrum. When combined with Yonsei-Yale stellar models, SME found
\teff~ = 5194 $\pm$ 73 K, \logg~ = 4.49 $\pm$ 0.05 and \feh~ =
0.17 $\pm$ 0.07. Final stellar parameters are listed in Table
~\ref{tab:stellar_pars_tbl}.

High-resolution imaging at WIYN using the speckle camera found no
stellar companions. Adaptive optics imaging at Mt. Palomar probed the
field of view beyond the 2$''$ field of view of speckle imaging,
revealing four companions within 6$''$ of the primary. Their
separations and brightnesses are: 2{\farcs}44 and delta magnitude of 4.9,
4{\farcs}32 away and delta magnitude of 3.3, 4{\farcs}39 and delta magnitude
7.3, and 5{\farcs}89 away delta magnitude of 7.0. This Palomar imaging
was conducted in the J-band and probes down to a $K p$ of
22.1 \citep{Adams2012}.  All four of these companion stars are
identifiable in the Keck guide camera and care was taken to keep these
stars from entering the slit during precise RV
observations.  However line bisector variations of a few observations
suggest that the nearest companion could have been present along the
slit on three occasions.  Strong limits on stellar companions were
imposed with Keck AO imaging on 2013 June 13/14 and 14/15 yielding a
FWHM of 0{\farcs}05.  No additional companions were seen. The
detectability curve from all imaging of Kepler-48 is shown in Figure
\ref{fig:koi148_fig1}.  The key result is a lack of stellar
companions within the inner 2$''$. We remain blind to stellar
companions within $\sim$0{\farcs}1 where Keck AO becomes increasingly
less sensitive.

Centroid analysis of the \ek images, suggests that the transits
occurred on the primary star, and not on any of the four companion
stars.  The centroid-based exclusion radii are less than 0{\farcs}3 
for all three planets, ruling out the four stars as sites of the dimming.
Figure~\ref{fig:koi148_fig1} (top left) shows a seeing limited image of
the field of view from the Keck guide
camera. Figure~\ref{fig:koi148_fig1} (middle) shows the detection
thresholds from the high-resolution imaging analysis.

We took high-resolution spectra of Kepler-48 with Keck-HIRES
starting in 2009 August (Figure \ref{fig:koi148_fig2}, top). The initial RV
epochs in summer 2009 would prove vital for detecting the
non-transiting planet in a nearly 3-year orbit. Further radial
velocity measurements will more tightly constrain the orbital
parameters of the non-transiting planet, which will likewise improve
the measurement of the masses of the transiting planets.

The 4.7 d  and 9.7d planets have been shown to be gravitationally
interacting, in a 2:1 mean motion resonance, as measured by their
TTVs \citep {Steffen2013_kepler_vii,  Wu2013, Xie2013}. 
These studies perform a stability analysis
to calculate the possibility of these two inner planets as false
positives, leading to a FPP of less than 0.001. This agrees with the
low false positive probability we find here, based on
\cite{Morton2012} analysis listed in Section \ref{sec:choose22} and
Table~\ref{tab:fpp_tbl}. The planet with the 43 d period does not
gravitationally interact with the inner planets in a way that leads to
a detectable TTV signal. The masses of the two interacting planets have
been measured with TTVs, but with different stellar parameters and
different planet radii, making a direct comparison here difficult.

These planet mass values are constrained with a TTV analysis which results in maximum
masses of 17.2 $\pm$ 3.9 and 10.1 $\pm$ 3.5 \mearthe, for 148.01 and
148.02, respectively.  RV measurements currently find the
mass of 3.9 $\pm$ 2.1 \mearth and 14.6 $\pm$ 2.3 \mearthe. Phase folded
RV curves for each planet are shown in Figure
\ref{fig:koi148_fig2} (bottom right). Radial velocities constrain the mass of .01
more tightly than TTV analysis. The mass determined for Kepler-48c is
consistent between the two methods within errors.  A joint analysis
using both TTVs and radial velocities would likely constrain the
masses even more.
 
 After the second observing season, a linear trend in the RVs was
 detected. After the third year of observing the RV trend turned over
 and the period of the non-transiting planet was known to within a
 factor of two. Only after the fourth year of observations and a full
 orbit of the non-transiting planet had occurred, was the period
 determined to better than 10\%.  We now find its period, $P=982
 \pm 8 $ d and \msini~= 657 $\pm$ 25 \mearthe (2.1 $\pm$ 0.1\mjupe).  As the orbital
 parameters, especially eccentricity, of the non-transiting planet are
 further constrained with more observations, the masses of the inner
 planets will be more tightly constrained.  The full set of RVs are listed, along with their \rphk~ activity values in Table ~\ref{tab:rvs_k00148}.

\subsection{KOI-153, Kepler-113}  

Kepler-113 was identified by \ek as having two transiting planet
candidates with orbital periods of 8.9 d and 4.7 d having radii of 2.2
and 1.8 \rearthe, respectively.  Recon-level spectra of this system
were taken at the 2.6m Nordic Optical Telescope and McDonald 2.7m in
August and 2009 September. Using SPC, stellar parameters of \teff~and
\logg~were confirmed to be within the errors of the KIC values, and no
RV scatter above 1 \kms was found. This star was found to be a slowly
rotating K3V star, making it amenable for precise RVs with
Keck-HIRES. The main challenge posed by this KOI is its faintness,
Kp = 13.5 mag, making high-resolution spectroscopy time consuming:
Keck-HIRES exposures of 45 minutes are required to achieve SNR=70 per
pixel. The final stellar parameters, determined with SME and
Yonsei-Yale stellar models, yielded a stellar radius 30\% less than
the KIC value, causing an equal decrease in the determination of the
planet radii.  The final stellar parameters are \teff~= 4725 $\pm$
74 K, \logg~= 4.64 $\pm$ 0.03 and \feh~= 0.05 $\pm$ 0.07, as listed in
Table~\ref{tab:stellar_pars_tbl}.

Speckle imaging at the WIYN telescope in 2010 June found no
companions. Adaptive optics imaging taken with ARIES in the 2009
observing season and with Palomar in 2010 found one companion within
6{\farcs}0. It was found in both the Ks and J bands at 5$''$.14 from
the primary with delta magnitudes of 8.1 and 8.3 respectively. Seeing
with ARIES was only 0{\farcs}4, worse than typical conditions, while
seeing with Palomar was 0{\farcs}1 -- 0{\farcs}2 for Ks and J
respectively.

Greater sensitivity to stellar companions was achieved with Keck AO imaging,
in $K'$ bandpass, on 2013 June 14/15 yielding a PSF FWHM of 0{\farcs}05, 
and again on 2013 July 5.   The image appears single in the
Keck AO image, except for the well known stellar companion located
5{\farcs}14 from the primary star.  This stellar neighbor cannot be the
cause of the dimmings because the astrometric exclusion distances are
only 0{\farcs}14 and 0{\farcs}09  for the two planets (see Table~\ref{tab:fpp_tbl}).  Figure
\ref{fig:koi153_fig1} at top left shows a seeing limited image of the
field of view of the keck-HIRES guide camera.  Figure
\ref{fig:koi153_fig1} (middle) shows the limiting magnitudes achieved
with each high-resolution imaging method, achieving sensitivity of
(and ruling out companions)
$\sim$7.5 magnitudes fainter than the primary star in near-IR bandpasses.

We obtained precise RVs spanning a time baseline of 832 days (Figure
\ref{fig:koi153_fig2}, top).  The RVs of Kepler-113b (P = 4.8 d)
phase with the transit-based ephemeris at the 2-$\sigma$ confidence
level, yielding a planet mass of 11.7 $\pm$ 4.2 \mearthe, and a density
of 10.7 $\pm$ 3.9 \gcc.  The mass and density are consistent with a
planet that is composed mostly of rock.  The 8.9 day transiting
planet is not detected in the RVs and we provide an upper limit to its
mass of 9 \mearthe.

We note that the upper limit of the mass for the 8.9 d planet is
difficult to interpret because the peak in the posterior distribution
is negative, at -4.0 \mearthe.  This upper limit is calculated by identifying
the 95th percentile of the posterior mass distribution.  The 95th percentile upper limit to
mass is thus 9 \mearth and unlikely to be much more massive.

We computed a periodogram of the RV residuals to the best fit.  That
periodogram exhibits peaks at  periods of 1.065 d (strongest), 16 d, 16.4, 0.984
d, and 0.515 d.  These periods are aliases of each other.  The
strongest peak is at 1.065 d and has a false alarm probability near
1\%.  This false alarm probability is difficult to determine
accurately because power is so clearly distributed among the aliases.
It appears likely that there is a non-transiting planet with a period
likely equal to one of the aliases quoted above.  The RMS of the 
RV residuals drops from 8.0 \ms to $\sim$4.9 \ms when adopting 
the 1.065 d period and a circular orbit as a ``real''
non-transiting planet.  The RMS decreases less for the other aliases,
making them less likely but not ruled out.  This reduction in the RV residuals
with only three additional parameters (for the circular orbit) indicates that one of
the periods is likely real.  Further, the lowest value of the reduced $\chi^2$
statistic in our fit of the RVs with a three-planet model occurred
with the non-transiting planet having $P$=1.0651 d, adding support to
that period.  A non-transiting planet with $\sim$1 d period would necessarily be
highly misaligned relative to the two transiting planets in the system. 
Among these four aliases we cannot be sure which
is real, if any, and which are the aliases.  Thus, we choose not to
report any of these periodicities as a definitive ``planet candidate''.
More RV measurements will be needed to
assess these periods.  The full set of RVs are listed, along with
their \rphk~activity values in Table~\ref{tab:rvs_k00153}.

\subsection{KOI-244, Kepler-25} 
\ek identified two transiting planets with orbital periods of 12.7 and
6.2 d, with planet radii of 5.2 and 2.7 \rearthe.  The RVs reveal a
non-transiting planet with an orbital period of 123 $\pm$ 2 d and
minimum mass of \msini = 90 $\pm$ 14 \mearthe, described below.  TTVs
for the system were published in \cite{Steffen2012_keplerIII},
validating the existence of both planets and providing names,
 Kepler-25b (P = 6.2 d) and Kepler-25c (P=12.7 d).  Note that the names
``b'' and ``c'' are placed in order of distance from star, unlike the
order of discovery.

Follow-up observations were begun with the acquisition of two recon
spectra at the McDonald Observatory 2.7m and with the Smithsonian
Astrophysical Observatory's Fred L. Whipple Observatory on Mt Hopkins
in Arizona, using the Tillinghast Reflector Echelle Spectrograph
(TRES), all in 2010 March. The stellar parameters determined from
these spectra agree, within errors, with the KIC values. This is a
non-typical RV target in the sense that its rotational Doppler
broadening is high, \vsini~=10 \kmse, and its surface temperature is
high, \teff~= 6270 K, implying few and washed-out absorption lines.
The star is rotating faster and has a higher temperature than we
normally choose.  But its brightness ($K p$ = 10.7) makes this a
suitable KOI to observe with high spectral resolution.  Also, its fast
rotation promotes the chance of measuring a Rossiter-McLaughlin
signal. The final stellar parameters, which agreed with the KIC
values, were obtained using SME analysis performed on a Keck-HIRES
spectrum. SME results were combined with asteroseismology analysis,
and the detection of solar like oscillations were used to determine
the stellar properties \citep{Huber2013}. The final stellar parameters
are \teff~= 6270 $\pm$ 79 K, \logg~= 4.28 $\pm$ 0.03, and 
\feh~= -0.04 $\pm$ 0.10, which are listed in Table ~\ref{tab:stellar_pars_tbl}.

Follow-up speckle and AO imaging of Kepler-25 conducted at the WIYN telescope and
at the Mt. Palomar 5-m found no companions, within limits, between 1$''$ to
6$''$ from the target.  The detection thresholds from Palomar are
extraordinary, 7.5/8 mag (J/K) at a separation of 1$''$ and $>$~9
mag for 2$''$ and beyond at J and K bands.  Keck AO imaging on 2013 June 13/14 and
14/15 was characterized by a PSF FWHM of 0{\farcs}05, offering tight
limits on any companions (none found) inward of 0{\farcs}5.  There was
early discussion on the CFOP about a possible false positive seen as
centroid motion of the primary star in and out of transit.  But
further analysis ruled out any such displacements to within 2{\farcs}0,
corresponding to half a \ek pixel. The confusion was due to this star
saturating the \ek CCD light detector. Figure \ref{fig:koi244_fig1}
(top) shows a seeing limited image of the field of view of the HIRES
guide camera, with Kepler-25 being the brighter of the two bright stars
in that image.  Figure~\ref{fig:koi244_fig1} (middle) shows the
limiting magnitudes achieved with each high-resolution imaging method, 
with a full detectability curve given in Figure
\ref{fig:koi244_fig1}.  No additional companions were seen.

Gravitational interactions between the two inner planets has been
revealed by TTVs as described by \cite{Steffen2012_keplerIII} and
\cite{Lithwick2012, Wu2013}.  \cite{Steffen2012_keplerIII} performs a dynamical
analysis of the planet motions based on the observed times of
transits, serving as a tool for false positive assessment. The false
positive probability from the dynamical analysis is $10^{-3}$.  Here,
employing the extraordinary high-resolution speckle and AO imaging, we
find a FPP of 0.0001 (see Section~\ref{sec:fpp1}), lower than from the
TTVs alone.  Thus, we support the reality of Kepler-25b.  The masses
are also constrained by TTVs to a similar level of accuracy
\citep{Lithwick2012, Wu2013}.

Extensive precise RV follow-up was carried out at Keck-HIRES from 2009
to 2012 (see Figure \ref{fig:koi244_fig2}, top), including two
separate measurements of the Rossiter-McLaughlin (RM) effect, in which
RVs were collected continuously while the planet was transiting the
host star. The RM results are summarized in \cite{Albrecht2013}
showing the stellar spin axis to be well aligned with the orbital
axis. \cite{Albrecht2013} find lambda, the projected angle between the
orbital plane of the transiting planet and the stellar equatorial
plane to be 2 $\pm$ 5 deg. The RVs taken while the planet was
transiting were removed for the orbital analysis presented here.

The two transiting planets in the system were confirmed with
TTVs, providing upper limits on the planet masses \citep{Lithwick2012, Wu2013}.  A
self-consistent two-planet model shows the RVs to phase well with the
transit ephemerides of both planets.  For Kepler-25c ($P$ = 12.7 d),
we find a mass = 24.6 $\pm$ 5.7 \mearthe, with a corresponding density
of 0.90 $\pm$ 0.21\gcc.  For Kepler-25b ($P$=6.2 d) we find a planet
mass of 9.6 $\pm$ 4.2 \mearth and a density of 2.5 $\pm$ 1.1 \gcc.  The RVs
reveal an additional non-transiting planet with of period 123 $\pm$ 2
d, with a lower limit to its mass of \msini~= 89.9 $\pm$ 13.7
\mearthe. The orbit of the non-transiting planet is slightly eccentric
($e$ = 0.18).  The densities of the two transiting planets are both so
low that they must consist of some light material, presumably H and He
\citep{Batygin2013}.

Spectra of Kepler-25 were also obtained with SOPHIE, the spectrograph
dedicated to high-precision RV measurements at the 1.93-m
telescope of the Haute-Provence Observatory, France
\citep{Bouchy2009}.  SOPHIE is a cross-dispersed, environmentally
stabilized echelle spectrograph fed by a set of two optical fibers and
calibrated in wavelength with thorium-argon lamps. Observations were
secured in \textit{high-resolution} mode (resolving power
$\lambda/\Delta\lambda=75000$). Twelve spectra were secured in August
- November 2011, with exposure times between 30 and 60 minutes
allowing signal-to-noise ratios per pixel at 550~nm between 30 and 70
to be reached.  The spectra were extracted using the SOPHIE pipeline,
and the radial velocities were measured from the weighted
cross-correlation with a numerical mask characteristic of a G2 star
\citep{Baranne1996}, together with the bisector of the
cross-correlation function.  All the spectral orders except the first
17 were used in the cross-correlation to reduce the dispersion of the
measurements.  The blue part of the spectra are particularly noisy and
using them degrades the accuracy of the radial-velocity measurement.
The error bars on the radial velocities were computed from the
cross-correlation using the method presented by \cite{Boisse2010};
they are typically of the order of $\pm$15 m s$^{-1}$.  Three spectra were
contaminated by moonlight.  Following the method described in
\cite{Hebrard2008} we estimated and corrected for the moonlight
contamination by using the second SOPHIE fiber aperture, which is
targeted on the sky while the first aperture points toward the
target. This typically results in RV corrections of a few
tens of \mse. The final SOPHIE measurements are reported in Table
~\ref{tab:rvs_k00244}.

The self-consistent three-planet model included RVs obtained with
SOPHIE, together with the HIRES RVs and \ek photometry, to provide the
final planet parameters for Kepler-25b, c, and d,
given in Table~\ref{tab:orbital_pars_tbl}.  All parameters for the two transiting planets agree at better than
$1\sigma$ with those obtained above without SOPHIE. This is not
surprising as the SOPHIE RV measurements here are about
four times less accurate than the HIRES ones. SOPHIE data alone do not
allow the two transiting planets to be significantly detected.  For
the outer, non-transiting planet Kepler-25d, the SOPHIE RVs alone favor an orbital
period of $93\pm2$~days. That value is an alias with a sampling of one
year of the 123~day period reported above for Kepler-25d. Additional
observations on a longer time span will allow the correct orbital
period to be established. This will be the object of a forthcoming
paper. Whether the period for Kepler-25d is 93 or 123~days has
no significant effect on the derived semi-amplitudes or masses measured for each
of the three planet masses for Kepler-25.

Finally, we studied the variations of the line bisectors obtained with
HIRES and SOPHIE. We found neither variations nor trends as a function
of RV.  The RV residuals considering only
one or two of the detected planets do not show either any correlations
with the bisectors. This reinforces the conclusion that each of the
three radial-velocity variations are due to planetary signals, and not
caused by spectral-line profile changes attributable to blends or
stellar activity.  The full set of RVs, minus the RVs used to measure
the Rossiter-McLaughlin effect, are listed, along with their \rphk~
activity values in Table ~\ref{tab:rvs_k00244}.

\subsection{KOI-245, Kepler-37}  

\ek identified three transiting planets having orbital periods of
13.36, 21.30, and 39.79 d with corresponding planet
radii of 0.32, 0.75 and 1.94 \rearthe, respectively. A fourth
candidate transiting planet
was reported in \cite{Batalha2013}, but it has since been deemed a 
false positive and it is not included in this study. A detailed
analysis of the \ek light curve, blend scenarios, and asteroseismic
analysis of this exceptionally bright ($K p$ = 9.7) KOI with a
sub-Mercury sized planet can be found in \cite{Barclay2013}. The
existences of all three planets were validated, yielding new names,
Kepler-37 b,c,d.  Here, we summarize the follow-up observations, and 
we place limits on the planet masses from the precise RV observations.

Recon spectra from the McDonald 2.7m and the Tillinghast 1.5m were
taken on 2010 March and 2010 April, respectively. These spectra
confirmed the stellar parameters from the KIC, and showed that \vsini~
is less than 2.0 \kmse. SME analysis of a Keck-HIRES spectrum also
agreed with the KIC parameters, and the final stellar parameters are
listed in Table ~\ref{tab:stellar_pars_tbl}, which were determined via
asteroseismology, with SME-based \teff, \logg, and \feh~used as inputs.
The final stellar parameters are \teff~= 5417 $\pm$ 75 K, \logg~= 4.57 $\pm$
0.05, and \feh~= -0.32 $\pm$ 0.07, and are consistent with those in \cite{Barclay2013}.
This KOI is the most dense star yet to reveal asteroseismic
oscillations, made possible by its brightness and lengthy coverage
in short cadence observations.

We present a seeing limited image of the field of view in Figure
\ref{fig:koi245_fig1} (top left).  There is a stellar neighbor located
7$''$ south of the primary star and about 4 magnitudes fainter in V
band.  Speckle imaging was acquired at the WIYN telescope and also at
the Gemini Telescope, showing Kepler-37 to be a single star within 2{\farcs}7
square box, with no companion detected to a threshold of 6 mag in R band
and 5.1 mag in V band. No neighboring star was seen.  Near infrared AO taken with 
ARIES also revealed no stellar companions. Extensive AO observations were
also taken with the Mt. Palomar 5m telescope using a Brackett-gamma
filter to avoid saturation of the IR detector.  No stellar companion
was found. A more thorough analysis of these AO observations is found
in \cite{Barclay2013} where the probability of background stars
falling, undetected, into the \ek aperture, is discussed.  Still we
obtained Keck AO images on 2013 June 14/15 to achieve
even better imaging of the region 0{\farcs}1 - 0{\farcs}5 from the
star.  The detectability curves from Keck AO and from all imaging are
shown in Figure \ref{fig:koi123_fig1}, showing detectability at
near-IR wavelengths 8 magnitudes fainter for all angular separations
beyond 0{\farcs}3.  No stellar companion was seen.
 
 The full set of RVs are listed, along with their \rphk~ activity
 values in Table ~\ref{tab:rvs_k00245}.  The precise RVs from
 Keck-HIRES are unable to determine the masses of the three transiting
 planets because their expected RV amplitudes are all below 1 \mse.
 We constrain the mass of the 40, 21, and 13 day planets to be less
 than 12.2, 12.0, and 10.0 \mearthe, respectively, based on the 95th
 percentile positive extent of the posterior distribution of planet
 mass.  The mass for the 40 d planet corresponds to a density upper
 limit of 8.7 \gcc, not useful to distinguish pure rocky from mostly
 volatile-rich compositions.  The 21 and 13 day planets similarly do not
 offer useful constraints on their densities. The 862 day baseline for
 the RV measurements finds no significant periods or trends above the
 noise.  Thus no non-transiting planets are detected.
 
\subsection{KOI-246, Kepler-68}   

\ek identified two transiting planets with orbital periods of 5.40 and
9.60 d and planet radii of 2.3 and 0.95 \rearthe, as described and
validated in \cite{Gilliland2013}, giving them names, Kepler-68 b and
c.  Our RVs indicate a non-transiting planet with an orbital period
of 625 $\pm$ 16 days and mass of 267 $\pm$ 16 \mearthe. Here, the planet
properties and orbital parameters are refined over those provided in
\cite{Gilliland2013}, with five additional RVs obtained in 2013.  For
a detailed summary of the light curve analysis, asteroseismic
analysis, RV analysis, and discussion of false positive scenarios with
{\em BLENDER}, see \cite{Gilliland2013}.  Recon spectra were taken at
the McDonald 2.7m on 2010 March 25 and 2010 March 28th. A spectrum was
acquired at Tillinghast 1.5m on 2010 March 25th. The near solar values
of temperature, \logg~ and \vsini~listed in the KIC were confirmed by
these recon spectra. SME combined with asteroseismology analysis was
then used to confirm the near solar values of stellar mass and stellar
radius.  The final stellar parameters are \teff~= 5793 $\pm$ 74 K,
\logg~= 4.28 $\pm$ 0.02, and \feh~= 0.12 $\pm$ 0.07, listed in Table
~\ref{tab:stellar_pars_tbl}.

Speckle imaging at WIYN taken in 2010 June and AO imaging
taken with ARIES, in summer 2010 found no companion stars that
could cause confusion or pollution in the light curve analysis. Figure
\ref{fig:koi246_fig1} (top left) shows a seeing limited image of the
field of view of the HIRES guide camera.  Figure
\ref{fig:koi246_fig1} (middle) shows the detection threshold
delta-magnitudes achieved with each high-resolution imaging method.
While Gilliland had already effectively ruled out false positives, we
obtained Keck AO images on 2013 June 14/15 yielding a FWHM of 0{\farcs}05, 
with a full detectability curve given in Figure
\ref{fig:koi246_fig1}, supporting and improving the lack of stellar
companions to within 0{\farcs}05 of the star.

 The precise RVs collected from 2009 to 2013 (see 
 Figure~\ref{fig:koi246_fig2}, top) determine the mass and density of the
 2.3 \rearthe, innermost transiting planet to be 5.97 \mearth and 2.60
 $\pm$ 0.74 \gcc, respectively.  The RVs also provide upper limits to
 the mass of the second transiting planet.  All current planet
 parameters for Kepler-68 are listed in
 Table~\ref{tab:orbital_pars_tbl}), and they differ from those in
 Gilliland et al. by typically $\sim$1-sigma.

The RVs clearly reveal a non-transiting planet with an orbital period
of 625 $\pm$ 16 d, with a minimum mass (\msini) of 0.84 $\pm$ 0.06
\mjupe. During the initial publication of the Kepler-68 results, the
period of the non-transiting planet had a high eccentricity alias
solution (Ian Crossfield, personal communication). The RVs 
obtained in 2013 show the true period to be that
quoted both here and in \cite{Gilliland2013}, not the alias.
The full set of RVs are listed, along with
their \rphk~ activity values in Table ~\ref{tab:rvs_k00246}.

The innermost planet with its density of 2.6 \gcc~ occupies an
important niche in exoplanet science.  This density is intermediate
between that of purely rocky planets and the mostly gaseous ones.
\cite{Lopez2012, Lopez2013} discusses the implications of such hybrid planets.
  
\subsection{KOI-261, Kepler-96}  

The \ek photometry revealed a single transiting planet in this system
with a period of 16.2 d and a radius of 2.67 \rearthe.  Three recon
spectra were taken at the McDonald 2.7m and Tillinghast Observatories
in 2010 March and 2010 April. Stellar parameters determined from these
spectra with SPC agreed in \teff, but the \logg~value was off by 0.4
dex from the KIC value. When the new \logg~was confirmed with SME
analysis of a Keck-HIRES template spectrum, Yonsei-Yale stellar models
were used to adjust the stellar radius from its KIC value of 1.9 to
1.02 $\pm$ 0.09 \rsune. The planet radius decreased from 6.2 to 2.7
\rearthe. Further evidence that the stellar radius is near solar, and
not near 1.9 \rsune, is the non-detection of stellar oscillations in
the asteroseismic analysis. A typical 1.9 \rsun star would have a
detectable asteroseismic signal, which was searched for and not
detected. The final stellar parameters are \teff~= 5690 $\pm$ 73 K,
\logg~= 4.42 $\pm$ 0.08, and \feh= 0.04 $\pm$ 0.07 and other stellar
values are listed in Table~\ref{tab:stellar_pars_tbl}.

High-resolution imaging with the WIYN speckle camera in 2010 September
detected no stellar companions, but MMT-ARIES observations in summer
2010 detected one nearby companion located 5{\farcs}4 northeast of the
primary star at PA=65.2 deg, and delta magnitude of 7.1 in J-band, and
6.8 in Ks band. The estimated \ek magnitude of the companion is 18.1.
Figure~\ref{fig:koi261_fig1} (at top left) shows a seeing limited
image of the field of view of the HIRES guide camera. Figure
\ref{fig:koi261_fig1} (middle) shows the detection threshold magnitudes achieved
with each high-resolution imaging method.  A detailed analysis of the
rotation period (from the \ek photometry) and the rotational Doppler
broadening, \vsini, indicates the star may be oriented nearly pole-on,
suggesting a spin-orbit misalignment \citep{Hirano2012}.  Tighter
limits on stellar companions were placed with Keck AO imaging on 2013
June 13/14, yielding a resolution of 0{\farcs}05 (FWHM), and all of the
detectability curves from all imaging attempts are given in Figure~\ref{fig:koi261_fig1}.  No
additional companions were seen besides that located 5{\farcs}4 away, noted above.

Precise RV follow-up was initiated in 2010 July and 25 RVs have been
acquired for a time baseline of 772 days (Figure
~\ref{fig:koi261_fig2}, top). The RVs show a weak correlation with the
\ek transit ephemeris. We constrain the RV $K$-amplitude to be 2.1
$\pm$ 0.8 \ms.  This translates to a planet mass of 8.46 $\pm$ 3.4
\mearthe.  The corresponding density is 2.26 $\pm$ 1.11 \gcc, which is
significant in that we constrain the density to be less than 3.4 \gcc
at the one-sigma level and 4.5 \gcc at the two-sigma level.  Such a
low density rules out a purely rocky composition, and requires some
contribution from low density materials such as water, H or He.  The
RVs show no sign of any non-transiting planet after 3 years at levels
of $\sim$ 3 \ms. The full set of RVs are listed, along with their
\rphk~ activity values in Table ~\ref{tab:rvs_k00261}.

As in the case of Kepler-25, Kepler-96 was observed with SOPHIE.  Six
60-min observations were secured in July - September 2011, allowing
signal-to-noise ratios of the order of 70 and RV accuracy
of $\pm$25 \ms reached on each exposure.  The SOPHIE RVs do not
allow the upper limit on the reflex motion due to Kepler-96b to be
significantly improved. 

\subsection{KOI-283, Kepler-131} 

\ek identified two transiting planets with orbital periods of 16.0 and
25.5 d and radii 2.4 and 0.8 \rearthe. The two recon spectra taken with the
McDonald 2.7m and the TRES spectrometer gave good agreement in \teff~but
the value of \logg~found with spectroscopy is 0.5 dex larger than the
KIC value, resulting in a decrease in stellar radius of 40\%. An SME
analysis of a Keck-HIRES spectrum combined with Yonsei-Yale stellar
models gave \teff~= 5685 $\pm$ 74 K, \logg~= 4.42 $\pm$ 0.08 and
\feh~= 0.12 $\pm$ 0.07, listed in Table~\ref{tab:stellar_pars_tbl}.

Speckle imaging from the WIYN found no stellar companions and the lone
stellar companion detected with AO imaging from
Mt. Palomar is 6$''$ away and 8 magnitudes fainter in J-band and
Ks-band. Such a stellar neighbor could not be responsible for the
dimming, as it would be easily detected with astrometric centroid
motion analysis in and out of transit in the \ek images.  Keck AO
imaging took place on 2013 June 13/14, yielding a FWHM of 0{\farcs}05,
with a full detectability curve given in Figure
\ref{fig:koi123_fig1}, revealing no additional companions.  Figure
\ref{fig:koi283_fig1} (top left) shows a seeing limited image of the
field of view of the HIRES guide camera.  Figure
\ref{fig:koi283_fig1} (middle) shows the limiting magnitudes achieved
with each high-resolution imaging method.

The RV measurements for Kepler-131 span 741 days (Figure
\ref{fig:koi283_fig2}, top) and are listed, along with their \rphk~
activity values in Table~\ref{tab:rvs_k00283}.  The results consist
of an RV detection of the 16 d planet giving a mass of 16.1 $\pm$ 3.5
\mearthe, and a density of 6.0 $\pm$ 2.0 \gcc.  We find a similar mass
and density whether we fit this planet by itself or with the second
planet simultaneously. The second planet shows up in the RVs only
marginally.  A self-consistent fit of RVs and photometry yields a peak
of the posterior mass distribution at $M$ = 8.25 $\pm$ 5.9 \mearth and
density of 78 $\pm$ 55 \gcc.  The density is unphysically large, and
hence the mass is too large, indeed a detection at less than
2$\sigma$.  We find an upper limit to the mass of Kepler-131c from the
95th percentile of the posterior mass distribution, yielding $M<$ 20.0
\mearth.  Certainly the mass of Kepler-131c remains highly uncertain.
Table~\ref{tab:orbital_pars_tbl} lists all of the best-fit planet
parameters.

\subsection{KOI-292, Kepler-97} 

The single transiting planet identified by \ek in this system has an
orbital period of 2.587 d and a radius of 1.5 \rearthe. Follow-up
observations began with recon spectra in 2010 March and 2010 April at
the McDonald 2.7m telescope. SPC analysis found this star to be a
slowly rotating main sequence star, an ideal target for Keck-HIRES
spectroscopy, the analysis of which using Yonsei-Yale stellar models
gave a final stellar radius 30\% smaller than that reported in the
KIC.  The final stellar parameters are \teff~= 5779 $\pm$ 74 K,
 \logg~= 4.43 $\pm$ 0.08, and \feh~= -0.20 $\pm$ 0.07, sub-solar
metallicity, with all stellar parameters in
Table~\ref{tab:stellar_pars_tbl}

Adaptive optics imaging at both Palomar and Keck-NIRC2 shows a
companion star with an angular separation of 0.37 $\pm$ 0{\farcs}01 from
the primary at a PA = 121.8 deg. The companion was measured to be 2.7
mag fainter in the optical (from speckle observations), 2.8 mag
fainter than the primary in the J-band, and 3.2 mag fainter in
$K'$-band.  These magnitude differences have uncertainties of 0.15 mag.
The Keck AO images revealed no additional neighboring stars (besides
the one 0{\farcs}37 away), down to delta magnitude of 8 mag ($K'$ band),
shown in detail in Figure~\ref{fig:koi292_fig1}.  Speckle imaging
revealed no companions down to a delta magnitude of 3 and 4 in the
R-band and V-band, respectively. The non-detection with speckle is
likely due to the companion being brighter in the infrared than in the
visible, relative to the primary. The companion is listed in
\cite{Adams2012}, and makes this a less than ideal target for precise
RVs due to the contamination of the companion in the spectrum of the
primary.  Figure~\ref{fig:koi292_fig1} (top left) shows a seeing
limited image of the field of view of the HIRES guide camera.  Figure
\ref{fig:koi292_fig1} (middle) shows the limiting magnitudes achieved
with each high-resolution imaging method.

The stellar companion located 0{\farcs}37  from the primary star
resides farther than the maximum exclusion radius determined from the
astrometric centroid measurements in and out of transit.  Thus, it is
unlikely that the neighboring star is responsible for the dimmings. If
it were, the centroid of light would appear to shift more than the
upper limit exclusion radius (see Table~\ref{tab:fpp_tbl}).  But it remains difficult
to calculate the possibility that the centroid measurements are not robust
enough to rule out this stellar companion.  We remain concerned that
while formally this stellar companion is ruled out as the cause of the
apparent transits, this assessment should be revisited.

The full set of RVs are listed, along with their \rphk~ activity
values in Table ~\ref{tab:rvs_k00283}.  The RVs exhibit a weak
downward trend of 5 \ms per year during three seasons (2 years).
Figure~\ref{fig:koi292_fig2} shows no deviations from a monotonic
downward trend, but more RVs are needed to confirm this trend. The
companion causing the linear trend likely has a an orbital period
longer than the twice the time baseline, $P>$ 4 yr and a mass greater
than 1 \mjupe.  The RV trend may be caused by the stellar companion
0{\farcs}37 away found with the AO imaging, but we have not pursued
this question in detail. 

Fitting for the linear trend, the RVs marginally correlate with the
ephemeris of the transiting planet.  The best-fit planet mass is
$M$~= 3.51 $\pm$1.9 \mearthe, corresponding to a density of 5.44 $\pm$
3.48 \gcc.  This is less than a 2$\sigma$ detection of mass and
density, meriting more RV measurements.  We adopt an upper limit on
the mass of the transiting planet to be 9.1 \mearth from the 95th
percentile of the posterior mass distribution.  This upper limit
corresponds to an upper limit for the planet density of 14 \gcc, which
seems unphysically high.  Table~\ref{tab:orbital_pars_tbl}) gives all of
the resulting planet parameters.  The phase folded RV curves are shown
in Figure~\ref{fig:koi292_fig2} (bottom right).

\subsection{KOI-299, Kepler-98} 

\ek identified a transiting planet around Kepler-98 with an orbital
period of 1.5 d and a radius of 2.0 \rearthe. Before precise RVs were
acquired, recon spectra were taken with the McDonald 2.7m in 2010
March and the Tillinghast 1.5m (TRES spectrometer) in 2010 June. The
stellar surface temperatures found by SPC from the recon spectra were
in fair agreement with KIC values.  We acquired a spectrum with
Keck-HIRES, giving final stellar parameters of \teff~= 5589 $\pm$
73 K, \logg~= 4.34 $\pm$ 0.10 and \feh~= 0.18 $\pm$ 0.07, found by
SME analysis and matching of spectral parameters to Yonsei-Yale
stellar models (see Table~\ref{tab:stellar_pars_tbl}).

Speckle imaging at the WIYN telescope conducted in 2010 September
detected no companions within its limits.  Keck AO imaging took place
on 2013 June 14, yielding a FWHM of 0{\farcs}05, with a full
detectability curve given in Figure~\ref{fig:koi299_fig1}.  No
stellar companions were detected at limits of 8 magnitues fainter 
(in K band) than the primary star.  The Keck guider image (Figure
\ref{fig:koi299_fig1}, top left) shows no companions from 1$"$ to
6$''$ from the primary, down to a delta magnitude of seven.

The full set of RVs are listed, along with their \rphk~ activity
values in Table ~\ref{tab:rvs_k00299}.  The RVs cover a time baseline
of 2 years and show only a marginal detection of the transiting planet
(Figure~\ref{fig:koi299_fig2}).  We find its mass to be 3.55 $\pm$ 1.6
\mearth with a density of 2.2 $\pm$ 1.2 \gcc. The 95th percentile of
the posterior mass distribution provides a useful upper limit to the
planet mass of 6.4 \mearth and a corresponding upper limit to density
of 3.9 \gcc.  This upper limit to the density of the planet indicates
a likely composition consistent with significant contributions of low
density material, presumably H, He, and/or water.
Table~\ref{tab:orbital_pars_tbl} gives the full set of planet
parameters.

\subsection{KOI-305, Kepler-99}  

The single transiting planet found by \ek in this system has an
orbital period of 4.6 d and a radius of 1.5 \rearthe.  Recon
spectroscopy of the target was initiated in 2010 March at the McDonald
2.7m. In 2010 May, a second recon spectrum was taken at the
Tillinghast 1.5m. Both spectra were used to determine \teff~and
\logg. The stellar parameters determined from each spectrum were
self-consistent.  These were in agreement with the KIC values in \teff,
but only in marginal agreement with the KIC in \logg. No large (over 1
\kms) RV variation was seen and the low rotational velocity of the
star led to collection of precise RVs. Once a Keck-HIRES template was
taken, SME was used with Yonsei-Yale stellar models to determine the
final stellar parameters of \teff~= 4782 $\pm$ 129 K, \logg~=
4.61$\pm$ 0.05, and \feh~= 0.18 $\pm$ 0.07.  The final stellar radius
is 30\% smaller than the KIC values (see
Table~\ref{tab:stellar_pars_tbl})

Speckle imaging taken at the WIYN telescope found no companions within
its limits and no further imaging is available.  Keck AO imaging took
place on 2013 June 15, yielding a FWHM of 0{\farcs}05 , with a full
detectability curve given in Figure~\ref{fig:koi305_fig1}, revealing
no companions.  When observing with Keck-HIRES, a guider image was
taken. No companions were detected from 2{\farcs}0 out to 4{\farcs}0 within
seven magnitudes of the primary star in the R-band. Figure
\ref{fig:koi305_fig1} (top left) shows a seeing limited image of the
field of view of the HIRES guide camera.

The full set of RVs are listed, along with their \rphk~ activity
values in Table ~\ref{tab:rvs_k00305}.  The RV baseline of 791 days
(Figure~\ref{fig:koi305_fig1}, top) shows no linear trends or
periodicities aside from the transiting planet. This single,
1.5 \rearth planet has a mass determined from RVs of 6.15 $\pm$
1.3 \mearthe.  The planet density is 10.9 $\pm$ 2.8 \gcc, suggesting the
planet has a rocky interior, similar in density to Kepler-10b and
CoRoT-7b (see Kepler-406 below).  The phase folded RV curve is shown in
Figure~\ref{fig:koi305_fig2} (bottom right).

\subsection{KOI-321, Kepler-406}  

The \ek mission identified two transiting planets around Kepler-406,
having orbital periods of 2.43 and 4.62 days with radii of 1.43 and
0.85 \rearthe, respectively.  Recon spectroscopy was taken at the
Tillinghast 1.5m (TRES), where \teff~was confirmed to be near the KIC
value. The slow rotational broadening of the stellar lines was seen
(\vsini$<$ 3 \kmse), and any RV variation was clearly below 500 \mse.
The \logg~ value found by recon, and later refined with SME and
asteroseismology analysis, is larger than that in the KIC, leading to
an increase in the stellar and planet radii of 30\%. The final stellar
parameters, although not published in \cite{Huber2013}, were found
using similar methods. The SME analysis of the Keck-HIRES spectrum
served as the starting point of the asteroseismology analysis that led
to \teff~= 5538 $\pm$ 44 K, \logg~= 4.41 $\pm$ 0.02, and \feh~= 0.18 $\pm$
0.04.  Table~\ref{tab:stellar_pars_tbl} contains the full set of
stellar parameters including mass and radius.

Imaging including Speckle imaging with WIYN in 2010 September and
AO imaging taken with the Lick 3m in 2011 September,
found no companions nearby that are contaminating the light curve.
Keck AO imaging took place on 2013 June 13, yielding a FWHM of 0{\farcs}05, 
with a full detectability curve given in Figure
\ref{fig:koi321_fig1}, also revealing no neighboring stars.  Figure
\ref{fig:koi321_fig1} (top left) shows a seeing limited image of the
field of view of the HIRES guide camera.
 
The full set of RVs are listed, along with their \rphk~ activity
values in Table ~\ref{tab:rvs_k00321}.  Precise RVs taken over 800
days, beginning in 2010 July, are shown in Figure
\ref{fig:koi321_fig2}, top.  The inner transiting planet appears
clearly in the RVs phased to the ephemeris from the transits.  We find
this 1.43 \rearth planet to have a mass of 6.35 $\pm$ 1.4 \mearth with
a corresponding density of 11.8 $\pm$ 2.7 \gcc. Such high density,
measured at the 4$\sigma$ confidence level, implies that this planet
is essentially made of rock and iron/nickel, perhaps compressed. 
{\em This planet represents one of the most convincing cases of a rocky planet
around a main sequence star, joining CoRoT-7b, Kepler-10, and Kepler-36b.}

The outer transiting planet shows only a marginal signal in the RVs,
yielding a mass of 2.71 $\pm$ 1.8 \mearth and a density of 24.4 $\pm$ 16
\gcc.  This density is unphysically high, indicating that the RVs
provide only a upper limit.  We adopt the 95th percentile of the
posterior mass distribution, giving an upper limit to its mass of 6.0
\mearthe.

\subsection{KOI-1442, Kepler-407} 

With a period of 0.67 days and a radius of 1.1 \rearthe,
Kepler-407b has a similar orbital period to the planet
Kepler-10b. However, the RVs provide only upper limits to the
transiting planet's mass, and a partial orbit of a non-transiting
planet. Follow-up observing began in 2011 March when recon spectra
were collected at the McDonald 2.7m. SPC analysis of the spectrum
confirmed the \teff~ and \logg~ from the KIC. After acquisition of the
Keck-HIRES template spectrum, SME analysis was run, and the final
stellar parameters were calculated by combining the SME result with
Yonsei-Yale isochrones. The final stellar parameters are \teff~= 5476
$\pm$ 75 K, \logg~= 4.43 $\pm$ 0.06 and \feh~= 0.04 $\pm$ 0.07.

Speckle observing at WIYN in 2010 September did not identify any
companion stars within the limits. There is one companion identified
in UKIRT images that is outside of the field of view of speckle. The
companion star is 2{\farcs}1 to the NE of the primary and roughly 5.3
magnitudes fainter in the \ek Bandpass. Centroid analysis of the
pixel level data constrains the centroid of the planet transit to be
within 0{\farcs}44 of the primary, ruling out the possibility that the
transits are actually occurring on a nearby star. Figure
\ref{fig:koi1442_fig1} (top left) shows a seeing limited image of the field of
view of the HIRES guide camera.  Figure ~\ref{fig:koi1442_fig1} (middle) shows
the limiting magnitudes achieved with speckle imaging.
Keck AO imaging took place on 2013 June 13/14, yielding a FWHM of 0{\farcs}05, 
with a full detectability curve given in Figure
\ref{fig:koi1442_fig1}.  We found no stellar companions, except for
the known companion 2{\farcs}1 away, thus greatly
limiting the false-positive probability, as discussed in Section \ref{sec:fpp1}.

 The most prominent signal in the RVs is a decrease of 300 \ms during
 the 2-year baseline, with clear curvature, as seen in Figure
\ref{fig:koi1442_fig2}, top. The slight curvature in the RVs is
 indicative of a massive body in a orbit of roughly 6-12 yr, with a
 minimum mass, \msini$\sim$5-10 \mjupe. With only roughly one quarter
 of this potential decade long orbit observed, the outer body remains
 poorly constrained in both orbital period and \msini.  Its mass could
 be above 10 \mjup and even above 80 \mjupe, for extremely face-on
 orbits.  Thus the non-transiting object is likely a planetary body of
 5-20 \mjupe, but could be classed as a brown dwarf or stellar,
 depending on the orbital inclination.  The RVs in the next few years
 will settle its \msini. The transiting planet and the non-transiting
 body are fit simultaneously. There are too few RV points to
 confidently measure the mass of the transiting planet and we place a
 one sigma upper limit on the mass of 1.7 \mearthe.  The phase folded
 RV curves are shown in Figure~\ref{fig:koi1442_fig2} (bottom right).
 We caution that the transiting planet could be orbiting that massive,
 non-transiting body, if luminous, rather than the primary star.  The
 full set of RVs are listed, along with their \rphk~activity values
 in Table~\ref{tab:rvs_k01442}.

\subsection{KOI-1612} 

The single planet in this system has a period of 2.5d and a radius of
0.82 \rearthe. This F-type star was first followed up from the ground
at the Tillinghast 1.5m where two spectra were taken. SPC determined
the stellar parameters to be consistent with the KIC and showed no
dramatic RV variation.  Before the asteroseismology analysis was
conducted in \cite{Huber2013}, \cite{Bruntt2012} identified this
bright star as having detectable asteroseismic oscillations.  We use
the SME-asteroseismology values in this work. The \teff~= 6104 $\pm$
74 K, \logg~= 4.29 $\pm$ 0.03, and \feh~= -0.20 $\pm$ 0.10.

 Speckle imaging taken at the WIYN in 2011 July found no companion
 stars within the limits. Figure
 \ref{fig:koi1612_fig1} (top left) shows a seeing limited image of the
 field of view of the HIRES guide camera.  Figure
 \ref{fig:koi1612_fig1} (middle) shows the limiting magnitudes
 achieved with speckle imaging.  Keck AO imaging took place on 2013
 June 13/14, yielding a FWHM of 0{\farcs}05, with a full detectability
 curve given in Figure~\ref{fig:koi1612_fig1}.  No stellar companions
 were found.  As described in Section \ref{sec:fpp1}, this stringent
 AO non-detection from Keck was able to limit the false-positive
 probability to 2.1\%, but not below 1\%.

 The first precise RVs were obtained shortly after the recon spectra
 were taken in 2011 May, resulting in an RV baseline of 477
 days (Figure~\ref{fig:koi1612_fig2}, top). The brightness of the star
 allowed us to obtain high signal to noise observations($SNR=200$),
 similar to those obtained for the standard CPS planet search
 stars. While we have not directly detected the planet, we place a one
 sigma upper limit on the mass of the planet equal to 4.3 \mearthe, but
 the limit on density is too uncertain to provide insight into
 composition. There are no RV signatures or periodicities
 that suggest the presence of a non-transiting planet in the system
 that would predict an RV amplitude greater than $~$ 4 \ms. A longer
 time baseline of observations will further constrain non-transiting
 planets. The posterior distribution of expected mass values is well
 behaved, and peaks near zero, as expected for RVs with well
 understood, photon limited uncertainties that provide only upper
 limits on the planet mass. The phase folded RV curve is shown in
 Figure~\ref{fig:koi1612_fig2} (bottom right).  The full set of RVs
 are plotted in Figure~\ref{fig:koi1612_fig2} (top) and listed, along
 with their \rphk~activity values, in Table~\ref{tab:rvs_k01612}.

\subsection{KOI-1925, Kepler-409} 

This KOI with a radius of 1.2 \rearth and period of 69 days was first
followed up in the 2012 observing season. Kepler-409 is one of
\ek's brightest KOIs (Kp = 9.44 mag) with a near earth sized planet. It
has well determined parameters from
asteroseismology\citep{Bruntt2012,Huber2013}. This analysis determined
that the final value of \logg~differed from the KIC value and
therefore the stellar radius was adjusted to only 40\% of the KIC
value.  Recon spectra from the Tillinghast 1.5m were taken and broadly
confirmed the analysis of asteroseismology. The final stellar
parameters are \teff~= 5460 $\pm$ 75 K, \logg~= 4.50 $\pm$ 0.03 and
\feh~= 0.08 $\pm$ 0.10. 

Careful Keck AO imaging took place on 2013 June 13/14, yielding a FWHM
of 0{\farcs}05, with a detectability curve given in Figure
\ref{fig:koi1925_fig1}.  We found no stellar companions, greatly
limiting the false-positive probability, as discussed in Section~
\ref{sec:fpp1}.

 Keck-HIRES RVs taken over 174 days in summer 2012 (Figure
 \ref{fig:koi1925_fig2}, top) all have high signal to noise ratios of
 200. This SNR is similar to the typical planet search stars, acquired
 in roughly ten minute exposures.  The planet's expected radial
 velocity amplitude, assuming an approximate mass-radius relationship
 tied to known low mass planets \citep{Weiss2013}, is below
 our detection threshold. Nonetheless, we limit the planet's mass to
 be less than 8.9 \mearth at the one sigma level. Such limits are
 insufficient to make conclusions about planet composition.  With 25
 RVs, showing an RMS of only 3.2 \ms we place upper limits on the mass
 of the transiting planet, and rule out any massive (\msini $>$ 30 \mearthe,
 expected K amplitude = 4.5 \ms~) planets in orbits interior to the
 transiting planet.

The full set of RVs are plotted in  Figure~\ref{fig:koi1925_fig2} (top)  
and listed, along with their \rphk~ activity values in  Table ~\ref{tab:rvs_k01925}.

\section{Discussion}           
\label{sec:discussion}

\subsection{Science Drivers and Selection Effects}

NASA's \ek mission has discovered over 4000 planet candidates to date,
offering unprecedented statistical information on the occurrence,
sizes, and orbits of planets, including over 400 multi-planet
systems\citep{Batalha2013, Fabrycky2012, Lissauer2012}.  Among the
remarkable results is that 30 - 50\% of Sun-like stars harbor planets
of size, 1-4 \rearthe, residing in orbits within 0.5 au
\citep{Batalha2013, Howard2012, Petigura2013, Fressin2013}.  The small
M dwarf stars also commonly harbor such small planets
\citep{Dressing2013}.

The nature of these 1-4 \rearth planets within 0.5 au remains poorly
understood, including their formation and internal composition.  They
may have formed as imagined for Neptune and Uranus, beyond the ice
line where they accreted gas and ices, notably H and He as well as
water.  Alternatively, 1-4 \rearth planets may be ``terrestrial''
planets that formed as did Earth, Venus, and Mars, but also acquired
varying, small amounts of H and He to puff their sizes above 1
\rearthe \citep{Batygin2013, Zeng_Seager08, Rogers2011, Valencia07}.

The nature of the 1-4 \rearth planets also bears on habitability, as
atmospheres of tiny mass yield modest pressures at the surface
allowing complex organic molecules to exist
\citep{Seager_Bains_Hu2013}.  But H and He ``atmospheres'' having over
1\% the mass of the planet will cause surface pressures above 1 Mbar
and temperatures above 1000 K, causing severe intermolecular Van der
Waals interactions that threaten the existence of complex organic molecules
such as RNA or DNA.

Here, we studied 42 small transiting planets and 7 non-transiting
planets that orbit 22 KOIs.  This KOI sample was selected from among
the candidate transiting planets identified by \ek
\citep{Batalha2013}, as described in Section~\ref{sec:vetting}. The
selection criteria stemmed from the goal of constraining the interior
structure of planets smaller than $\sim$4 \rearthe, with planets
smaller than 2.5 \rearth being higher priority.  Measuring
small-planet masses with RV measurements required orbital periods
short enough that the reflex motion of the star might be detectable by
precise Doppler measurements.  However, we had no prior knowledge of
the planet masses in advance, and thus there was no selection effect
in mass, for a given planet radius.  Planets having photometrically
determined radii could have had any mass prior to our RV measurements.

With extensive observations and modeling of false-positive channels,
we found a false positive probability under 1\% for all of the
transiting planets except two which have only a $\sim$2\% false
positive probability (see Section~\ref{sec:fpp1} and Table 3). The
likely existence of all 42 transiting planets allows them to be
employed in statistical studies, notably of their radii, masses,
densities, orbits, and host stars.

\subsection{Properties of the Host Stars}

The stellar properties were derived from high-resolution spectroscopy
and (for 11 KOIs) from asteroseismology, providing accurate values of
stellar mass, radius, metallicity, and age (see Table~\ref{tab:stellar_pars_tbl} and
Section~\ref{sec:starchar}).  The 22 host stars have effective
temperatures in the range, \teff~= 4700--6300 K, i.e. mid-K-type stars
to late F stars.  The surface gravities are in the range, \logg~=
4.1--4.6, i.e. on the main sequence.  The 22 metallicities are solar
within 0.3 dex, with comparable numbers of slightly metal-rich and
metal-poor stars, similar to the solar neighborhood.  The ages of the
stars (from stellar structure models) are distributed between 1--8.5
Gyr, with one (Kepler-25) nominally 11 Gyr old, consistent with the
distribution of ages in the solar neighborhood.  {\em Thus, the 22
  KOIs in the stellar sample presented here are apparently similar to
  the distribution of FGK stars in the solar neighborhood.}

\subsection{Properties of the Planets and Host Stars}

The planet properties were derived from modeling the transit light curve and the
RVs, yielding measurements of the planet radii, masses, and bulk
densities for the 42 transiting planets, listed in Table~\ref{tab:orbital_pars_tbl} and
described in Section~\ref{sec:planetchar}.  We found
16 transiting planets that have masses and densities measured at a
confidence level of 2$\sigma$ or better, useful for characterizing the
population of the planets statistically.  Another 14 planets emerged
having uncertainties in density less than 6.5 \gcc, offering
constraints on the light element content.

\begin{figure}  			
\epsscale{1.2}
\plotone{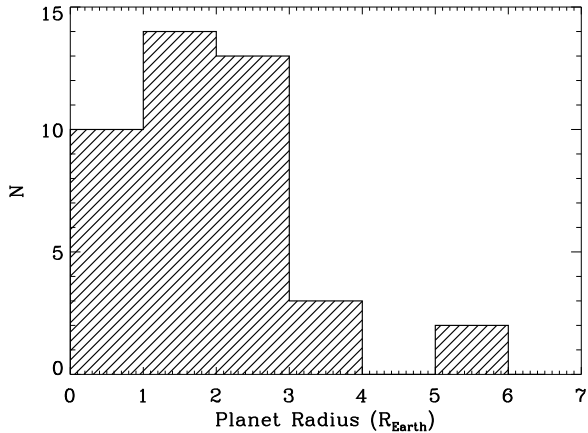}  
\caption{Histogram of the radii of the 42 transiting planets studied
  here, most of which are smaller than 3 \rearthe, with 6 greater than
  3 \rearthe.  The radii were determined from full models of the
  transit light curves, and stellar radii coming from analysis of high
  resolution spectra and, for 11 cases, asteroseismological analysis.}
\label{fig:rp_hist}
\end{figure}

Figure \ref{fig:rp_hist} shows a histogram of the 42 planet radii in
the present sample of transiting planets.  By selection, the radii are
mostly smaller than 4 \rearth with most being 1--3 \rearthe.  This
range of planet radii corresponds to the most common size of planets
(above 1 \rearthe) revealed by \ek \citep{Howard2012, Fressin2013, Petigura2013}.
The 22 KOIs, harboring 42 planets, were selected by the radii of the
planet candidates, favoring 1-3 \rearthe, without knowledge of
their masses or densities.

\begin{figure}  			
\epsscale{1.0}
\plotone{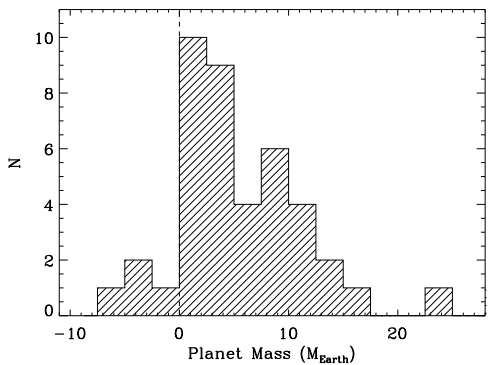}
\caption{Histogram of the masses of the 42 transiting planets studied
  here. The mass determinations stem from the multiple RV measurements
  of the host star and the constrained orbital period and measured
  time of transit from the \ek photometry.  Circular orbits were
  assumed in all cases.  Mass determinations represent the peak of the
  posterior mass distribution from the MCMC analysis of the RVs.  For
  four planets, fluctuations (errors) in the RVs happen to yield a
  peak in the posterior mass distribution that resides at a negative
  mass ($K$ is negative, given the frozen orbital phase).  While
  unphysical, the negative masses offer an unbiased statistical
  balance against the fluctuations that yield overly positive masses.
  The masses span a range up to 25 \mearthe.  See
  Fig~\ref{fig:mass_limvs_radius} for the well determined planet
  masses.  }
\label{fig:mp_hist}
\end{figure}

Figure \ref{fig:mp_hist} shows the histogram of 42 planet masses measured
here. The mass of each planet is defined arbitrarily as the peak mass (mode)
of the posterior distribution of masses from the MCMC analysis and
planet-model of the light curve and RVs.  The planet masses span a range up
to 25 \mearthe.  Four planets have a peak in the posterior distribution
at negative mass, due to errors in the RVs or ``noise'' from other
planets (see Section~\ref{sec:planetchar}), and four planets have
densities unphysically high, namely Kepler-102b, Kepler-37c, Kepler-37b, Kepler-103c,
due to long orbital periods.

\begin{figure}  			
\epsscale{1.1}
\plotone{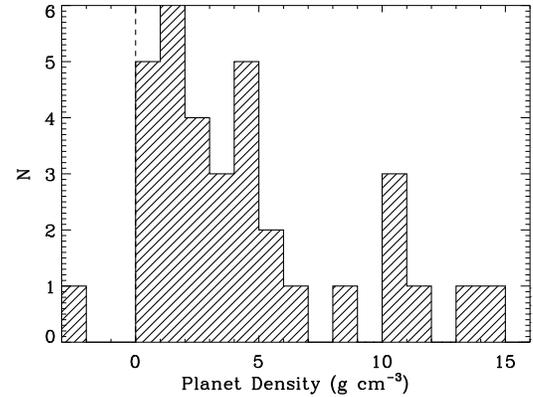}
\caption{Histogram of the densities of the 42 transiting planets studied
  here. Four values of density are negative, due to negative masses
  (See Fig.~\ref{fig:mp_hist}), three being offscale.  The physical
  planet densities span a range from below unity to 15 \gcc.  As each
  density is the mode of a broad posterior distribution, caution should be
  exercised in interpreting the histogram.  See
  Figure \ref{fig:rho_vs_radius} for the well-measured density measurements.
}
\label{fig:rhop_hist}
\end{figure}

Figure \ref{fig:rhop_hist} shows the histogram of planet densities for
the 42 transiting planets.  The densities span a range from near zero
to 15 \gcc.  Planets with low density are presumably composed of
considerable amounts of H and He gas (by volume) and planets with densities above
6 \gcc are likely composed mostly of rock and iron-nickel \citep{Batygin2013, Zeng_Seager08, 
Rogers2011, Valencia07}.  While negative mass and negative density are not physical,
bulk densities of 10-12 \gcc may be composed of solid or compressed iron. 
Several planets have densities out of the 
range of the plot due to poorly determined masses (See Table~\ref{tab:orbital_pars_tbl}).

\begin{figure}  			
\epsscale{1.1}
\plotone{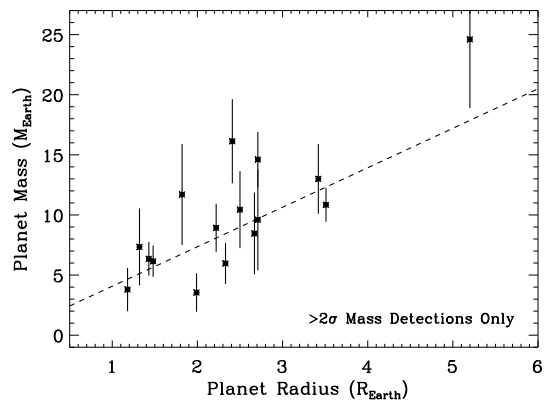}  
\caption{Planet mass vs radius for the 16 transiting planets studied
  here having a mass measured at the 2$\sigma$ confidence level or
  better.  Planet mass increases systematically with increasing radius.
  The weighted, linear fit to the planet masses and radii is, $M =
  3.28 R + 0.79$ (with $M$ and $R$ in Earth units), for planet radii
  in the range 1--5.5 \rearthe.  The monotonic relationship is
  consistent with a predominantly rocky composition for planets
  smaller than $\sim$ 2 \rearthe, and an increasing contribution of
  low density material (probably H, He, or light molecules) for larger
  planets.  The present data do not warrant more than a linear
  relationship, but see \cite{Weiss2013b} and Rogers et
  al. (2013).}
\label{fig:mass_detected_vs_radius}
\end{figure}

Figure \ref{fig:mass_detected_vs_radius} shows planet mass 
as a function of planet radius, for the 16 planets having a
mass measured here at the 2$\sigma$ level or better.  The planet
masses represent the peak of the posterior mass distribution, and the
uncertainties represent the mass 34\% in area from the peak in
both directions.  Planet mass is correlated with radius, but
the available radii and masses and their scatter warrant only a linear fit.  A weighted
linear fit (dotted line) has the relation: $M = 3.28
R + 0.79$ with M and R representing the planet mass and radius in
Earth units.  The scatter of the masses about the linear fit is
roughly consistent with the uncertainties (error bars), leaving unclear
whether some of the scatter in mass at a given radius is intrinsic to
the planets.  For radii above 4 \rearthe, the mass-radius relation
curves upward, as is easily seen by placing gas giant planets on the
plot at $R\approx$11 \rearth and $M$ = 318 \mearthe. Two companion papers
will examine this relationship more closely (\cite{Weiss2013b} and 
Rogers et al. 2013,  in prep).

\begin{figure}  			
\epsscale{1.1}
\plotone{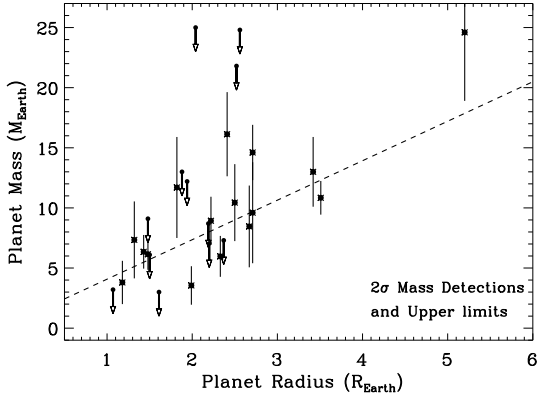}  
\caption{Planet mass vs radius for both the 16 transiting planets
  having a 2$\sigma$ mass measurement, along with 14 upper limits to
  the masses of the remaining planets.  The best-fit linear relation
  between planet mass and radius is, $M = 3.28 R + 0.79$  (with $M$
  and $R$ in Earth units), for planet
  radii in the range 1--5.5 \rearthe.    The upper limits are
  consistent with the linear trend in M vs R found from the 16
  detections and their scatter.  These upper limits are consistent
  with mass increasing systematically with radius, as in
  Fig.~\ref{fig:mass_detected_vs_radius} Some planets appear to have
  masses below the linear fit at the 2-$\sigma$ level, indicating that
  some of them contain more volatile material than average for their
  radius, i.e. diversity of composition for a given radius.  See \cite{Weiss2013b} 
  and Rogers et al. (2013) for further statistical
  treatment.  The 12 planets with an uncertainty in density greater
  than 6.5 \gcc were excluded from this plot.}
\label{fig:mass_limvs_radius}
\end{figure}

Figure \ref{fig:mass_limvs_radius} shows planet mass vs planet radius
for the 16 planets with 2$\sigma$ mass detections, as in Figure
\ref{fig:mass_detected_vs_radius}.  Here, we include the 14 additional
planets that do not meet the 2-sigma criterion, i.e., have a poorer
mass determination.  For them, we plot only an upper limit as a
downward arrow.  The upper limit is defined to be the planet mass at
the 95th percentile of the posterior distribution.  Including the
upper limits provides 14 additional masses and radii, which are
apparently consistent with the linear fit to the 16 planets with
2$\sigma$ masses.  Two upper limits in mass reside below the linear
fit, but not by more than the errors.  Thus the monotonic linear
relation between mass and radius is not refuted by inclusion of the
upper limits in planet mass.  We exclude from this figure the 12
planets having mass determinations so poor that the density is
uncertain by more than 6.5 \gcc.

\begin{figure}  			
\epsscale{1.1}
\plotone{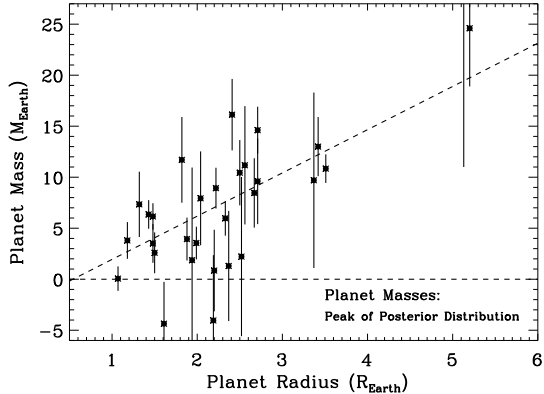}  
   \caption{Planet mass vs radius for 30 transiting planets measured
     here.  The plotted masses represent the peak of the posterior
     distribution of the planet masses.  Most peaks reside at positive
     masses, but RV fluctuations occassionaly yield a peak at negative
     masses, clearly representing non-detections.  We retain the
     negative masses to avoid Lutz-Kelker-type bias toward positive
     masses. The best-fit linear relation is $M = 4.24 R - 2.29$ (in Earth units) for
     planet radii in the range 1--5.5 \rearthe, including negative
     masses in the fit. Of the 42 transiting planets studied here, we
     excluded from this plot the 12 transiting planets having a
     density uncertainty greater than 6.5 \gcc (but included all
     mass measurements, even if less than a 2-sigma detection).}
\label{fig:mass_vs_radius}
\end{figure}

Figure \ref{fig:mass_vs_radius} shows planet mass vs planet radius for
all 30 transiting planets.  (We exclude the 12 planets with
uncertainties in density greater than 6.5 \gcc.)  Planet mass
increases monotonically with radius, as seen in
Figure~\ref{fig:mass_limvs_radius}.  Here, masses less secure than
2$\sigma$, and even negative masses, are included in the plot.  We
specifically allowed our MCMC analysis to include negative planet
masses in the solution.  Among the 42 transiting planets, 4 of them
had a peak in the posterior mass distribution that was negative (see
Table~\ref{tab:orbital_pars_tbl}). Such unphysical masses are a normal outcome of fluctuations
in the RV measurements from both RV errors and unknown planets, which
will be mostly uncorrelated in orbital phase with the known planets,
aside from mean motion resonances. Negative masses arise naturally
from RV uncertainties of \ms for planets that induce RV
semi-amplitudes, $K$, that are smaller (due to small planet mass
and/or long period).  RV errors may cause measured RVs to be high when
the actual RV is low, and vice-versa, due to those errors.  For the
low mass planets, such RV non-detections are expected, yielding equal
numbers of apparently positive and ``negative'' masses.  

The ``negative'' masses are useful for any statistical treatment of masses
or densities of the planet population as an ensemble.  Because planet
masses are certainly positive, the ensemble of planets (in a given
range of planet radii) will, on average, yield positive values of
mass.  As the KOI target selection was based on planet radius not
mass, we may measure the average planet mass as a function of planet radius.
We report in Table~\ref{tab:orbital_pars_tbl}, column 5, the planet
masses corresponding to the peak in the MCMC posterior distribution,
four of which are negative.  For the uncertainty, we report the mass
corresponding to 34\% of the integrated area of the distribution on
either side, representing the 1$\sigma$ departure from the peak mass.

Figure \ref{fig:rho_vs_radius} shows planet density vs radius for the
30 transiting planets having an uncertainty in density less than 6.5
\gcc.  Planet density {\em decreases with increasing planet radius}
from 1.5 to 5 \rearthe.  The densities of planets smaller than 1.5 \rearth
are systematically greater than $\sim$5 \gcc, similar to that of Earth
(5.5 \gcc), indicating that these planets have mostly rocky interiors.
Planets with radii larger than 2 \rearth have densities predominatly
and increasingly less than 5 \gcc, indicating increasing amounts by volume of light material,
presumably H and He \citep{Batygin2013, Zeng_Seager08, Rogers2011, 
Valencia07}. Thus the transition from rocky to gas-dominated
planets (by volume) occurs at roughly 2 \rearthe (\cite{Weiss2013b}, 
and Rogers et al. (2013, in prep)).

\begin{figure}  			
\epsscale{1.1}
\plotone{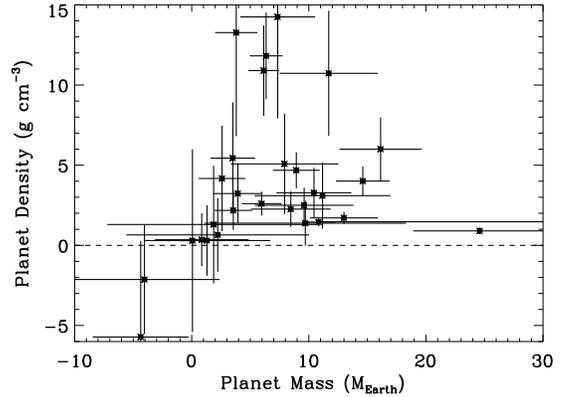}  
\caption{Planet density vs planet mass for the 30 transiting planets
  measured here.  Planet density appears to increase from 1 to $\sim$6
  \mearthe, and then decline from $\sim$6--25 \rearthe, with large
  scatter in density.  This behavior is consistent with planets of
  0--6 \mearth having a mostly rocky interior experiencing increased
  gravitational compression (or greater iron/nickel content).  The
  decline in density for planet masses, $M > 6$ \mearth indicates
  increasing amounts of low density material, presumably H, He, and
  light molecules, with increasing planet mass.  The scatter in planet
  density, at a given planet mass, indicates a diversity of chemical
  composition at a given mass, with different admixtures of iron,
  rock, and light elements.}
\label{fig:rho_vs_mass}
\end{figure}

Figure \ref{fig:rho_vs_mass} shows planet density vs planet mass for
30 transiting planets measured here.  Planet density is not a strong
function of planet mass in the domain of 0--30 \mearthe.  There is
weak evidence in the plot of a rise in density for masses from 0--6
\mearthe, and a decline in density for planet masses 6--25 \mearthe.
This behavior of the density is consistent with increasing
gravitational compression of rocky planet interiors with increasing
mass from 1-6 \mearthe.  The decline in density from 6--25 \mearth is
consistent with an increasing contribution from light elements, H and
He, toward Neptune-like envelopes.
 
\begin{figure}  			
\epsscale{1.1}
\plotone{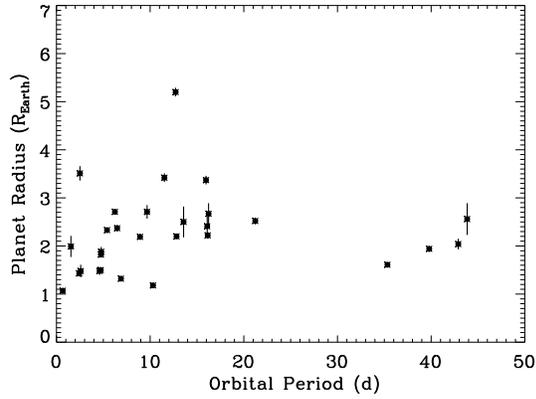}  
\caption{Planet radius vs orbital period for 30 well-measured
  transiting planets.  Planet radii are not strongly correlated with
  orbital period in the present sample.}
\label{fig:radius_vs_period}
\end{figure}

\begin{figure}  			
\epsscale{1.1}
\plotone{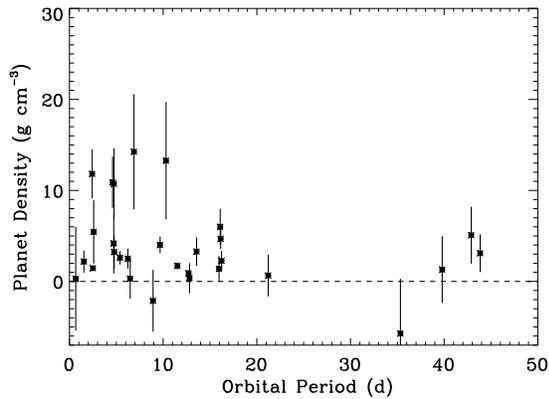}  
\caption{Planet density vs orbital period for 30 well-measured
  transiting planets.  Planet density is not strongly correlated with
  orbital period. There is no convincing evidence that the density,
  and hence planet composition, is a strong function of orbital
  distance, out to periods of 45 days. But mass detection biases may
  affect this interpretation.}
\label{fig:rho_vs_period}
\end{figure}

 Figure \ref{fig:radius_vs_period} shows planet radius vs orbital
period for 30 well-measured transiting planets.   The planet radii are
not a strong function of orbital period.  It is difficult to interpret
this lack of correlation securely.  Selection effects may contribute
to the sizes, and hence anticipated masses, of planets farther from
the host star where Doppler reflex velocities are lower.
Figure \ref{fig:rho_vs_period} shows planet density vs orbital period
for 30 well-measured transiting planets, which shows no clear
dependence.  As with  Figure \ref{fig:radius_vs_period}, the
interpretation of no correlation between period and bulk density is
not clear, due to possible selection effects with orbital period.

\begin{figure}  			
\epsscale{1.1}
\plotone{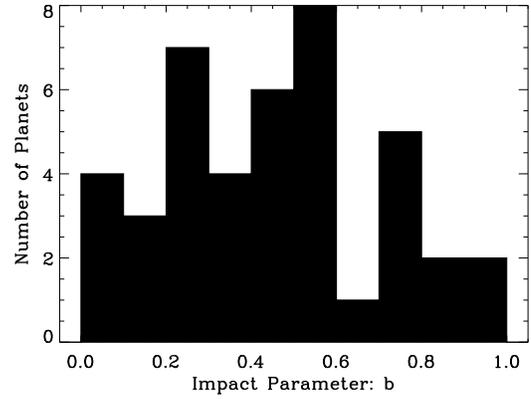}  
\caption{Histogram of impact parameters, $b$, for all 42 transiting
  planets studied here.  The impact parameter is derived from a
  self-consistent transiting-planet model fit to the photometry and
  RVs, with constraints on the stellar radius from spectroscopy and
  asteroseismology.  The roughly uniform distribution of impact
  parameters, from 0 to 1, is consistent with the geometrical
  expectation of a random distribution of orbital tilt angles relative
  to our line of sight, leading to a uniform distribution of impact
  parameters, $b$.  
}
\label{fig:b_hist}
\end{figure}

Figure \ref{fig:b_hist} shows a histogram of impact parameters, $b$,
for all 42 transiting planets.  The impact parameter is mostly
decoupled from the RV and mass detections. The values of $b$ offer a
test of the integrity of the light curve fitting.  The roughly uniform
distribution of $b$ from 0--1 is consistent with the expectation of
randomly oriented orbital planes leading to a uniform distribution of
impact parameters, $b$. We note that the impact parameter distribution 
in \cite{Batalha2013} appears not to be uniform, raising concerns about bias there. 

\subsection{Conclusions about Planet Composition and Formation}
 
In general, the distribution of planet masses for a
given planet radius may be a function of orbital period and
the type of host star, stemming from the complex processes of planet
formation in a protoplanetary disk.  The distribution of planet masses
surely depends on planet radius, stellar mass, orbital semi-major axis
and eccentricity, and on the chemical and thermodynamic properties of
the protoplanetary region where they form. Thus, the measured planet
masses and radii here inform only one plane of a multi-dimensional
space that characterizes planet properties.

\begin{figure}  			
\epsscale{1.1}
\plotone{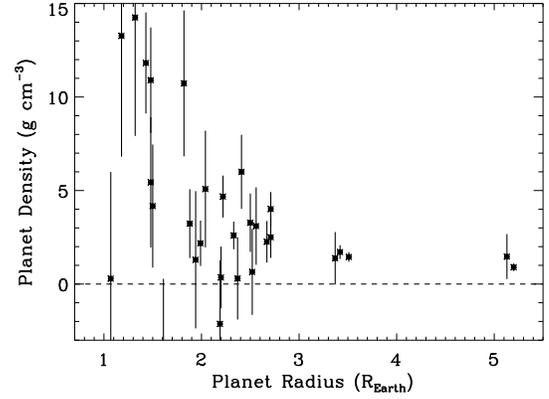}  
\caption{Planet density vs radius for all 30 transiting
  planets having an uncertainty in density less than 6.5 \gcc.  Planet
  density declines systematically with increasing planet radius in
  the domain of 1.0--5.5 \rearthe.  Increasing planet radius apparently
  is associated with increasing amounts of
  low density material, presumably H, He, and light molecules. For an
  analysis of density vs radius for all known planets, see \cite{Weiss2013b} 
 and Rogers et al. 2014 (in prep). }
\label{fig:rho_vs_radius}
\end{figure}

Planet density may decrease with increasing planet radius due to an
increasing admixture of light building material (H, He, and water)
\citep{Lopez2012, Lopez2013, Rogers2011, Rogers_Seager2010b,
  Zeng_Seager08,Weiss2013b} and see also Rogers
et al. (2013, in prep).  Indeed this decline in bulk density with
increasing radius is seen clearly in Figure~\ref{fig:rho_vs_radius}.
This supports previous analyses, e.g., \cite{Weiss2013}, that
similarly noted a declining density with increasing radius.  This
density behavior with radius strongly supports the suggestion that
planets with sizes increasing above 2 \rearth have interiors with an
increasing contribution from light elements, H and He, as described
theoretically by \citep{Batygin2013, Zeng_Seager08, Rogers2011, Valencia07}.

Interestingly, average planet masses may remain constant or even
decrease with increasing planet radii for radii near the transition
from rocky to volatile-rich planets near 2 \rearthe.  For example,
planets of 1.8 \rearth may, on average, be only slightly more massive
(or even less massive) than those of 1.5 \rearth if they contain large
differences in the amount of light elements.

The unknown composition and density of planets having radii, $R<3$ \rearth
motivated the RV measurements here, but left great uncertainty
about whether RV signals would be detectable.  In retrospect, the low
densities near 1 \gcc~for the planets with $R>2.5$\rearth certainly
led to many of the non-detections of RV signals we found.  We simply
didn't know, at the start, if $\sim$2.5 \rearth planets were mostly
rocky or mostly volatile-rich. It is worth emphasizing that for each
planet candidate only the radius was known, leaving the resulting
mass and density unbiased for the given planet radius.

The primary goal of this paper is to provide a statistically useful
and unbiased set of planet masses and densities as a function of
planet radii. However, these planet masses and densities pertain only
to the stellar masses of 0.8--1.1 \msun and to orbital periods ($P<$25
d) represented in this work.  The population of planet masses and
densities, as a function of radius, may be a function of orbital
period and stellar mass, confining the conclusions here to such stars
and orbital distances.

This work benefited greatly from the transit ephemeris for each
candidate that came directly from the \ek photometry.  That ephemeris
provided a prediction of the times (i.e. orbital phase) when the RVs
were expected to be maxima or minima.  We optimized the times of
observations based on those predictions, to more efficiently detect
and measure the RV semiamplitude, $K$, and hence planet mass.
Moreover, for circular orbits, only the RV amplitude is unknown prior
to taking RV measurements, thus concentrating the RV information into
the measurement of planet mass, rather than having to constrain other
orbital parameters too.  The only caveat stems from the unknown
orbital eccentricities.

Why were no clear false positives detected?  This program began by studying
42 transiting planet ``candidates''.  The \ek photometry,
planet-detection pipeline, and associated ``data validation'' (DV)
efforts conferred ``KOI'' status to these planet ``candidates''.  The
\ek Follow-up Observation Team (FOP) had obtained low signal-to-noise
spectra for each of the 22 KOIs, rendering the stars and prospective
planets suitable for precise RV follow-up.  The false-positive
probabilities were not known for the 22 KOIs and their associated 42
transiting planets.  However we carried out careful vetting of the
planets by detailed light curve models, astrometric ``centroid''
analysis, multiple high-resolution imaging, and sophisticated
modelling of the false-positive probability (Table~\ref{tab:fpp_tbl}).  After all
that, {\em none of the 42 transiting planets were revealed to be false
  positives} (but one has an FPP of $\sim$2\%).  

This absence of false positives is noteworthy, as
\cite{Fressin2013, Morton2012} predict a false positive probability of
$\sim$10\%.  We might have expected 4 false positives from 42
transiting \ek planets, but we found none.  The implication is that
the preliminary recon spectroscopy, seeing-limited imaging, and the
centroid analysis were effective in diminishing the false positive
rate.  These KOIs are also much brighter than the typical KOI, these
being magnitude 10--13 mag, and hence are closer than the typical \ek
KOI.  Typical distances are only 100--200 pc, making the various
imaging techniques more efficient at detecting angularly nearby stars
that can harbor the false positive.  The \ek pipeline and DV efforts
seem to be performing well in keeping the false positive rate of
small planets to no more than the 10\% rate that one predicts
simplistically without imaging and spectroscopy.   Moreover, many of
the KOIs studied here are multi-planet transiting systems which have
been shown to be mostly real planets \citep{Lissauer2012}.  Based on these 42
transiting planets, the \ek pipeline and vetting appears to have a
false positive rate under 10\% for planets less than 4 \rearthe.
This conclusion deserves a more careful analysis, including
quantitative assessment of early vetting of these candidates.

\subsection{Rocky Planets and Interiors}

Six of the planets studied here have 2-$\sigma$ densities between
5.0--12.0 \gcc.  They are marked in bold in column 7 of Table~\ref{tab:orbital_pars_tbl}.
These six planets are likely to have compositions that consist mostly
of rocky or iron/nickel material.  Those six rocky planets are
Kepler-100b, Kepler-102e, Kepler-102d, Kepler-48c, Kepler-113b, Kepler-131b, Kepler-99b, and Kepler-406b.
Most of these planets have densities
secure only at the 2--3 $\sigma$ level. They should be considered only
``candidate'' rocky planets, requiring further RVs to secure their
masses and densities.

Several of these 6 high-density planets have densities greater than
that of Earth, making them plausibly composed nearly entirely of rock
or iron/nickel.  These planets join previously detected planets that
have measured masses and radii implying rocky compositions, namely
CoRoT-7b, Kepler-10b and Kepler-36b.

Detailed models of planet interiors, including possible chemical
compositions, stratified differentiation, and equations of state are
needed to predict the plausible bulk densities associated with planets
with a given mass.  Recent work on the interiors of rocky
planets have been carried out by \cite{Rogers_Seager2010a,
  Rogers_Seager2010b, Rogers2011, Zeng_Sasselov2013,
  Lopez2013} and Rogers et al. (2013, in prep).  

Discussions have ensued about whether iron-rich planets such as
Mercury, might form from giant collisions that strip the silicate and
volatile envelopes or instead whether such compositional oddities
might result from processes intrinsic to planet formation.  Recent
work by \cite{Wurm2013} provides an interesting mechanism of
photophoretic separation of metals from silicates.  Meanwhile, models of
photoevaporation seem promising in predicting the statistical behavior
of these close-in rocky planets \citep{Lopez2012, Lopez2013}.
 
Future observations of RVs and TTVs are needed to provide more masses
and densities for small planets.   It is highly desirable to measure
masses and densities for 1-3 \rearth planets that orbit farther from
the host star, to determine whether the transition size between rocky and gaseous
planets is different at 1 au than at 0.1 au as measured here.

\section{Acknowledgements}

\acknowledgments{\ek was competitively selected as the tenth NASA
  Discovery mission. Funding for this mission is provided by the NASA
  Science Mission Directorate. Some of the data presented herein were
  obtained at the W. M. Keck Observatory, which is operated as a
  scientific partnership among the California Institute of Technology,
  the University of California, and the National Aeronautics and Space
  Administration. The Keck Observatory was made possible by the
  generous financial support of the W. M. Keck Foundation.  Some of
  the asteroseismology analysis was performed by the Stellar
  Astrophysics Centre which is funded by the Danish National Research
  Foundation (Grant DNRF106). The research is supported by the ASTERISK project
  (ASTERoseismic Investigations with SONG and \ek) funded by the
  European Research Council (Grant 267864).  D.H. is supported by an
  appointment to the NASA Postdoctoral Program at Ames Research
  Center, administered by Oak Ridge Associated Universities through a
  contract with NASA. W.J.C., Y.E., T.L.C., G.R.D, R.H and
  A.M. acknowledge financial support from the UK Science and
  Technology Facilities Council (STFC). S.B. acknowledges NSF grant
  AST-1105930.  Funding for the Stellar Astrophysics Centre is
  provided by The Danish National Research Foundation (Grant agreement
  no.: DNRF106). The research is supported by the ASTERISK project
  (ASTERoseismic Investigations with SONG and \ek) funded by the
  European Research Council (Grant agreement no.
  267864). S.H. acknowledges financial support from the Netherlands
  organisation for Scientific Research (NWO).  The research leading to
  the presented results has received funding from the European
  Research Council under the European Community's Seventh Framewrok
  Programme (FP7/2007-2013)/ERC grant agreement no. 338251
  (StellarAges).  W. F. Welsh and J. A. Orosz acknowledge support from
  NASA through the \ek Participating Scientist Program and from the
  NSF via grant AST-1109928. D. Fischer acknowledges support from NASA
  ADAP12-0172. O. R. Sanchis-Ojeda \& J. N. Winn are supported by the
  \ek Participating Scientist Program (PSP) through grant
  NNX12AC76G. E. Ford is partially supported by NASA PSP grants
  NNX08AR04G \& NNX12AF73G. Eric Agol acknowledges NSF Career grant
  AST-0645416.  R.L.G. has been partially supported by NASA
  co-operative agreement: NNX09AG09A.  A. Santerne acknowledges the
  support by the European Research Council/European Community under
  the FP7 through Starting Grant agreement number 239953.  L.A.R 
  acknowledges support provided by NASA through Hubble Fellowship 
  grant \#HF-51313.01 awarded by the Space Telescope Science Institute, 
  which is operated by the Association of Universities for Research in 
  Astronomy, Inc., for NASA, under contract NAS 5-26555The authors
  would like to thank the many people who gave so generously of their
  time to make this Mission a success.  All \ek data products are
  available to the public at the Mikulski Archive for Space Telescopes
  {\url{http://stdatu.stsci.edu/kepler}} and the spectra and their
  products are made available at the NExSci Exoplanet Archive and its
  CFOP website: {\url{http://exoplanetarchive.ipac.caltech.edu}} We
  thank the many observers who contributed to the measurements
  reported here.  We gratefully acknowledge the efforts and dedication
  of the Keck Observatory staff, especially Scott Dahm, Hien Tran, and
  Grant Hill for support of HIRES and Greg Wirth for support of remote
  observing.  This work made use of the SIMBAD database (operated at CDS,
Strasbourg, France) and NASA's Astrophysics Data System Bibliographic
Services.  This research has made use of the NASA Exoplanet Archive,
which is operated by the California Institute of Technology, under
contract with the National Aeronautics and Space Administration under
the Exoplanet Exploration Program.  Finally, the authors wish to
extend special thanks to those of Hawai`ian ancestry on whose sacred
mountain of Mauna Kea we are privileged to be guests.  Without their
generous hospitality, the Keck observations presented herein would not
have been possible.}


\clearpage  %

\bibliographystyle{apj}
\bibliography{h22} 

\clearpage  %


\clearpage
\begin{figure*}   [htp]
\epsscale{0.9}
\plotone{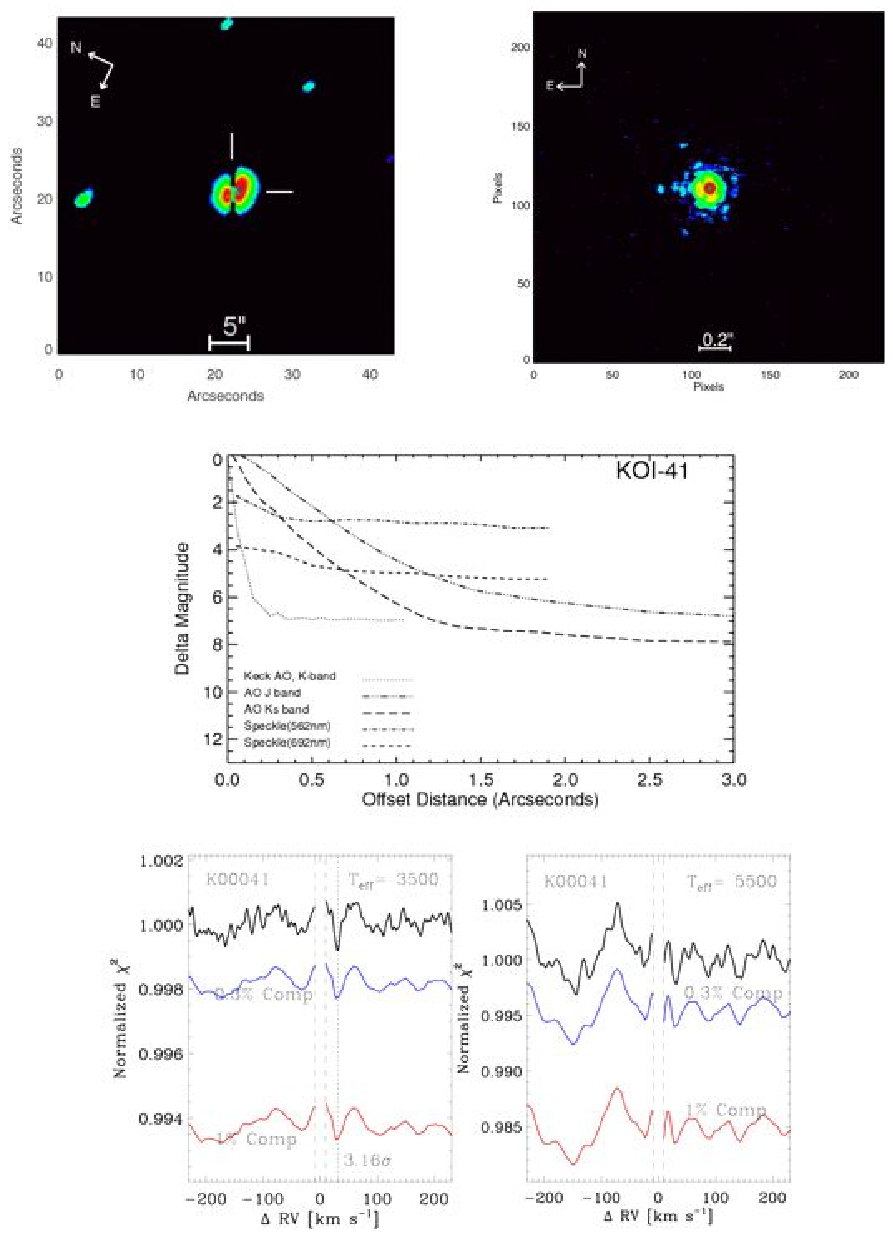}
\caption{ Upper left) Seeing limited image for Kepler-100, (KOI-41). Upper right) Adaptive 
Optics image. Middle) Limiting magnitudes of companion stars found with 
Speckle and AO imaging. Lower left and lower right)  
Diagnostic plots to detect the presence of absorption lines from any 
secondary star at a different radial velocity relative to the primary star.
The plot gives the reduced chi-square statistic (black line) of the fit to 
the residuals (after subtracting the primary star spectrum) for representative 
secondary spectra for stars having \teff=3500 K (left panel) and \teff=5500 K 
(right panel), as a function of relative RV.    Injection of such mock companion 
stars, at different relative RVs, was performed for a relative brightness of 0.3\% 
(blue line) and 1\% (red line).    The lack of drops below those two color lines 
indicates the  lack of companions at those thresholds.  For Kepler-100, no 
companion star is apparent.  Companions brighter than 0.3\% (optical, V-band) 
would have been detected.    Companions separated in velocity by less than 
10 \kms would not be detected as the absorption lines mostly overlap.
}
\label{fig:koi41_fig1}
\end{figure*}

\begin{figure*}    
\epsscale{1.0}
\plotone{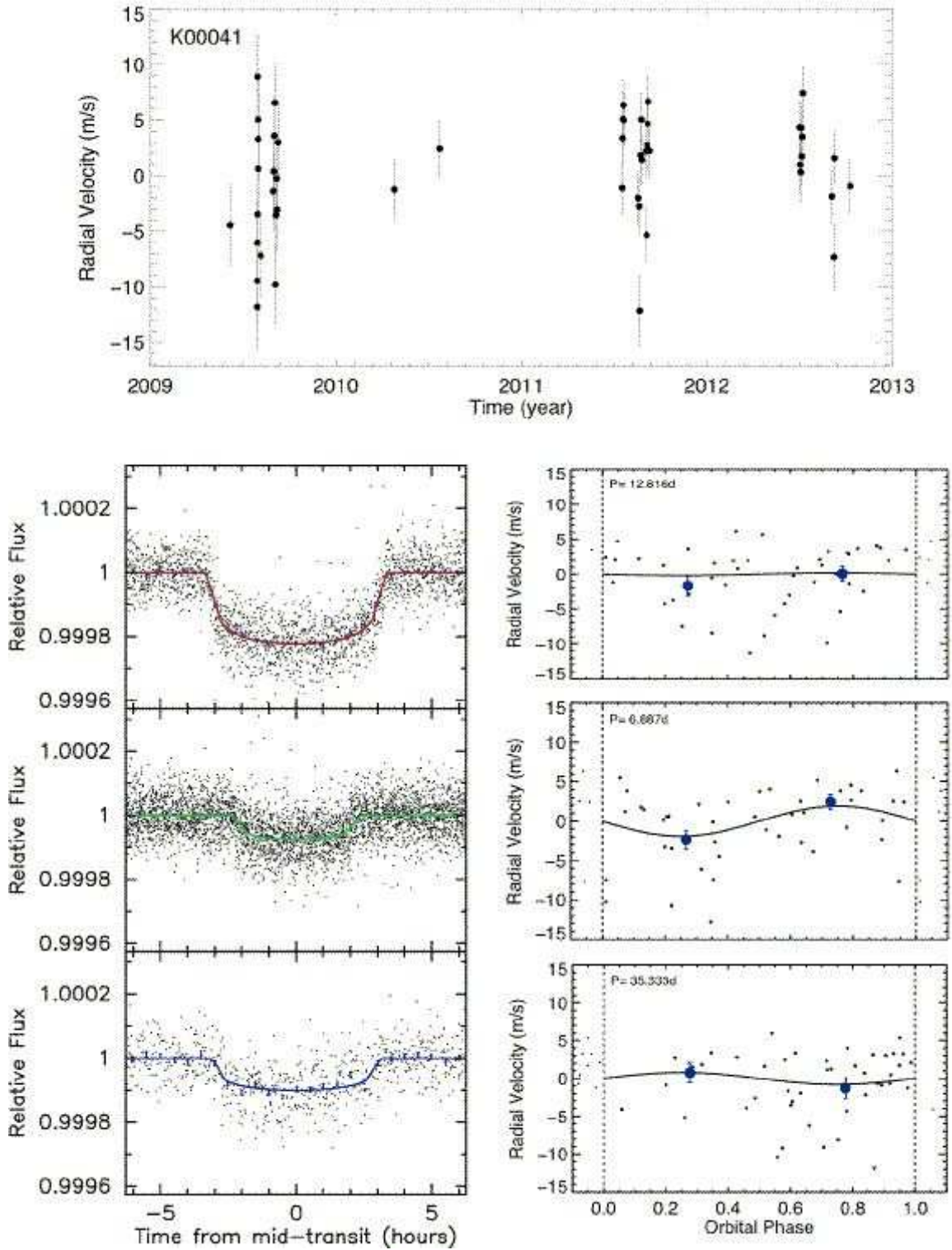}
\caption{ Top) RV versus time for Kepler-100 (KOI-41). Lower left) Phase folded \ek photometry for each planet. Lower Right) Phase folded radial velocities for each planet. 
The blue points represent RVs binned into the quadrature intervals, 0.25 $\pm$ 0.125 and 
0.75 $\pm$ 0.125. The amplitude of the blue point is the weighted average in
 each bin. The error of the blue point is the standard deviation of the RVs 
 in the bin, divided by the square root of the number of RVs included.
Kepler-100c: Rp = $2.20 \pm 0.05$ \rearthe, Mp = $0.85 \pm 4.0$ \mearthe.
Kepler-100b: Rp = $1.32 \pm 0.04$ \rearthe, Mp = $7.34 \pm 3.2$ \mearthe.
Kepler-100d: Rp = $1.61 \pm 0.05$ \rearthe, Mp = $-4.36 \pm 4.1$\mearthe.
}
\label{fig:koi41_fig2}
\end{figure*}

\begin{figure*}   
\epsscale{1.0}
\plotone{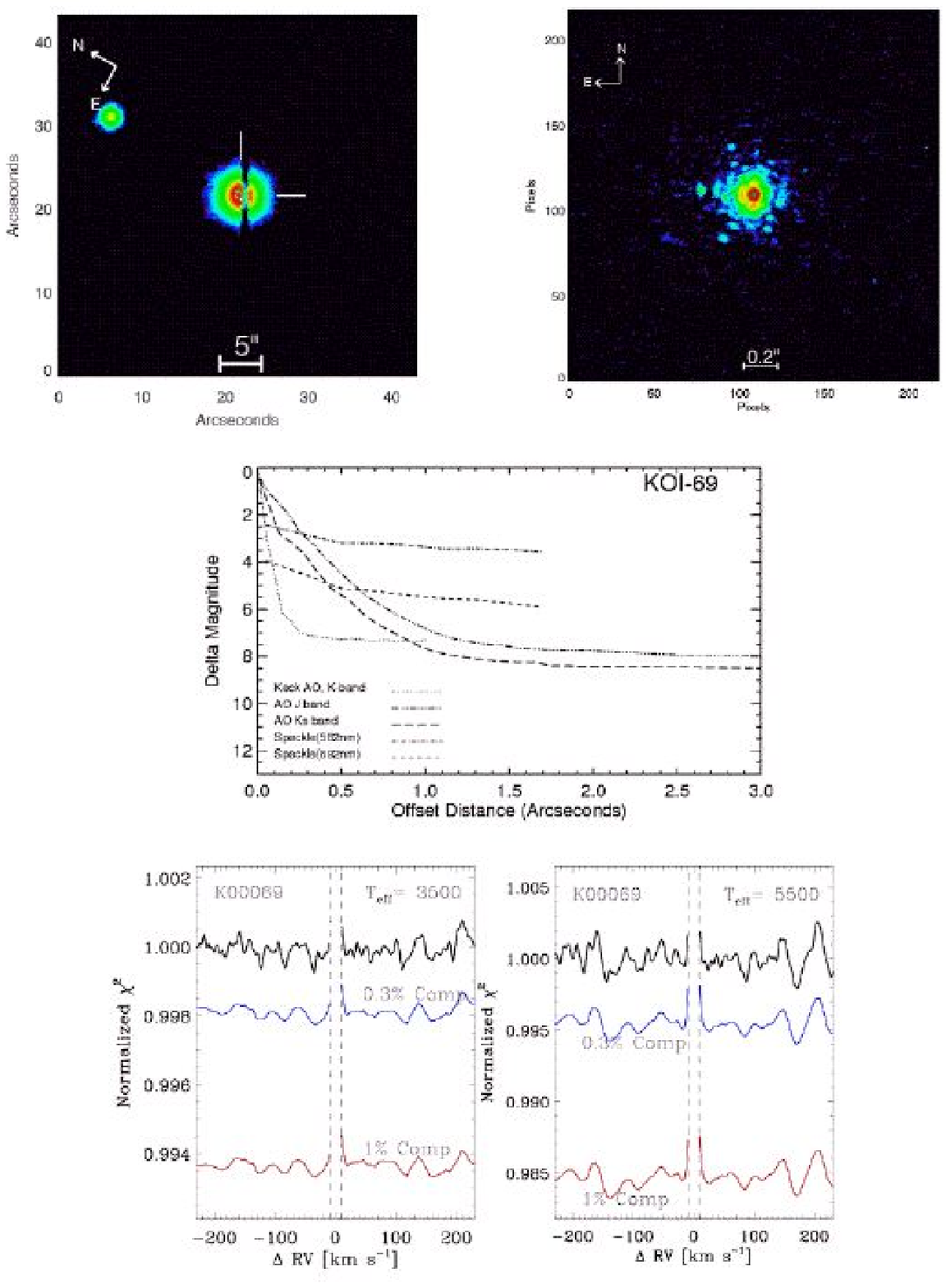}
\caption{ Upper left) Seeing limited image for Kepler-93 (KOI-69). Upper right) Adaptive Optics image. Middle) Limiting magnitudes of companion stars found with Speckle and AO imaging. Lower left and lower right)  Same as Figure 1, but for Kepler-93. No companions are detected. Any companion brighter than 0.3\% the brightness(V-band) of the primary would have been detected.
}
\label{fig:koi69_fig1}
\end{figure*}

\begin{figure*}    
\epsscale{1.0}
\plotone{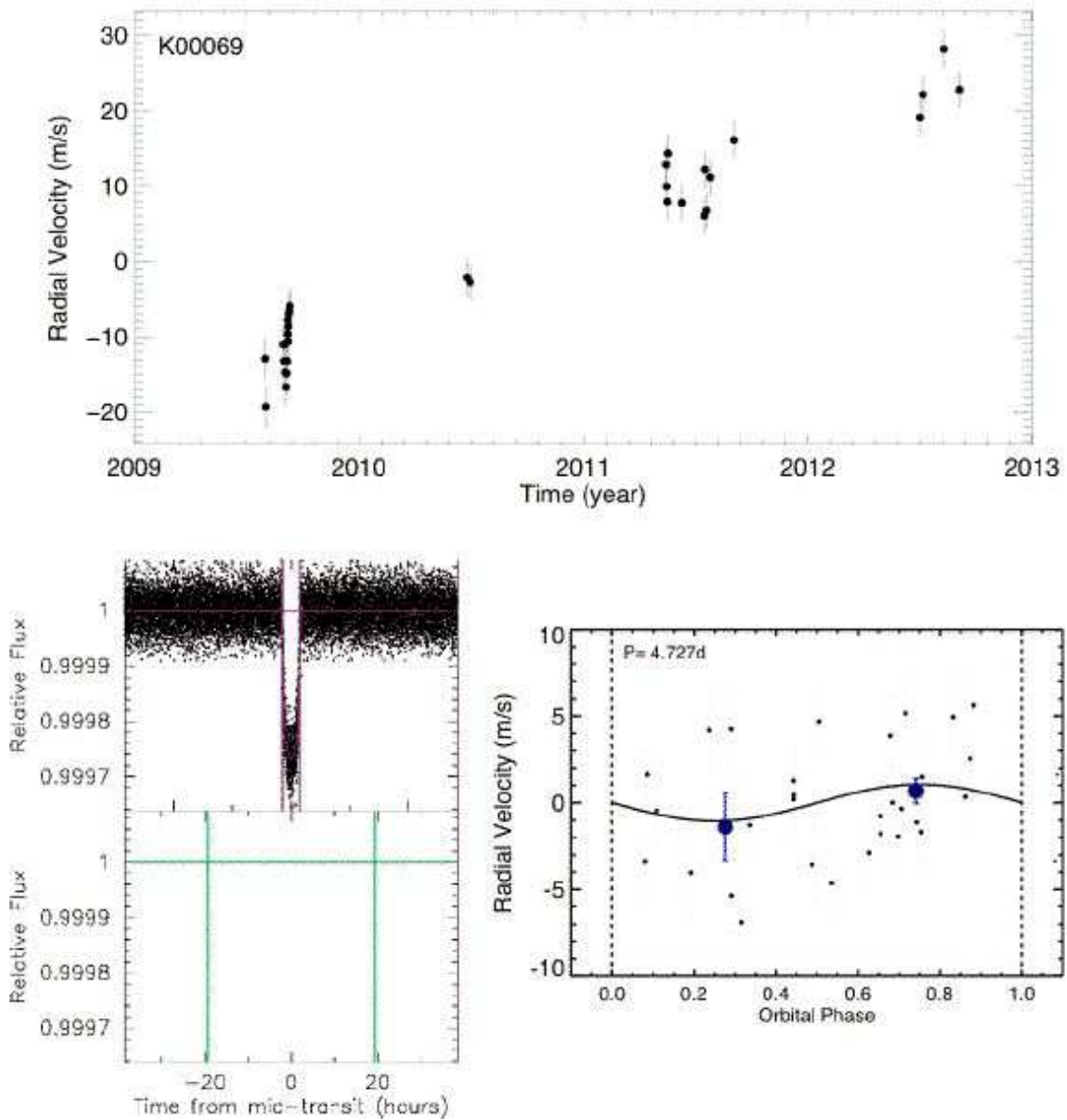}
\caption{ Top) Radial Velocity versus time for Kepler-93 (KOI-69). Lower left) Phase folded \ek photometry for each planet. Lower Right) Phase folded radial velocities for each planet. 
The blue points represent binned RVs near quadrature, same as Figure  \ref{fig:koi41_fig2}.
Kepler-93b: Rp = $1.50 \pm 0.03$ \rearthe, Mp = $2.59 \pm 2.0$ \mearthe.
Kepler-93c: Rp = NA, Mp  $>$954  \mearthe, Period $>$ 1460d.
}
\label{fig:koi69_fig2}
\end{figure*}

\begin{figure*}   
\epsscale{1.0}
\plotone{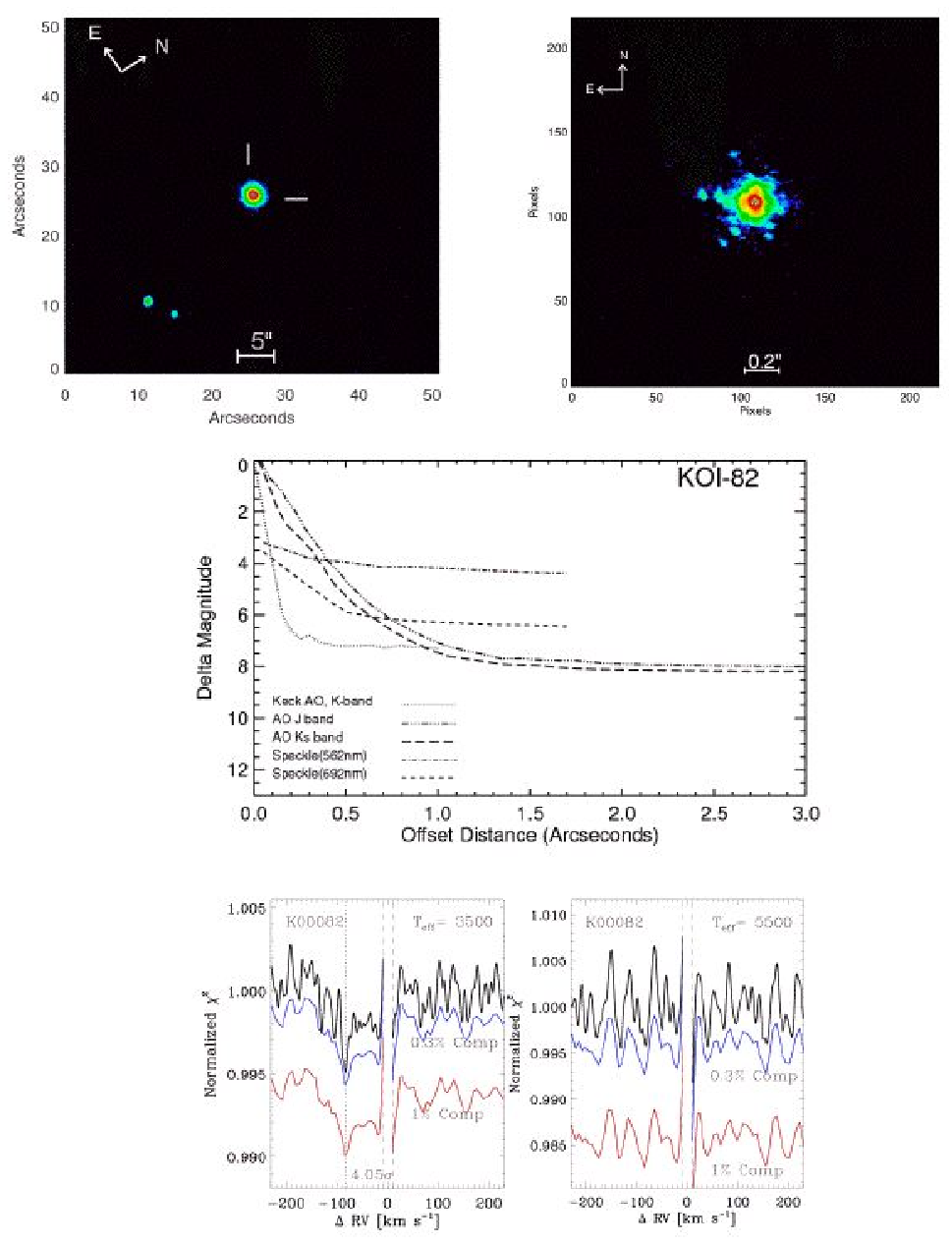}
\caption{ Upper left) Seeing limited image for Kepler-102 (KOI-82). Upper right) Adaptive Optics image. Middle) Limiting magnitudes of companion stars found with Speckle and AO imaging. Lower left and lower right)  Same as Figure 1, but for Kepler-102. No companions are detected. Any companion brighter than 0.5\% the brightness(V-band) of the primary would have been detected.
}
\label{fig:koi82_fig1}
\end{figure*}

\begin{figure*}    
\epsscale{0.9}
\plotone{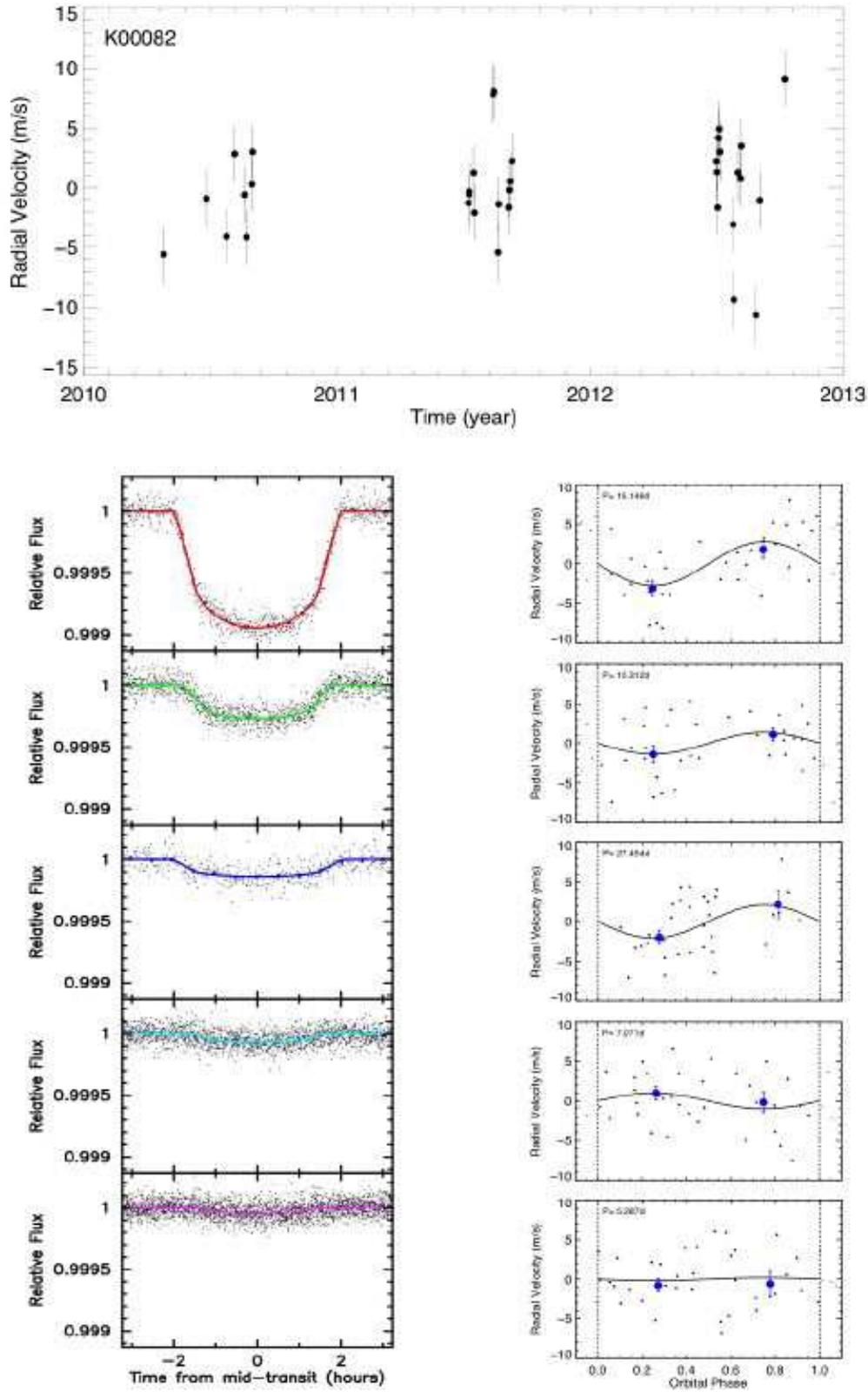}
\caption{ Top) RV versus time for Kepler-102 (KOI-82). Lower left) Phase folded \ek 
photometry for each planet. Lower Right) Phase folded radial velocities for each planet. 
The blue points represent binned RVs near quadrature, same as Figure  \ref{fig:koi41_fig2}.
Kepler-102e: Rp = $2.22 \pm 0.07$ \rearthe, Mp = $8.93 \pm 2.0$ \mearthe.
Kepler-102d: Rp = $1.18 \pm 0.04$ \rearthe, Mp = $3.80 \pm 1.8$ \mearthe.
Kepler-102f: Rp = $0.88 \pm 0.03$ \rearthe, Mp = $0.62 \pm 3.3$ \mearthe.
Kepler-102c: Rp = $0.58 \pm 0.02$ \rearthe, Mp = $-1.58 \pm 2.0$\mearthe.
Kepler-102b: Rp = $0.47 \pm 0.02$ \rearthe, Mp = $0.41 \pm 1.6$ \mearthe.
}
\label{fig:koi82_fig2}
\end{figure*}

\begin{figure*}   
\epsscale{1.0}
\plotone{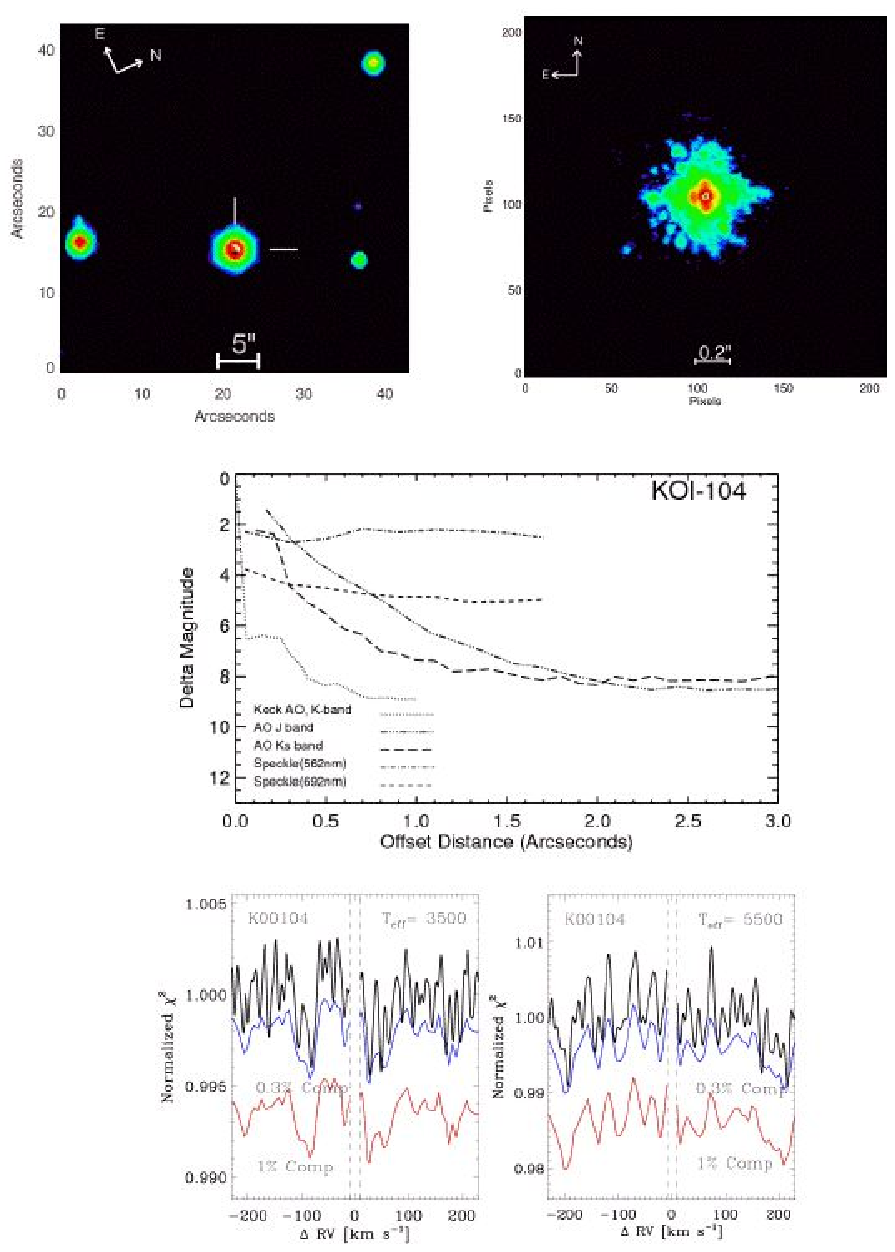}
\caption{ Upper left) Seeing limited image for Kepler-94 (KOI-104). Upper right) Adaptive Optics image. Middle) Limiting magnitudes of companion stars found with Speckle and AO imaging. Lower left and lower right)  Same as Figure 1, but for Kepler-94. No companions are detected. Any companion brighter than 0.5\% the brightness(V-band) of the primary would have been detected.
}
\label{fig:koi104_fig1}
\end{figure*}

\begin{figure*}    
\epsscale{1.0}
\plotone{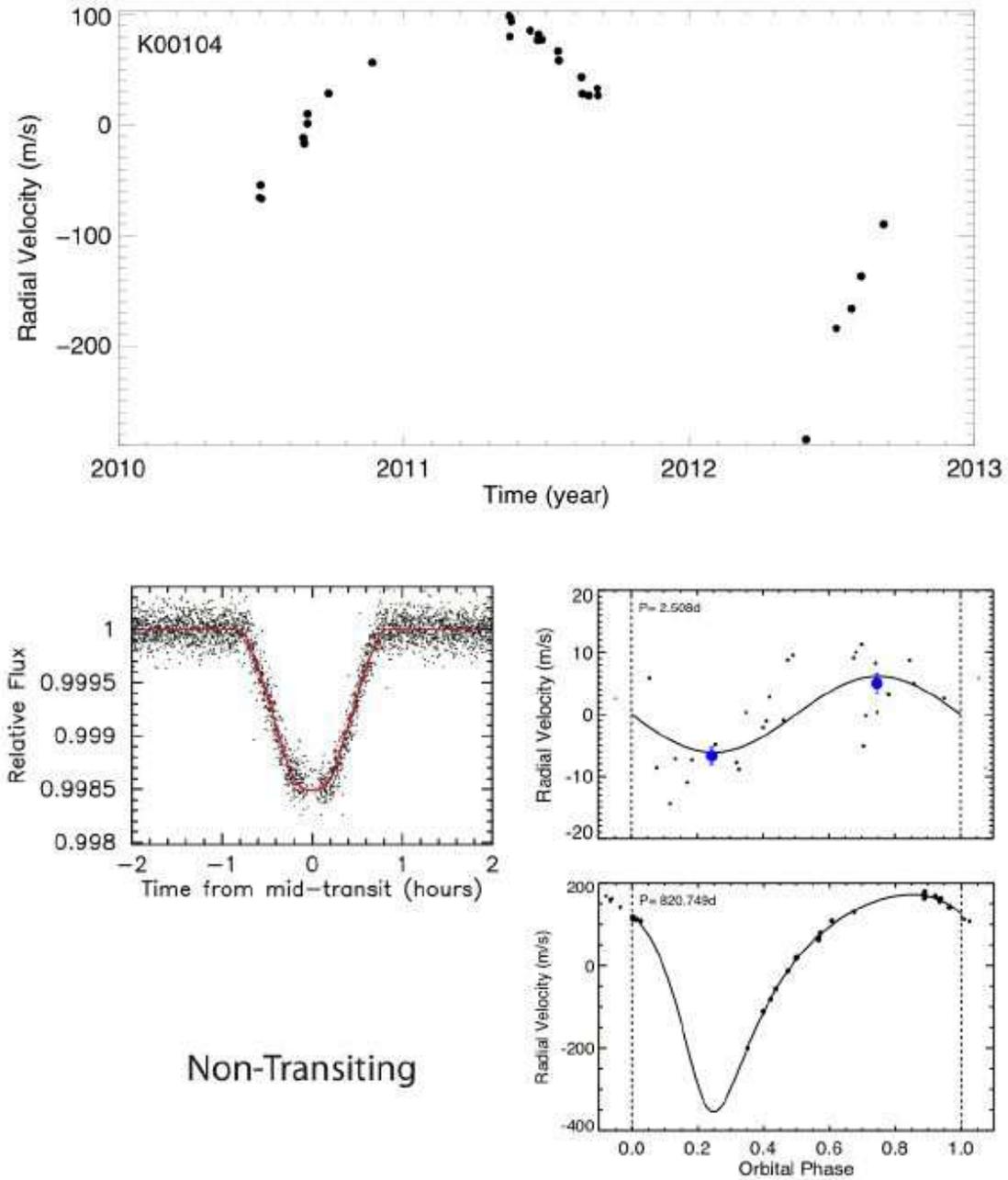}
\caption{ Top) RV versus time for Kepler-94 (KOI-104). Lower left) Phase folded \ek 
photometry for the transiting planet. The non-transiting planet has no light curve.  
Lower Right) Phase folded radial velocities for each planet. 
The blue points represent binned RVs near quadrature, same as Figure  \ref{fig:koi41_fig2}.
Kepler-94b: Rp = $3.51 \pm 0.15$ \rearthe, Mp = $10.84 \pm 1.4$\mearthe.
Kepler-94c: Rp = NA, Mp = $3126 \pm 202$ \mearthe, Period = 820 $\pm$ 3 d.
}
\label{fig:koi104_fig2}
\end{figure*}

\begin{figure*}   
\epsscale{1.0}
\plotone{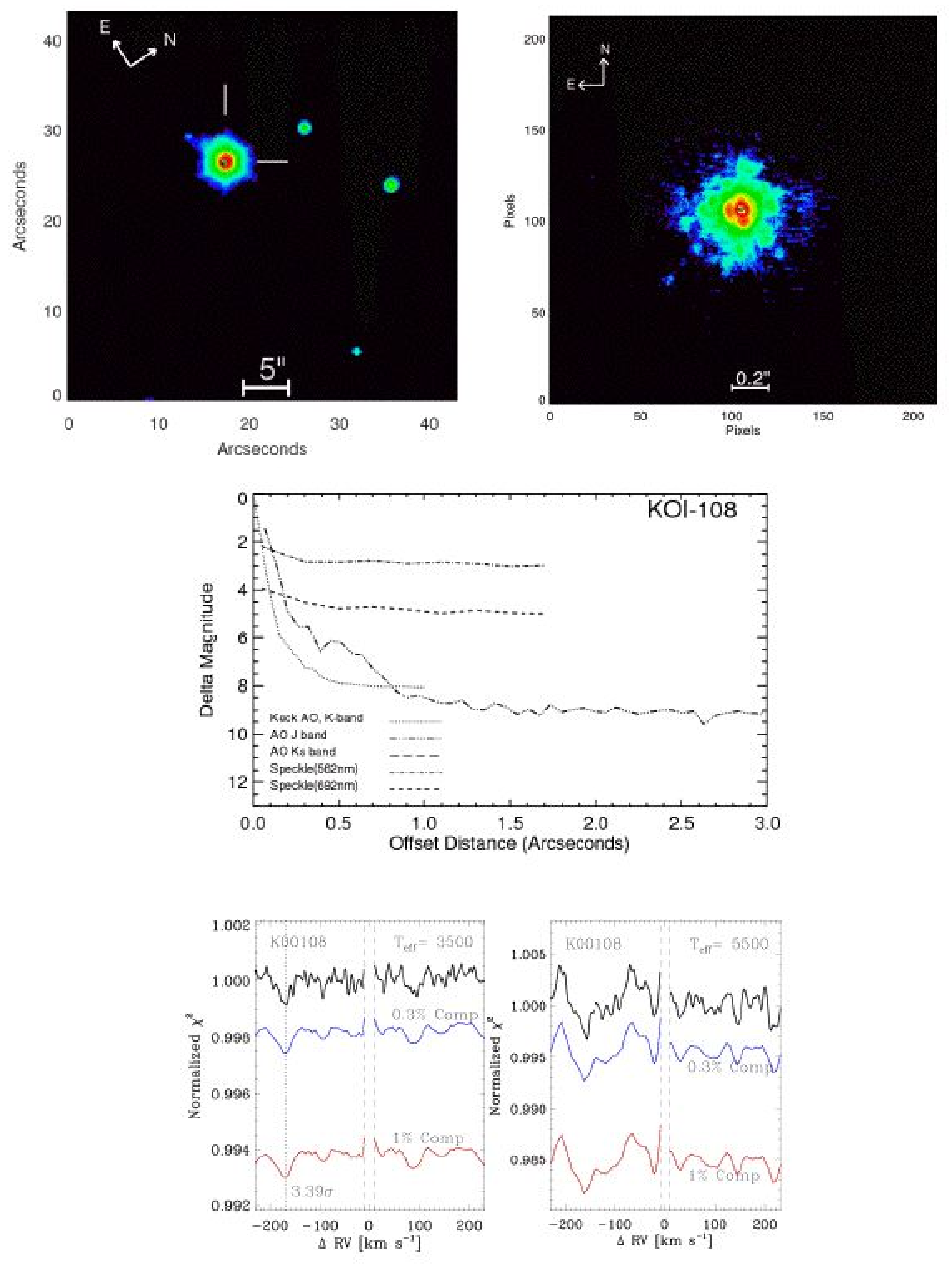}
\caption{ Upper left) Seeing limited image for Kepler-103 (KOI-108). Upper right) Adaptive Optics image. Middle) Limiting magnitudes of companion stars found with Speckle and AO imaging. Lower left and lower right)  Same as Figure 1, but for Kepler-103. No companions are detected. Any companion brighter than 0.3\% the brightness(V-band) of the primary would have been detected.
}
\label{fig:koi108_fig1}
\end{figure*}

\begin{figure*}     
\epsscale{1.0}
\plotone{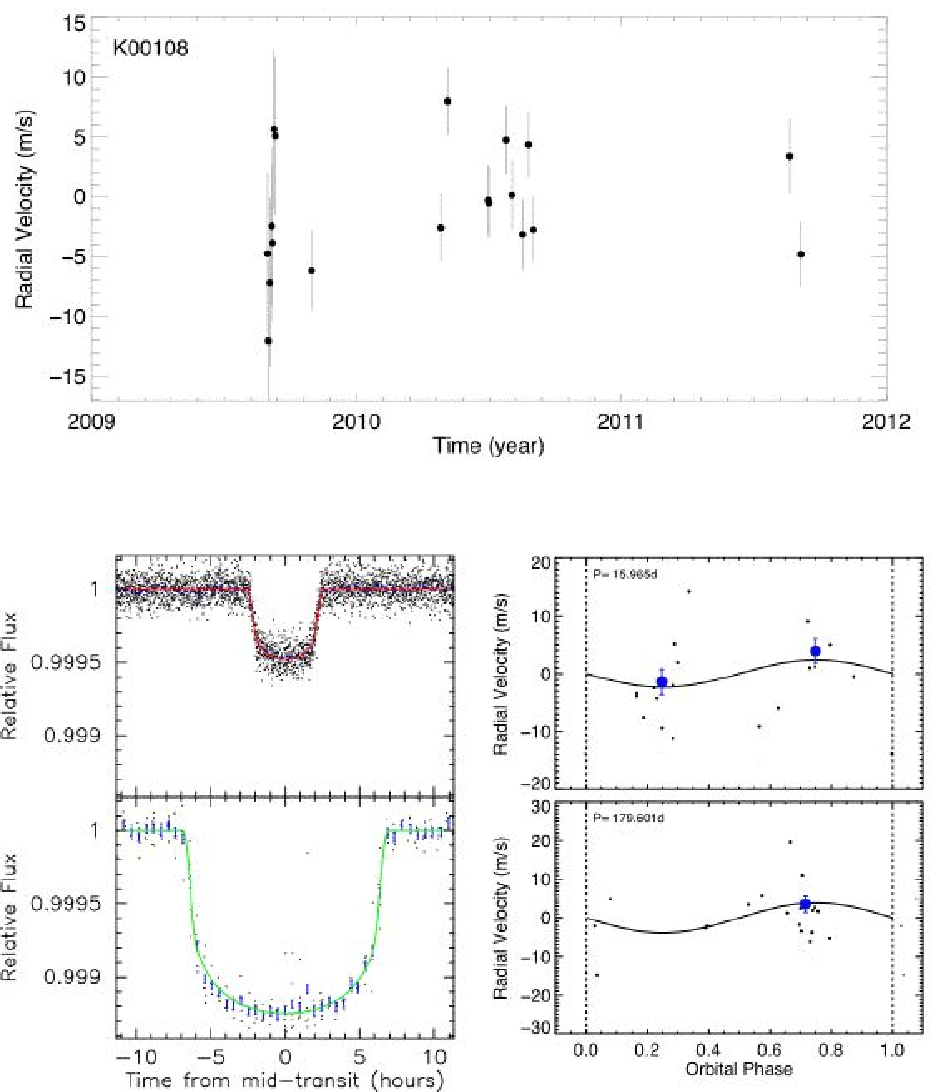}
\caption{ Top) RV versus time for Kepler-103 (KOI-108). Lower left) Phase folded \ek 
photometry for each planet. Lower Right) Phase folded radial velocities for each planet. 
The blue points represent binned RVs near quadrature, same as Figure  \ref{fig:koi41_fig2}.
Kepler-103b: Rp = $3.37 \pm 0.09$ \rearthe, Mp = $9.7 \pm 8.6$\mearthe.
Kepler-103c: Rp = $5.14 \pm 0.14$ \rearthe, Mp = $36.1 \pm 25.2$\mearthe.
}
\label{fig:koi108_fig2}
\end{figure*}

\begin{figure*}   
\epsscale{1.0}
\plotone{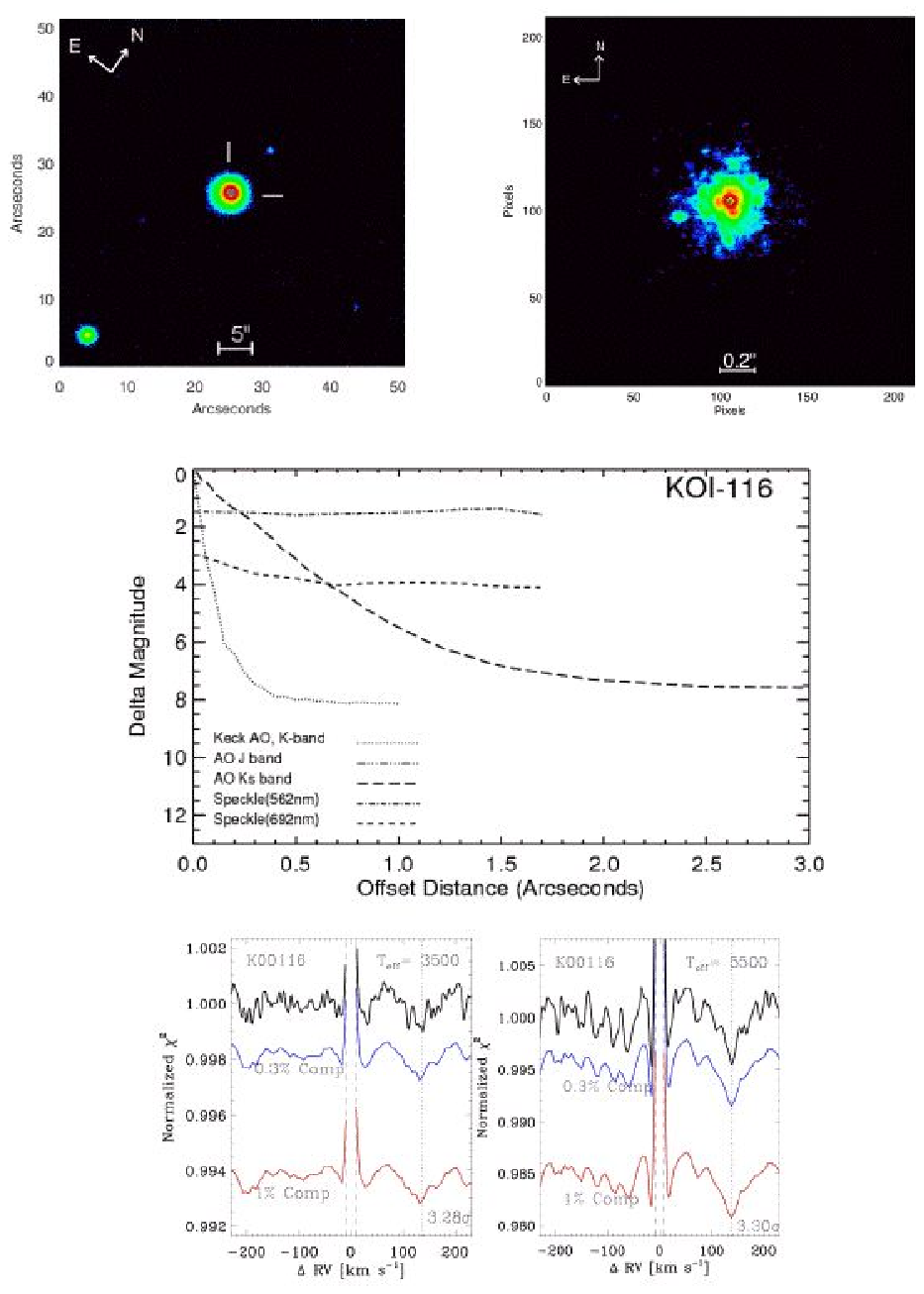}
\caption{ Upper left) Seeing limited image for Kepler-106 (KOI-116). Upper right) Adaptive Optics image. Middle) Limiting magnitudes of companion stars found with Speckle and AO imaging. Lower left and lower right)  Same as Figure 1, but for Kepler-106. No companions are detected. Any companion brighter than 0.3\% the brightness(V-band) of the primary would have been detected.
}
\label{fig:koi116_fig1}
\end{figure*}

\begin{figure*}  
\epsscale{1.0}
\plotone{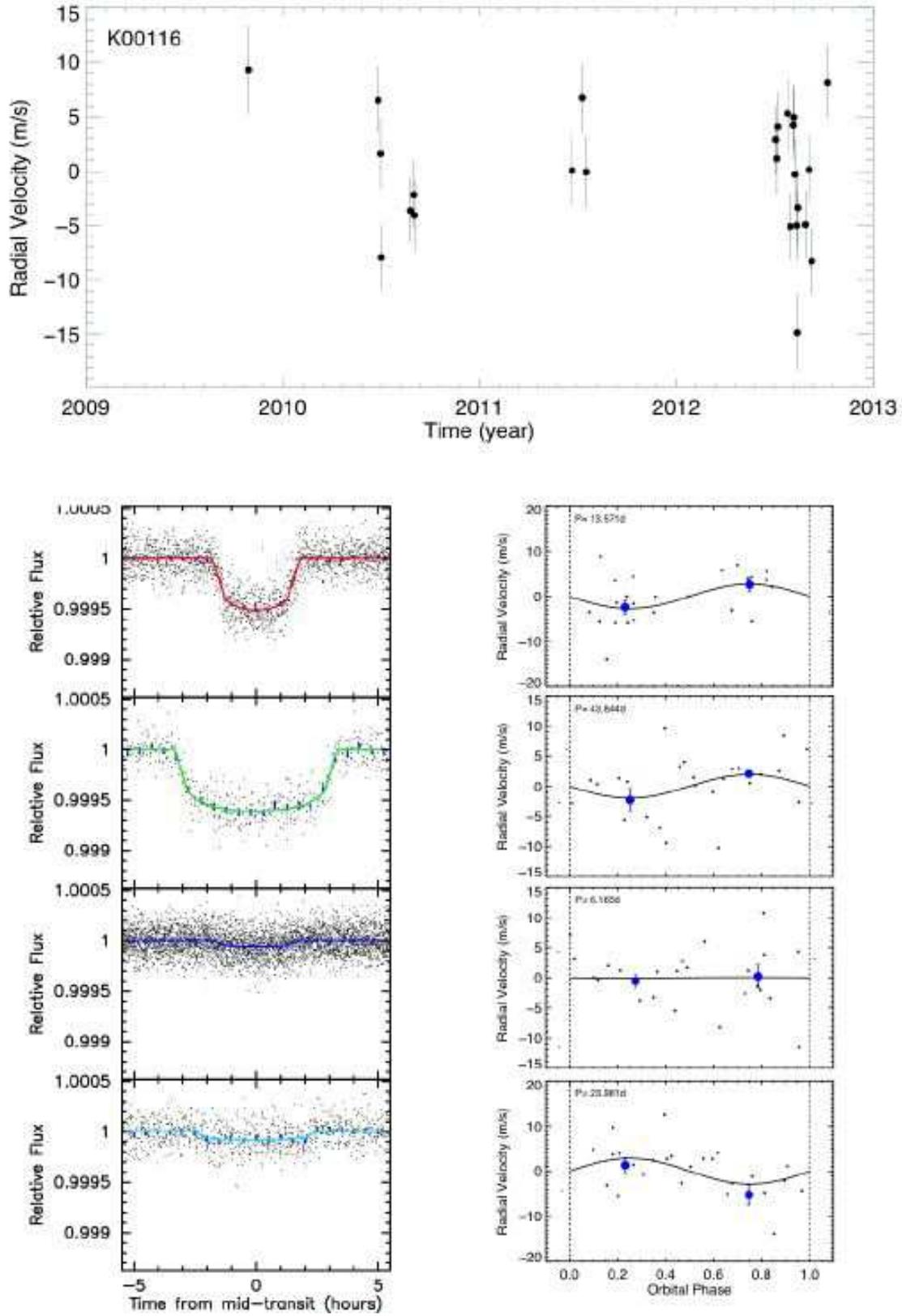}
\caption{ Top) RV versus time for Kepler-106 (KOI-116). Lower left) Phase folded \ek 
photometry for each planet. Lower Right) Phase folded radial velocities for each planet. 
The blue points represent binned RVs near quadrature, same as Figure  \ref{fig:koi41_fig2}.
Kepler-106c: Rp = $2.50 \pm 0.32$ \rearthe, Mp = $10.44 \pm 3.2$\mearthe.
Kepler-106e: Rp = $2.56 \pm 0.33$ \rearthe, Mp = $11.17 \pm 5.8$\mearthe.
Kepler-106b: Rp = $0.82 \pm 0.11$ \rearthe, Mp = $0.15 \pm 2.8$ \mearthe.
Kepler-106d: Rp = $0.95 \pm 0.13$ \rearthe, Mp = $-6.39 \pm 7.0$\mearthe.
}
\label{fig:koi116_fig2}
\end{figure*}

\begin{figure*}   
\epsscale{1.0}
\plotone{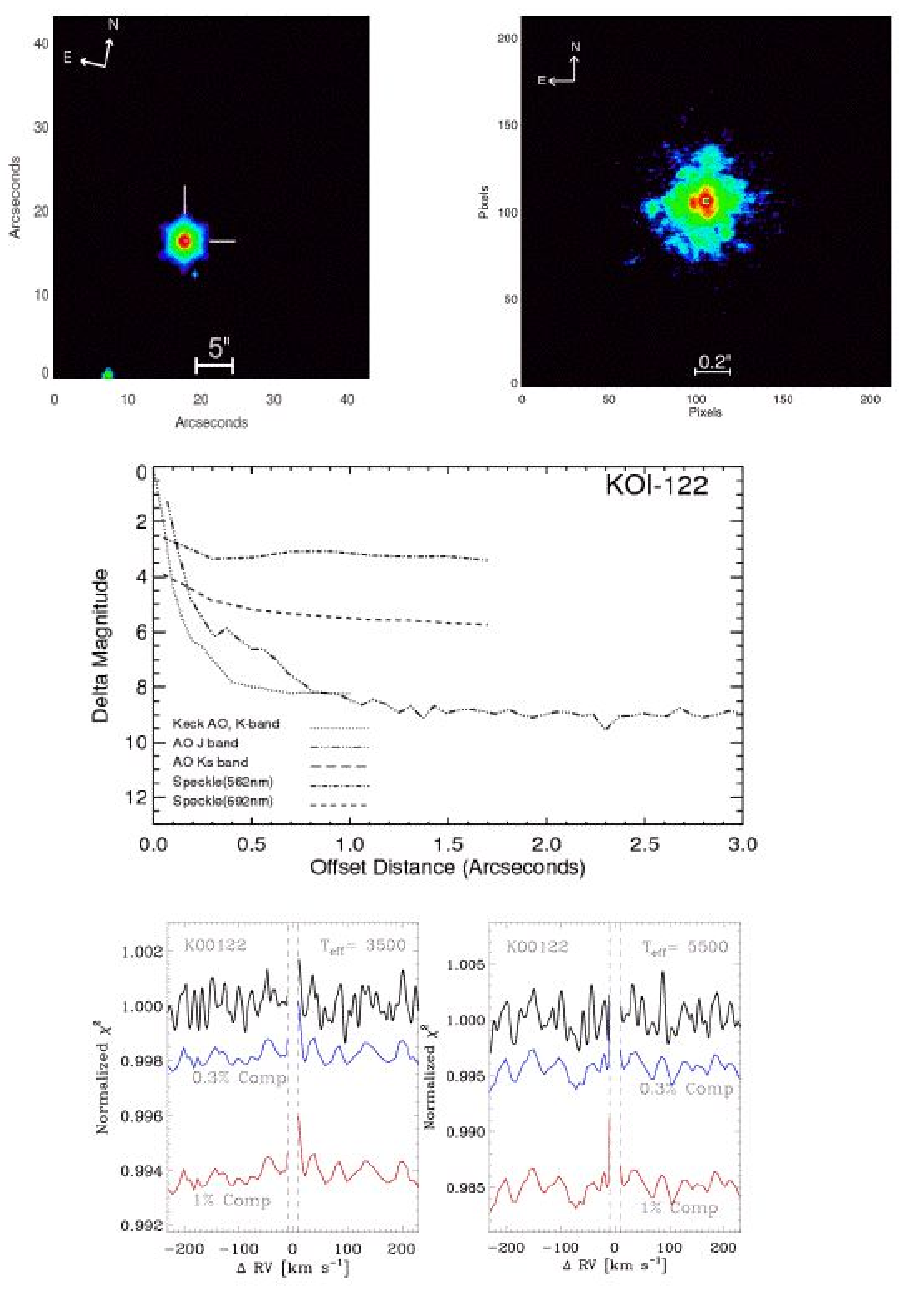}
\caption{ Upper left) Seeing limited image for Kepler 95 (KOI-122). Upper right) Adaptive Optics image. Middle) Limiting magnitudes of companion stars found with Speckle and AO imaging. Lower left and lower right)  Same as Figure 1, but for Kepler 95. No companions are detected. Any companion brighter than 0.3\% the brightness(V-band) of the primary would have been detected.
}
\label{fig:koi122_fig1}
\end{figure*}

\begin{figure*} 
\epsscale{1.0}
\plotone{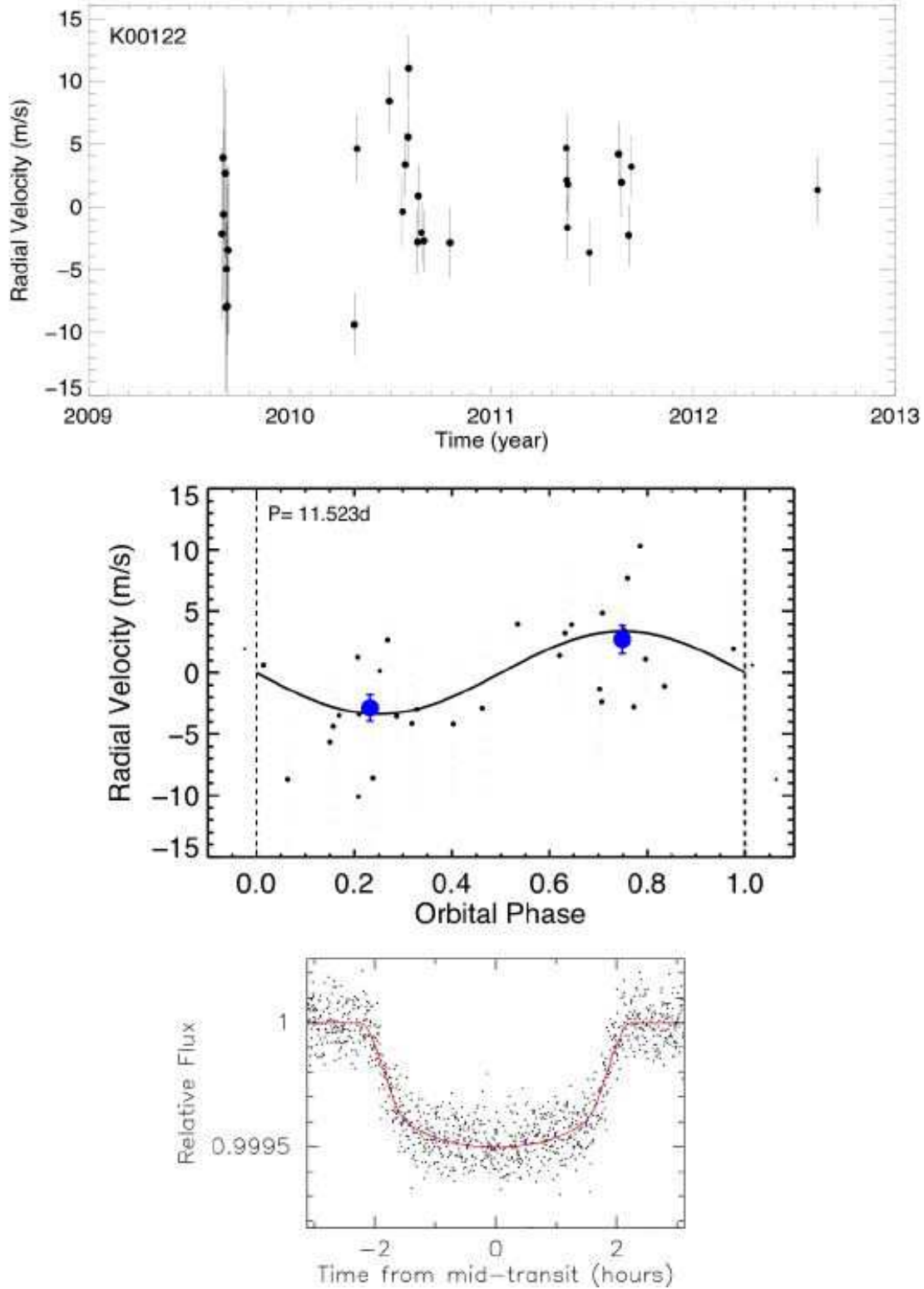}
\caption{ Top) RV versus time for KOI-122. Lower left) Phase folded \ek 
photometry for each planet. Lower Right) Phase folded radial velocities for each planet. 
The blue points represent binned RVs near quadrature, same as Figure  \ref{fig:koi41_fig2}.
Kepler 95b: Rp = $3.42 \pm 0.09$ \rearthe, Mp = $13.0 \pm 2.9$ \mearthe.
}
\label{fig:koi122_fig2}
\end{figure*}

\clearpage
\begin{figure*}   
\epsscale{1.0}
\plotone{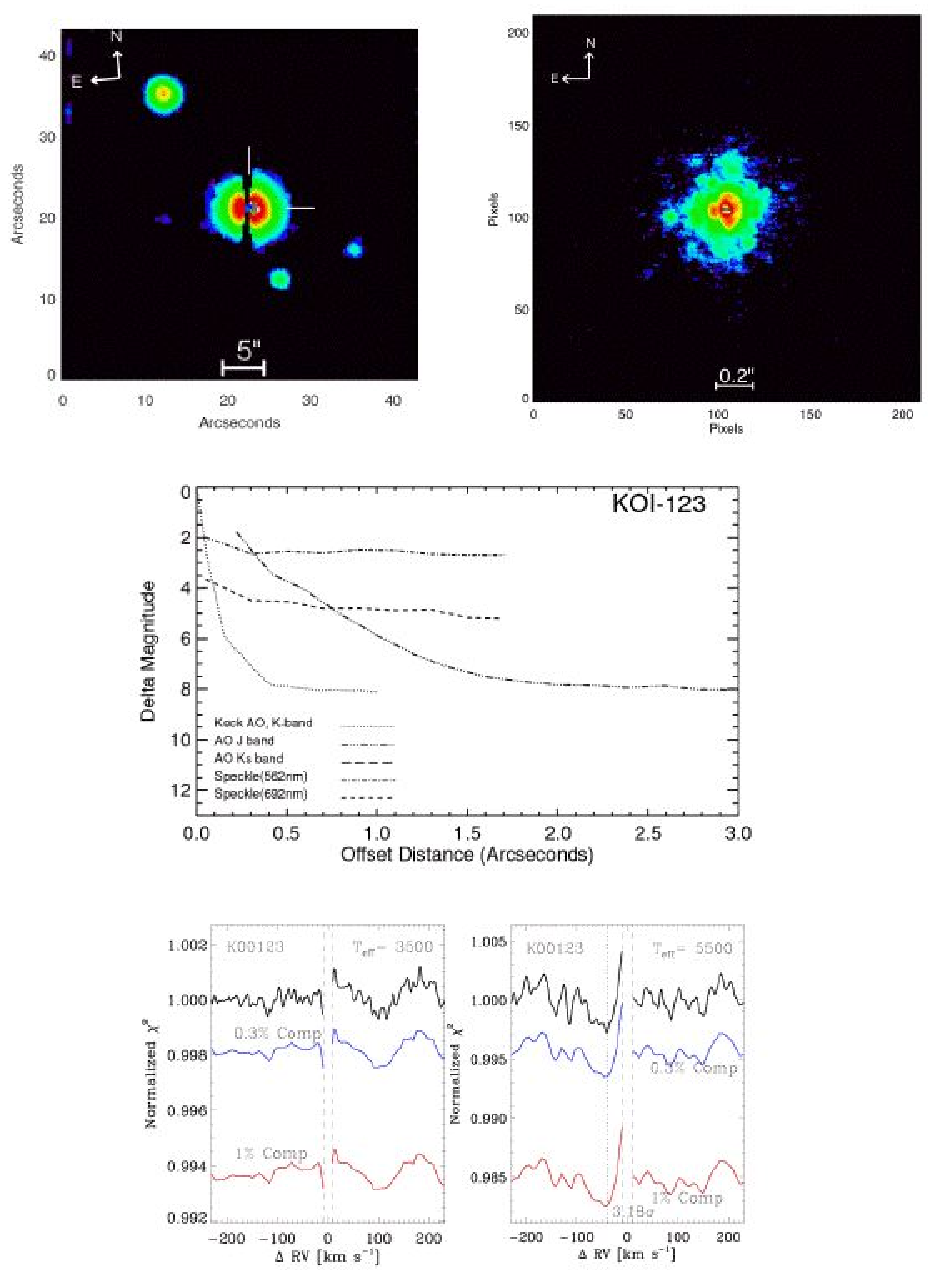}
\caption{ Upper left) Seeing limited image for Kepler-109 (KOI-123). Upper right) Adaptive Optics image. Middle) Limiting magnitudes of companion stars found with Speckle and AO imaging. Lower left and lower right)  Same as Figure 1, but for Kepler-109. No companions are detected. Any companion brighter than 0.3\% the brightness(V-band) of the primary would have been detected.
}
\label{fig:koi123_fig1}
\end{figure*}

\begin{figure*}    
\epsscale{1.0}
\plotone{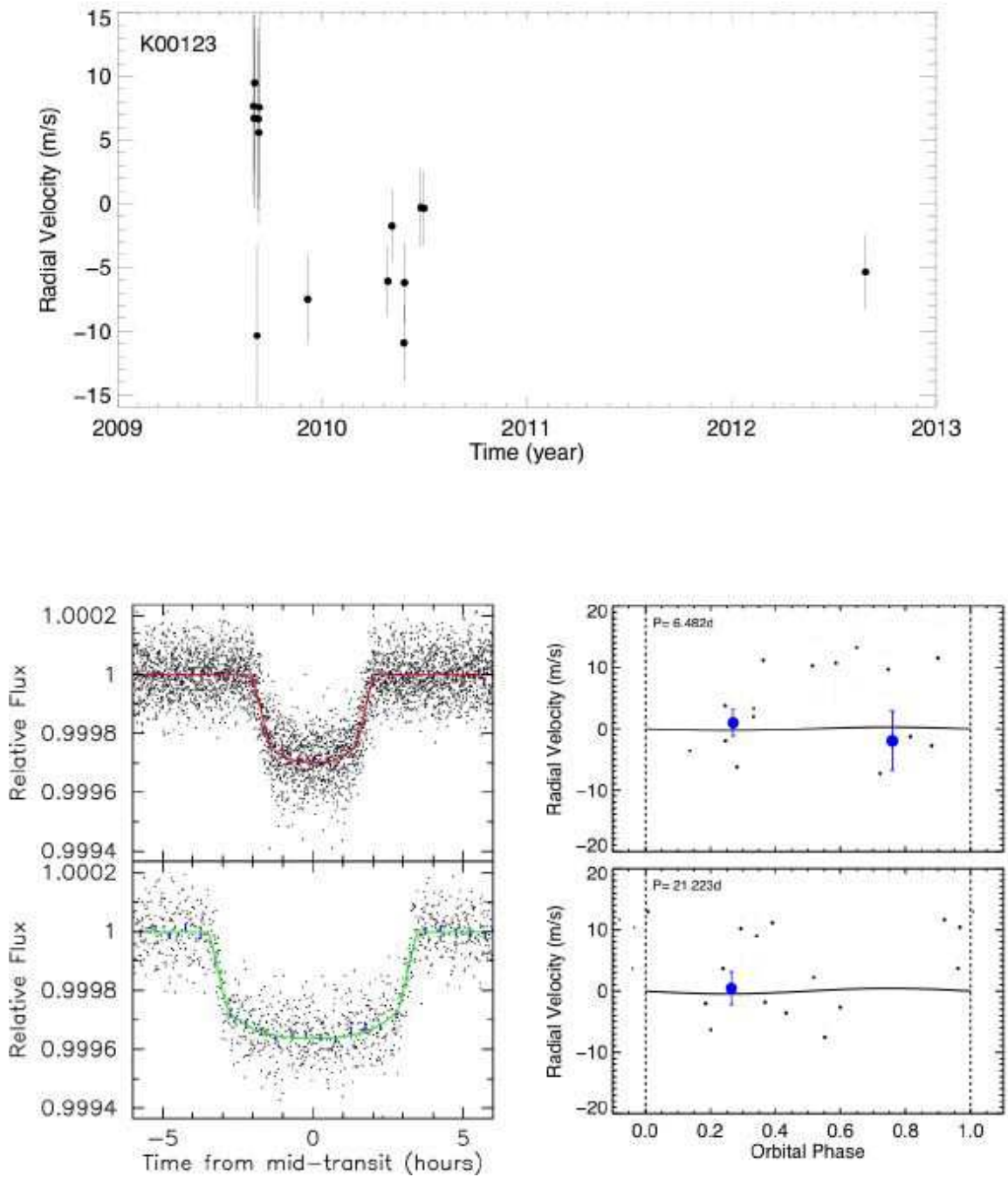}
\caption{ Top) RV versus time for Kepler-109 (KOI-123). Lower left) Phase folded \ek 
photometry for each planet. Lower Right) Phase folded radial velocities for each planet. 
The blue points represent binned RVs near quadrature, same as Figure  \ref{fig:koi41_fig2}.
Kepler-109b: Rp = $2.37 \pm 0.07$ \rearthe, Mp = $1.3 \pm 5.4$ \mearthe.
Kepler-109c: Rp = $2.52 \pm 0.07$ \rearthe, Mp = $2.22 \pm 7.8$ \mearthe.
}
\label{fig:koi123_fig2}
\end{figure*}

\begin{figure*}   
\epsscale{1.0}
\plotone{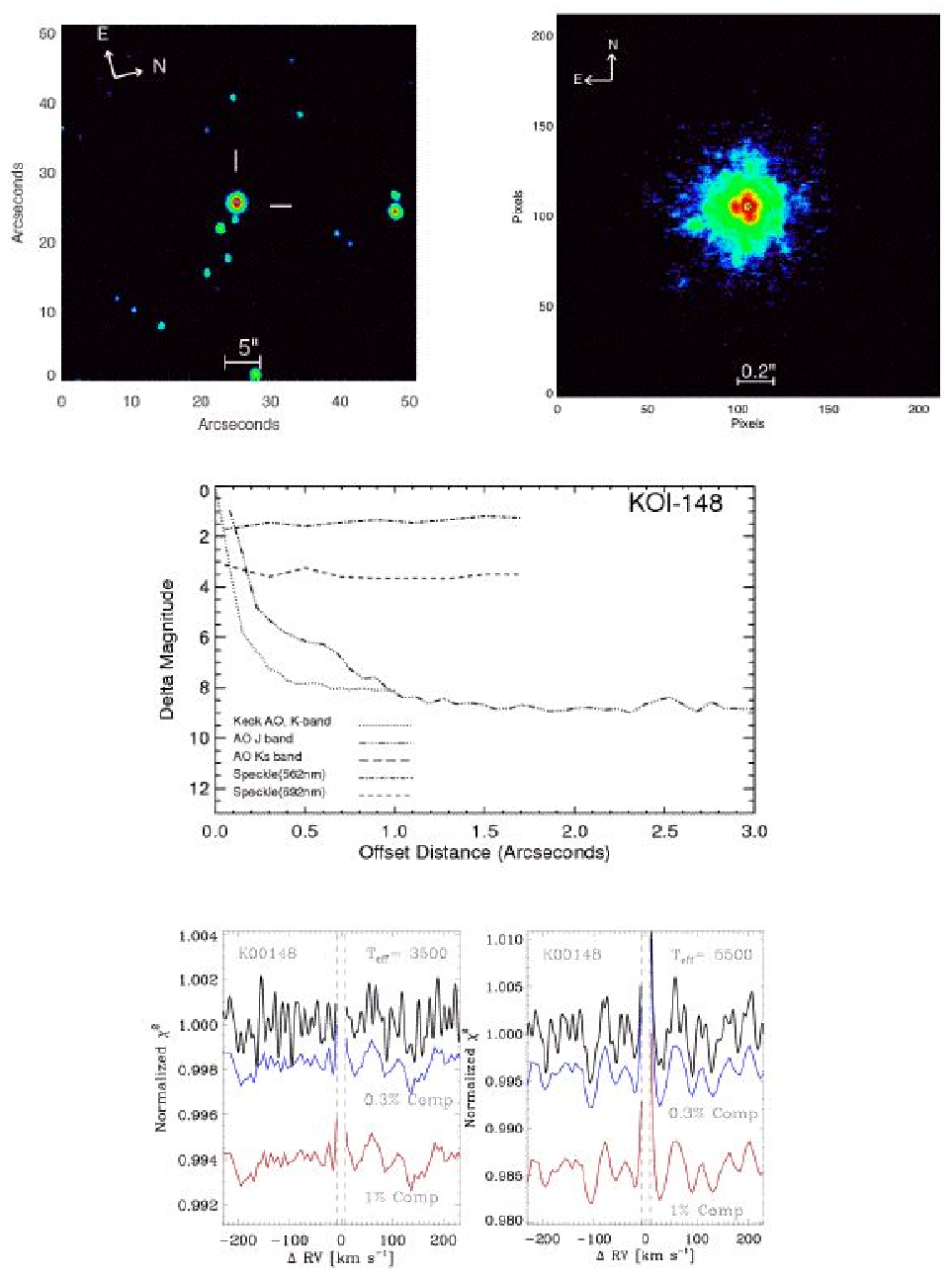}
\caption{ Upper left) Seeing limited image for Kepler-48 (KOI-148). Upper right) Adaptive Optics image. Middle) Limiting magnitudes of companion stars found with Speckle and AO imaging. Lower left and lower right)  Same as Figure 1, but for Kepler-48. No companions are detected. Any companion brighter than 0.5\% the brightness(V-band) of the primary would have been detected.
}
\label{fig:koi148_fig1}
\end{figure*}

\begin{figure*}  
\epsscale{1.0}
\plotone{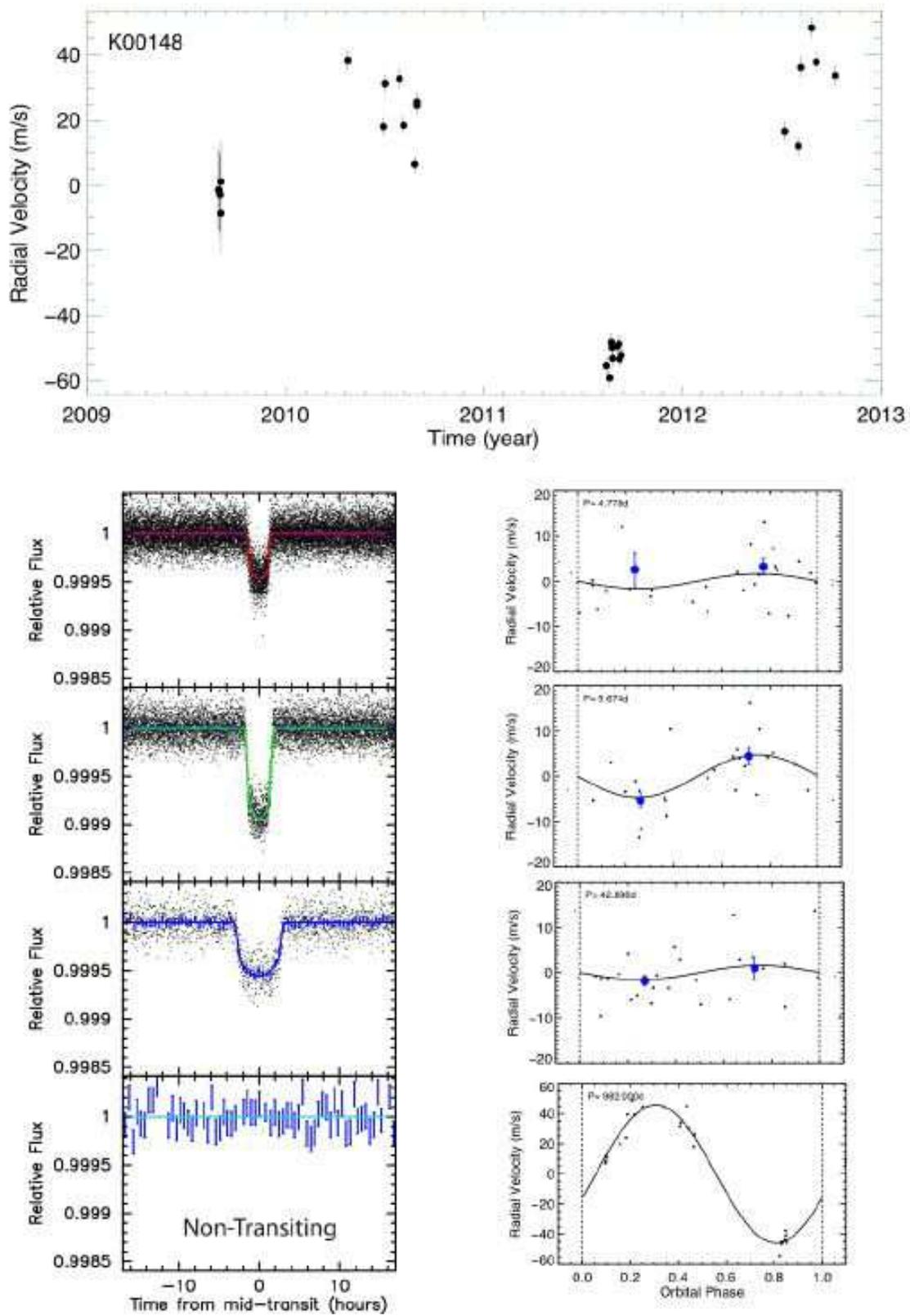}
\caption{ Top) RV versus time for Kepler-48 (KOI-148). Lower left) Phase folded \ek 
photometry for each planet. Lower Right) Phase folded radial velocities for each planet. 
The blue points represent binned RVs near quadrature, same as Figure  \ref{fig:koi41_fig2}.
Kepler-48b: Rp = $1.88 \pm 0.10$ \rearthe, Mp = $3.94 \pm 2.1$ \mearthe.
Kepler-48c: Rp = $2.71 \pm 0.14$ \rearthe, Mp = $14.61 \pm 2.3$\mearthe.
Kepler-48d: Rp = $2.04 \pm 0.11$ \rearthe, Mp = $7.93 \pm 4.6$ \mearthe.
Kepler-48e: Rp = NA, Mp = $657 \pm 25$  \mearthe, Period = $982 \pm 8$d.
}
\label{fig:koi148_fig2}
\end{figure*}

\begin{figure*}   
\epsscale{1.0}
\plotone{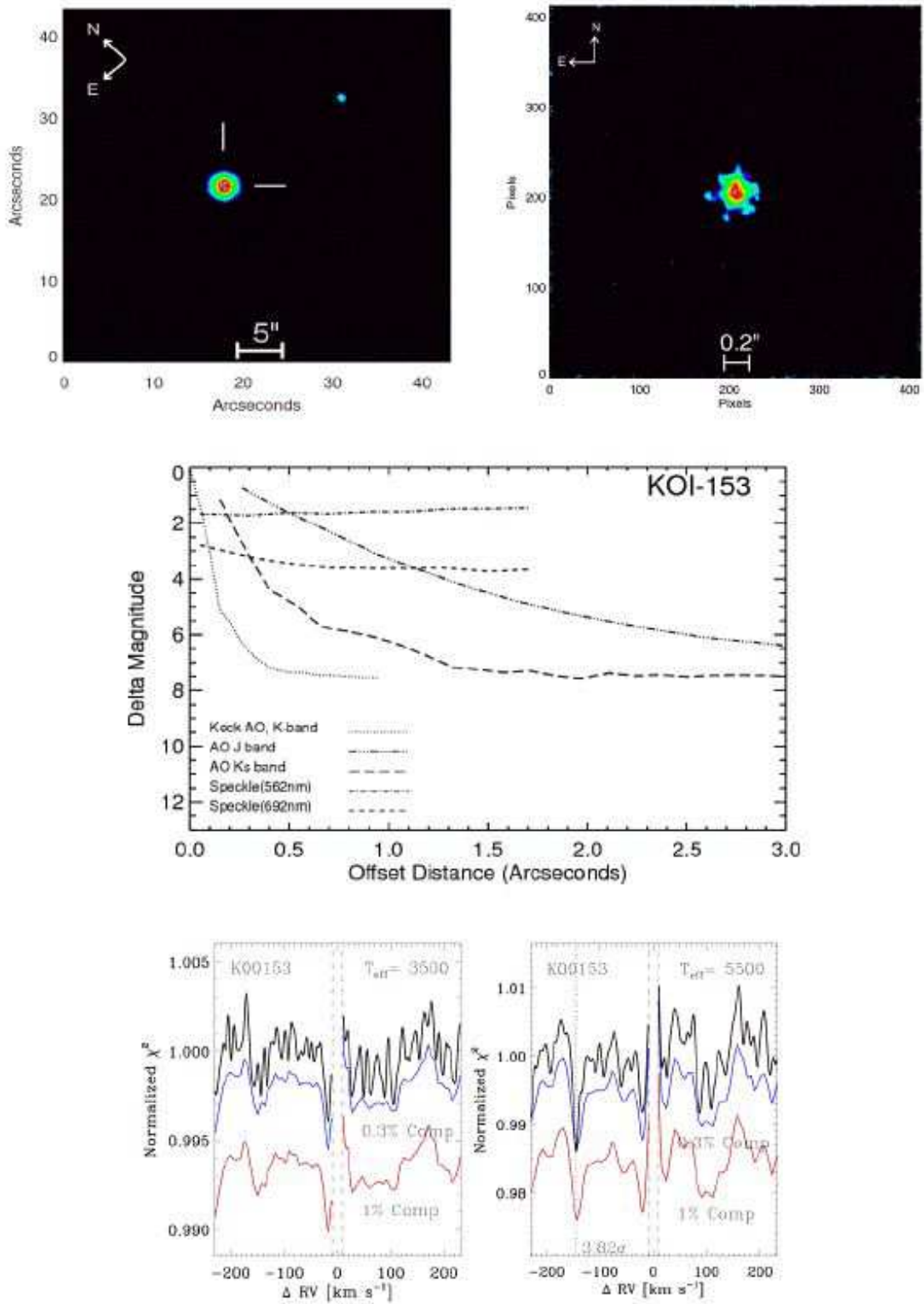}
\caption{ Upper left) Seeing limited image for Kepler-113 (KOI-153). Upper right) Keck-NIRC2 adaptive Optics image.  Middle) Limiting magnitudes of companion stars found with Speckle and AO imaging. Lower left and lower right)  Same spectroscopic search for secondary lines as described in Figure 1, but for Kepler-113. No companions are detected. Any companion brighter than 0.5\% the brightness(V-band) of the primary would have been detected.  
}
\label{fig:koi153_fig1}
\end{figure*}

\begin{figure*}     
\epsscale{1.0}
\plotone{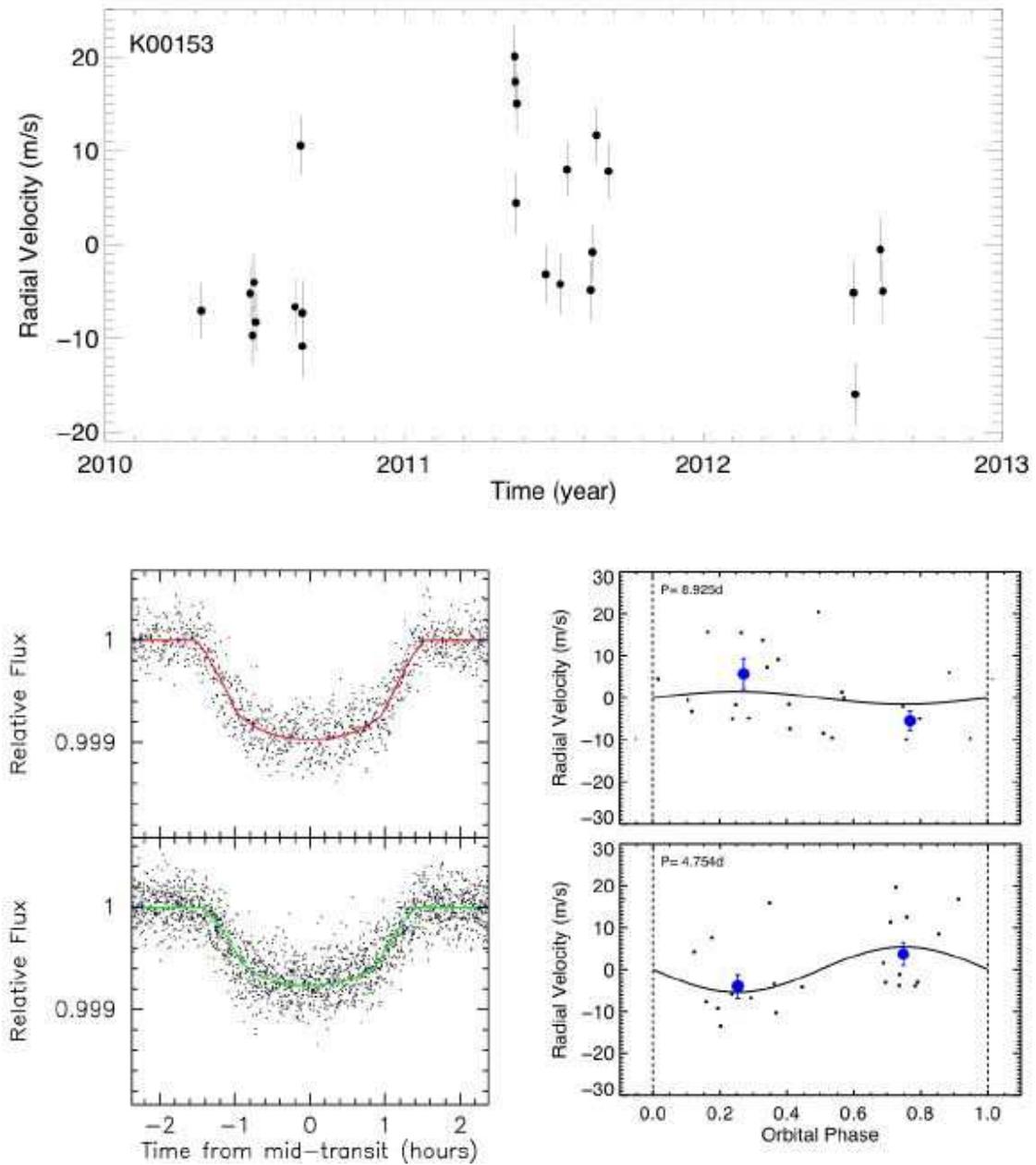}
\caption{ Top) RV versus time for Kepler-113 (KOI-153). Lower left) Phase folded \ek 
photometry for each planet. Lower Right) Phase folded radial velocities for each planet. 
The blue points represent binned RVs near quadrature, same as Figure  \ref{fig:koi41_fig2}.
Kepler-113c: Rp = $2.18 \pm 0.06$ \rearthe, Mp = $-4.04 \pm 6.4$ \mearthe.
Kepler-113b: Rp = $1.82 \pm 0.05$ \rearthe, Mp = $11.7 \pm 4.2$ \mearthe.
}
\label{fig:koi153_fig2}
\end{figure*}

\begin{figure*}   
\epsscale{1.0}
\plotone{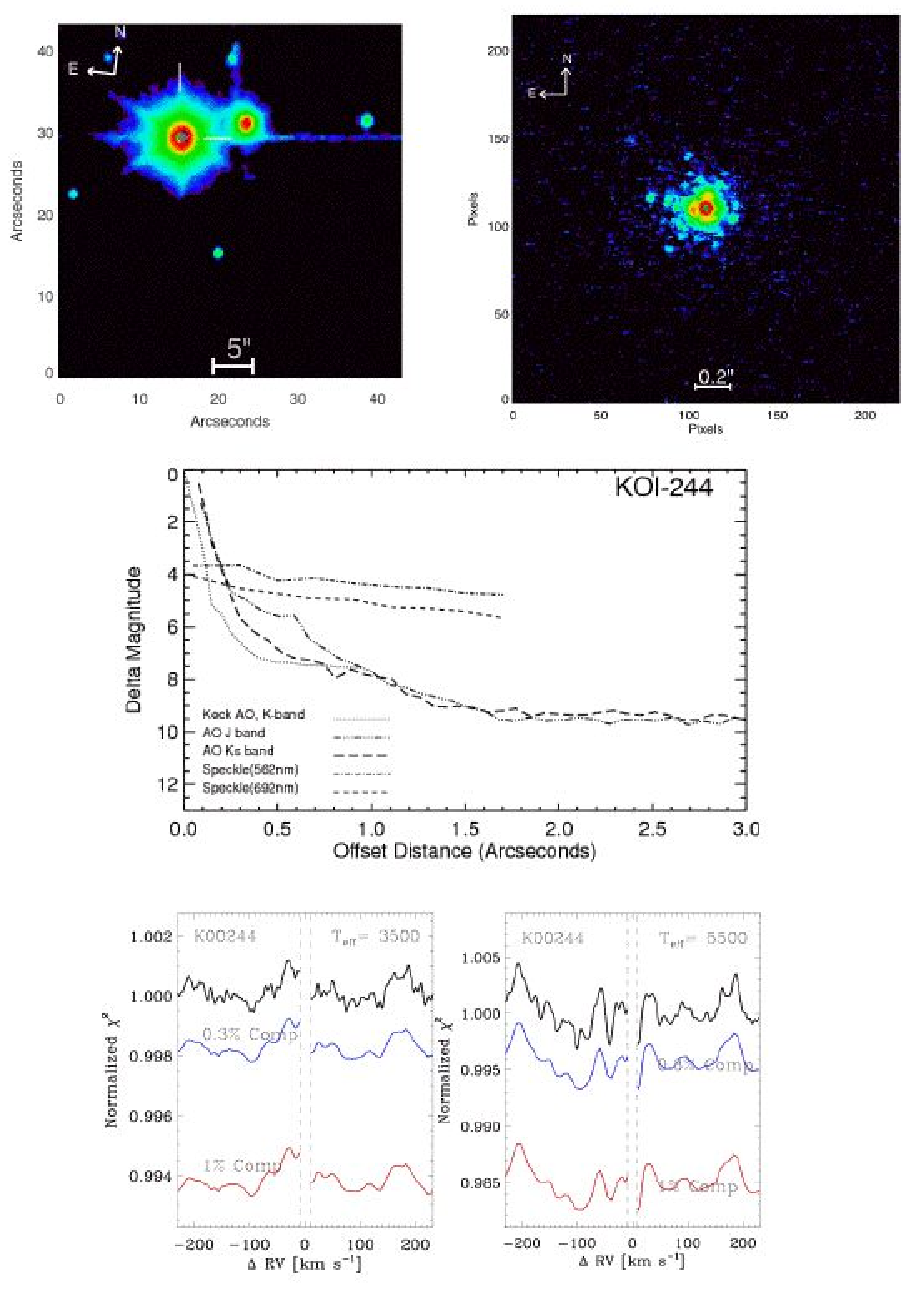}
\caption{ Upper left) Seeing limited image for Kepler-25 (KOI-244). Upper right) Adaptive Optics image. Middle) Limiting magnitudes of companion stars found with Speckle and AO imaging. Lower left and lower right)  Same as Figure 1, but for Kepler-25. No companions are detected. Any companion brighter than 0.3\% the brightness(V-band) of the primary would have been detected.
}
\label{fig:koi244_fig1}
\end{figure*}

\begin{figure*}   
\epsscale{1.0}
\plotone{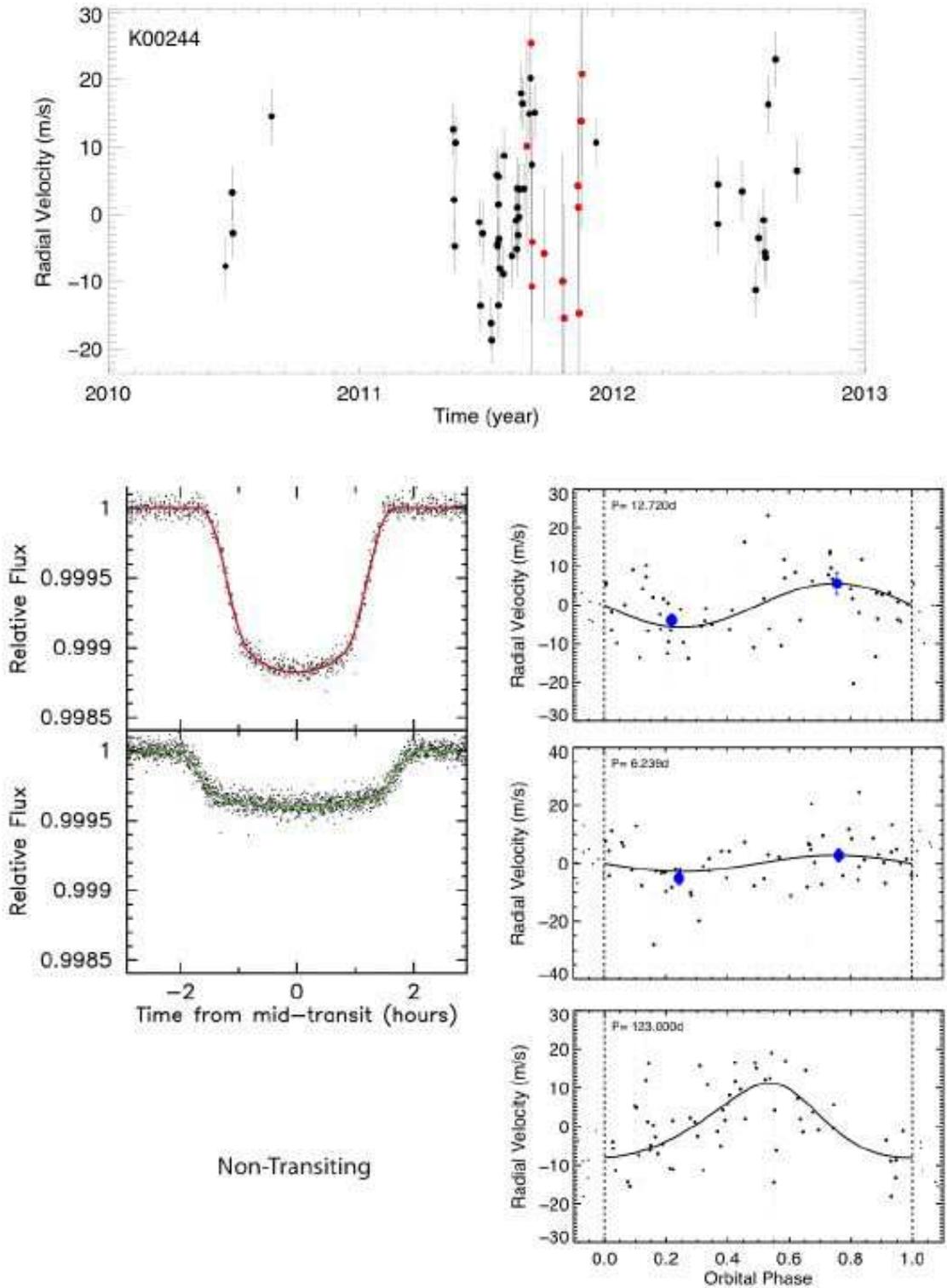}
\caption{ Top) RV versus time for Kepler-25 (KOI-244). The red points
  are from the SOPHIE spectrometer at Obs. de Haute-Provence. Lower left) Phase folded \ek 
photometry for each planet. The non-transiting planet has no light curve. Lower Right) Phase folded radial velocities for each planet. 
The blue points represent binned RVs near quadrature, same as Figure  \ref{fig:koi41_fig2}.
Kepler-25c: Rp = $5.20 \pm 0.09$ \rearthe, Mp = $24.60 \pm 5.7$\mearthe.
Kepler-25b: Rp = $2.71 \pm 0.05$ \rearthe, Mp = $9.6 \pm 4.2$ \mearthe.
Kepler-25d: Rp = NA, Mp = $89.90 \pm 13.7$\mearthe, period = $123\pm 2$ d. The 123d period has an alias at 93 days. Each period provides an equally good fit.
}
\label{fig:koi244_fig2}
\end{figure*}

\begin{figure*}   
\epsscale{1.0}
\plotone{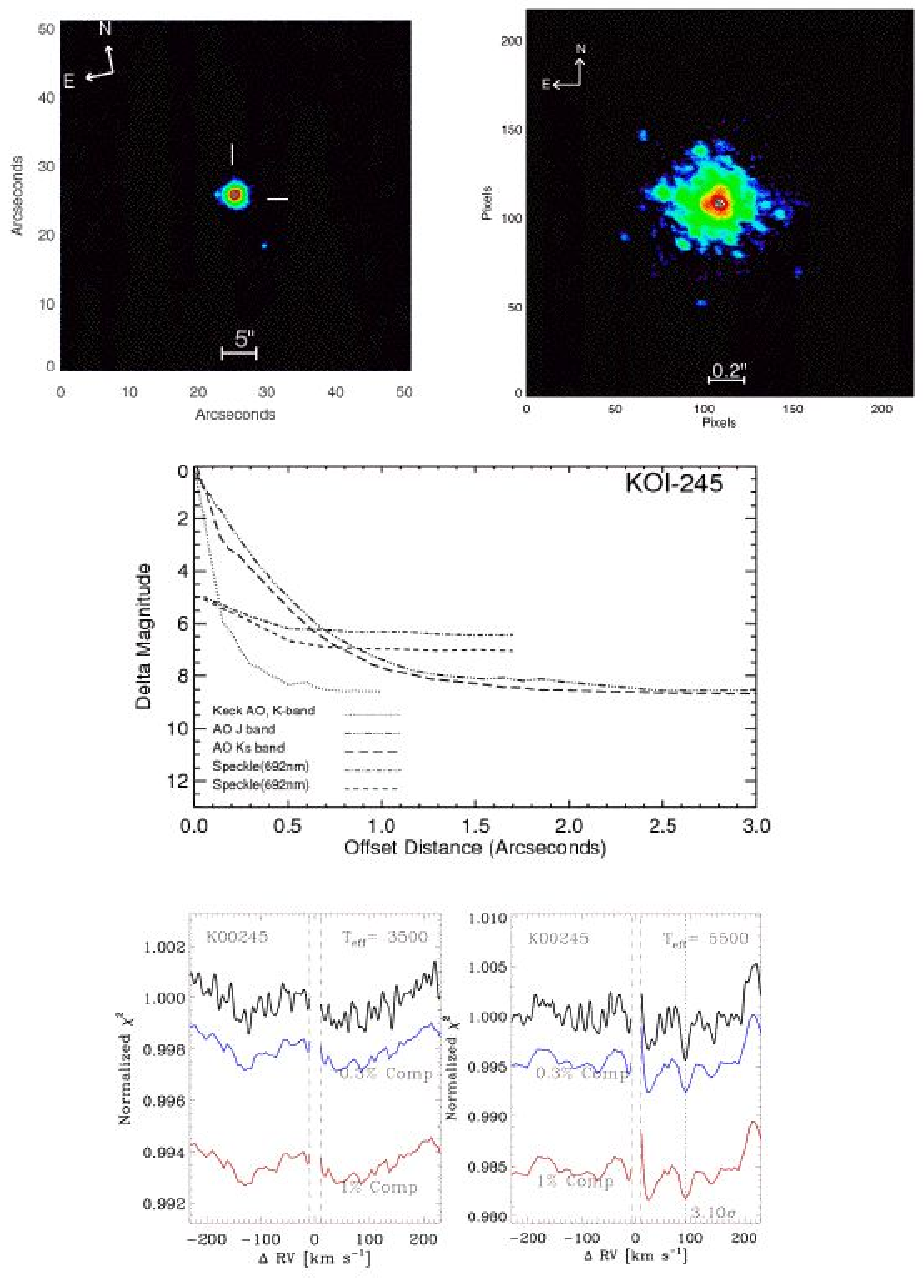}
\caption{ Upper left) Seeing limited image for Kepler-37 (KOI-245). Upper right) Adaptive Optics image. Middle) Limiting magnitudes of companion stars found with Speckle and AO imaging. Lower left and lower right)  Same as Figure 1, but for Kepler-37. No companions are detected. Any companion brighter than 0.3\% the brightness(V-band) of the primary would have been detected.
}
\label{fig:koi245_fig1}
\end{figure*}

\begin{figure*}   
\epsscale{1.0}
\plotone{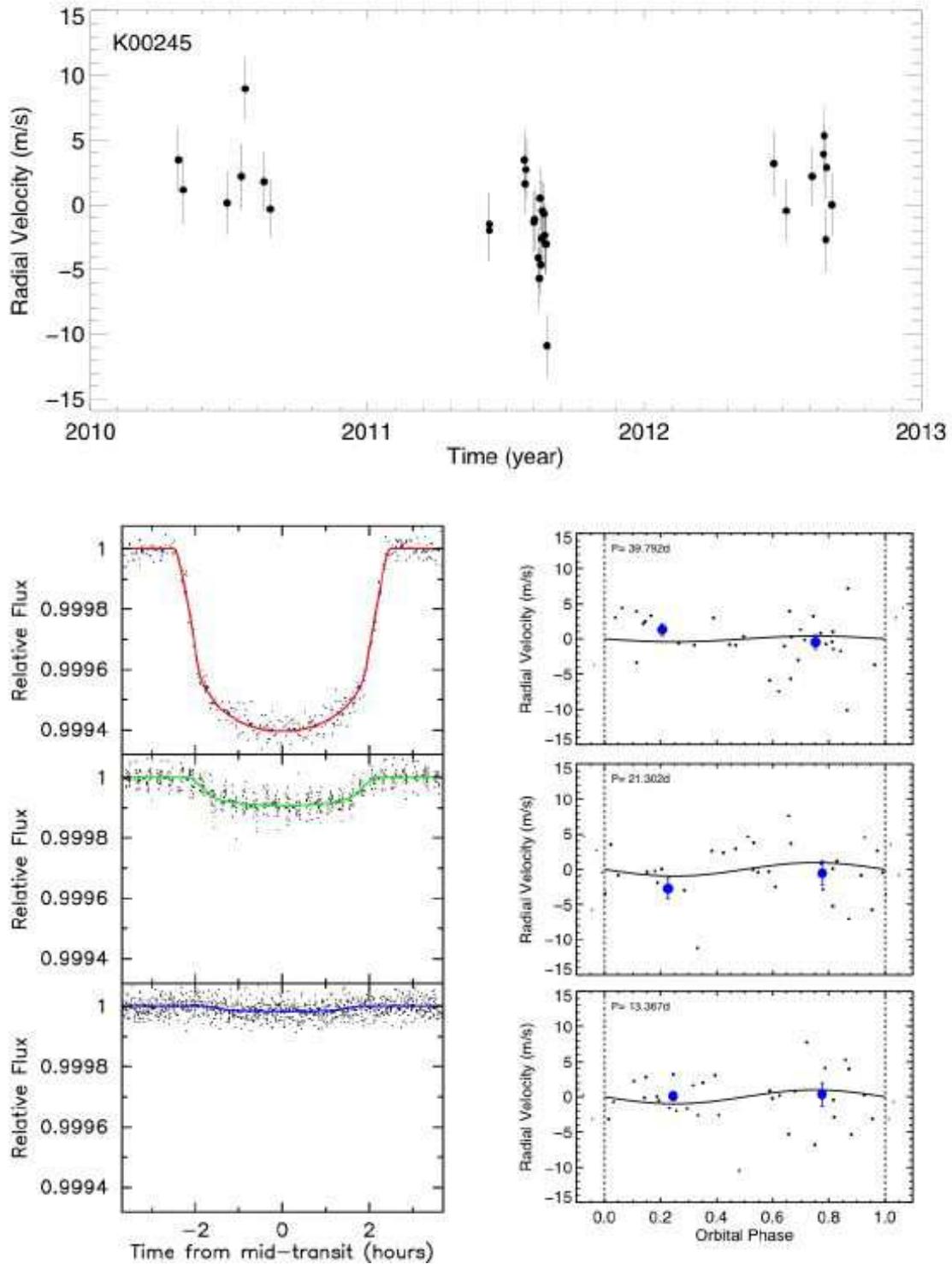}
\caption{ Top) RV versus time for Kepler-37 (KOI-245). Lower left) Phase folded \ek 
photometry for each planet. Lower Right) Phase folded radial velocities for each planet. 
The blue points represent binned RVs near quadrature, same as Figure  \ref{fig:koi41_fig2}.
Kepler-37d: Rp = $1.94 \pm 0.06$ \rearthe, Mp = $1.87 \pm 9.08$\mearthe.
Kepler-37c: Rp = $0.75 \pm 0.03$ \rearthe, Mp = $3.35 \pm 4.0$ \mearthe.
Kepler-37b: Rp = $0.32 \pm 0.02$ \rearthe, Mp = $2.78 \pm 3.7$\mearthe.
}
\label{fig:koi245_fig2}
\end{figure*}

\begin{figure*}   
\epsscale{1.0}
\plotone{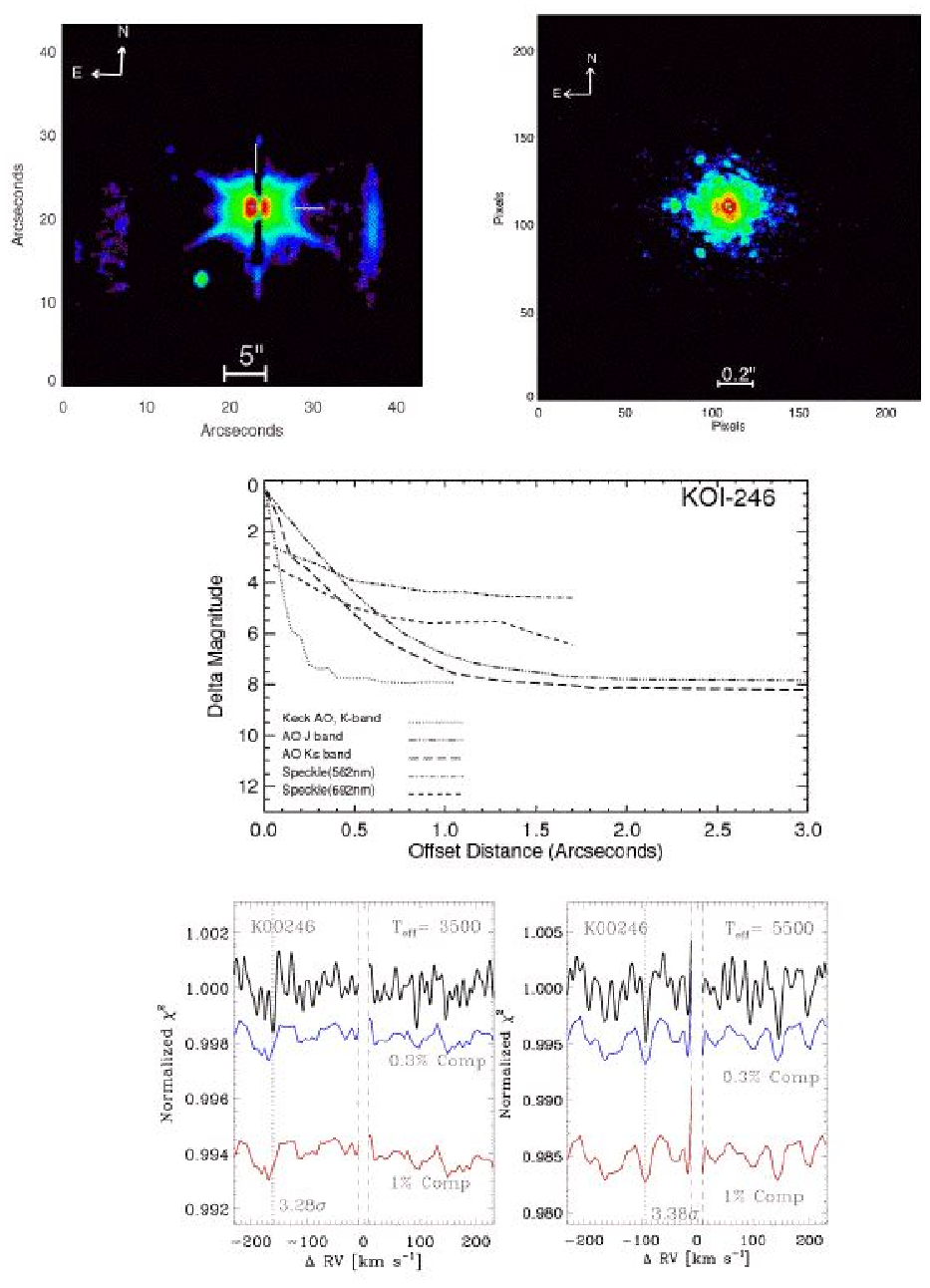}
\caption{ Upper left) Seeing limited image for Kepler-68 (KOI-246). Upper right) Adaptive Optics image. Middle) Limiting magnitudes of companion stars found with Speckle and AO imaging. Lower left and lower right)  Same as Figure 1, but for Kepler-68. No companions are detected. Any companion brighter than 0.5\% the brightness(V-band) of the primary would have been detected.
}
\label{fig:koi246_fig1}
\end{figure*}

\begin{figure*}   
\epsscale{1.0}
\plotone{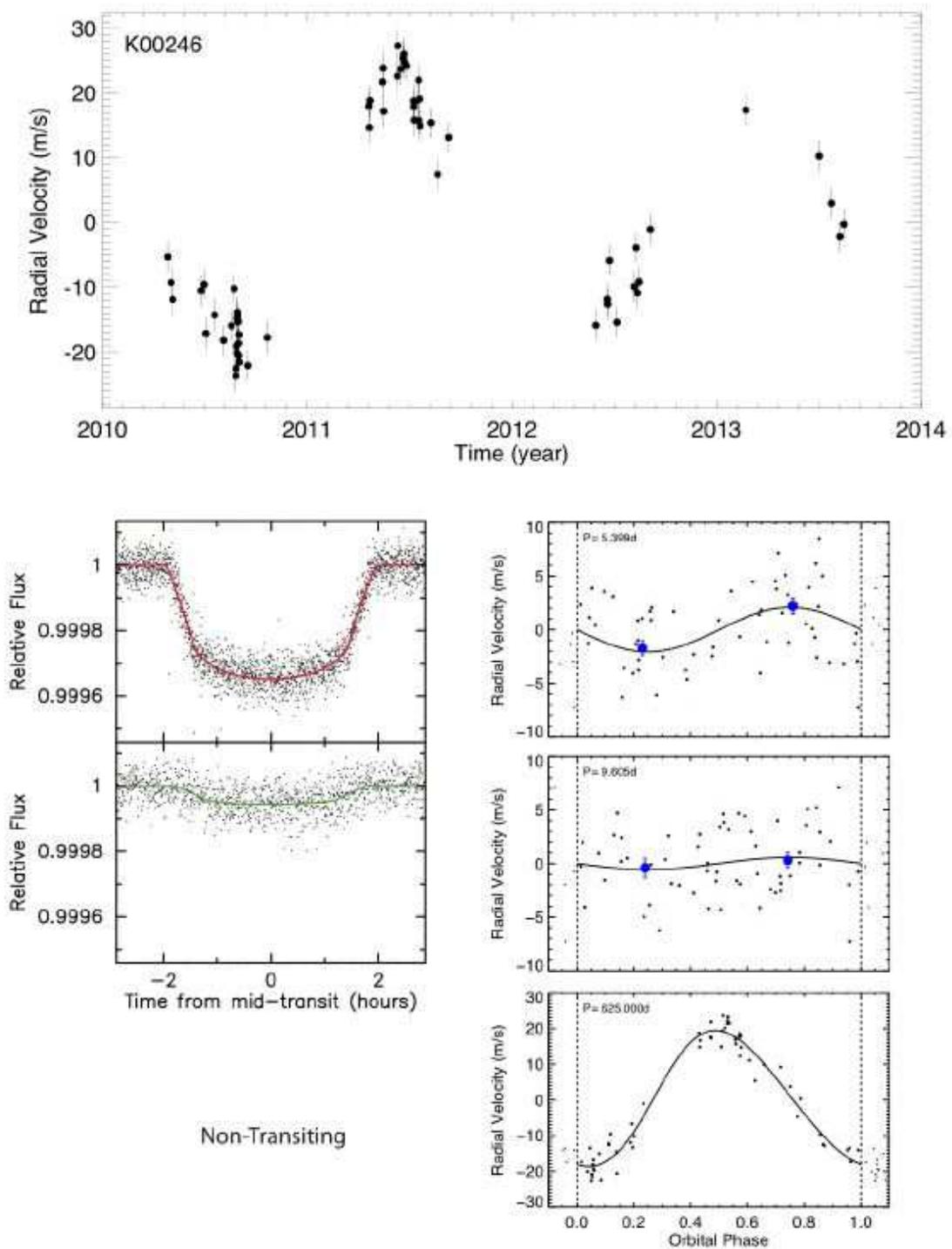}
\caption{ Top) RV versus time for Kepler-68 (KOI-246). Lower left) Phase folded \ek photometry for each planet. Lower Right) Phase folded radial velocities for each planet. 
The blue points represent binned RVs near quadrature, same as Figure  \ref{fig:koi41_fig2}.
Kepler-68b: Rp = $2.33 \pm 0.02$ \rearthe, Mp = $5.97 \pm 1.7$ \mearthe.
Kepler-68c: Rp = $1.00 \pm 0.02$ \rearthe, Mp = $2.18 \pm 3.5$ \mearthe.
Kepler-68d: Rp = NA, Mp = $267 \pm 16$  \mearthe, period = $625 \pm16$.
}
\label{fig:koi246_fig2}
\end{figure*}
\clearpage

\begin{figure*}   
\epsscale{1.0}
\plotone{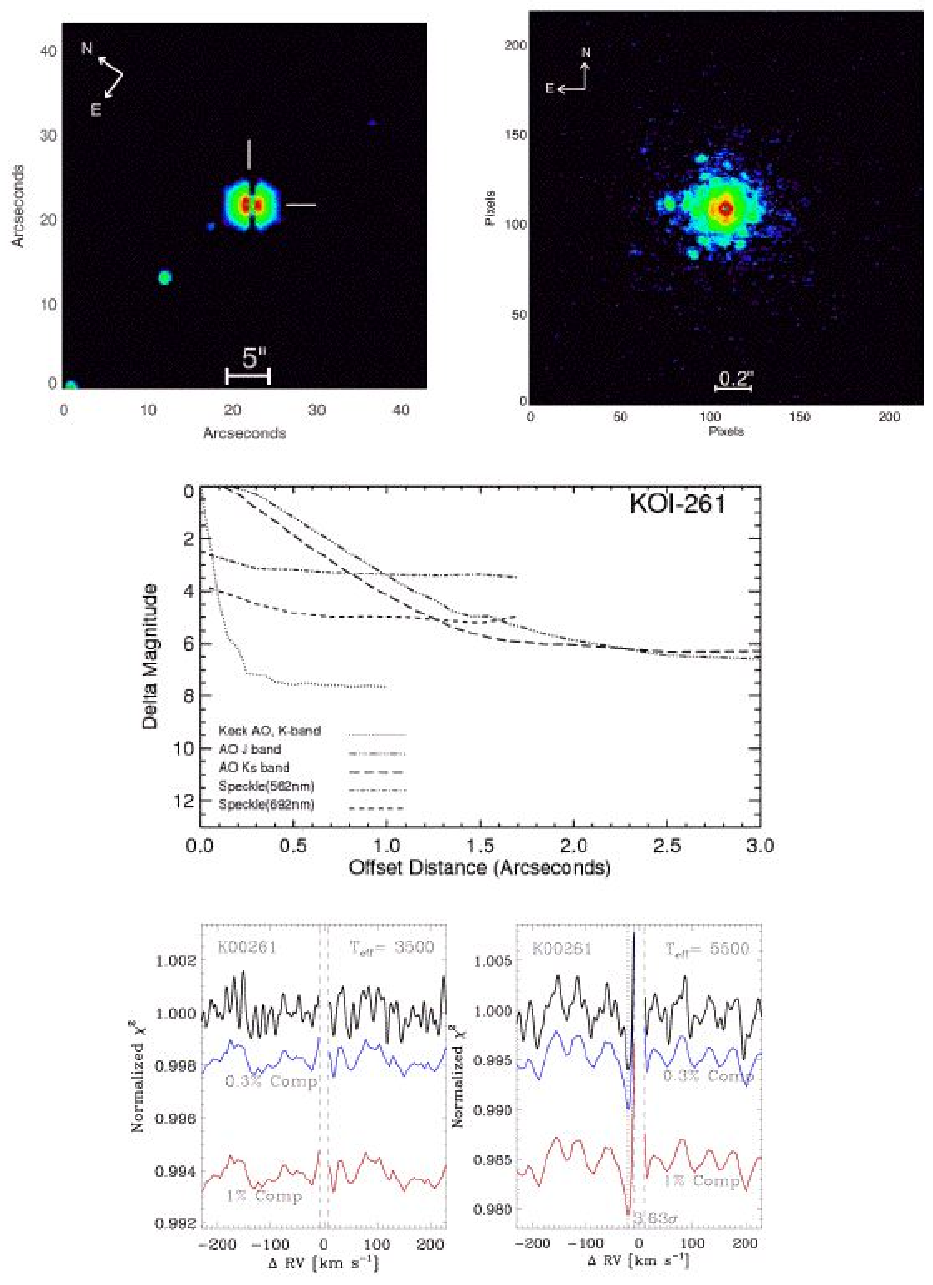}
\caption{ Upper left) Seeing limited image for Kepler-96 (KOI-261). Upper right) Adaptive Optics image. Middle) Limiting magnitudes of companion stars found with Speckle and AO imaging. Lower left and lower right)  Same as Figure 1, but for Kepler-96. No companions are detected. Any companion brighter than 0.5\% the brightness(V-band) of the primary would have been detected.
}
\label{fig:koi261_fig1}
\end{figure*}

\begin{figure*}    
\epsscale{1.0}
\plotone{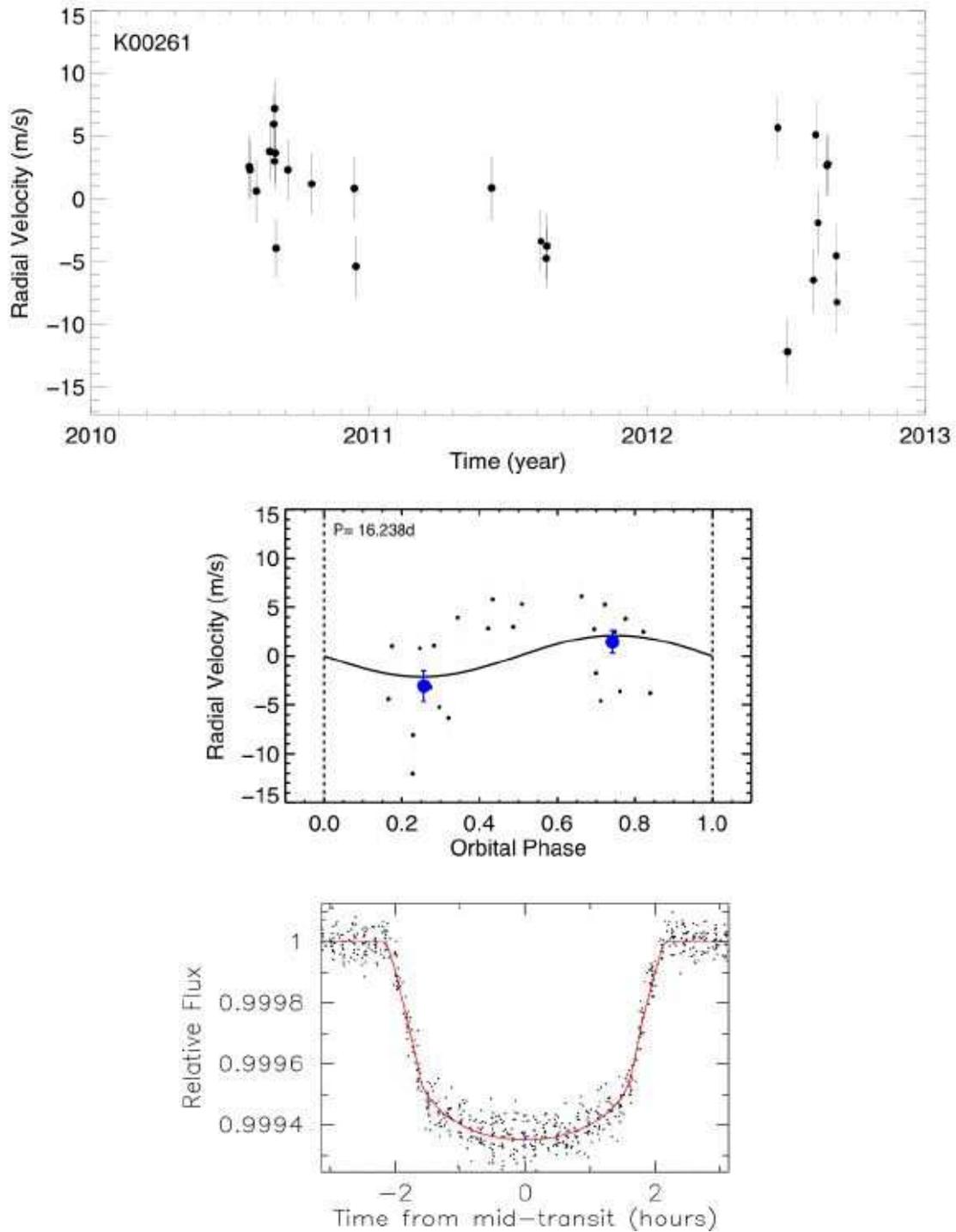}
\caption{ Top) RV versus time for Kepler 96 (KOI-261).  Middle) Phase folded radial 
velocities for each planet. Bottom) Phase folded \ek photometry for the planet.
The blue points represent binned RVs near quadrature, same as Figure  \ref{fig:koi41_fig2}.
Kepler 96b: Rp = $2.67 \pm 0.22$ \rearthe, Mp = $8.46 \pm 3.4$ \mearthe.
}
\label{fig:koi261_fig2}
\end{figure*}

\begin{figure*}   
\epsscale{1.0}
\plotone{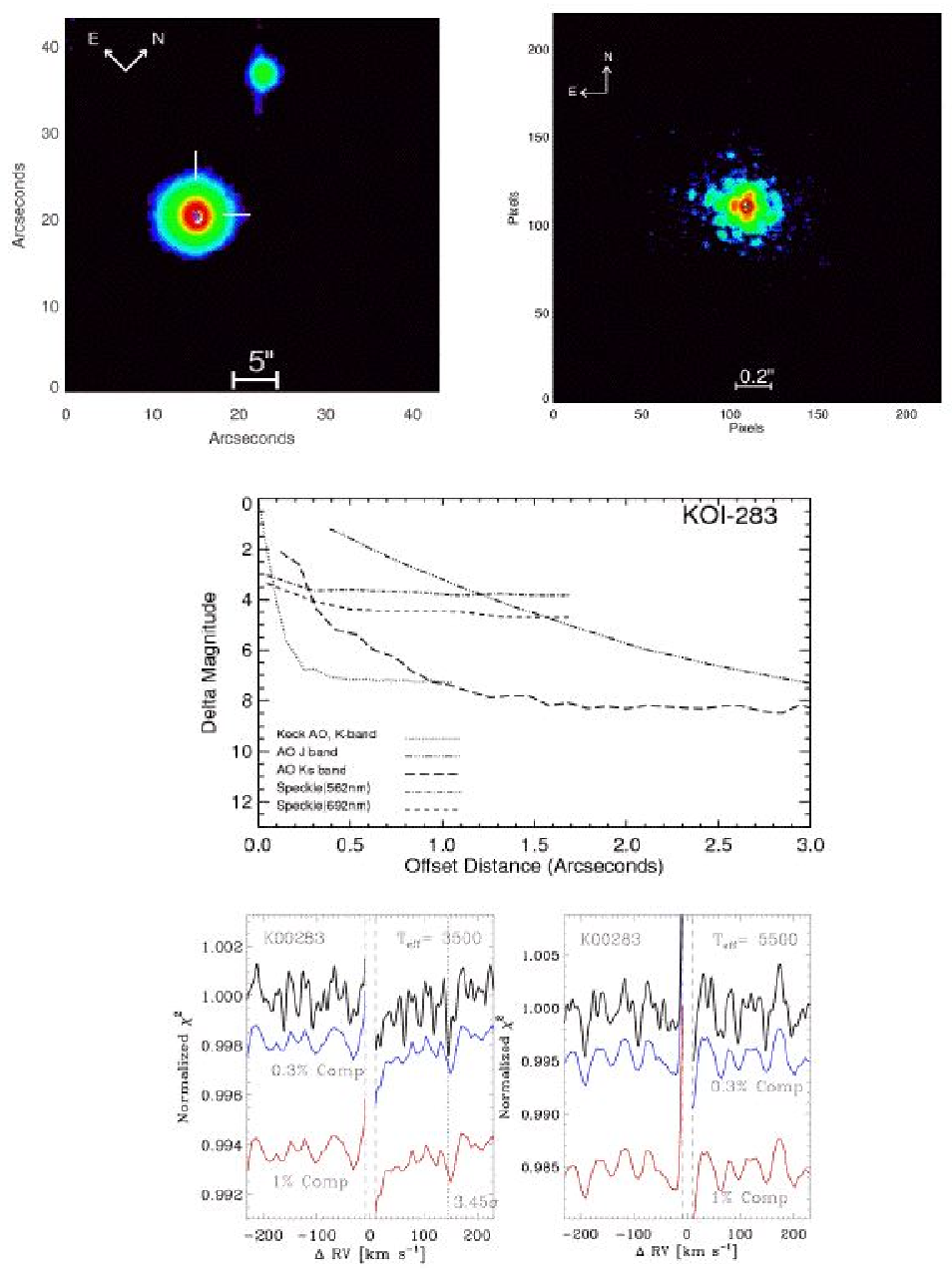}
\caption{ Upper left) Seeing limited image for Kepler-131. Upper right) Adaptive Optics image. Middle) Limiting magnitudes of companion stars found with Speckle and AO imaging. Lower left and lower right)  Same as Figure 1, but for Kepler-131. No companions are detected. Any companion brighter than 0.5\% the brightness(V-band) of the primary would have been detected.
}
\label{fig:koi283_fig1}
\end{figure*}

\begin{figure*}  
\epsscale{1.0}
\plotone{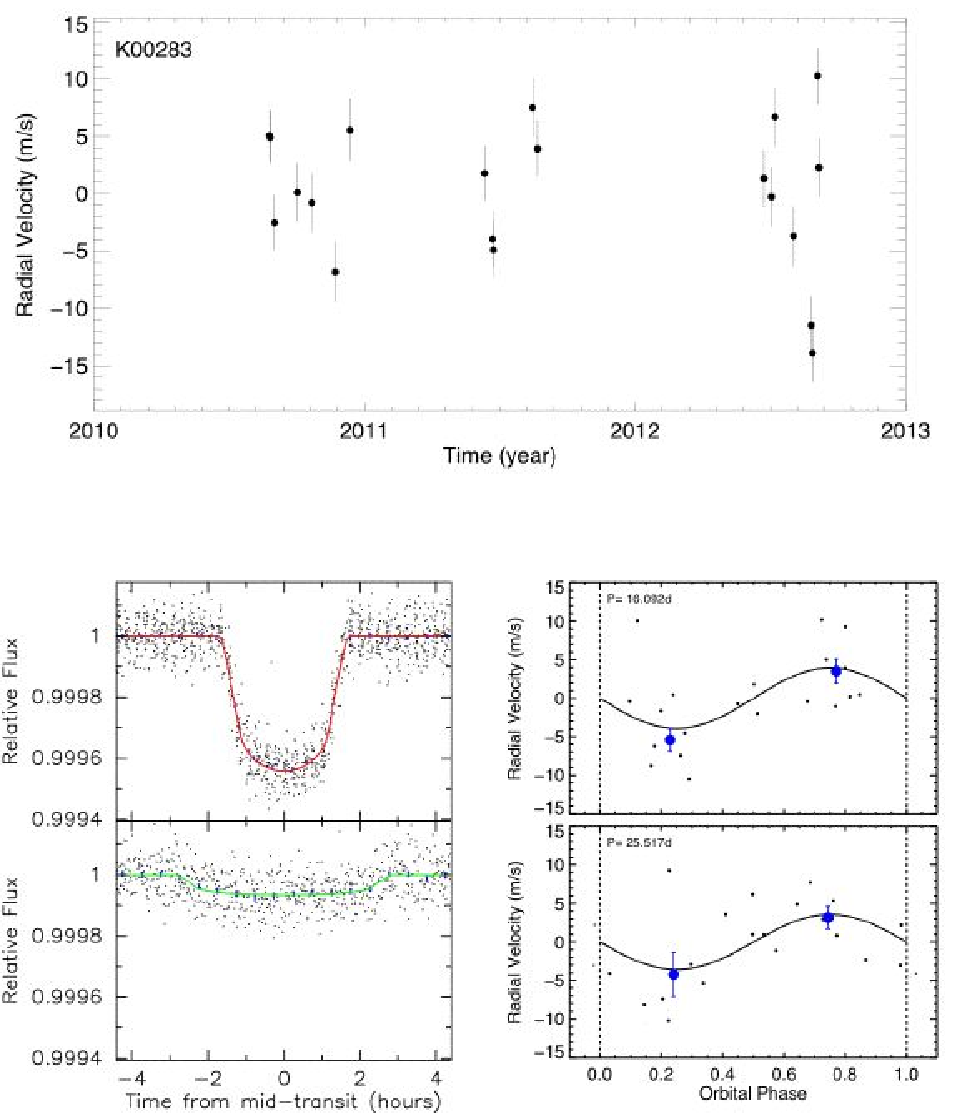}
\caption{ Top) RV versus time for Kepler-131 (KOI-283). Lower left) Phase folded \ek 
photometry for each planet. Lower Right) Phase folded radial velocities for each planet. 
The blue points represent binned RVs near quadrature, same as Figure  \ref{fig:koi41_fig2}.
Kepler-131b: Rp = $2.41 \pm 0.20$ \rearthe, Mp = $16.13 \pm 3.5$\mearthe.
Kepler-131c: Rp = $0.84 \pm 0.07$ \rearthe, Mp = $8.25 \pm 5.9$\mearthe.
}
\label{fig:koi283_fig2}
\end{figure*}

\begin{figure*}   
\epsscale{1.0}
\plotone{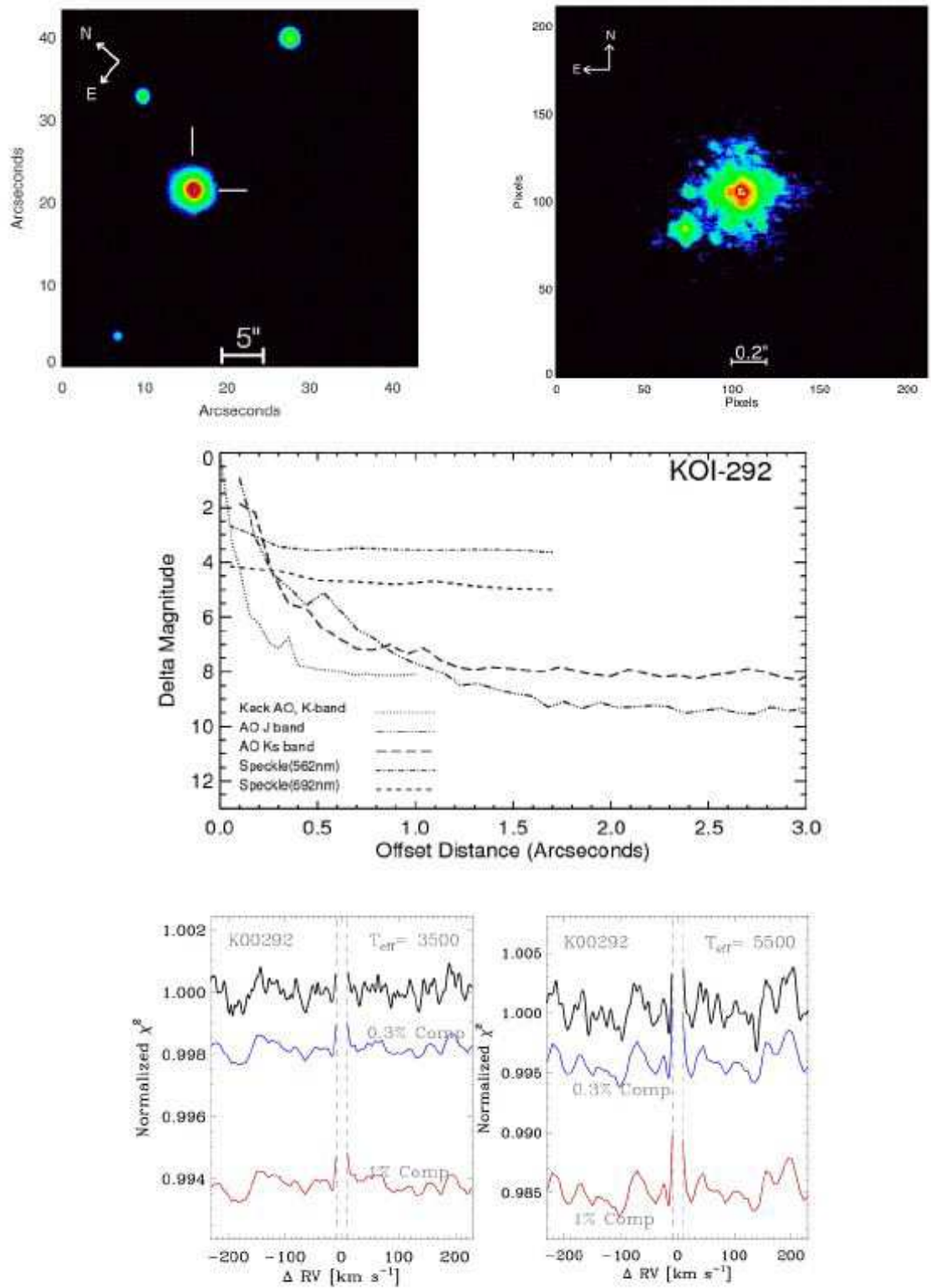}
\caption{ Upper left) Seeing limited image for Kepler-97 (KOI-292). Upper right) Adaptive Optics image. Note companion 0{\farcs}38 to the SE, 3.2mag fainter in K-band (2.2$\mu$).  The transiting planet could orbit the primary or secondary star.  But lack of photo-center displacement suggests the planet orbits the primary star.  Middle) Limiting magnitudes of companion stars found with Speckle and AO imaging. Lower left and lower right)  Same as Figure 1, but for Kepler-97. No companions are detected. Any companion brighter than 0.5\% the brightness(V-band) of the primary would have been detected.
}
\label{fig:koi292_fig1}
\end{figure*}

\begin{figure*}   
\epsscale{1.0}
\plotone{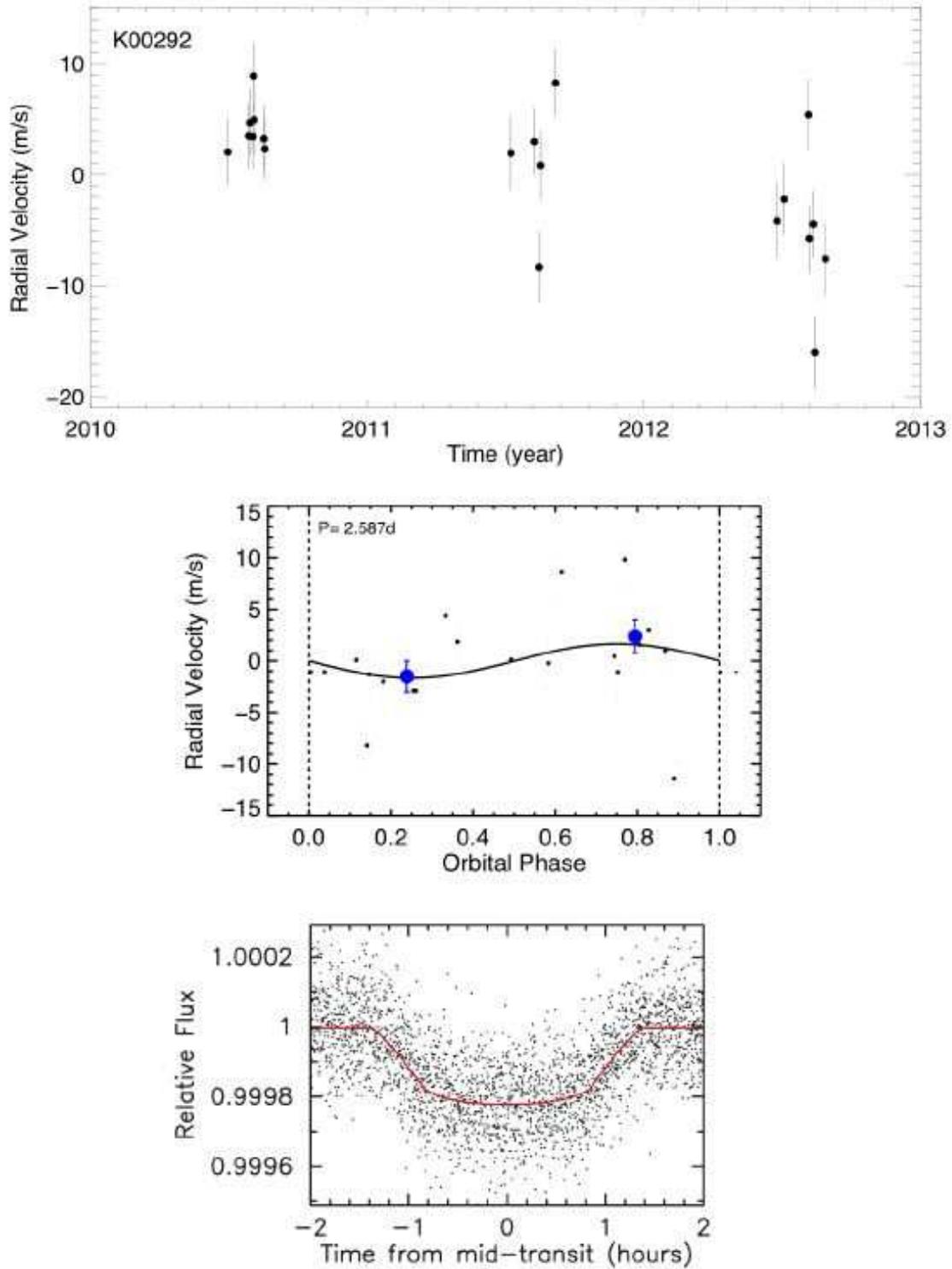}
\caption{ Top) RV versus time for Kepler-97 (KOI-292). MIddle)Phase folded radial velocities for each planet. 
Bottom) Phase folded \ek photometry for each planet. 
The blue points represent binned RVs near quadrature, same as Figure  \ref{fig:koi41_fig2}.
Kepler-97b: Rp = $1.48 \pm 0.13$ \rearthe, Mp = $3.51 \pm 1.9$ \mearthe.
KOI-292.10: Rp = NA, Mp = $>$789  \mearthe, period $>$ 344.
}
\label{fig:koi292_fig2}
\end{figure*}

\begin{figure*}   
\epsscale{1.0}
\plotone{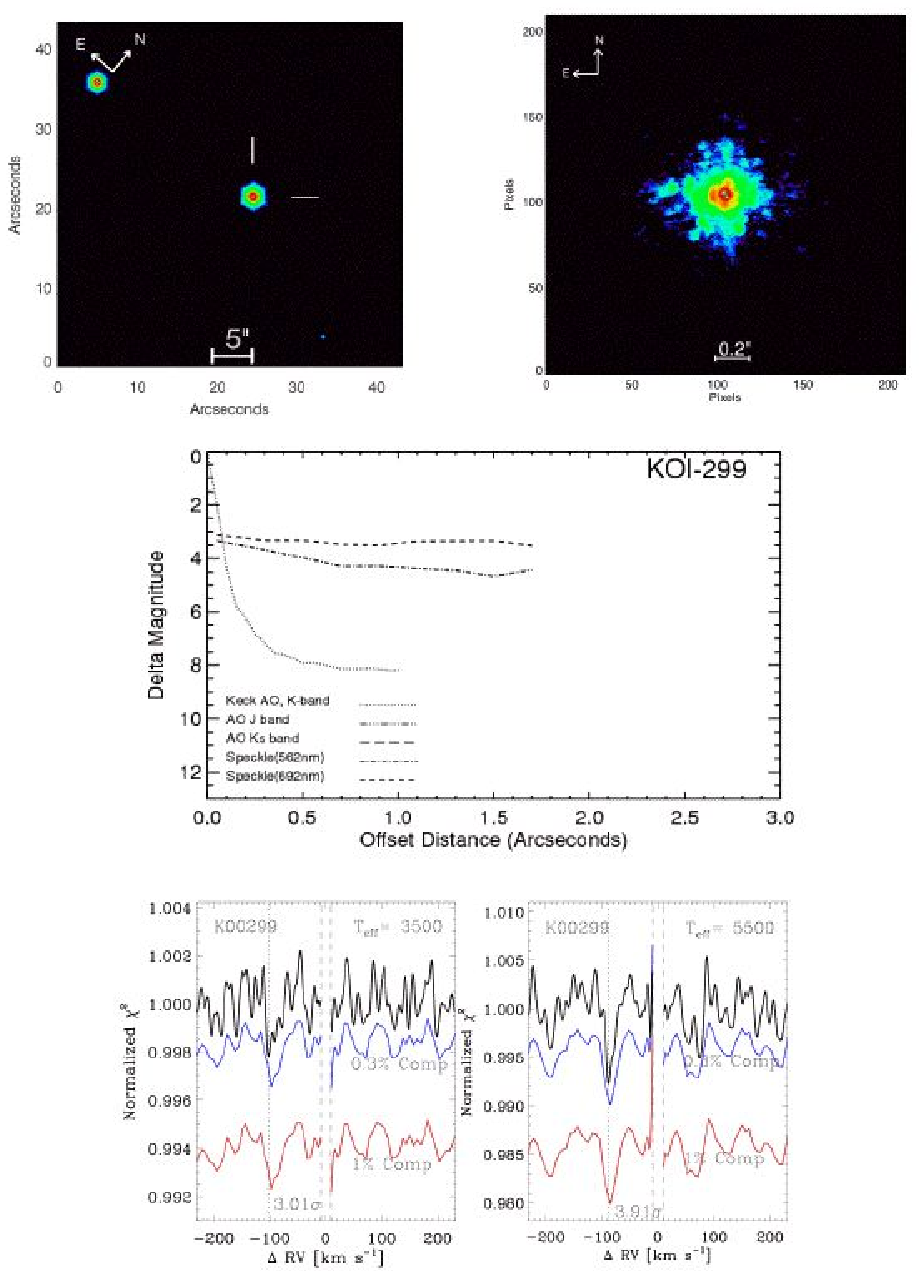}
\caption{ Upper left) Seeing limited image for Kepler-98 (KOI-299). Upper right) Adaptive Optics image. Middle) Limiting magnitudes of companion stars found with Speckle and AO imaging. Lower left and lower right)  Same as Figure 1, but for Kepler-98. No companions are detected. Any companion brighter than 0.5\% the brightness(V-band) of the primary would have been detected.
}
\label{fig:koi299_fig1}
\end{figure*}

\begin{figure*}   
\epsscale{1.0}
\plotone{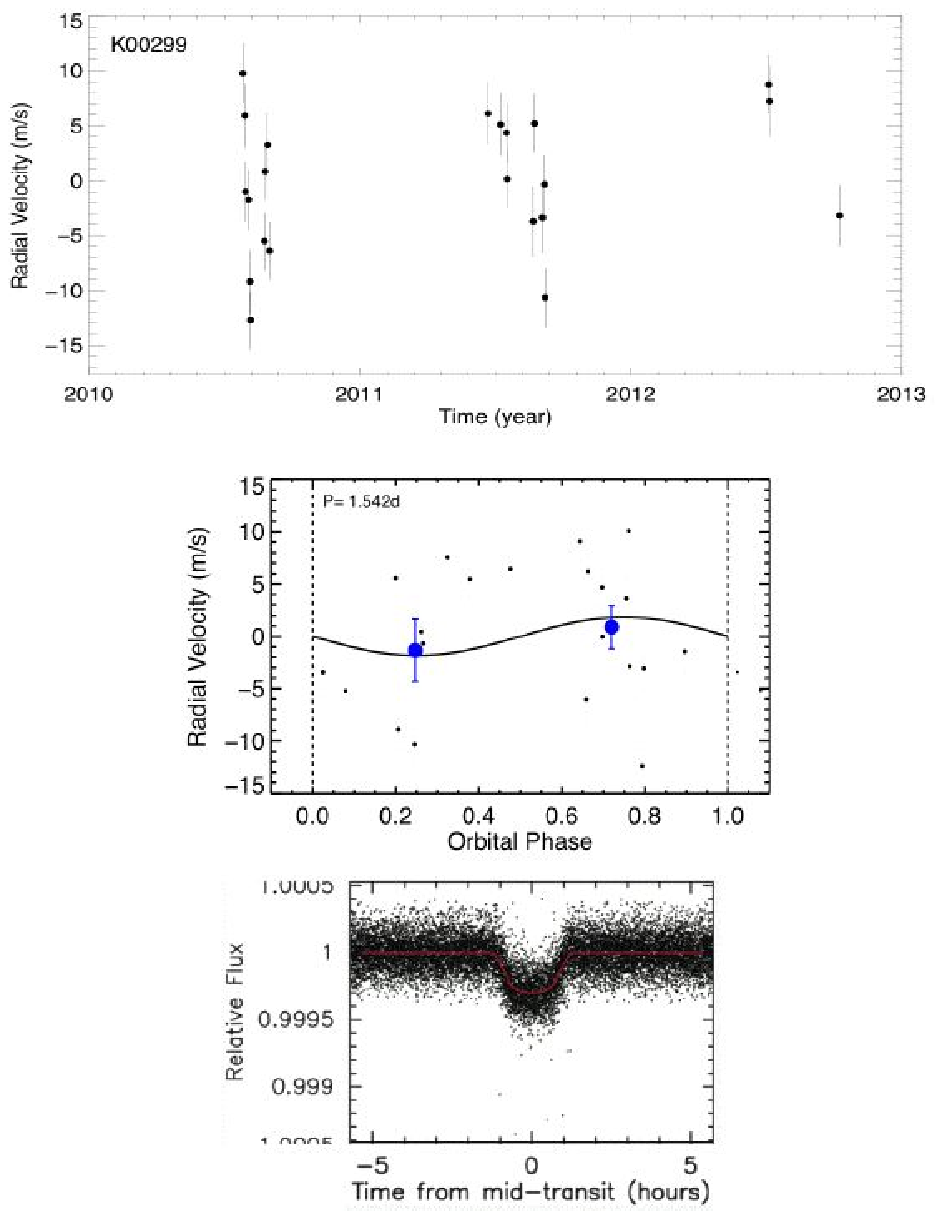}
\caption{ Top) RV versus time for Kepler-98 (KOI-299).  Middle) Phase folded radial 
velocities for each planet. Bottom) Phase folded \ek photometry for the transiting planet.
The blue points represent binned RVs near quadrature, same as Figure  \ref{fig:koi41_fig2}.
Kepler-98b: Rp = $1.99 \pm 0.22$ \rearthe, Mp = $3.55 \pm 1.6$ \mearthe.
}
\label{fig:koi299_fig2}
\end{figure*}

\begin{figure*}   
\epsscale{1.0}
\plotone{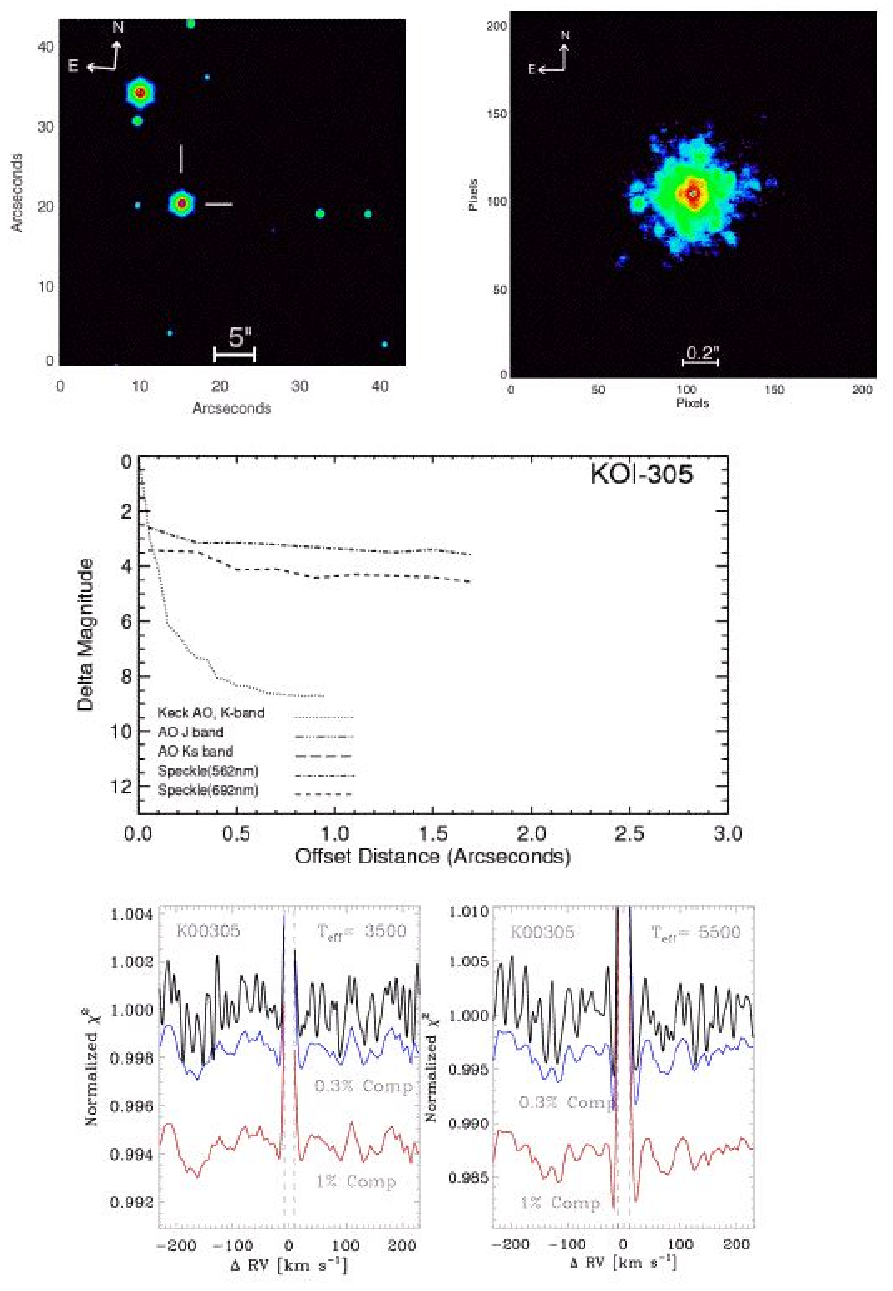}
\caption{ Upper left) Seeing limited image for Kepler-99 (KOI-305). Upper right) Adaptive Optics image. Middle) Limiting magnitudes of companion stars found with Speckle and AO imaging. Lower left and lower right)  Same as Figure 1, but for Kepler-99. No companions are detected. Any companion brighter than 0.5\% the brightness(V-band) of the primary would have been detected.
}
\label{fig:koi305_fig1}
\end{figure*}

\begin{figure*}    %
\epsscale{1.0}
\plotone{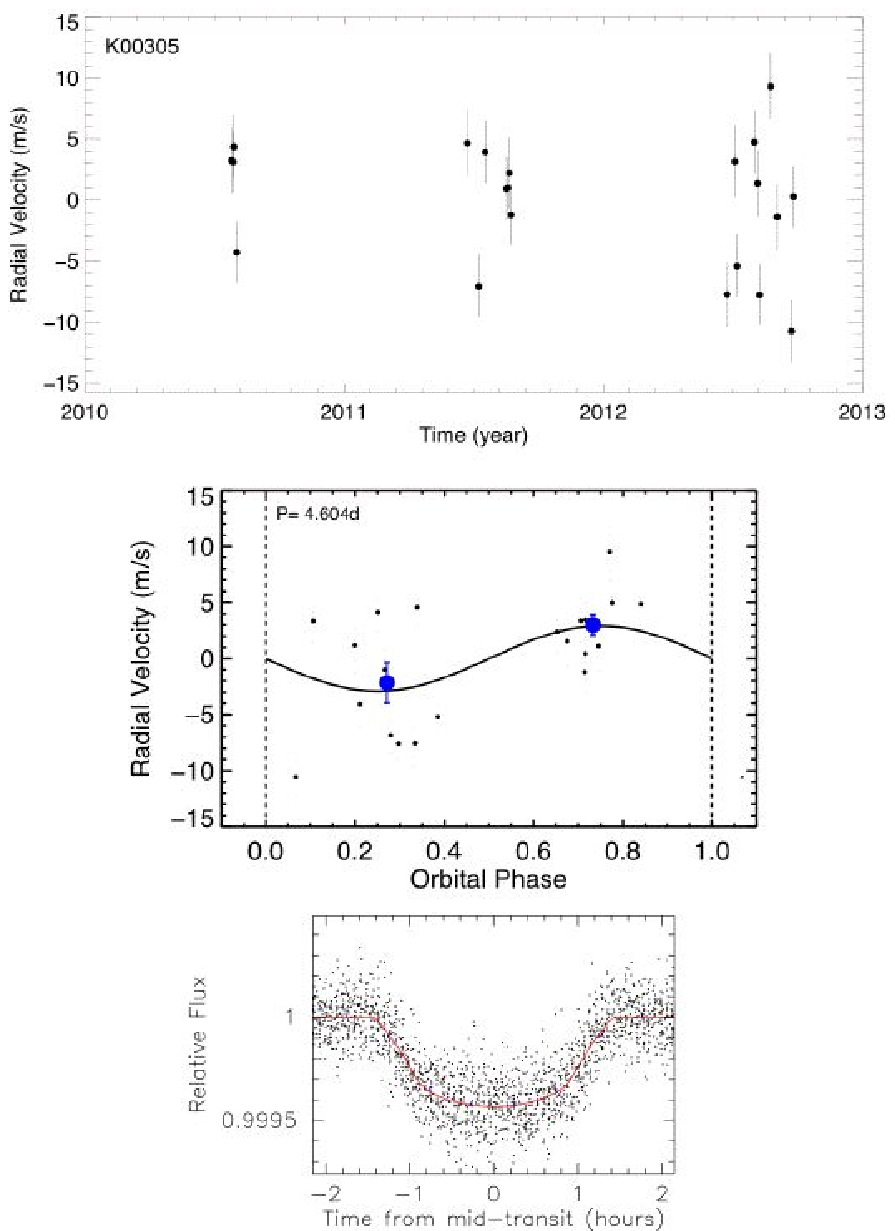}
\caption{ Top) RV versus time for Kepler-99 (KOI-305). Middle) Phase folded radial 
velocities for each planet.  Bottom) Phase folded \ek photometry for the planet. 
The blue points represent binned RVs near quadrature, same as Figure  \ref{fig:koi41_fig2}.
Kepler-99b: Rp = $1.48 \pm 0.08$ \rearthe, Mp = $6.15 \pm 1.3$ \mearthe.
}
\label{fig:koi305_fig2}
\end{figure*}

\begin{figure*}   
\epsscale{1.0}
\plotone{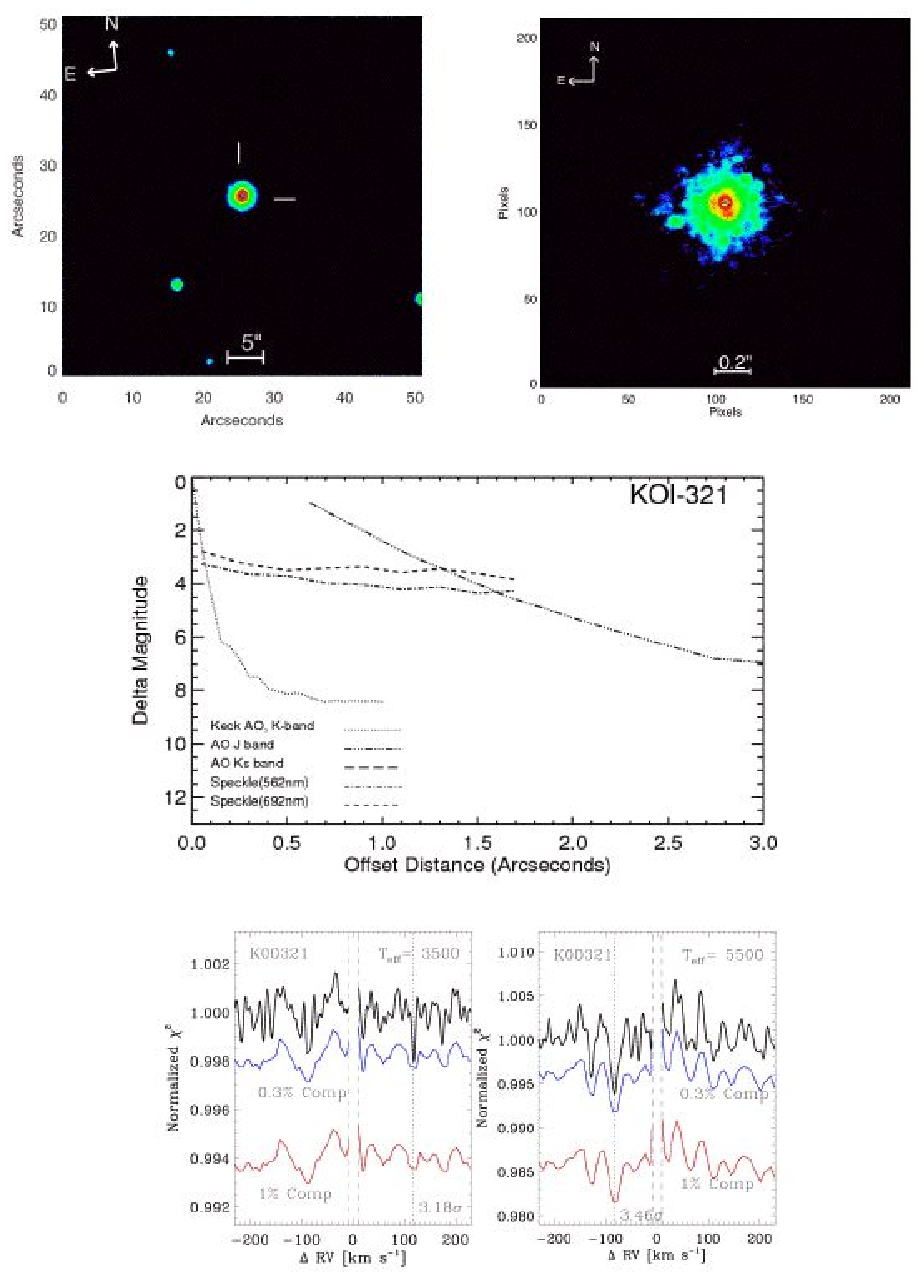}
\caption{ Upper left) Seeing limited image for Kepler-406 (KOI-321). Upper right) Adaptive Optics image. Middle) Limiting magnitudes of companion stars found with Speckle and AO imaging. Lower left and lower right)  Same as Figure 1, but for Kepler-406. No companions are detected. Any companion brighter than 0.5\% the brightness(V-band) of the primary would have been detected.
}
\label{fig:koi321_fig1}
\end{figure*}

\begin{figure*}    %
\epsscale{1.0}
\plotone{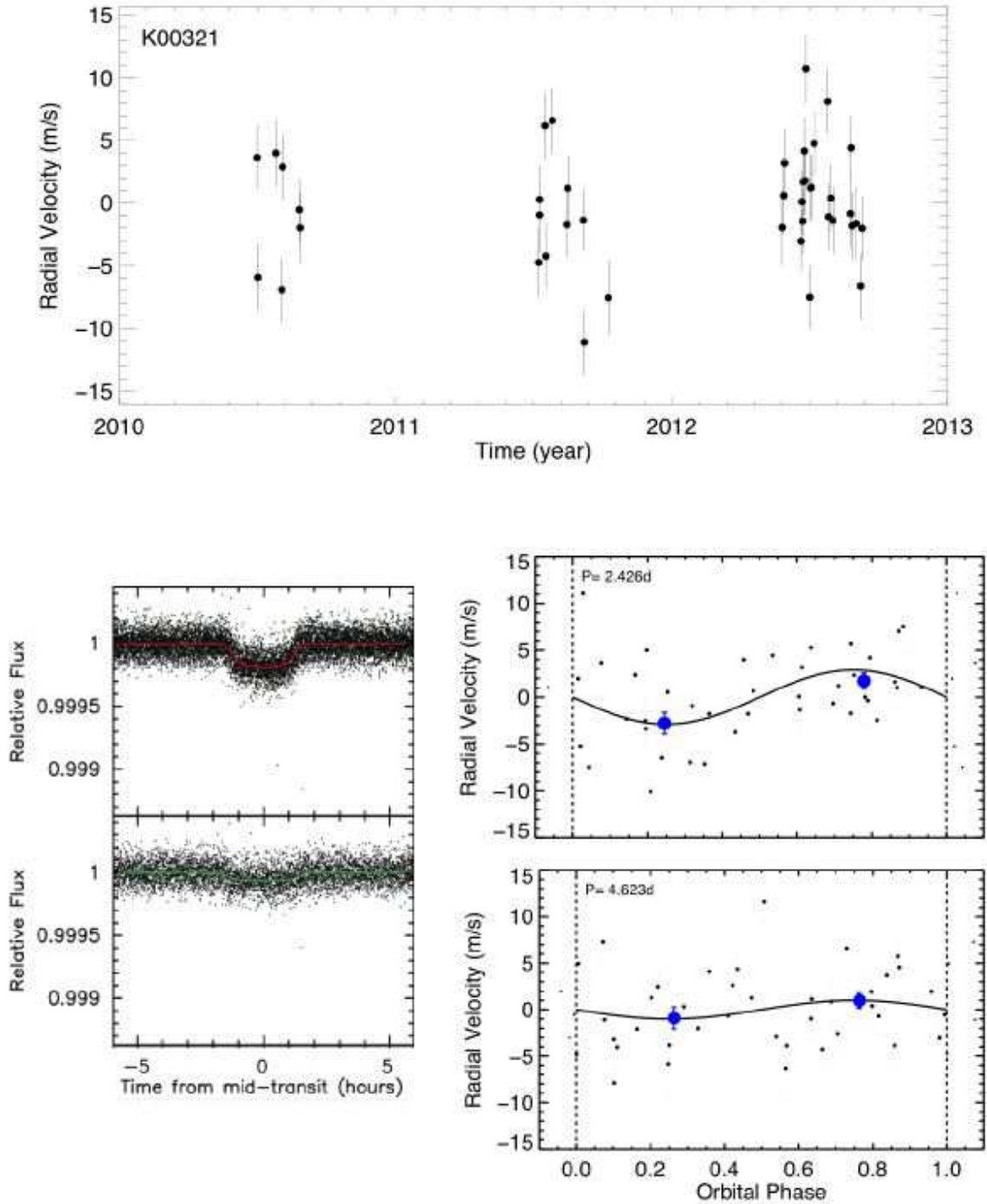}
\caption{ Top) RV versus time for Kepler-406 (KOI-321). Lower left) Phase folded \ek 
photometry for each planet. Lower Right) Phase folded radial velocities for each planet. 
The blue points represent binned RVs near quadrature, same as Figure  \ref{fig:koi41_fig2}.
Kepler-406b: Rp = $1.43 \pm 0.03$ \rearthe, Mp = $6.35 \pm 1.4$ \mearthe.
Kepler-406c: Rp = $0.85 \pm 0.03$ \rearthe, Mp = $2.71 \pm 1.8$ \mearthe.
}
\label{fig:koi321_fig2}
\end{figure*}

\clearpage

\begin{figure*}   
\epsscale{1.0}
\plotone{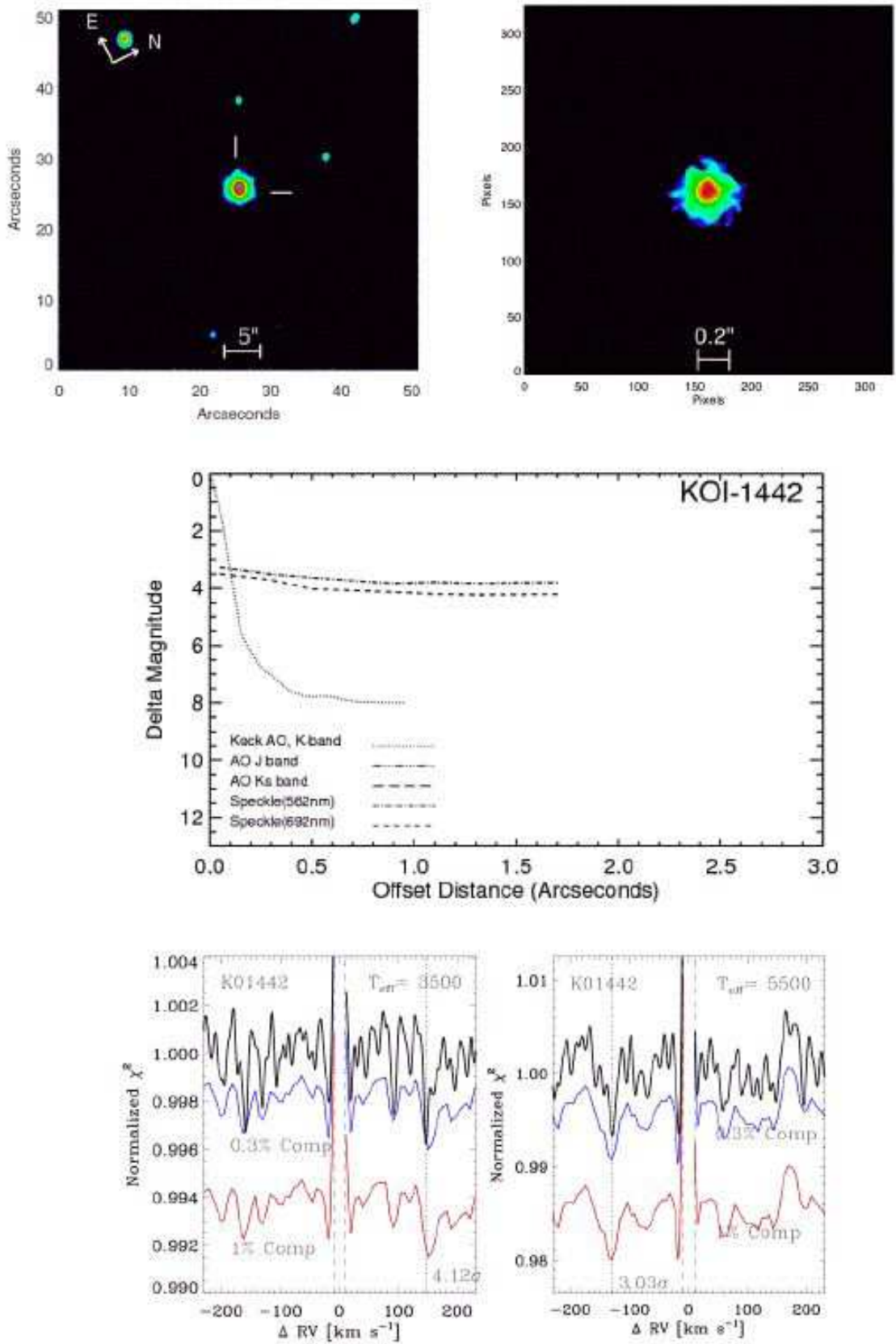}
\caption{ Upper left) Seeing limited image for Kepler-407 (KOI-1442). Upper right) Adaptive Optics image from Keck NIRC2. Middle) Limiting magnitudes of companion stars from Speckle and AO imaging. Lower left and lower right)  Same as Figure 1, but for Kepler-407. No secondary lines from a companion are detected . Any companion brighter than 0.5\% the brightness(V-band) of the primary would have been detected. }
\label{fig:koi1442_fig1}
\end{figure*}

\begin{figure*}    
\epsscale{1.0}
\plotone{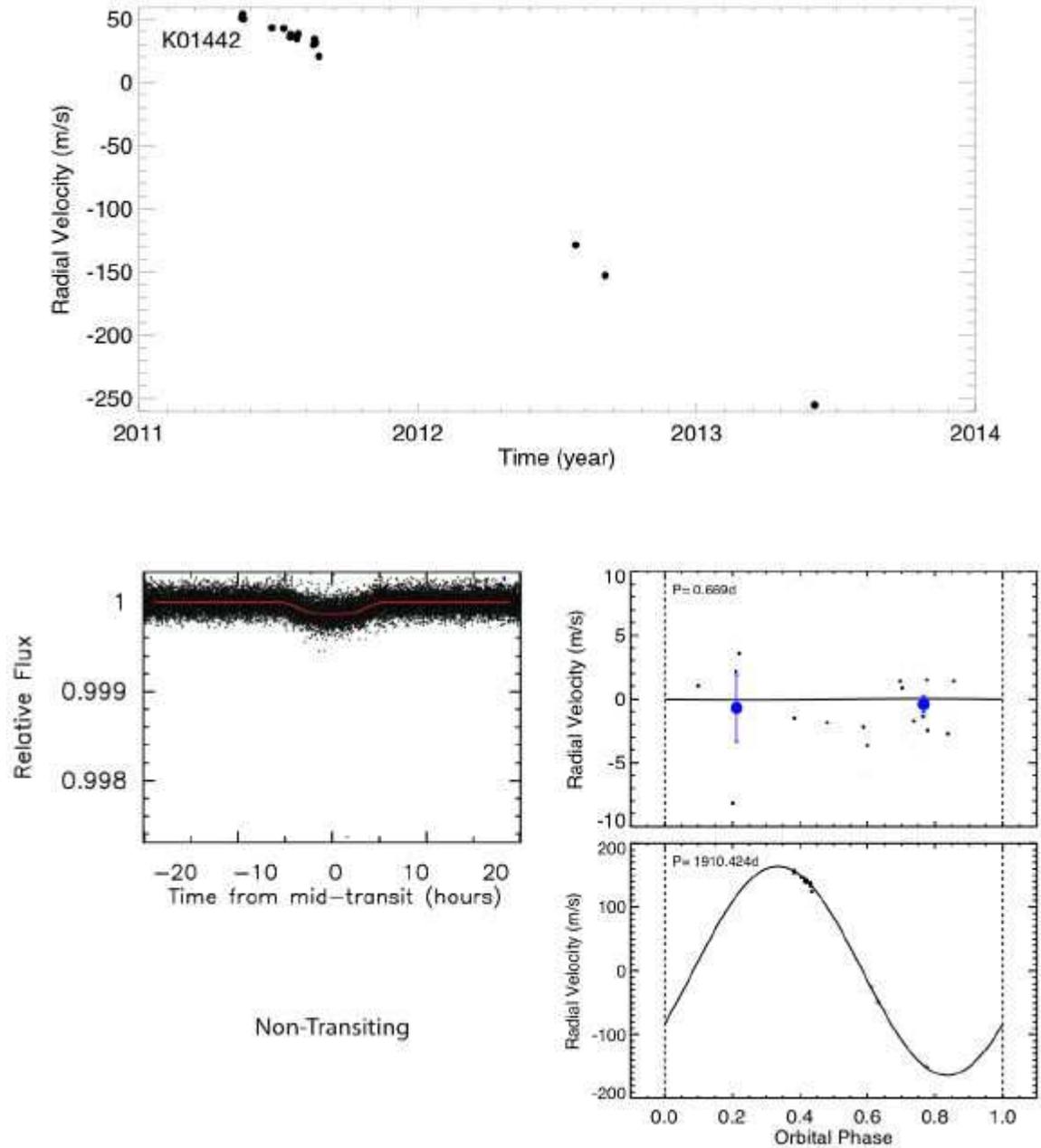}
\caption{ Top) RV versus time for Kepler-407 (KOI-1442). Lower left) Phase folded \ek 
photometry for the transiting planet.  Lower Right) Phase folded radial velocities for each planet. 
The blue points represent binned RVs near quadrature, same as Figure  \ref{fig:koi41_fig2}.
Kepler-407b: Rp = $1.07 \pm 0.02$ \rearthe, Mp = $0.06 \pm 1.2$ \mearthe.
Kepler-407c: Rp = NA, Mp = $4000\pm 2000 $\mearthe, period = $3000 \pm
500$d.  Kepler-407 could be a planet, brown dwarf, or a star,
depending on the inclination of the orbit.
}
\label{fig:koi1442_fig2}
\end{figure*}

\begin{figure*}   
\epsscale{1.0}
\plotone{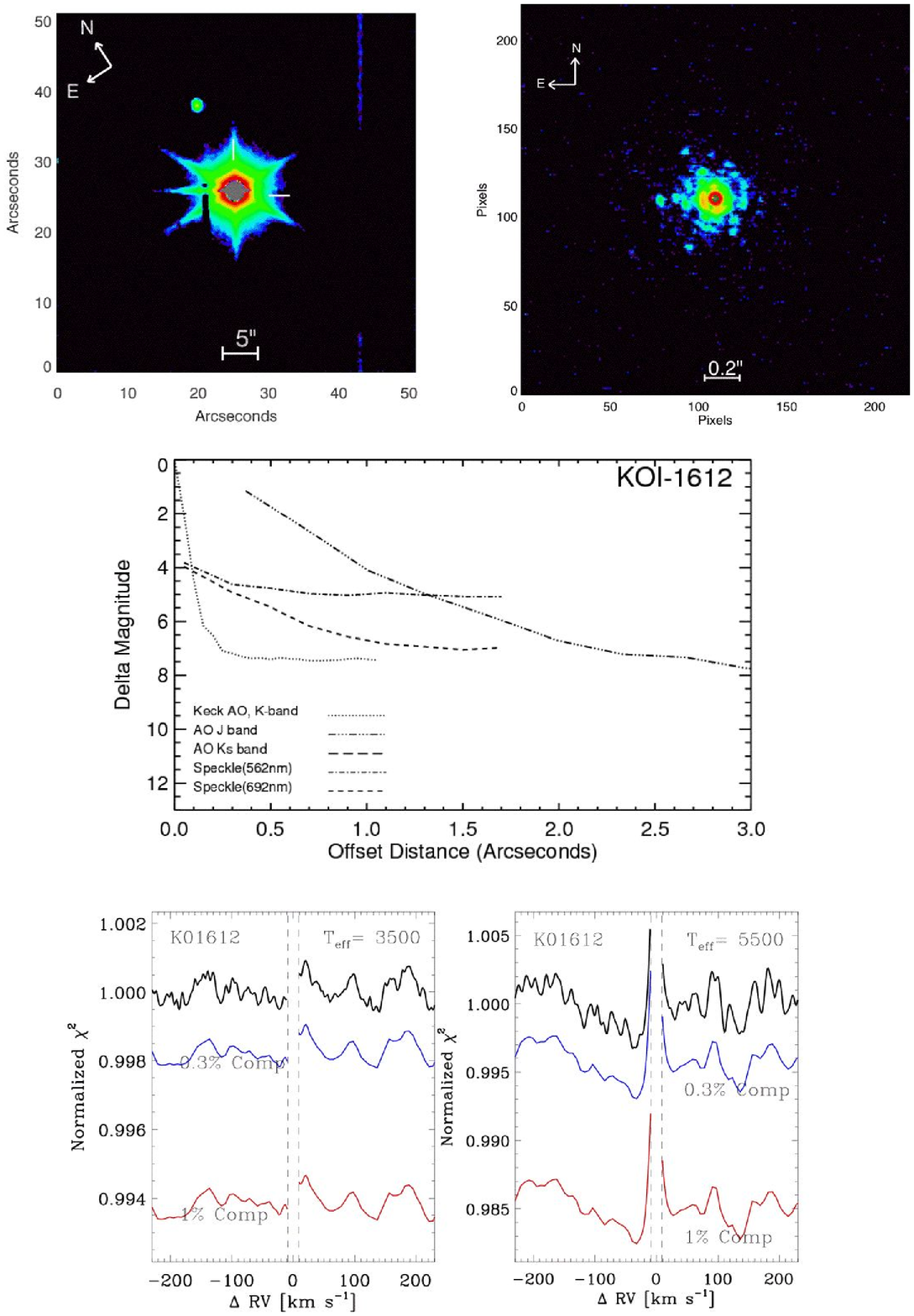}
\caption{ Upper left) Seeing limited image for KOI-1612. Upper right) Adaptive Optics image. Middle) Limiting magnitudes of companion stars found with Speckle and AO imaging. Lower left and lower right)  Same as Figure 1, but for KOI-1612. No companions are detected. Any companion brighter than 0.5\% the brightness(V-band) of the primary would have been detected.
}
\label{fig:koi1612_fig1}
\end{figure*}

\begin{figure*}    
\epsscale{1.0}
\plotone{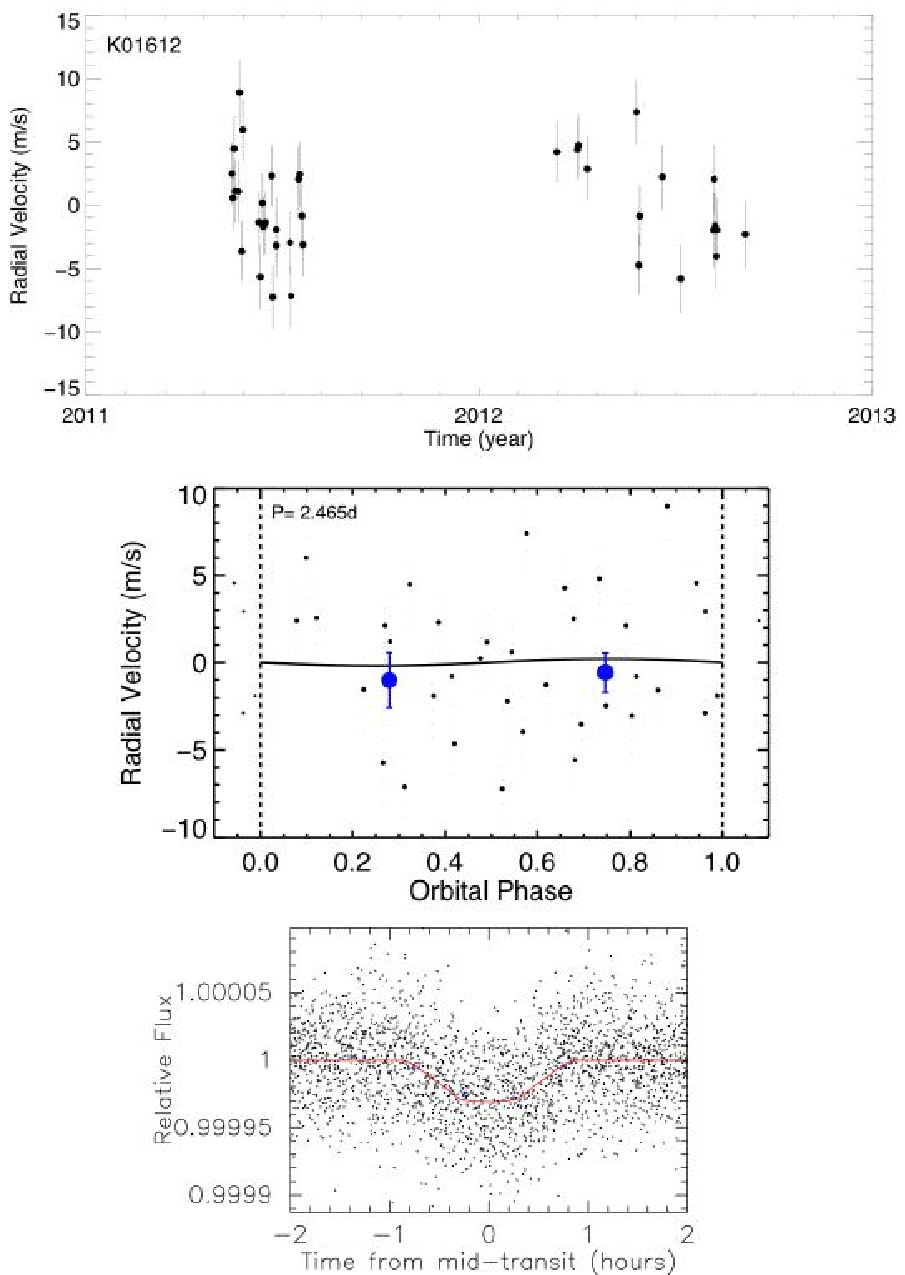}
\caption{ Top) RV versus time for KOI-1612. Middle) Phase folded radial 
velocities for each planet.  Bottom) Phase folded \ek photometry for the planet. 
The blue points represent binned RVs near quadrature, same as Figure  \ref{fig:koi41_fig2}.
KOI-1612.01: Rp = $0.82 \pm 0.03$ \rearthe, Mp = $0.48 \pm 3.2$ \mearthe.
}
\label{fig:koi1612_fig2}
\end{figure*}

\begin{figure*}   
\epsscale{1.0}
\plotone{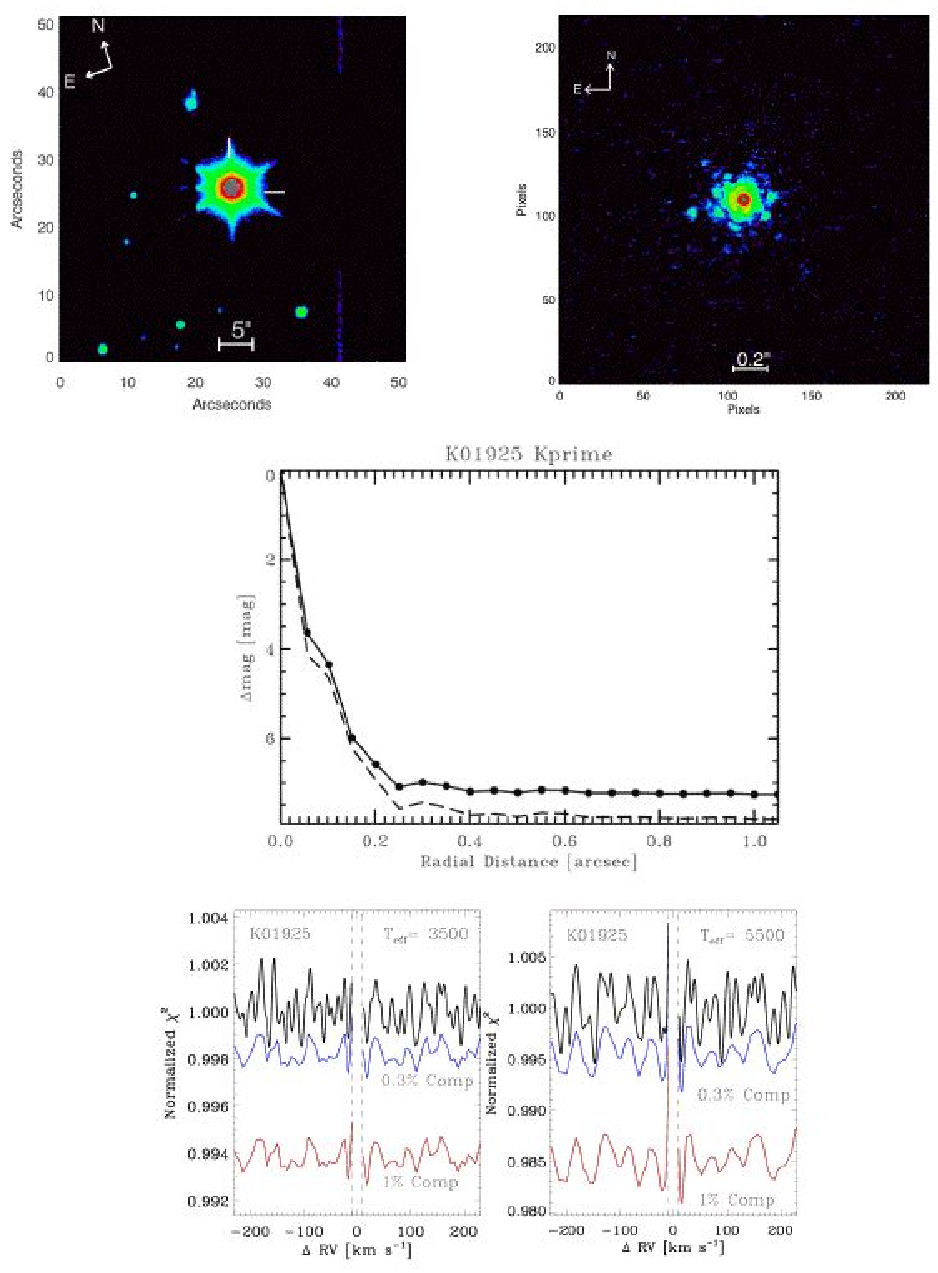}
\caption{ Upper left) Seeing limited image for Kepler-409b (KOI-1925). Upper right) Adaptive Optics image. Middle) Limiting magnitudes of companion stars found with Speckle and AO imaging. Lower left and lower right)  Same as Figure 1, but for Kepler-409. No companions are detected. Any companion brighter than 0.5\% the brightness(V-band) of the primary would have been detected.
}
\label{fig:koi1925_fig1}
\end{figure*}

\begin{figure*}    
\epsscale{1.0}
\plotone{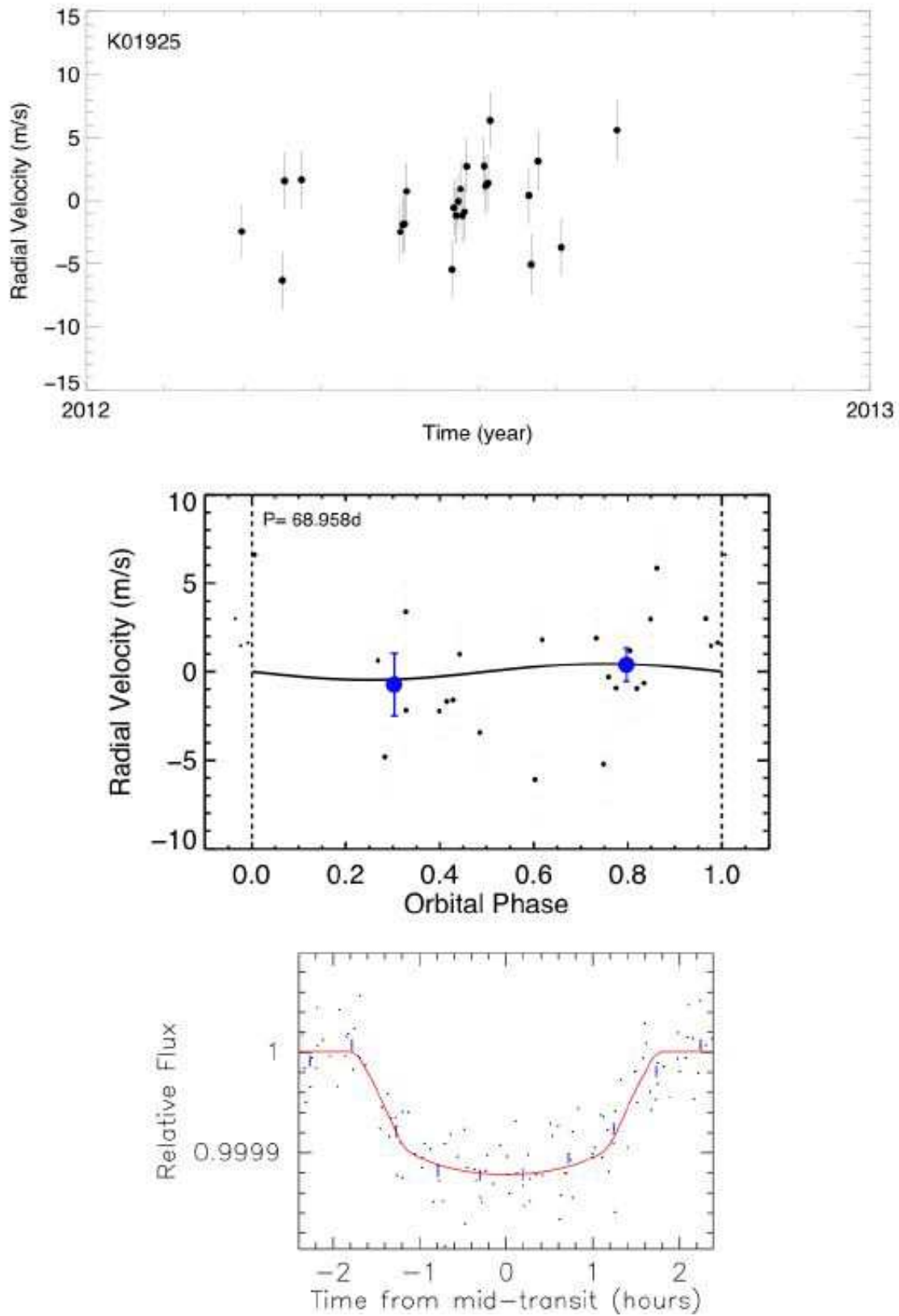}
\caption{ Top) RV versus time for Kepler-409 (KOI-1925). Middle) Phase folded radial 
velocities for each planet.  Bottom) Phase folded \ek photometry for each planet. 
The blue points represent binned RVs near quadrature, same as Figure  \ref{fig:koi41_fig2}.
Kepler-409b: Rp = $1.19 \pm 0.03$ \rearthe, Mp = $2.69 \pm 6.2$ \mearthe.
}
\label{fig:koi1925_fig2}
\end{figure*}

 \clearpage


\clearpage

\LongTables  

 \begin{landscape}  
\begin{deluxetable}{llllllllllllll} 
\tabletypesize{\footnotesize}
\tablecaption{Stellar Parameters
\label{tab:stellar_pars_tbl}}
\tablewidth{0pt}
\tablehead{
\colhead{Kepler \#} & \colhead{KOI}    & \colhead{KIC } & \colhead{RA} & \colhead{Dec} &  
  \colhead{$T_{\rm eff}$}   & \colhead{Stellar}      & \colhead{\feh} & 
 \colhead{$M_*$} & \colhead{$R_*$}   & \colhead{\vsinie} &  \colhead{\ek}   & 
 \colhead{Age} & \colhead{Source$^a$}  \\
\colhead{} & \colhead{} & \colhead{} &  \colhead{} &  \colhead{} & 
  \colhead{(K)} & \colhead{\logg} &  \colhead{} &
  \colhead{(\msun)} &  \colhead{(\rsun)} &  \colhead{({\rm \kms})} &  \colhead{Magn.} &  
  \colhead{(Gyr)} &  \colhead{MCMC+}   
}
\tabcolsep=0.00cm  %
\startdata
Kepler-100 & 41 &  6521045 & 19:25:32.6 & 41:59:24 &  $5825 \pm 75$ & $4.125 \pm 0.03$ &  $+0.02 \pm 0.10$ &  $1.08 \pm 0.06$ &  $1.49 \pm 0.04$ & 3.7 & 11.20 &  6.46 & AS  \\ 
Kepler-93  & 69 &  3544595 & 19:25:40.3 & 38:40:20 &  $5669 \pm 75$ & $4.468 \pm 0.03$ & $-0.18 \pm 0.10$ &  $0.91 \pm 0.06$ &  $0.92 \pm 0.02$ & 0.5 &  9.93 &  5.05 & AS  \\ 
Kepler-102  & 82 & 10187017 & 18:45:55.8 & 47:12:28 &  $4903 \pm 74$ & $4.607 \pm 0.03$ &  $+0.08 \pm 0.07$ &  $0.80 \pm 0.06$ &  $0.74 \pm 0.02$ & 0.5 & 11.49 &  1.41 & SME  \\ 
Kepler-94  & 104 & 10318874 & 18:44:46.7 & 47:29:49 &  $4781 \pm 98$ & $4.590 \pm 0.04$ &  $+0.34 \pm 0.07$ &  $0.81 \pm 0.06$ &  $0.76 \pm 0.03$ & 0.5 & 12.90 &  1.41 & SME  \\ 
Kepler-103  & 108 &  4914423 & 19:15:56.2 & 40:03:52 &  $5845 \pm 88$ & $4.162 \pm 0.04$ &  $+0.07 \pm 0.11$ &  $1.09 \pm 0.07$ &  $1.44 \pm 0.04$ & 2.5 & 12.29 &  5.70 & AS  \\ 
Kepler-106  & 116 &  8395660 & 20:03:27.3 & 44:20:15 &  $5858 \pm 114$ & $4.407 \pm 0.14$ & $-0.12 \pm 0.11$ &  $1.00 \pm 0.06$ &  $1.04 \pm 0.17$ & 0.3 & 12.88 &  4.83 & SME  \\ 
Kepler-95  & 122 &  8349582 & 18:57:55.7 & 44:23:52 &  $5699 \pm 74$ & $4.171 \pm 0.04$ &  $+0.30 \pm 0.10$ &  $1.08 \pm 0.08$ &  $1.41 \pm 0.04$ & 0.7 & 12.35 &  5.63 & AS  \\ 
Kepler-109  & 123 &  5094751 & 19:21:34.2 & 40:17:05 &  $5952 \pm 75$ & $4.211 \pm 0.04$ & $-0.08 \pm 0.10$ &  $1.04 \pm 0.06$ &  $1.32 \pm 0.04$ & 1.0 & 12.36 &  5.73 & AS  \\ 
Kepler-48  & 148 &  5735762 & 19:56:33.4 & 40:56:56 &  $5194 \pm 73$ & $4.487 \pm 0.05$ &  $+0.17 \pm 0.07$ &  $0.88 \pm 0.06$ &  $0.89 \pm 0.05$ & 0.5 & 13.04 &  3.14 & SME  \\ 
Kepler-113  & 153 & 12252424 & 19:11:59.4 & 50:56:39 &  $4725 \pm 74$ & $4.636 \pm 0.03$ &  $+0.05 \pm 0.07$ &  $0.75 \pm 0.06$ &  $0.69 \pm 0.02$ & 0.4 & 13.46 &  6.89 & SME  \\ 
Kepler-25  & 244 &  4349452 & 19:06:33.2 & 39:29:16 &  $6270 \pm 79$ & $4.278 \pm 0.03$ & $-0.04 \pm 0.10$ &  $1.19 \pm 0.06$ &  $1.31 \pm 0.02$ & 9.5 & 10.73 & 11.00 & AS  \\ 
Kepler-37  & 245 &  8478994 & 18:56:14.2 & 44:31:05 &  $5417 \pm 75$ & $4.567 \pm 0.05$ & $-0.32 \pm 0.07$ &  $0.80 \pm 0.07$ &  $0.77 \pm 0.03$ & 0.5 &  9.70 &  5.66 & AS  \\ 
Kepler-68  & 246 & 11295426 & 19:24:07.7 & 49:02:24 & $5793 \pm  74$ & $4.282 \pm 0.02$ &  $+0.12 \pm 0.07$ &  $1.08 \pm 0.05$ &  $1.24 \pm 0.02$ & 0.5 & 10.00 &  6.30 & AS  \\ 
Kepler-96  & 261 &  5383248 & 19:48:16.7 & 40:31:30 &  $5690 \pm 73$ & $4.421 \pm 0.08$ &  $+0.04 \pm 0.07$ &  $1.00 \pm 0.06$ &  $1.02 \pm 0.09$ & 0.5 & 10.30 &  2.34 & SME  \\ 
Kepler-131  & 283 &  5695396 & 19:14:07.4 & 40:56:32 &  $5685 \pm 74$ & $4.417 \pm 0.08$ &  $+0.12 \pm 0.07$ &  $1.02 \pm 0.06$ &  $1.03 \pm 0.10$ & 0.4 & 11.52 &  3.66 & SME  \\ 
Kepler-97  & 292 & 11075737 & 19:09:18.3 & 48:40:24 &  $5779 \pm 74$ & $4.430 \pm 0.08$ &  $-0.20 \pm 0.07$ &  $0.94 \pm 0.06$ &  $0.98 \pm 0.09$ & 0.5 & 12.87 &  8.42 & SME  \\ 
Kepler-98  & 299 &  2692377 & 19:02:38.8 & 37:57:52 &  $5539 \pm 73$ & $4.341 \pm 0.10$ &  $+0.18 \pm 0.07$ &  $0.99 \pm 0.06$ &  $1.11 \pm 0.12$ & 0.5 & 12.90 &  2.79 & SME  \\ 
Kepler-99  & 305 &  6063220 & 19:49:24.9 & 41:18:00 & $4782 \pm 129$ & $4.605 \pm 0.05$ &  $+0.18 \pm 0.07$ &  $0.79 \pm 0.06$ &  $0.73 \pm 0.04$ & 0.5 & 12.97 &  1.47 & SME  \\ 
Kepler-406  & 321 &  8753657 & 19:27:23.5 & 44:58:05 &  $5538 \pm 75$ & $4.409 \pm 0.02$ &  $+0.18 \pm 0.07$ &  $1.07 \pm 0.06$ &  $1.07 \pm 0.02$ & 0.4 & 12.52 &  5.84 & AS  \\ 
Kepler-407  & 1442 & 11600889 & 19:04:08.7 & 49:36:52 &  $5476 \pm 75$ & $4.426 \pm 0.06$ & $+0.33 \pm 0.07$ &  $1.00 \pm 0.06$ &  $1.01 \pm 0.07$ & 2.0 & 12.52 &  7.47 & SME  \\ 
\nodata  & 1612 & 10963065 & 18:59:08.6 & 48:25:23 &  $6104 \pm 74$ & $4.294 \pm 0.03$ & $-0.20 \pm 0.10$ &  $1.08 \pm 0.07$ &  $1.23 \pm 0.03$ & 3.1 &  8.77 &  6.68 & AS  \\ 
Kepler-409  & 1925 &  9955598 & 19:34:43.0 & 46:51:09 &  $5460 \pm 75$ & $4.499 \pm 0.03$ & $+0.08 \pm 0.10$ &  $0.92 \pm 0.06$ &  $0.89 \pm 0.02$ & 2.0 &  9.44 &  6.80 & AS  \\ 
\hline
\enddata

\end{deluxetable}
\clearpage 
\end{landscape} 

\clearpage  
$^a$ ``AS'': Asteroseismology provided the input stellar values of \teff and
  \logg \ to the MCMC analysis of the planet-transit model that further
  constrained stellar parameters.  LTE
  spectrum synthesis, SME, yielded initial measurements of  \teff, \logg, and
  metallicity, used as input values and priors for the
  asteroseismology analysis.
``SME'': the SME code performed LTE spectrum analysis
  to yield \teff, \logg, and metalicity, used as input to the MCMC
  analysis of the planet
  transit model. Iteration with Yonsei-Yale stellar models constrained
  all stellar values.
\clearpage  %

 \begin{landscape}  

\begin{deluxetable}{l cccccc ccccc c }         
\tabletypesize{\footnotesize} 
\tablecaption{Planet Properties and Orbital Parameters 
\label{tab:orbital_pars_tbl}}
\tablehead{
\colhead{Kepler \#} &\colhead{KOI}     & \colhead{Period$^a$}   & \colhead{Radius$^b$} & \colhead{Mass(peak)$^c$} &\colhead{Mass$^d$}  & \colhead{Planet Density$^e$ } &  \colhead{$K^f$} &
\colhead{Stellar density} & \colhead{Impact} & \colhead{$R_{\rm P}/R_{*}$} & \colhead{Midtransit $E$} & \colhead{Reduced Chi$^2$}  \\
\colhead{} & \colhead{} & \colhead{(days)} & \colhead{(\rearth)} & \colhead{(\mearth)} & \colhead{95\%(\mearth)}& \colhead{(\gcc)}  & \colhead{(\ms)} & \colhead{(\gcc)} & \colhead{Param. ($b$)} & \colhead{ }  & \colhead{(BJD-2454900)} & \colhead{ } \\
}
\startdata
Kepler-100c  &     41.01 &                  12.8159 &   $2.20 \pm 0.05$ &         $0.85 \pm 4.0$ &  7.05 &    $0.35 \pm 1.65$ &         $0.18 \pm 0.8$ &        $0.44 \pm 0.01$ &        $0.05 \pm 0.04$ &                    0.013550 &               55.94713 & 1.185   \\
Kepler-100b  &     41.02 &                  6.88705 &   $1.32 \pm 0.04$ &         $7.34 \pm 3.2$ &  10.0 &     ${\bf 14.25 \pm 6.33}$ &         $1.90 \pm 0.8$ &        $0.44 \pm 0.01$ &        $0.54 \pm 0.02$ &                    0.008094 &                  66.17797 & 1.110   \\
Kepler-100d  &     41.03 &                  35.3331 &   $1.61 \pm 0.05$ &        $-4.36 \pm 4.1$ &  3.0  &     $-5.72 \pm 6.00$ &        $-0.79 \pm 1.0$ &        $0.44 \pm 0.01$ &        $0.75 \pm 0.01$ &                    0.009926 &                  86.98031 &    1.261  \\
Kepler-93b  &     69.01 &                  4.72674 &   $1.50 \pm 0.03$ &         $2.59 \pm 2.0$ &  6.1  &     $4.17 \pm 3.29$ &         $1.05 \pm 0.8$ &        $1.64 \pm 0.01$ &        $0.20 \pm 0.02$ &                    0.014927 &                  67.92580 &   1.017   \\
Kepler-93c  &     69.10$^g$ &                  $>$1460 &        \nodata    &                 $>$954 &  \nodata &           \nodata    &                  $>$60 &             \nodata    &             \nodata    &                  \nodata    &               \nodata     &    \nodata  \\
Kepler-102e  &     82.01 &                  16.1457 &   $2.22 \pm 0.07$ &        $ 8.93 \pm 2.0$ &  11.7 &     ${ 4.68 \pm 1.12}$ &         $2.77 \pm 0.6$ &        $2.76 \pm 0.09$ &        $0.30 \pm 0.03$ &                    0.027639 &                  67.75384 &   1.026  \\
Kepler-102d  &     82.02 &                  10.3117 &   $1.18 \pm 0.04$ &         $3.80 \pm 1.8$ &  6.6 &      ${\bf 13.27 \pm 6.46}$ &         $1.37 \pm 0.6$ &        $2.76 \pm 0.09$ &        $0.22 \pm 0.05$ &                    0.014682 &                  67.07920 &  0.950   \\
Kepler-102f  &     82.03 &                  27.4536 &   $0.88 \pm 0.03$ &         $0.62 \pm 3.3$ &  5.2 &     $ 4.92 \pm 24.5$ &         $2.11 \pm 0.8$ &        $2.76 \pm 0.09$ &        $0.57 \pm 0.03$ &                    0.010963 &                  78.02565 &   1.404  \\
Kepler-102c  &     82.04 &                  7.07142 &   $0.58 \pm 0.02$ &        $-1.58 \pm 2.0$ &  3.0 &     $-44.24 \pm 56.7$ &        $-1.00 \pm 0.6$ &        $2.76 \pm 0.09$ &        $0.06 \pm 0.05$ &                    0.007228 &                  72.98486 &  0.883 \\
Kepler-102b  &     82.05 &                  5.28696 &   $0.47 \pm 0.02$ &         $0.41 \pm 1.6$ &  4.3 &     $23.29 \pm 94.93$ &         $0.19 \pm 0.7$ &        $2.76 \pm 0.09$ &        $0.06 \pm 0.06$ &                    0.005800 &                  68.84920 &   0.958  \\
Kepler-94b  &  104.01 &                  2.50806 &   $3.51 \pm 0.15$ &        $10.84 \pm 1.4$ &  16.5 &    $1.45 \pm 0.26$ &         $6.11 \pm 0.8$ &        $2.90 \pm 0.23$ &        $0.85 \pm 0.01$ &                    0.042380 &                  67.99980 &   1.076  \\
Kepler-94c  &  104.10$^h$ &              $820.3 \pm 3$ &        \nodata    &         $3126 \pm 200$ &  4321 &     \nodata    &      $262.7 \pm 13.9$ &             \nodata    &             \nodata    &                  \nodata    &               \nodata     &   \nodata  \\
Kepler-103b  &    108.01 &                  15.9654 &   $3.37 \pm 0.09$ &        $9.7 \pm 8.6$ &  30.0 &    $1.38 \pm 1.4$ &         $2.32 \pm 2.1$ &        $0.52 \pm 0.02$ &        $0.72 \pm 0.01$ &                    0.021500 &                  75.17614 &   0.866\\
Kepler-103c  &    108.02 &                  179.612 &   $5.14 \pm 0.14$ &      $ 36.10 \pm 25.2$ &  95 &     $1.47 \pm 1.2$ &        $3.85 \pm 2.7$ &        $0.52 \pm 0.02$ &        $0.44 \pm 0.02$ &                    0.032766 &                 228.32581 &   0.824  \\
Kepler-106c  &    116.01 &                  13.5708 &   $2.50 \pm 0.32$ &     $10.44 \pm 3.2$ &     18.8 &    $3.28 \pm 1.56$ &         $2.71 \pm 0.8$ &        $1.38 \pm 0.15$ &        $0.69 \pm 0.03$ &                    0.022113 &                  69.27837 &  1.000 \\
Kepler-106e  &    116.02 &                  43.8445 &   $2.56 \pm 0.33$ &     $11.17 \pm 5.8$ &     24.8 &    $3.10 \pm 2.07$ &         $1.95 \pm 1.0$ &        $1.38 \pm 0.15$ &        $0.28 \pm 0.13$ &                    0.022676 &                  84.93360 &  0.933 \\
Kepler-106b  &    116.03 &                  6.16486 &   $0.82 \pm 0.11$ &         $0.15 \pm 2.8$ &  5.3 &     $1.26 \pm 26.26$ &         $0.05 \pm 0.9$ &        $1.38 \pm 0.15$ &        $0.39 \pm 0.15$ &                    0.007258 &                  68.64035 &  0.992  \\
Kepler-106d  &    116.04 &                  23.9802 &   $0.95 \pm 0.13$ &       $-6.39 \pm 7.0$ &  8.1  &    $-41.00 \pm 39.75$ &        $-2.84 \pm 0.9$ &        $1.38 \pm 0.15$ &        $0.45 \pm 0.11$ &                    0.008362 &                  80.53263 &  1.051 \\
Kepler-95b  &    122.01 &                  11.5231 &   $3.42 \pm 0.09$ &         $13.0 \pm 2.9$ &  16.7 &      $1.71 \pm 0.37$ &         $3.36 \pm 0.6$ &        $0.54 \pm 0.02$ &        $0.74 \pm 0.01$ &                    0.022166 &                  64.96841 & 0.891  \\
Kepler-109b  &    123.01 &                  6.48163 &   $2.37 \pm 0.07$ &          $1.3 \pm 5.4$ &  7.3 &       $0.3 \pm 2.2$ &         $0.29 \pm 1.8$ &        $0.65 \pm 0.02$ &        $0.58 \pm 0.02$ &                    0.016434 &                  55.97755 &   1.000  \\  
Kepler-109c  &    123.02 &                  21.2227 &   $2.52 \pm 0.07$ &         $2.22 \pm 7.8$ &  21.8 &      $0.65 \pm 2.30$ &         $0.43 \pm 1.5$ &        $0.65 \pm 0.02$ &        $0.16 \pm 0.08$ &                    0.017432 &                  70.57250 &  0.975  \\
Kepler-48b  &    148.01 &                  4.77800 &   $1.88 \pm 0.10$ &         $3.94 \pm 2.1$ &  13.0 &      $3.23 \pm 1.84$ &         $1.62 \pm 0.9$ &        $1.98 \pm 0.13$ &        $0.17 \pm 0.11$ &                    0.019318 &                  57.06113 &  0.933  \\
Kepler-48c  &    148.02 &                  9.67395 &   $2.71 \pm 0.14$ &     $14.61 \pm 2.3$ &     22.7 &     ${4.01 \pm 0.91}$ &         $4.74 \pm 0.7$ &        $1.98 \pm 0.13$ &        $0.38 \pm 0.05$ &                    0.027892 &                  58.33925 &   0.992  \\
Kepler-48d  &    148.03 &                  42.8961 &   $2.04 \pm 0.11$ &         $7.93 \pm 4.6$ &  25.0 &      $5.08 \pm 3.12$ &         $1.57 \pm 0.9$ &        $1.98 \pm 0.13$ &        $0.20 \pm 0.11$ &                    0.021021 &                  79.06554 &  0.976 \\
Kepler-48e  &    148.10 &              $982 \pm 8$ &        \nodata    &           $657 \pm 25$ &  727 &           \nodata    &        $45.83 \pm 0.8$ &             \nodata    &             \nodata    &                  \nodata    &               \nodata     &  \nodata   \\
Kepler-113c  &    153.01$^i$ &                  8.92507 &   $2.18 \pm 0.06$ &         $-4.04 \pm 6.4$ &  8.7 &       $-2.13 \pm 3.4$ &        $-1.52 \pm 2.4$ &        $2.71 \pm 0.08$ &        $0.56 \pm 0.02$ &                    0.029135 &                  72.71374 &  0.950   \\
Kepler-113b  &    153.02$^i$ &                  4.75400 &   $1.82 \pm 0.05$ &     $11.7 \pm 4.2$ &      20.  &   ${\bf 10.73 \pm 3.9}$ &         $5.4 \pm 1.9$ &        $2.71 \pm 0.08$ &        $0.08 \pm 0.06$ &                    0.024190 &                  61.54630 &   0.950  \\
Kepler-25c  &   244.01 &                  12.7204 &   $5.20 \pm 0.09$ &        $24.60 \pm 5.7$ &  32.4 &      $0.90 \pm 0.21$ &         $5.63 \pm 1.3$ &        $0.75 \pm 0.01$ &        $0.88 \pm 0.00$ &                    0.036409 &                 111.52730 & 1.127  \\
Kepler-25b  &   244.02 &                   6.2385 &   $2.71 \pm 0.05$ &          $9.6 \pm 4.2$ &  16.0 &      $2.50 \pm 1.10$ &         $2.80 \pm 1.2$ &        $0.75 \pm 0.01$ &        $0.54 \pm 0.01$ &                    0.018950 &                 104.70541 &  1.051   \\
Kepler-25d  &   244.10$^h$ &              $123 \pm 2$ &        \nodata    &       $89.90 \pm 13.7$ &  101 &           \nodata    &         $9.67 \pm 1.5$ &             \nodata    &             \nodata    &                  \nodata    &               \nodata     & \nodata  \\
Kepler-37d  &    245.01 &                  39.7922 &   $1.94 \pm 0.06$ &        $1.87 \pm 9.08$ &  12.2 &     $1.30 \pm 3.67$ &        $0.40 \pm 1.2$ &        $2.46 \pm 0.04$ &        $0.53 \pm 0.01$ &                    0.023068 &                 108.24950 & 1.160  \\
Kepler-37c  &    245.02 &                  21.3020 &   $0.75 \pm 0.03$ &         $3.35 \pm 4.0$ &  12.0 &    $44.33 \pm 53.60$ &         $0.92 \pm 1.1$ &        $2.46 \pm 0.04$ &        $0.43 \pm 0.02$ &                    0.008909 &                 124.83685 &  1.227   \\
Kepler-37b  &    245.03 &                  13.3675 &   $0.32 \pm 0.02$ &        $ 2.78 \pm 3.7$ &  10.0 &  $548.8 \pm 700.0$ &        $0.95 \pm 1.1$ &        $2.46 \pm 0.04$ &        $0.48 \pm 0.13$ &                    0.003828 &                 117.04171 &   0.950  \\
Kepler-68b  &   246.01 &                  5.39875 &   $2.33 \pm 0.02$ &         $5.97 \pm 1.7$ &  9.3 &      $2.60 \pm 0.74$ &         $2.07 \pm 0.59$ &        $0.79 \pm 0.01$ &        $0.48 \pm 0.01$ &                    0.017383 &                 106.85783 &  1.076  \\
Kepler-68c  &   246.02 &                  9.60504 &   $1.00 \pm 0.02$ &         $2.18 \pm 3.5$ &  7.2 &    $10.77 \pm 17.29$ &         $0.57 \pm 0.9$ &        $0.79 \pm 0.01$ &        $0.77 \pm 0.01$ &                    0.007455 &                  69.38025 &  1.084  \\
Kepler-68d  &   246.10$^h$ &             $625 \pm 16$ &        \nodata    &           $267 \pm 16$ &  283 &          \nodata    &        $19.06 \pm 0.58$ &             \nodata    &             \nodata    &                  \nodata    &               \nodata     &  \nodata  \\
Kepler-96b  &    261.01 &                  16.2385 &   $2.67 \pm 0.22$ &         $8.46 \pm 3.4$ &  12.0 &      $2.26 \pm 1.11$ &         $2.10 \pm 0.8$ &        $1.54 \pm 0.25$ &        $0.54 \pm 0.07$ &                    0.023967 &                 104.01897 &  1.219  \\
Kepler-131b  &    283.01 &                  16.0920 &   $2.41 \pm 0.20$ &        $16.13 \pm 3.5$ &  19.4 &      ${\bf 6.00 \pm 1.98}$ &         $3.95 \pm 0.8$ &        $1.44 \pm 0.21$ &        $0.79 \pm 0.02$ &                    0.021263 &                 103.59795 &  1.051  \\
Kepler-131c  &    283.02 &                  25.5169 &   $0.84 \pm 0.07$ &        $8.25 \pm 5.9$ &  20.0 &   $77.7 \pm 55.$ &         $3.58 \pm 0.9$ &        $1.44 \pm 0.21$ &        $0.27 \pm 0.16$ &                    0.007394 &                  87.42342 &  0.891  \\
Kepler-97b  &    292.01 &                  2.58664 &   $1.48 \pm 0.13$ &         $3.51 \pm 1.9$ &  9.1 &      $5.44 \pm 3.48$ &         $1.65 \pm 0.9$ &        $1.53 \pm 0.31$ &        $0.41 \pm 0.16$ &                    0.013803 &                 104.84121 &  0.950   \\
Kepler-97c  &    292.10$^g$ &                   $>$789 &        \nodata    &                 $>$344 &  \nodata &           \nodata    &                  $>$25 &             \nodata    &             \nodata    &                  \nodata    &               \nodata     &  \nodata   \\
Kepler-98b  &    299.01 &                  1.54168 &   $1.99 \pm 0.22$ &         $3.55 \pm 1.6$ &  6.4 &      $2.18 \pm 1.21$ &         $1.82 \pm 0.8$ &        $1.12 \pm 0.34$ &        $0.56 \pm 0.14$ &                    0.016377 &                 103.54328 & 0.950  \\
Kepler-99b  &    305.01 &                  4.60358 &   $1.48 \pm 0.08$ &         $6.15 \pm 1.3$ &  9.7 &      ${\bf 10.90 \pm 2.82}$ &         $2.91 \pm 0.6$ &        $2.86 \pm 0.23$ &        $0.21 \pm 0.13$ &                    0.018475 &                 104.83893 &  0.992   \\
Kepler-406b  &   321.01 &                  2.42629 &   $1.43 \pm 0.03$ &         $6.35 \pm 1.4$ &  8.1 &     ${\bf 11.82 \pm 2.70}$ &         $2.89 \pm 0.6$ &        $1.12 \pm 0.02$ &        $0.10 \pm 0.06$ &                    0.012285 &                 103.45680 &  1.093 \\
Kepler-406c  &   321.02 &                  4.62332 &   $0.85 \pm 0.03$ &         $2.71 \pm 1.8$ &  6.0 &    $24.39 \pm 16.13$ &         $1.00 \pm 0.6$ &        $1.12 \pm 0.02$ &        $0.36 \pm 0.06$ &                    0.007261 &                  65.37438 &  0.883  \\
Kepler-407b  &   1442.01 &                 0.669310 &   $1.07 \pm 0.02$ &         $0.06 \pm 1.2$ &  3.2 &      $0.29 \pm 5.70$ &         $0.05 \pm 0.9$ &        $1.38 \pm 0.16$ &        $0.25 \pm 0.15$ &                    0.010414 &                  67.13162 &  0.958  \\
Kepler-407c  &   1442.10$^g$ &           $3000 \pm 500$ &        \nodata    &         $4000 \pm 2000$ &  8000 &           \nodata    &           $164 \pm 20$ &             \nodata    &             \nodata    &                  \nodata    &               \nodata     &  \nodata  \\
\nodata  &   1612.01 &                  2.46502 &   $0.82 \pm 0.03$ &         $0.48 \pm 3.2$ &  5.0 &     $4.42 \pm 29.82$ &         $0.20 \pm 1.3$ &        $0.82 \pm 0.01$ &        $0.92 \pm 0.01$ &                    0.006136 &                  65.67928 & 0.849  \\
Kepler-409b  &   1925.01 &                  68.9584 &   $1.19 \pm 0.03$ &         $2.69 \pm 6.2$ &  22.0  &     $8.88 \pm 20.60$ &         $0.45 \pm 1.0$ &        $1.81 \pm 0.00$ &        $0.90 \pm 0.01$ &                    0.012223 &                 112.08151 & 1.059  \\
\enddata

\end{deluxetable}
\clearpage
\end{landscape} 

\clearpage

$^a$ Each planet's orbital period is determined using only the \ek 
photometry. 

$^b$ Each planet's radius is measured using the combined 
RV/photometry analysis, consistent with the best values found
for the stellar mass and radius. Errors are dominated by uncertainties
in the stellar radius.

$^c$ Each planet mass is calculated using the combined RV/photometry 
analysis. The stellar parameters are those found in Table 1. The peak of 
the posterior distribution for the planet mass is chosen as the best value.
The peak can have either a positive or negative value.

$^d$ We choose the 95th percentile of the MCMC distribution as the 
upper limit to the mass. We choose to list this value for all planets, but
these values are more meaningful when the mass detection is marginal
or the mass is poorly constrained.

$^e$  Planets with measurements of density
greater than 5 \gcc are typed with bold font, including only those with RV 
mass detections at the $2\sigma$ level or better.   These planets likely have 
interior compositions that are mostly rock and iron/nickel.  We caution that planets having
measured densities only 2--3 $\sigma$ above zero require more RV
measurements to secure their rocky nature.  Two other planets with
densities between 4-5 \gcc are also likely rocky, namely Kepler-102e
and Kepler-48c.

$^f$ The RV amplitude listed is the peak of the posterior distribution for
the combined RV/photometry fit. Uncertainties are determining by
integrating the posterior distribution out from the peak value until
68\% of the values in the MCMC chain are included.

$^g$Long term variation in RVs indicates a long period orbiting object, which may not be a planet \citep{Marcy2001}.

$^h$ A floating orbital eccentricity was allowed in the model for the
outer, non-transiting planet, as indicated by apparent non-sinusoidal
long-term variation in the
RVs, for Kepler-94, Kepler-25, and Kepler-68.  The best-fit eccentricities are  
0.38$\pm$0.05, 0.18$\pm$0.10, and 0.10$\pm$0.04,
respectively.   See notes in Section \ref{sec:obs}
on the individual KOIs for more information.

$^i$  For Kepler-113, the RV residuals to the fit to the two transiting
planets indicate the presence of a third, non-transiting planet.  The
highest periodogram peak is at a period, P = 1.065 d.  But peaks appear
at aliases, namely at 16 d, 0.984 d, and 0.515 d.   We are not able to
determine which among this family of alias periods is the real period,
if any, thus we are not able to assuredly suggest the period or existence of
this prospective non-transiting planet around Kepler-113.

\clearpage 

\begin{deluxetable*}{llllllllll}
\tabletypesize{\footnotesize}
\tablecaption{False Positive Probabilities 
\label{tab:fpp_tbl}}
\tablehead{
\colhead{Kepler \#}  &  \colhead{KOI}     & \colhead{$r_{\rm Excl.}$ [``]$^a$}   &
\colhead{Comp. sep.[``]$^b$} & \colhead{P$_{\rm EB}$}   & \colhead{P$_{\rm HEB}$} &  \colhead{P$_{\rm BGEB}$}   & \colhead{P$_{\rm BGPL}$} & \colhead{{\rm Prior}$_{\rm PL}^c$}  &  \colhead{FPP$^d$} \\
}
\startdata
Kepler-100c &  41.01 & 4-SAT & single & $<$1e-4  & $<$1e-4 &  $<$1e-4  & $<$1e-4 & 0.56 & $<$1e-4   \\
Kepler-100b &  41.02 & 4-SAT & single  & $<$1e-4  & $<$1e-4  & $<$1e-4  & $<$1e-4 &  0.25  & $<$1e-4  \\
Kepler-100d &  41.03 & 4-SAT & single  & $<$1e-4 & $<$1e-4 &  $<$1e-4 & $<$1e-4 & 0.37  & $<$1e-4  \\
Kepler-93b &  69.01 & 4-SAT  & single  &  $<$1e-4 &  $<$1e-4 &  $<$1e-4  & $<$1e-4  &  0.33 &  $<$1e-4 \\
Kepler-102e &  82.01 & 4-SAT & single  & $<$1e-4 &  $<$1e-4 &  $<$1e-4 & $<$1e-4 &  0.56 &  $<$1e-4   \\
Kepler-102d &  82.02 & 4-SAT & single    & $<$1e-4  & $<$1e-4 &  $<$1e-4 &  $<$1e-4  & 0.18 &  $<$1e-4 \\
Kepler-102f &  82.03 & 4-SAT & single    &  $<$1e-4  & $<$1e-4 &  $<$1e-4  & $<$1e-4 &  0.07 &  $<$1e-4 \\
Kepler-102c &   82.04 & 4-SAT & single    &  $<$1e-4 &  $<$1e-4 &  2.5e-4  & 0.89e-4 &  0.01 &  3.4e-4 \\
Kepler-102b  &  82.05 & 4-SAT & single    &  $<$1e-4  & $<$1e-4 &  7.6e-4  & 0.2e-4 &  0.00 &  7.8e-4 \\
Kepler-94b &  104.01 & 0.035 & single  & $<$1e-4 & $<$1e-4 & $<$1e-4  & $<$1e-4 & 0.34 & $<$1e-4    \\
Kepler-103b &  108.01 & 0.009 & 2.44,4.87 & $<$1e-4  & $<$1e-4 &  $<$1e-4 &  $<$1e-4  & 0.37 & $<$1e-4  \\
Kepler-103c &  108.02 & 0.04  & 2.44,4.87 & $<$1e-4  & 1e-4 & $<$1e-4 &  $<$1e-4 &  0.13  & {\bf 1e-4} \\
Kepler-106c &  116.01 & 0.23 & single  &  $<$1e-4 &  $<$1e-4  & $<$1e-4 &  $<$1e-4 &  0.55  & $<$1e-4 \\
Kepler-106e &  116.02 & 0.49 & single  & $<$1e-4 &  $<$1e-4 &  $<$1e-4  & $<$1e-4 &  0.53 &  $<$1e-4 \\
Kepler-106b &  116.03 & 0.83 & single  & $<$1e-4 &  $<$1e-4 &  $<$1e-4 &  $<$1e-4 &  0.05 & $<$1e-4  \\
Kepler-106d &  116.04 & 2.1  & single  & $<$1e-4  & $<$1e-4 & 0.2e-4 &  1.4e-4 &  0.09  & 1.7e-4 \\
Kepler-95b &  122.01 & 0.12 & 4.1 & $<$1e-4 &  $<$1e-4  & $<$1e-4  & $<$1e-4  & 0.37 &  $<$1e-4 \\
Kepler-109b &  123.01 & 0.11 & 2.03,5.3 &  $<$1e-4 &  $<$1e-4 & $<$1e-4  & $<$1e-4 &  0.56 &  $<$1e-4 \\
Kepler-109c &  123.02 & 0.09 & 2.03,5.3 & $<$1e-4  & $<$1e-4  & $<$1e-4  & $<$1e-4 &  0.55 &  $<$1e-4 \\
Kepler-48b &  148.01 & 0.16 & 2.44$^e$ & $<$1e-4  & $<$1e-4 &  $<$1e-4  & $<$1e-4  & 0.54  & $<$1e-4 \\
Kepler-48c &  148.02 & 0.26 & 2.44$^e$ &  $<$1e-4  & $<$1e-4 &  $<$1e-4 &  $<$1e-4 &  0.45 &  $<$1e-4 \\
Kepler-48d &  148.03 & 0.28 & 2.44$^e$ &  $<$1e-4  & $<$1e-4 &  $<$1e-4  & $<$1e-4  & 0.56  & $<$1e-4 \\
Kepler-113c &  153.01 & 0.14 & 5.14 & $<$1e-4  & $<$1e-4 &  $<$1e-4 &  $<$1e-4  & 0.56 &  $<$1e-4 \\
Kepler-113d &  153.02 & 0.09 & 5.14 &  $<$1e-4 & $<$1e-4  &  $<$1e-4 &  $<$1e-4  & 0.47 & $<$1e-4 \\
Kepler-25c &  244.01 & 4-SAT & single  &  $<$1e-4  & $<$1e-4  & $<$1e-4 &  $<$1e-4  & 0.12 &  $<$1e-4 \\
Kepler-25b &  244.02 & 4-SAT & single  &  $<$1e-4  & $<$1e-4 &  $<$1e-4  & $<$1e-4 &  0.51  & $<$1e-4 \\
Kepler-37d &  245.01 & 4-SAT & single  & $<$1e-4 & $<$1e-4 &  $<$1e-4 &  $<$1e-4 &  0.50 &  $<$1e-4  \\
Kepler-37c &  245.02 & 4-SAT & single  & $<$1e-4  & $<$1e-4  & $<$1e-4 &  $<$1e-4  & 0.04  & $<$1e-4 \\
Kepler-37b & 245.03 & 4-SAT & single  & $<$1e-4  & $<$1e-4  & 0.001 & $<$1e-4  & 0.01  &   4e-4$^g$ \\
Kepler-68b &  246.01 & 4-SAT & single  & $<$1e-4  & $<$1e-4 &  $<$1e-4  & $<$1e-4  & 0.55 &  $<$1e-4 \\
Kepler-68c &  246.02 & 4-SAT & single  & $<$1e-4  & $<$1e-4 &  6.2e-4  & $<$1e-4  & 0.15 &  6.3e-4 \\
Kepler-96b &  261.01 & 4-SAT & 5.4 & $<$1e-4 & $<$1e-4 & $<$1e-4 & $<$1e-4 & 0.53 &  $<$1e-4 \\
Kepler-131b &  283.01 & 4-SAT  & 6 & $<$1e-4 &  $<$1e-4 & $<$1e-4  & $<$1e-4  & 0.55 &  $<$1e-4 \\
Kepler-131c &  283.02 & 4-SAT  & 6 & $<$1e-4  & $<$1e-4 &  9.8e-04 &  $<$1e-4 &  0.06 &  9.8e-4 \\
Kepler-97b &  292.01 & 0.20 & 0.38$^f$ & $<$1e-4  & $<$1e-4 & $<$1e-4 &  $<$1e-4 &  0.31  & $<$1e-4 \\
Kepler-98b &  299.01 & 0.15 & single & $<$1e-4 &  $<$1e-4 &  $<$1e-4 &  $<$1e-4 &  0.52 &  $<$1e-4 \\
Kepler-99b &  305.01 & 0.18 & single  & $<$1e-4  & $<$1e-4 & $<$1e-4  & $<$1e-4  & 0.31  & $<$1e-4 \\
Kepler-406b &  321.01 & 0.24 & single  & $<$1e-4  & $<$1e-4 &  $<$1e-4 &  $<$1e-4  & 0.30  & $<$1e-4 \\
Kepler-406c &  321.02 & 0.20 & single  &  $<$1e-4  & $<$1e-4  & $<$1e-4  & $<$1e-4  & 0.06  & $<$1e-4 \\
Kepler-407b &  1442.01 & 0.33 & 2.1 & $<$1e-4  & $<$1e-4 &  $<$1e-4  & $<$1e-4  & 0.14 & $<$1e-4 \\
\nodata &  1612.01 & 4-SAT & single & $<$1e-4  & $<$1e-4  & 2.1e-2  & 1.9e-4  & 0.05  & {\bf 0.021} \\
Kepler-409b &  1925.01 & 4-SAT & single &  $<$1e-4  & $<$1e-4  & 3.2e-4  & $<$1e-4 &  0.18 &  3.2e-4 \\
\enddata
\end{deluxetable*}

\clearpage 

$^a$ The exclusion radius is the maximum angular distance to any
neighboring star that might cause the periodic dimming. It is computed
from the upper limit to the displacement of the photo-center during
times of ``transit.'' This upper limit yields a maximum angular distance
to any star that could cause the apparent transit.  Stars farther away
can't be responsible. The note {\it 4-SAT} denotes saturated images
from \ek, limiting astrometric accuracy to 4$''$ (one pixel).

$^b$ Detected companions within 6$''$ of the target star, found by
AO or speckle observations.  Keck adaptive optics imaging was performed on all 22 KOIs, giving the highest resolution (0{\farcs}050) and best contrast detectability (8 mag at $K'$ bandpass, i.e., 2.2 $\mu$).  ``Single'' denotes no such companions.
Otherwise we list the angular separation in arcsec, (see Sec
\ref{sec:obs}).  The detection thresholds from Keck Adaptive Optics
and from all AO and speckle observations for each KOI are
described in \ref{sec:obs}, and plotted in Figures \ref{fig:koi41_fig1} - \ref{fig:koi1925_fig2}.

$^c$ The planet prior represents the probability for a star to host a planet of that approximate size (between 2/3 and 4/3 the candidate's radius), used in the FPP calculation.  This number is based on an approximate occurrence rate estimate using the \ek candidates, corrected for transit probability, but not for individual detectability.  For cases in which this planet prior is very small (e.g., for Kepler-102c and Kepler-102b), there are very few detected candidates in this radius range, and the true planet prior should be higher.

$^d$ The False-Positive Probability for the Planet: The sum of the probabilities of all false-positive scenarios divided by the sum of probabilities of all scenarios including that the planet is real.

$^e$ For Kepler-48, there are four neighboring stars with separations
and delta magnitudes (J-band): (2{\farcs}44,4.9), (4{\farcs}32,3.3), (4{\farcs}39,7.3),
(5{\farcs}89,7.0) \citep{Adams2012}.

$^f$ Kepler-97: The neighbor (0{\farcs}36) is farther than the
astrometric maximum radius (0{\farcs}2).   Thus, the planet probably doesn't orbit neighbor.

$^g$ Careful false positive analysis by \cite{Barclay2013} shows FPP = 0.0004.

\clearpage 


\begin{deluxetable}{ccccc}
\tabletypesize{\footnotesize}
\tablecaption{Radial Velocities for KOI-41, Kepler-100
\label{tab:rvs_k00041}}
\tablewidth{0pt}
\tablehead{
\colhead{}         & \colhead{Radial Velocity}     & \colhead{Uncertainty}  & \colhead{\rphk}  \\
\colhead{BJD -- 2450000}   & \colhead{(\mse)}  & \colhead{(\mse)} & \colhead{} 
}
\startdata
    4989.010954 &    -4.46 &     3.73 &   -5.028 \\
    5042.015300 &   -11.82 &     3.81 &   -5.034 \\
    5042.022823 &    -9.46 &     3.86 &   -5.028 \\
    5042.030068 &    -6.02 &     3.77 &   -5.041 \\
    5042.842452 &    -3.45 &     3.63 &   -5.041 \\
    5042.849778 &     8.90 &     3.75 &   -5.014 \\
    5044.044453 &     5.07 &     3.73 &   -5.034 \\
    5044.051871 &     3.30 &     3.77 &   -5.014 \\
    5044.059302 &     0.62 &     3.73 &   -5.028 \\
    5048.909978 &    -7.17 &     3.93 &   -5.014 \\
    5073.846868 &    -1.39 &     3.70 &   -5.041 \\
    5074.844066 &     0.39 &     3.59 &   -5.034 \\
    5075.868856 &     3.61 &     3.67 &   -5.028 \\
    5076.756808 &     6.56 &     3.65 &   -5.021 \\
    5077.894672 &    -9.78 &     3.64 &   -5.048 \\
    5078.934772 &    -3.54 &     3.71 &   -5.014 \\
    5079.962974 &    -0.23 &     3.71 &   -5.041 \\
    5081.002505 &    -3.08 &     3.55 &   -4.953 \\
    5082.982668 &     2.99 &     3.77 &   -5.041 \\
    5312.060754 &    -1.25 &     2.78 &   -5.041 \\
    5400.916085 &     2.42 &     2.62 &   -5.109 \\
    5760.064528 &    -1.11 &     2.51 &   -5.041 \\
    5760.958645 &     3.39 &     2.56 &   -5.041 \\
    5761.876117 &     5.12 &     2.48 &   -5.048 \\
    5762.912248 &     6.33 &     2.42 &   -5.041 \\
    5763.870182 &     5.01 &     2.45 &   -5.034 \\
    5791.871645 &    -1.99 &     2.47 &   -5.048 \\
    5793.996566 &    -2.77 &     2.55 &   -5.070 \\
    5795.012741 &   -12.15 &     3.13 &   -5.085 \\
    5796.793068 &     1.87 &     2.42 &   -5.021 \\
    5797.853511 &     5.06 &     2.41 &   -5.041 \\
    5798.773192 &     1.43 &     2.37 &   -5.048 \\
    5807.741761 &     2.21 &     2.55 &   -5.070 \\
    5808.840217 &    -5.33 &     2.61 &   -5.028 \\
    5809.749926 &     2.75 &     2.45 &   -5.048 \\
    5810.749216 &     4.68 &     2.42 &   -5.041 \\
    5811.865319 &     6.67 &     2.40 &   -5.041 \\
    5814.757203 &     2.23 &     2.47 &   -5.048 \\
    6109.979987 &     4.30 &     2.46 &   -5.055 \\
    6110.856162 &     0.99 &     2.59 &   -5.055 \\
    6111.867315 &     0.32 &     2.56 &   -5.055 \\
    6112.888295 &     4.29 &     2.49 &   -5.055 \\
    6113.851194 &     1.74 &     2.65 &   -5.117 \\
    6114.900100 &     3.51 &     2.56 &   -5.034 \\
    6115.897107 &     7.42 &     2.53 &   -5.048 \\
    6173.889875 &    -1.85 &     2.47 &   -5.055 \\
    6177.929359 &    -7.30 &     3.07 &   -5.048 \\
    6178.805849 &     1.60 &     2.44 &   -5.048 \\
    6209.725678 &    -0.93 &     2.55 &   -5.063 \\

\enddata
\end{deluxetable}

\begin{deluxetable}{ccccc}
\tabletypesize{\footnotesize}
\tablecaption{Radial Velocities for KOI-69, Kepler-93
\label{tab:rvs_k00069}}
\tablewidth{0pt}
\tablehead{
\colhead{}         & \colhead{Radial Velocity}     & \colhead{Uncertainty}  & \colhead{\rphk}  \\
\colhead{BJD -- 2450000}   & \colhead{(\mse)}  & \colhead{(\mse)} & \colhead{} 
}
\startdata
    5044.067242 &   -12.90 &     2.73 &   -4.983 \\
    5045.039518 &   -19.29 &     2.85 &   -4.972 \\
    5074.010142 &   -11.01 &     2.44 &   -4.972 \\
    5075.008297 &   -13.22 &     2.45 &   -4.983 \\
    5075.985757 &   -11.06 &     2.43 &   -4.972 \\
    5077.017498 &   -14.77 &     2.48 &   -4.994 \\
    5078.016299 &   -16.70 &     2.47 &   -4.972 \\
    5078.943562 &   -14.89 &     2.44 &   -4.972 \\
    5079.935707 &   -13.23 &     2.39 &   -4.972 \\
    5080.753597 &    -7.78 &     2.38 &   -4.972 \\
    5080.792010 &    -9.66 &     2.44 &   -4.966 \\
    5081.769317 &    -8.65 &     2.59 &   -4.966 \\
    5081.776018 &   -10.64 &     2.71 &   -4.977 \\
    5082.741078 &    -6.95 &     2.47 &   -4.966 \\
    5083.753822 &    -6.48 &     2.43 &   -4.966 \\
    5084.751762 &    -5.96 &     2.39 &   -4.961 \\
    5373.033115 &    -2.16 &     2.41 &   -4.977 \\
    5377.938646 &    -2.78 &     2.38 &   -4.966 \\
    5696.963643 &    12.83 &     2.57 &   -4.977 \\
    5697.938827 &     9.94 &     2.42 &   -4.977 \\
    5698.937161 &     7.95 &     2.45 &   -4.972 \\
    5700.017234 &    14.34 &     2.39 &   -4.972 \\
    5723.039490 &     7.80 &     2.47 &   -4.977 \\
    5759.820878 &     6.04 &     2.51 &   -4.983 \\
    5760.863548 &    12.19 &     2.43 &   -4.972 \\
    5762.926837 &     6.74 &     2.40 &   -4.983 \\
    5768.842459 &    11.17 &     2.44 &   -4.983 \\
    5807.767167 &    16.08 &     2.47 &   -5.005 \\
    6110.033656 &    19.10 &     2.50 &   -4.972 \\
    6115.033413 &    22.14 &     2.45 &   -4.977 \\
    6148.820507 &    28.18 &     2.56 &   -4.972 \\
    6174.837273 &    22.79 &     2.36 &   -4.972 \\
\enddata
\end{deluxetable}

\begin{deluxetable}{ccccc}
\tabletypesize{\footnotesize}
\tablecaption{Radial Velocities for KOI-82, Kepler-102
\label{tab:rvs_k00082}}
\tablewidth{0pt}
\tablehead{
\colhead{}         & \colhead{Radial Velocity}     & \colhead{Uncertainty}  & \colhead{\rphk}  \\
\colhead{BJD -- 2450000}   & \colhead{(\mse)}  & \colhead{(\mse)} & \colhead{} 
}
\startdata
    5312.078536 &    -5.59 &     2.39 &   -4.586 \\
    5373.830100 &    -0.92 &     2.40 &   -4.618 \\
    5402.922434 &    -4.08 &     2.31 &   -4.658 \\
    5414.824561 &     2.83 &     2.30 &   -4.625 \\
    5428.776770 &    -0.60 &     2.33 &   -4.644 \\
    5431.766490 &    -4.13 &     2.30 &   -4.644 \\
    5439.774551 &     0.28 &     2.27 &   -4.484 \\
    5440.884419 &     3.01 &     2.31 &   -4.667 \\
    5752.082671 &    -1.28 &     2.41 &   -4.573 \\
    5752.923634 &    -0.57 &     2.54 &   -4.594 \\
    5752.940613 &    -0.33 &     2.60 &   -4.587 \\
    5759.888028 &     1.25 &     2.33 &   -4.586 \\
    5760.825411 &    -2.10 &     2.32 &   -4.601 \\
    5787.782634 &     7.80 &     2.32 &   -4.598 \\
    5788.796321 &     8.08 &     2.32 &   -4.604 \\
    5794.989286 &    -5.42 &     2.47 &   -4.561 \\
    5795.870970 &    -1.38 &     2.32 &   -4.563 \\
    5809.768343 &    -1.66 &     2.31 &   -4.633 \\
    5810.766141 &    -0.21 &     2.29 &   -4.636 \\
    5811.846830 &     0.54 &     2.28 &   -4.617 \\
    5814.775163 &     2.23 &     2.31 &   -4.594 \\
    6110.049088 &     2.21 &     2.41 &   -4.626 \\
    6110.877659 &     1.30 &     2.33 &   -4.604 \\
    6111.885201 &    -1.67 &     2.32 &   -4.603 \\
    6112.910601 &     4.16 &     2.36 &   -4.581 \\
    6113.892412 &     4.91 &     2.38 &   -4.610 \\
    6115.097881 &     3.01 &     2.60 &   -4.624 \\
    6133.877339 &    -3.08 &     2.34 &   -4.620 \\
    6134.913293 &    -9.37 &     2.37 &   -4.624 \\
    6141.045454 &     1.27 &     2.40 &   -4.630 \\
    6144.835181 &     0.78 &     2.37 &   -4.636 \\
    6145.835817 &     3.49 &     2.27 &   -4.630 \\
    6166.822306 &   -10.67 &     2.46 &   -4.671 \\
    6172.780889 &    -1.06 &     2.42 &   -4.641 \\
    6208.797480 &     9.06 &     2.41 &   -4.580 \\
\enddata
\end{deluxetable}

\begin{deluxetable}{ccccc}
\tabletypesize{\footnotesize}
\tablecaption{Radial Velocities for KOI-104, Kepler-94
\label{tab:rvs_k00104}}
\tablewidth{0pt}
\tablehead{
\colhead{}         & \colhead{Radial Velocity}     & \colhead{Uncertainty}  & \colhead{\rphk}  \\
\colhead{BJD -- 2450000}   & \colhead{(\mse)}  & \colhead{(\mse)} & \colhead{} 
}
\startdata
    5377.841906 &   -65.50 &     2.75 &   -4.638 \\
    5378.934287 &   -54.36 &     2.96 &   -4.616 \\
    5379.961656 &   -66.73 &     3.07 &   -4.701 \\
    5433.761242 &   -11.72 &     2.60 &   -4.685 \\
    5434.927884 &   -16.61 &     2.80 &   -4.656 \\
    5438.852111 &     1.55 &     3.20 &   -4.857 \\
    5438.881509 &     9.99 &     2.98 &   -4.756 \\
    5465.829854 &    28.56 &     3.26 &   -4.424 \\
    5521.729851 &    56.90 &     3.14 &   -4.966 \\
    5697.049110 &    99.24 &     2.93 &   -4.601 \\
    5698.037021 &    80.44 &     2.80 &   -4.686 \\
    5699.034699 &    97.68 &     3.26 &   -4.671 \\
    5699.997226 &    93.70 &     2.79 &   -4.560 \\
    5723.980007 &    85.68 &     2.97 &   -4.577 \\
    5733.960824 &    77.25 &     2.90 &   -4.567 \\
    5734.920790 &    82.12 &     2.76 &   -4.591 \\
    5738.849308 &    77.70 &     2.87 &   -4.471 \\
    5759.912267 &    66.97 &     3.05 &   -4.873 \\
    5760.882081 &    58.78 &     2.79 &   -4.654 \\
    5789.922472 &    43.45 &     2.89 &   -4.946 \\
    5790.939908 &    28.25 &     2.81 &   -4.635 \\
    5798.969403 &    26.80 &     2.79 &   -4.600 \\
    5809.956676 &    33.38 &     2.82 &   -4.730 \\
    5810.845889 &    26.81 &     2.67 &   -4.586 \\
    6077.025088 &  -285.15 &     2.79 &   -4.576 \\
    6115.875911 &  -184.39 &     2.73 &   -4.547 \\
    6135.066858 &  -166.59 &     3.17 &   -4.581 \\
    6147.946195 &  -136.85 &     3.03 &   -4.585 \\
    6176.762087 &   -90.03 &     2.85 &   -4.573 \\
\enddata
\end{deluxetable}

\begin{deluxetable}{ccccc}
\tabletypesize{\footnotesize}
\tablecaption{Radial Velocities for KOI-108, Kepler-103
\label{tab:rvs_k00108}}
\tablewidth{0pt}
\tablehead{
\colhead{}         & \colhead{Radial Velocity}     & \colhead{Uncertainty}  & \colhead{\rphk}  \\
\colhead{BJD -- 2450000}   & \colhead{(\mse)}  & \colhead{(\mse)} & \colhead{} 
}
\startdata
    5073.968560 &    -4.76 &     6.71 &   -4.999 \\
    5074.938879 &   -12.07 &     6.64 &   -4.999 \\
    5077.005648 &    -7.20 &     7.06 &   -5.054 \\
    5079.991586 &    -2.46 &     6.66 &   -5.054 \\
    5080.982908 &    -3.87 &     6.66 &   -4.836 \\
    5082.902081 &     5.64 &     6.67 &   -4.999 \\
    5084.913610 &     5.06 &     6.63 &   -4.986 \\
    5134.787135 &    -6.19 &     3.33 &   -5.012 \\
    5313.104615 &    -2.59 &     2.84 &   -5.006 \\
    5321.997726 &     7.95 &     2.89 &   -5.006 \\
    5377.875117 &    -0.31 &     2.90 &   -5.006 \\
    5378.838317 &    -0.52 &     3.00 &   -5.026 \\
    5402.987989 &     4.71 &     2.91 &   -5.026 \\
    5411.034645 &     0.12 &     2.90 &   -5.006 \\
    5425.916614 &    -3.17 &     2.93 &   -5.012 \\
    5433.844314 &     4.34 &     2.78 &   -4.986 \\
    5440.822218 &    -2.76 &     2.70 &   -5.006 \\
    5794.030378 &     3.36 &     3.14 &   -5.046 \\
    5809.900772 &    -4.82 &     2.70 &   -4.999 \\
\enddata
\end{deluxetable}

\begin{deluxetable}{ccccc}
\tabletypesize{\footnotesize}
\tablecaption{Radial Velocities for KOI-116, Kepler-106
\label{tab:rvs_k00116}}
\tablewidth{0pt}
\tablehead{
\colhead{}         & \colhead{Radial Velocity}     & \colhead{Uncertainty}  & \colhead{\rphk}  \\
\colhead{BJD -- 2450000}   & \colhead{(\mse)}  & \colhead{(\mse)} & \colhead{} 
}
\startdata
    5133.897351 &     9.29 &     4.06 &   -4.783 \\
    5373.958416 &     6.52 &     3.03 &   -4.948 \\
    5379.001449 &     1.61 &     3.29 &   -4.966 \\
    5380.025601 &    -7.92 &     2.98 &   -4.960 \\
    5433.918565 &    -3.61 &     3.02 &   -4.920 \\
    5439.876430 &    -2.17 &     3.07 &   -4.960 \\
    5441.023521 &    -4.05 &     3.23 &   -4.978 \\
    5734.085379 &     0.06 &     3.14 &   -4.960 \\
    5752.894726 &     6.75 &     3.20 &   -4.960 \\
    5759.992485 &    -0.07 &     3.27 &   -4.978 \\
    6111.910098 &     2.89 &     3.16 &   -4.948 \\
    6113.987149 &     1.15 &     3.32 &   -4.990 \\
    6115.949113 &     4.10 &     3.11 &   -4.960 \\
    6133.944504 &     5.29 &     3.24 &   -4.966 \\
    6138.981896 &    -5.09 &     3.04 &   -4.960 \\
    6144.975318 &     4.26 &     3.35 &   -4.954 \\
    6145.858923 &     4.94 &     3.06 &   -4.954 \\
    6147.834465 &    -0.29 &     3.30 &   -4.960 \\
    6151.043480 &    -4.99 &     3.19 &   -4.948 \\
    6152.041410 &   -14.82 &     3.46 &   -5.003 \\
    6153.049571 &    -3.34 &     3.56 &   -5.074 \\
    6166.790534 &    -4.91 &     3.15 &   -4.948 \\
    6173.825374 &     0.14 &     3.12 &   -4.966 \\
    6178.784524 &    -8.28 &     3.07 &   -4.966 \\
    6207.805559 &     8.16 &     3.34 &   -4.948 \\
\enddata
\end{deluxetable}

\begin{deluxetable}{ccccc}
\tabletypesize{\footnotesize}
\tablecaption{Radial Velocities for KOI-122, Kepler-95
\label{tab:rvs_k00122}}
\tablewidth{0pt}
\tablehead{
\colhead{}         & \colhead{Radial Velocity}     & \colhead{Uncertainty}  & \colhead{\rphk}  \\
\colhead{BJD -- 2450000}   & \colhead{(\mse)}  & \colhead{(\mse)} & \colhead{} 
}
\startdata
    5073.994534 &    -2.19 &     6.81 &   -5.002 \\
    5075.946624 &     3.90 &     6.80 &   -4.991 \\
    5076.769704 &    -0.63 &     6.72 &   -4.974 \\
    5079.919778 &     2.64 &     6.81 &   -5.014 \\
    5080.930639 &    -8.02 &     6.81 &   -5.014 \\
    5081.929904 &    -4.97 &     6.79 &   -4.996 \\
    5082.942536 &    -7.91 &     6.78 &   -5.002 \\
    5083.865170 &    -3.44 &     6.76 &   -5.014 \\
    5084.842177 &    -3.48 &     6.71 &   -4.968 \\
    5313.064246 &    -9.40 &     2.50 &   -4.991 \\
    5318.098919 &     4.62 &     2.67 &   -4.991 \\
    5377.020314 &     8.40 &     2.53 &   -4.985 \\
    5400.944047 &    -0.41 &     2.80 &   -5.008 \\
    5405.939014 &     3.36 &     2.57 &   -4.996 \\
    5411.002552 &     5.55 &     2.49 &   -4.974 \\
    5411.896640 &    11.00 &     2.65 &   -4.979 \\
    5427.839158 &    -2.81 &     2.59 &   -5.008 \\
    5428.800667 &     0.84 &     2.49 &   -4.996 \\
    5434.798117 &    -2.08 &     2.44 &   -4.996 \\
    5439.834670 &    -2.72 &     2.42 &   -5.008 \\
    5486.806230 &    -2.87 &     2.86 &   -5.026 \\
    5697.081519 &     4.67 &     2.68 &   -5.039 \\
    5698.069331 &     2.09 &     2.60 &   -5.033 \\
    5699.066159 &    -1.67 &     2.63 &   -5.045 \\
    5700.096380 &     1.79 &     2.65 &   -5.033 \\
    5738.817453 &    -3.66 &     2.67 &   -5.014 \\
    5791.763490 &     4.17 &     2.48 &   -4.963 \\
    5797.013255 &     1.95 &     2.77 &   -5.045 \\
    5809.927350 &    -2.30 &     2.47 &   -5.014 \\
    5814.898557 &     3.18 &     2.52 &   -5.026 \\
    6152.011007 &     1.33 &     2.64 &   -4.985 \\
\enddata
\end{deluxetable}

\begin{deluxetable}{ccccc}
\tabletypesize{\footnotesize}
\tablecaption{Radial Velocities for KOI-123, Kepler-109
\label{tab:rvs_k00123}}
\tablewidth{0pt}
\tablehead{
\colhead{}         & \colhead{Radial Velocity}     & \colhead{Uncertainty}  & \colhead{\rphk}  \\
\colhead{BJD -- 2450000}   & \colhead{(\mse)}  & \colhead{(\mse)} & \colhead{} 
}
\startdata
    5074.992468 &     7.66 &     7.12 &   -5.014 \\
    5075.969200 &     6.70 &     7.11 &   -5.007 \\
    5076.849395 &     9.48 &     7.05 &   -4.994 \\
    5080.956368 &   -10.35 &     7.11 &   -5.021 \\
    5082.920913 &     6.64 &     7.13 &   -4.994 \\
    5083.964085 &     5.57 &     7.21 &   -5.000 \\
    5084.949365 &     7.55 &     7.11 &   -4.994 \\
    5170.755814 &    -7.51 &     3.49 &   -5.000 \\
    5314.059619 &    -6.09 &     2.80 &   -5.007 \\
    5321.104302 &    -1.74 &     2.85 &   -5.021 \\
    5343.073868 &   -10.93 &     2.97 &   -5.028 \\
    5344.098191 &    -6.20 &     3.10 &   -5.028 \\
    5372.963165 &    -0.29 &     3.02 &   -5.007 \\
    5378.870526 &    -0.34 &     2.97 &   -5.021 \\
    6166.844942 &    -5.35 &     2.91 &   -4.994 \\
\enddata
\end{deluxetable}

\begin{deluxetable}{ccccc}
\tabletypesize{\footnotesize}
\tablecaption{Radial Velocities for KOI-148, Kepler-48
\label{tab:rvs_k00148}}
\tablewidth{0pt}
\tablehead{
\colhead{}         & \colhead{Radial Velocity}     & \colhead{Uncertainty}  & \colhead{\rphk}  \\
\colhead{BJD -- 2450000}   & \colhead{(\mse)}  & \colhead{(\mse)} & \colhead{} 
}
\startdata
    5074.026507 &    -1.31 &    12.26 &   -4.879 \\
    5075.032090 &    -1.48 &    12.21 &   -4.859 \\
    5076.799857 &    -2.81 &    12.08 &   -4.873 \\
    5077.955169 &     1.14 &    12.19 &   -4.865 \\
    5077.975134 &    -8.53 &    12.17 &   -4.879 \\
    5312.106882 &    38.33 &     3.16 &   -4.873 \\
    5377.908091 &    18.04 &     2.79 &   -4.848 \\
    5381.054562 &    31.24 &     2.83 &   -4.876 \\
    5407.078286 &    32.63 &     2.97 &   -4.820 \\
    5414.941524 &    18.40 &     2.73 &   -4.813 \\
    5434.897144 &     6.44 &     2.55 &   -4.840 \\
    5438.783842 &    24.56 &     2.77 &   -4.810 \\
    5438.811770 &    25.55 &     2.92 &   -4.813 \\
    5787.843659 &   -55.28 &     2.86 &   -4.818 \\
    5793.775124 &   -59.11 &     2.85 &   -4.833 \\
    5796.821157 &   -48.12 &     2.63 &   -4.810 \\
    5797.947000 &   -49.70 &     2.66 &   -4.828 \\
    5799.028618 &   -53.15 &     2.90 &   -4.818 \\
    5807.829071 &   -49.37 &     2.92 &   -4.835 \\
    5810.977776 &   -48.64 &     2.89 &   -4.833 \\
    5811.953784 &   -53.35 &     2.60 &   -4.843 \\
    5814.865587 &   -52.15 &     2.75 &   -4.843 \\
    6115.912076 &    16.62 &     3.13 &   -4.791 \\
    6141.025441 &    12.07 &     3.22 &   -4.840 \\
    6145.890599 &    36.27 &     3.17 &   -4.851 \\
    6164.829375 &    48.32 &     3.11 &   -4.825 \\
    6173.848569 &    37.85 &     2.91 &   -4.848 \\
    6208.750960 &    33.64 &     3.07 &   -4.810 \\
\enddata
\end{deluxetable}

\begin{deluxetable}{ccccc}
\tabletypesize{\footnotesize}
\tablecaption{Radial Velocities for KOI-153, Kepler-113
\label{tab:rvs_k00153}}
\tablewidth{0pt}
\tablehead{
\colhead{}         & \colhead{Radial Velocity}     & \colhead{Uncertainty}  & \colhead{\rphk}  \\
\colhead{BJD -- 2450000}   & \colhead{(\mse)}  & \colhead{(\mse)} & \colhead{} 
}
\startdata
    5314.091538 &    -7.08 &     2.97 &   -5.018 \\
    5373.884706 &    -5.21 &     3.08 &   -5.060 \\
    5376.896191 &    -9.70 &     3.07 &   -5.175 \\
    5378.901697 &    -4.04 &     3.05 &   -5.028 \\
    5381.020292 &    -8.27 &     3.16 &   -5.133 \\
    5428.828387 &    -6.66 &     3.00 &   -5.084 \\
    5435.804603 &    10.57 &     3.16 &   -5.257 \\
    5437.846080 &    -7.31 &     3.55 &   -4.945 \\
    5437.882874 &   -10.84 &     3.39 &   -5.098 \\
    5697.110582 &    20.09 &     3.43 &   -4.815 \\
    5698.004879 &    17.39 &     3.05 &   -4.830 \\
    5699.002293 &     4.43 &     3.27 &   -4.731 \\
    5700.066664 &    15.06 &     3.20 &   -4.949 \\
    5735.002572 &    -3.16 &     3.13 &   -5.084 \\
    5752.830942 &    -4.22 &     3.23 &   -5.013 \\
    5761.058504 &     8.01 &     2.99 &   -5.065 \\
    5789.952972 &    -4.86 &     3.20 &   -5.023 \\
    5792.017851 &    -0.80 &     2.95 &   -5.052 \\
    5796.872421 &    11.64 &     3.06 &   -5.076 \\
    5811.816797 &     7.82 &     3.02 &   -5.079 \\
    6110.983363 &    -5.14 &     3.32 &   -5.149 \\
    6112.973425 &   -15.97 &     3.34 &   -5.098 \\
    6144.037413 &    -0.53 &     3.27 &   -5.152 \\
    6147.014037 &    -4.98 &     3.29 &   -4.965 \\
\enddata
\end{deluxetable}

\begin{deluxetable}{ccccc}
\tabletypesize{\footnotesize}
\tablecaption{Radial Velocities for KOI-244, Kepler-25
\label{tab:rvs_k00244}}
\tablewidth{0pt}
\tablehead{
\colhead{}         & \colhead{Radial Velocity}     & \colhead{Uncertainty}  & \colhead{\rphk}  \\
\colhead{BJD -- 2450000}   & \colhead{(\mse)}  & \colhead{(\mse)} & \colhead{} 
}
\startdata
    5367.102834 &    -7.68 &     4.28 &   -5.183 \\
    5376.963016 &     3.26 &     3.89 &   -5.195 \\
    5377.950450 &    -2.78 &     3.73 &   -5.172 \\
    5433.941912 &    14.54 &     4.22 &   -5.258 \\
    5696.949358 &    12.64 &     3.97 &   -5.207 \\
    5697.952294 &     2.17 &     4.29 &   -5.219 \\
    5698.948910 &    -4.71 &     3.98 &   -5.219 \\
    5700.027670 &    10.59 &     4.03 &   -5.207 \\
    5734.052698 &    -1.17 &     3.67 &   -5.195 \\
    5735.986640 &   -13.52 &     3.84 &   -5.207 \\
    5739.049628 &    -2.81 &     4.63 &   -5.195 \\
    5751.935799 &   -16.11 &     3.97 &   -5.195 \\
    5752.793676 &   -18.65 &     3.70 &   -5.183 \\
    5759.966063 &     5.85 &     3.59 &   -5.195 \\
    5760.809525 &    -4.73 &     3.78 &   -5.183 \\
    5761.102651 &    -4.29 &     3.67 &   -5.183 \\
    5762.100290 &     5.63 &     4.10 &   -5.207 \\
    5762.110117 &     1.44 &     3.86 &   -5.219 \\
    5762.118219 &   -13.47 &     5.07 &   -5.161 \\
    5762.891494 &    -3.59 &     3.28 &   -5.207 \\
    5763.840742 &    -8.05 &     3.75 &   -5.195 \\
    5768.852800 &    -8.83 &     3.90 &   -5.219 \\
    5769.943894 &     8.72 &     4.09 &   -6.789 \\
    5782.065146 &    -6.14 &     4.58 &   -5.301 \\
    5787.762931 &    -0.92 &     4.28 &   -5.232 \\
    5788.938209 &    -5.17 &     3.93 &   -5.219 \\
    5789.798064 &     1.01 &     4.21 &   -5.245 \\
    5789.804939 &     3.83 &     4.57 &   -5.245 \\
    5790.755601 &    -3.04 &     3.81 &   -5.232 \\
    5791.933168 &    -0.39 &     3.94 &   -5.245 \\
    5792.767270 &     3.70 &     3.76 &   -5.245 \\
    5794.925518 &    17.95 &     4.26 &   -5.272 \\
    5796.893200 &    16.45 &     3.96 &   -5.245 \\
    5798.759251 &     3.79 &     3.96 &   -5.258 \\
    5806.789430 &    14.91 &     3.97 &   -5.272 \\
    5808.797455 &    20.19 &     3.80 &   -5.258 \\
    5810.734832 &     7.34 &     4.10 &   -5.258 \\
    5814.927264 &    15.10 &     4.13 &   -5.272 \\
    5903.699358 &    10.63 &     4.06 &   -5.258 \\
    6079.872621 &    -1.44 &     4.36 &   -5.333 \\
    6080.109747 &     4.40 &     4.07 &   -5.207 \\
    6114.852234 &     3.40 &     4.35 &   -5.161 \\
    6134.929891 &   -11.23 &     3.95 &   -5.219 \\
    6139.028357 &    -3.47 &     4.17 &   -5.207 \\
    6145.944136 &    -0.85 &     4.72 &   -5.207 \\
    6147.806840 &    -5.67 &     4.06 &   -5.219 \\
    6148.839914 &    -6.37 &     4.43 &   -5.232 \\
    6152.962700 &    16.32 &     4.35 &   -5.245 \\
    6163.794337 &    22.99 &     4.15 &   -5.232 \\
    6194.822303 &     6.48 &     4.58 &   -5.258 \\
    5803.3364  & 10.11    &   16.00    &  0 \\
    5809.5034  & 25.41    &   19.00    &  0 \\
    5810.4814  & -10.69   &   25.00    &  0 \\
    5811.3940  & -4.09    &   12.00    &  0 \\
    5828.2935  & -5.79    &   10.00    &  0 \\
    5855.2837  & -9.89    &   19.00    &  0 \\
    5857.2627  & -15.39   &   17.00    &  0 \\
    5877.6963  &  4.21    &   16.00    &  0 \\
    5878.2544  &  1.01    &   16.00    &  0 \\
    5879.2661  & -14.69   &   16.00    &  0 \\
    5882.3022  & 13.81    &   16.00    &  0 \\
    5883.2764  & 20.81    &   15.00    &  0 \\
\enddata
\end{deluxetable}

\begin{deluxetable}{ccccc}
\tabletypesize{\footnotesize}
\tablecaption{Radial Velocities for KOI-245, Kepler-37
\label{tab:rvs_k00245}}
\tablewidth{0pt}
\tablehead{
\colhead{}         & \colhead{Radial Velocity}     & \colhead{Uncertainty}  & \colhead{\rphk}  \\
\colhead{BJD -- 2450000}   & \colhead{(\mse)}  & \colhead{(\mse)} & \colhead{} 
}
\startdata
    5313.086400 &     3.47 &     2.46 &   -4.972 \\
    5319.116525 &     1.15 &     2.61 &   -4.980 \\
    5376.949842 &     0.14 &     2.45 &   -5.008 \\
    5396.019973 &     2.18 &     2.53 &   -4.990 \\
    5400.968136 &     8.98 &     2.39 &   -5.116 \\
    5425.934247 &     1.79 &     2.27 &   -5.023 \\
    5433.932956 &    -0.32 &     2.28 &   -5.013 \\
    5723.052494 &    -1.50 &     2.46 &   -4.959 \\
    5723.100134 &    -1.99 &     2.41 &   -4.954 \\
    5768.862142 &     3.46 &     2.29 &   -5.013 \\
    5769.817056 &     1.64 &     2.35 &   -5.013 \\
    5770.858798 &     2.72 &     2.36 &   -4.976 \\
    5782.046825 &    -1.33 &     2.38 &   -5.038 \\
    5782.900841 &    -1.14 &     2.38 &   -5.003 \\
    5787.754366 &    -4.12 &     2.34 &   -5.013 \\
    5788.994542 &    -5.71 &     2.40 &   -5.028 \\
    5789.882547 &     0.50 &     2.40 &   -5.023 \\
    5790.747253 &    -4.64 &     2.34 &   -5.023 \\
    5791.748658 &    -2.62 &     2.39 &   -5.023 \\
    5792.759449 &    -0.45 &     2.34 &   -5.013 \\
    5794.938166 &    -0.70 &     2.37 &   -5.028 \\
    5795.758577 &    -2.35 &     2.47 &   -5.043 \\
    5796.781113 &    -3.02 &     2.38 &   -5.003 \\
    5797.783521 &    -3.02 &     2.32 &   -5.003 \\
    5798.751068 &   -10.90 &     2.40 &   -4.999 \\
    6098.081597 &     3.20 &     2.52 &   -4.938 \\
    6115.057869 &    -0.46 &     2.46 &   -4.976 \\
    6148.956270 &     2.18 &     2.29 &   -4.959 \\
    6163.786571 &     3.90 &     2.38 &   -4.918 \\
    6164.745854 &     5.34 &     2.36 &   -4.907 \\
    6166.810249 &    -2.70 &     2.46 &   -4.910 \\
    6168.007627 &     2.90 &     2.50 &   -4.938 \\
    6174.952383 &     0.01 &     2.40 &   -4.907 \\
\enddata
\end{deluxetable}

\begin{deluxetable}{ccccc}
\tabletypesize{\footnotesize}
\tablecaption{Radial Velocities for KOI-246, Kepler-68
\label{tab:rvs_k00246}}
\tablewidth{0pt}
\tablehead{
\colhead{}         & \colhead{Radial Velocity}     & \colhead{Uncertainty}  & \colhead{\rphk}  \\
\colhead{BJD -- 2450000}   & \colhead{(\mse)}  & \colhead{(\mse)} & \colhead{} 
}
\startdata
    5313.081849 &    -5.32 &     2.49 &   -5.144 \\
    5319.109464 &    -9.27 &     2.64 &   -5.190 \\
    5322.050550 &   -11.90 &     2.34 &   -5.162 \\
    5372.982740 &   -10.49 &     2.32 &   -5.162 \\
    5377.929044 &    -9.56 &     2.33 &   -5.153 \\
    5381.000211 &   -17.12 &     2.39 &   -5.162 \\
    5396.962763 &   -14.26 &     2.50 &   -5.144 \\
    5412.922534 &   -18.20 &     2.33 &   -5.171 \\
    5426.913366 &   -15.93 &     2.35 &   -5.119 \\
    5431.784060 &   -10.20 &     2.32 &   -5.144 \\
    5434.870088 &   -22.55 &     2.33 &   -5.111 \\
    5434.876488 &   -23.68 &     2.33 &   -5.111 \\
    5435.931255 &   -19.07 &     2.41 &   -5.162 \\
    5436.968463 &   -14.79 &     2.62 &   -5.190 \\
    5436.974551 &   -20.13 &     2.65 &   -5.230 \\
    5437.940359 &   -14.42 &     2.42 &   -5.119 \\
    5437.949792 &   -13.86 &     2.38 &   -5.103 \\
    5438.991420 &   -15.34 &     2.38 &   -5.153 \\
    5439.002623 &   -15.28 &     2.42 &   -5.136 \\
    5439.923524 &   -18.65 &     2.34 &   -5.111 \\
    5439.931556 &   -20.54 &     2.39 &   -5.127 \\
    5440.971227 &   -17.33 &     2.34 &   -5.119 \\
    5440.980301 &   -21.55 &     2.37 &   -5.136 \\
    5455.809962 &   -22.09 &     2.36 &   -5.080 \\
    5490.829645 &   -17.75 &     2.44 &   -5.088 \\
    5672.026425 &    18.02 &     2.27 &   -5.162 \\
    5672.998402 &    14.69 &     2.38 &   -5.171 \\
    5673.995774 &    18.86 &     2.42 &   -5.153 \\
    5696.973926 &    21.76 &     2.39 &   -5.144 \\
    5697.964434 &    23.88 &     2.46 &   -5.144 \\
    5698.962303 &    17.22 &     2.37 &   -5.153 \\
    5722.994969 &    22.64 &     2.42 &   -5.171 \\
    5724.033632 &    27.33 &     2.48 &   -5.171 \\
    5728.900947 &    23.73 &     2.48 &   -5.299 \\
    5734.064071 &    25.42 &     2.40 &   -5.171 \\
    5734.951047 &    26.01 &     2.38 &   -5.162 \\
    5735.974739 &    24.78 &     2.37 &   -5.162 \\
    5739.034181 &    24.27 &     2.46 &   -5.171 \\
    5751.796956 &    18.81 &     2.47 &   -5.162 \\
    5752.104703 &    17.96 &     2.40 &   -5.200 \\
    5752.778961 &    15.83 &     2.42 &   -5.162 \\
    5759.975172 &    18.93 &     2.40 &   -5.153 \\
    5761.076467 &    22.02 &     2.30 &   -5.153 \\
    5761.841845 &    15.82 &     2.35 &   -5.200 \\
    5763.032747 &    19.13 &     2.36 &   -5.220 \\
    5763.851041 &    14.98 &     2.36 &   -5.153 \\
    5782.907862 &    15.44 &     2.40 &   -5.171 \\
    5795.024254 &     7.43 &     2.53 &   -5.153 \\
    5814.735957 &    13.15 &     2.34 &   -5.095 \\
    6077.045184 &   -15.87 &     2.41 &   -5.181 \\
    6098.093799 &   -11.87 &     2.46 &   -5.171 \\
    6098.828990 &   -12.59 &     2.44 &   -5.144 \\
    6102.007905 &    -5.88 &     2.37 &   -5.162 \\
    6114.871876 &   -15.41 &     2.37 &   -5.153 \\
    6145.875413 &    -9.89 &     2.45 &   -5.136 \\
    6148.928774 &    -3.85 &     2.33 &   -5.111 \\
    6151.060872 &   -10.87 &     2.41 &   -5.127 \\
    6153.983044 &    -9.16 &     2.36 &   -5.095 \\
    6174.826576 &    -1.05 &     2.45 &   -5.088 \\
    6345.155209 &    17.40 &     2.44 &   -5.162 \\
    6475.843586 &    10.26 &     2.45 &   -5.162 \\
    6498.070164 &     3.00 &     2.43 &   -5.162 \\
    6513.044113 &    -2.18 &     2.45 &   -5.162 \\
    6519.919555 &    -0.24 &     2.39 &   -5.162 \\
\enddata
\end{deluxetable}

\begin{deluxetable}{ccccc}
\tabletypesize{\footnotesize}
\tablecaption{Radial Velocities for KOI-261, Kepler-96
\label{tab:rvs_k00261}}
\tablewidth{0pt}
\tablehead{
\colhead{}         & \colhead{Radial Velocity}     & \colhead{Uncertainty}  & \colhead{\rphk}  \\
\colhead{BJD -- 2450000}   & \colhead{(\mse)}  & \colhead{(\mse)} & \colhead{} 
}
\startdata
    5405.030168 &     2.55 &     2.47 &   -4.755 \\
    5405.889943 &     2.32 &     2.34 &   -4.745 \\
    5413.976523 &     0.62 &     2.48 &   -4.739 \\
    5431.794557 &     3.78 &     2.30 &   -4.765 \\
    5436.986914 &     5.95 &     2.35 &   -4.793 \\
    5437.960924 &     3.00 &     2.26 &   -4.765 \\
    5437.971745 &     7.19 &     2.30 &   -4.762 \\
    5438.829298 &     3.66 &     2.31 &   -4.758 \\
    5439.855631 &    -3.95 &     2.31 &   -4.758 \\
    5455.819605 &     2.30 &     2.38 &   -4.758 \\
    5486.833994 &     1.19 &     2.42 &   -4.775 \\
    5542.714646 &     0.82 &     2.46 &   -4.769 \\
    5544.703444 &    -5.41 &     2.55 &   -4.576 \\
    5723.092577 &     0.86 &     2.58 &   -4.663 \\
    5787.903358 &    -3.39 &     2.50 &   -4.758 \\
    5795.035974 &    -4.74 &     2.41 &   -4.755 \\
    5795.835975 &    -3.78 &     2.57 &   -4.775 \\
    6099.033843 &     5.65 &     2.59 &   -4.708 \\
    6111.925286 &   -12.19 &     2.59 &   -4.742 \\
    6145.907599 &    -6.48 &     2.52 &   -4.782 \\
    6148.981644 &     5.12 &     2.64 &   -4.779 \\
    6152.067220 &    -1.91 &     2.53 &   -4.797 \\
    6163.805762 &     2.65 &     2.43 &   -4.782 \\
    6164.846196 &     2.79 &     2.45 &   -4.786 \\
    6175.877276 &    -4.53 &     2.47 &   -4.797 \\
    6176.894658 &    -8.24 &     2.49 &   -4.786 \\
\enddata
\end{deluxetable}

\begin{deluxetable}{ccccc}
\tabletypesize{\footnotesize}
\tablecaption{Radial Velocities for KOI-283, Kepler-131
\label{tab:rvs_k00283}}
\tablewidth{0pt}
\tablehead{
\colhead{}         & \colhead{Radial Velocity}     & \colhead{Uncertainty}  & \colhead{\rphk}  \\
\colhead{BJD -- 2450000}   & \colhead{(\mse)}  & \colhead{(\mse)} & \colhead{} 
}
\startdata
    5433.868391 &     5.08 &     2.30 &   -4.824 \\
    5434.854201 &     4.86 &     2.31 &   -4.828 \\
    5440.950440 &    -2.54 &     2.37 &   -4.860 \\
    5471.813137 &     0.10 &     2.56 &   -4.856 \\
    5490.808418 &    -0.81 &     2.63 &   -4.856 \\
    5522.775891 &    -6.82 &     2.61 &   -4.864 \\
    5542.700150 &     5.54 &     2.67 &   -4.886 \\
    5724.012916 &     1.75 &     2.42 &   -4.928 \\
    5734.972723 &    -3.99 &     2.43 &   -4.904 \\
    5736.009887 &    -4.90 &     2.46 &   -4.900 \\
    5788.893783 &     7.50 &     2.58 &   -4.860 \\
    5795.941507 &     3.92 &     2.39 &   -4.844 \\
    6101.059779 &     1.30 &     2.50 &   -4.895 \\
    6110.960155 &    -0.31 &     2.58 &   -4.975 \\
    6115.924507 &     6.65 &     2.58 &   -4.891 \\
    6140.911995 &    -3.73 &     2.61 &   -4.895 \\
    6164.891469 &   -11.47 &     2.52 &   -4.975 \\
    6166.889764 &   -13.88 &     2.53 &   -4.928 \\
    6173.867998 &    10.24 &     2.46 &   -4.856 \\
    6175.860356 &     2.27 &     2.55 &   -4.840 \\
\enddata
\end{deluxetable}

\begin{deluxetable}{ccccc}
\tabletypesize{\footnotesize}
\tablecaption{Radial Velocities for KOI-292, Kepler-97
\label{tab:rvs_k00292}}
\tablewidth{0pt}
\tablehead{
\colhead{}         & \colhead{Radial Velocity}     & \colhead{Uncertainty}  & \colhead{\rphk}  \\
\colhead{BJD -- 2450000}   & \colhead{(\mse)}  & \colhead{(\mse)} & \colhead{} 
}
\startdata
    5377.974537 &     2.05 &     3.02 &   -5.104 \\
    5405.869250 &     3.50 &     2.94 &   -5.120 \\
    5407.043984 &     4.70 &     3.08 &   -5.136 \\
    5410.952610 &     3.43 &     2.90 &   -5.120 \\
    5411.803603 &     8.88 &     2.95 &   -5.112 \\
    5412.865682 &     4.95 &     3.11 &   -5.128 \\
    5425.822700 &     3.24 &     3.03 &   -5.088 \\
    5426.931969 &     2.33 &     2.87 &   -5.088 \\
    5751.875319 &     1.94 &     3.30 &   -5.096 \\
    5782.969047 &     2.99 &     3.00 &   -5.120 \\
    5788.957075 &    -8.30 &     3.19 &   -5.096 \\
    5790.834867 &     0.86 &     3.14 &   -4.906 \\
    5810.874922 &     8.26 &     3.04 &   -5.096 \\
    6103.079865 &    -4.15 &     3.45 &   -5.154 \\
    6112.853787 &    -2.20 &     3.28 &   -5.136 \\
    6144.947724 &     5.39 &     3.18 &   -5.145 \\
    6145.924744 &    -5.74 &     3.05 &   -5.136 \\
    6151.014731 &    -4.42 &     3.03 &   -5.120 \\
    6153.016068 &   -15.95 &     3.28 & ***e2147 \\
    6166.911312 &    -7.58 &     3.21 &   -5.104 \\
\enddata
\end{deluxetable}

\begin{deluxetable}{ccccc}
\tabletypesize{\footnotesize}
\tablecaption{Radial Velocities for KOI-299, Kepler-98
\label{tab:rvs_k00299}}
\tablewidth{0pt}
\tablehead{
\colhead{}         & \colhead{Radial Velocity}     & \colhead{Uncertainty}  & \colhead{\rphk}  \\
\colhead{BJD -- 2450000}   & \colhead{(\mse)}  & \colhead{(\mse)} & \colhead{} 
}
\startdata
    5404.011500 &     9.78 &     2.76 &   -4.788 \\
    5406.943814 &     5.91 &     2.90 &   -4.835 \\
    5407.871779 &    -1.02 &     2.75 &   -4.794 \\
    5411.927391 &    -1.76 &     2.74 &   -4.846 \\
    5413.947119 &    -9.22 &     2.91 &   -4.876 \\
    5414.852128 &   -12.73 &     2.66 &   -4.853 \\
    5433.792757 &    -5.53 &     2.61 &   -4.811 \\
    5434.771547 &     0.81 &     2.61 &   -4.831 \\
    5437.918288 &     3.25 &     2.92 &   -4.842 \\
    5440.852924 &    -6.40 &     2.67 &   -4.849 \\
    5735.034887 &     6.12 &     2.81 &   -4.748 \\
    5751.843073 &     5.10 &     2.87 &   -4.797 \\
    5760.040755 &     4.36 &     2.69 &   -4.785 \\
    5760.909806 &     0.13 &     2.67 &   -4.801 \\
    5796.002983 &    -3.74 &     3.22 &   -4.860 \\
    5797.816289 &     5.21 &     2.65 &   -4.831 \\
    5807.988113 &    -3.40 &     3.20 &   -4.320 \\
    5810.917683 &    -0.34 &     2.67 &   -4.811 \\
    5811.762281 &   -10.68 &     2.68 &   -4.804 \\
    6113.004738 &     8.75 &     2.69 &   -4.782 \\
    6114.053151 &     7.21 &     3.26 &   -4.966 \\
    6208.773499 &    -3.19 &     2.78 &   -4.814 \\
\enddata
\end{deluxetable}

\begin{deluxetable}{ccccc}
\tabletypesize{\footnotesize}
\tablecaption{Radial Velocities for KOI-305, Kepler-99
\label{tab:rvs_k00305}}
\tablewidth{0pt}
\tablehead{
\colhead{}         & \colhead{Radial Velocity}     & \colhead{Uncertainty}  & \colhead{\rphk}  \\
\colhead{BJD -- 2450000}   & \colhead{(\mse)}  & \colhead{(\mse)} & \colhead{} 
}
\startdata
    5404.043451 &     3.25 &     2.69 &   -4.414 \\
    5405.840133 &     3.15 &     2.47 &   -4.610 \\
    5406.910870 &     4.38 &     2.58 &   -4.779 \\
    5410.923078 &    -4.28 &     2.49 &   -4.676 \\
    5736.069898 &     4.67 &     2.69 &   -4.712 \\
    5751.907376 &    -7.05 &     2.59 &   -4.624 \\
    5760.980986 &     3.95 &     2.55 &   -4.842 \\
    5790.878287 &     0.93 &     2.57 &   -4.648 \\
    5792.968210 &     1.01 &     2.57 &   -4.663 \\
    5795.056867 &     2.20 &     2.90 &   -4.644 \\
    5797.876148 &    -1.21 &     2.52 &   -4.628 \\
    6102.027398 &    -7.74 &     2.62 &   -4.590 \\
    6112.943513 &     3.17 &     2.97 &   -4.666 \\
    6116.071830 &    -5.38 &     2.58 &   -4.591 \\
    6140.890738 &     4.76 &     2.59 &   -4.579 \\
    6145.029741 &     1.35 &     2.72 &   -4.589 \\
    6147.892714 &    -7.78 &     2.50 &   -4.631 \\
    6163.884725 &     9.32 &     2.70 &   -4.531 \\
    6172.829726 &    -1.39 &     2.69 &   -4.593 \\
    6192.868451 &   -10.73 &     2.57 &   -4.573 \\
    6195.854351 &     0.27 &     2.51 &   -4.559 \\
\enddata
\end{deluxetable}

\begin{deluxetable}{ccccc}
\tabletypesize{\footnotesize}
\tablecaption{Radial Velocities for KOI-321, Kepler-406
\label{tab:rvs_k00321}}
\tablewidth{0pt}
\tablehead{
\colhead{}         & \colhead{Radial Velocity}     & \colhead{Uncertainty}  & \colhead{\rphk}  \\
\colhead{BJD -- 2450000}   & \colhead{(\mse)}  & \colhead{(\mse)} & \colhead{} 
}
\startdata
    5379.033747 &     3.62 &     2.57 &   -4.945 \\
    5380.109093 &    -5.92 &     2.66 &   -4.980 \\
    5403.979414 &     3.99 &     2.68 &   -5.007 \\
    5411.834541 &    -6.94 &     2.54 &   -5.029 \\
    5412.898763 &     2.87 &     2.63 &   -5.012 \\
    5434.825372 &    -0.53 &     2.52 &   -5.085 \\
    5435.909173 &    -1.98 &     2.82 &   -5.085 \\
    5751.813315 &    -4.74 &     2.76 &   -5.001 \\
    5752.852880 &    -0.94 &     2.96 &   -4.996 \\
    5752.868470 &     0.29 &     2.73 &   -4.975 \\
    5759.839971 &     6.17 &     2.66 &   -4.950 \\
    5760.936394 &    -4.23 &     2.58 &   -4.965 \\
    5769.859474 &     6.60 &     2.56 &   -5.041 \\
    5788.839547 &    -1.72 &     2.59 &   -5.059 \\
    5789.984420 &     1.17 &     2.63 &   -5.066 \\
    5810.791430 &    -1.37 &     2.51 &   -5.078 \\
    5811.920824 &   -11.11 &     2.59 &   -5.126 \\
    5843.811883 &    -7.55 &     3.01 &   -5.148 \\
    6073.925432 &    -1.96 &     3.05 &   -4.960 \\
    6075.917333 &     0.60 &     3.09 &   -5.001 \\
    6076.992302 &     3.17 &     2.75 &   -4.960 \\
    6098.855454 &    -3.04 &     2.44 &   -4.918 \\
    6099.985448 &     0.10 &     2.62 &   -4.941 \\
    6101.029734 &    -1.44 &     2.59 &   -4.950 \\
    6101.872764 &     1.69 &     2.61 &   -4.936 \\
    6103.050119 &     4.15 &     2.68 &   -4.941 \\
    6104.059437 &     1.79 &     2.72 &   -4.985 \\
    6105.064136 &    10.72 &     2.80 &   -4.980 \\
    6109.955028 &    -7.50 &     2.49 &   -4.970 \\
    6111.022468 &     1.20 &     2.73 &   -4.965 \\
    6111.941178 &     1.24 &     2.53 &   -4.905 \\
    6115.999606 &     4.75 &     2.51 &   -5.001 \\
    6133.832918 &     8.11 &     2.57 &   -5.035 \\
    6134.888536 &    -1.09 &     2.63 &   -5.035 \\
    6138.008915 &     0.39 &     2.66 &   -5.035 \\
    6140.863836 &    -1.40 &     2.69 &   -5.041 \\
    6163.843324 &    -0.84 &     2.72 &   -5.085 \\
    6164.774988 &     4.40 &     2.57 &   -5.085 \\
    6166.003960 &    -1.82 &     2.71 &   -5.059 \\
    6171.983117 &    -1.63 &     2.94 &   -5.085 \\
    6177.836688 &    -6.64 &     2.67 &   -5.078 \\
    6179.763596 &    -2.02 &     2.56 &   -5.119 \\
\enddata
\end{deluxetable}

\begin{deluxetable}{ccccc}
\tabletypesize{\footnotesize}
\tablecaption{Radial Velocities for KOI-1442, Kepler-407
\label{tab:rvs_k01442}}
\tablewidth{0pt}
\tablehead{
\colhead{}         & \colhead{Radial Velocity}     & \colhead{Uncertainty}  & \colhead{\rphk}  \\
\colhead{BJD -- 2450000}   & \colhead{(\mse)}  & \colhead{(\mse)} & \colhead{} 
}
\startdata
    5697.018217 &    50.75 &     2.93 &   -5.038 \\
    5698.101339 &    53.47 &     2.96 &   -5.049 \\
    5699.097179 &    50.07 &     2.95 &   -4.969 \\
    5736.037333 &    42.70 &     3.03 &   -5.108 \\
    5752.051431 &    42.24 &     2.82 &   -5.072 \\
    5760.015457 &    35.72 &     2.73 &   -5.084 \\
    5760.776641 &    37.46 &     2.79 &   -5.090 \\
    5768.875006 &    34.26 &     2.70 &   -5.141 \\
    5769.798374 &    36.20 &     2.80 &   -5.176 \\
    5770.841452 &    37.96 &     2.76 &   -5.128 \\
    5790.914061 &    29.22 &     2.85 &   -5.078 \\
    5791.889570 &    33.87 &     2.69 &   -5.090 \\
    5792.876920 &    31.36 &     2.74 &   -5.072 \\
    5797.899896 &    20.24 &     2.84 &   -5.102 \\
    6133.912806 &  -128.73 &     3.07 &   -5.066 \\
    6172.849950 &  -152.83 &     2.97 &   -5.044 \\
    6446.924398 &  -255.23 &     3.22 &   -5.044 \\
\enddata
\end{deluxetable}

\begin{deluxetable}{ccccc}
\tabletypesize{\footnotesize}
\tablecaption{Radial Velocities for KOI-1612
\label{tab:rvs_k01612}}
\tablewidth{0pt}
\tablehead{
\colhead{}         & \colhead{Radial Velocity}     & \colhead{Uncertainty}  & \colhead{\rphk}  \\
\colhead{BJD -- 2450000}   & \colhead{(\mse)}  & \colhead{(\mse)} & \colhead{} 
}
\startdata
    5698.087742 &     2.49 &     2.50 &   -4.992 \\
    5699.127102 &     0.56 &     2.49 &   -4.972 \\
    5700.112050 &     4.47 &     2.48 &   -5.006 \\
    5700.943498 &     1.12 &     2.50 &   -5.013 \\
    5703.924719 &     1.10 &     2.47 &   -5.035 \\
    5704.886679 &     8.90 &     2.53 &   -5.013 \\
    5706.894189 &    -3.62 &     2.39 &   -5.028 \\
    5707.890606 &     5.94 &     2.47 &   -5.051 \\
    5723.104987 &    -1.35 &     2.43 &   -5.020 \\
    5724.114372 &    -5.66 &     2.54 &   -5.028 \\
    5726.075420 &     0.17 &     2.36 &   -5.067 \\
    5727.023738 &    -1.65 &     2.36 &   -5.051 \\
    5728.890788 &    -1.36 &     2.51 &   -5.178 \\
    5734.956730 &     2.31 &     2.45 &   -5.051 \\
    5736.051585 &    -7.29 &     2.52 &   -5.051 \\
    5739.066041 &    -3.17 &     2.55 &   -5.043 \\
    5739.075902 &    -1.94 &     2.59 &   -5.059 \\
    5751.924655 &    -2.95 &     2.50 &   -5.059 \\
    5752.784580 &    -7.19 &     2.53 &   -5.059 \\
    5760.075266 &     2.04 &     2.57 &   -5.035 \\
    5761.082387 &     2.43 &     2.56 &   -5.059 \\
    5762.898826 &    -0.86 &     2.54 &   -5.075 \\
    5763.857299 &    -3.11 &     2.56 &   -5.075 \\
    6000.139475 &     4.20 &     2.43 &   -5.043 \\
    6019.033769 &     4.40 &     2.46 &   -5.020 \\
    6020.045352 &     4.72 &     2.52 &   -5.013 \\
    6028.005633 &     2.87 &     2.48 &   -4.985 \\
    6073.886610 &     7.36 &     2.51 &   -5.051 \\
    6075.968379 &    -4.72 &     2.40 &   -5.075 \\
    6076.937976 &    -0.86 &     2.49 &   -5.059 \\
    6098.068823 &     2.23 &     2.53 &   -5.067 \\
    6115.023931 &    -5.82 &     2.70 &   -5.075 \\
    6144.877583 &    -1.98 &     2.95 &   -5.101 \\
    6145.901440 &     2.03 &     2.77 &   -5.092 \\
    6146.966014 &    -1.61 &     2.64 &   -5.101 \\
    6147.817037 &    -4.03 &     2.64 &   -5.119 \\
    6148.853184 &    -1.97 &     2.55 &   -5.110 \\
    6174.848128 &    -2.30 &     2.67 &   -5.138 \\
\enddata
\end{deluxetable}

\begin{deluxetable}{ccccc}
\tabletypesize{\footnotesize}
\tablecaption{Radial Velocities for KOI-1925, Kepler-409
\label{tab:rvs_k01925}}
\tabletypesize{\scriptsize}
\tablehead{
\colhead{}         & \colhead{Radial Velocity}     & \colhead{Uncertainty}  & \colhead{\rphk}  \\
\colhead{BJD -- 2450000}   & \colhead{(\mse)}  & \colhead{(\mse)} & \colhead{} 
}
\startdata
    6000.146032 &    -2.45 &     2.20 &   -5.061 \\
    6019.055308 &    -6.34 &     2.29 &   -5.045 \\
    6020.099174 &     1.57 &     2.23 &   -5.024 \\
    6028.041949 &     1.66 &     2.20 &   -5.029 \\
    6073.949745 &    -2.48 &     2.23 &   -5.019 \\
    6075.094428 &    -1.92 &     2.25 &   -5.024 \\
    6075.977029 &    -1.84 &     2.23 &   -5.024 \\
    6076.946404 &     0.74 &     2.27 &   -5.014 \\
    6098.057089 &    -5.45 &     2.30 &   -5.096 \\
    6098.838256 &    -0.55 &     2.20 &   -5.061 \\
    6099.909720 &    -1.19 &     2.21 &   -5.061 \\
    6100.930048 &    -0.06 &     2.16 &   -5.061 \\
    6101.982553 &     0.93 &     2.20 &   -5.056 \\
    6102.956964 &    -1.21 &     2.21 &   -5.050 \\
    6103.963911 &    -0.88 &     2.22 &   -5.050 \\
    6104.981934 &     2.73 &     2.18 &   -5.061 \\
    6113.020955 &     2.76 &     2.21 &   -5.034 \\
    6113.831179 &     1.20 &     2.23 &   -5.045 \\
    6114.839891 &     1.39 &     2.24 &   -5.029 \\
    6115.860558 &     6.37 &     2.21 &   -5.045 \\
    6133.923304 &     0.39 &     2.31 &   -5.061 \\
    6134.962073 &    -5.06 &     2.39 &   -5.078 \\
    6138.035933 &     3.13 &     2.32 &   -5.067 \\
    6148.866212 &    -3.70 &     2.30 &   -5.056 \\
    6174.857241 &     5.59 &     2.36 &   -5.045 \\
\enddata
\end{deluxetable}

\clearpage

\enddocument